\documentclass{ieeeaccess}
\usepackage{cite}
\usepackage{amsmath,amssymb,amsfonts}
\usepackage{algorithmic}
\usepackage{graphicx}
\usepackage{textcomp}

\usepackage{siunitx}
\DeclareSIUnit{\belmilliwatt}{Bm}
\DeclareSIUnit{\dBm}{\deci\belmilliwatt}

\usepackage{braket}
\usepackage{hyperref} 
\usepackage{cleveref} 
\crefname{appsec}{appendix}{appendices}
\Crefname{appsec}{Appendix}{Appendices}

\usepackage{color}
\usepackage{multirow}
\usepackage{booktabs}
\usepackage{threeparttable}

\usepackage{array}         

\newcommand{\helv}{\fontfamily{phv}\selectfont}
\DeclareTextFontCommand{\leg}{\helv\fontsize{8pt}{9pt}\selectfont}
\def\BibTeX{{\rm B\kern-.05em{\sc i\kern-.025em b}\kern-.08em
    T\kern-.1667em\lower.7ex\hbox{E}\kern-.125emX}}
\begin{document}
\history{Date of publication xxxx 00, 0000, date of current version xxxx 00, 0000.}
\doi{xx.xxxx/XXX.xxxx.DOI}

\title{Revisiting Thermal Scalability for Large-Scale Superconducting Quantum Systems}

\author{
\uppercase{Shaswot Shresthamali}\authorrefmark{1},
\uppercase{Ilkwon Byun\authorrefmark{1}, \IEEEmembership{Member, IEEE}},
\uppercase{Teruo Tanimoto\authorrefmark{1}, \IEEEmembership{Member, IEEE}, 
\uppercase{Yoshinori Uzawa\authorrefmark{4,5},
\uppercase{Kunihiro Inomata\authorrefmark{3}},
\uppercase{Tsuyoshi Yamamoto\authorrefmark{2,3}},
and 
Koji Inoue}\authorrefmark{1},\IEEEmembership{Member, IEEE}}
}
\address[1]{Faculty of Information Science and Electrical Engineering, Kyushu University, Fukuoka, Fukuoka 819-0395, Japan}
\address[2]{Secure System Platform Research Laboratories, NEC Corporation, Kawasaki, Kanagawa 211-0011, Japan}
\address[3]{Global Research and Development Center for Business by Quantum-AI Technology (G-QuAT), National Institute of Advanced Industrial Science and Technology (AIST), Tsukuba, Ibaraki 305-8568, Japan}
\address[4]{Advanced Technology Center, National Astronomical Observatory of Japan (NAOJ), Tokyo 181-8588, Japan}
\address[5]{Graduate University for Advanced Studies (SOKENDAI), Tokyo 181-8588, Japan}

\tfootnote{This work was supported in part by Japan Science and Technology Agency through Moonshot R\&D program under Grant JPMJMS2067 and JSPS KAKENHI Grant Number JP24K20843.\newline The complete implementation code is available at \url{https://github.com/shaswot/SCQ_heatmodel}.}

\markboth
{Shresthamali \headeretal: Revisiting Thermal Scalability for Large-Scale Superconducting Quantum Systems}
{Shresthamali \headeretal: Revisiting Thermal Scalability for Large-Scale Superconducting Quantum Systems}

\corresp{Corresponding author: Shaswot Shresthamali (email: shaswot@fujitsu.com).}

\begin{abstract}
The readout amplification chain imposes a critical thermal scalability bottleneck in large-scale superconducting quantum systems. This happens through three mechanisms: amplifier dissipation, passive conduction through bias wiring and Joule heating within that same wiring.
These terms are absent or only partially represented in several prior system-level thermal-scalability models, leading to bottleneck misidentification and scalability overestimation. 
In this work, we improve upon previous system-level heat estimation models by fully accounting for the major heat sources in modern cryogenic quantum systems including the active dissipation, passive conduction, and Joule heating in the readout amplifier module.
Our analysis demonstrates that amplifier-associated heat emerges as the dominant thermal bottleneck that fundamentally alters the thermal landscape of modern large-scale cryogenic systems. We explore various technology options and their tradeoffs to identify configurations that reduce this critical heat load and improve scalability. 
Finally, we evaluate forward-looking system configurations, including larger refrigeration platforms and optical approaches, and analyze forward-looking pathways toward single-fridge 10k-qubit cryogenic systems.
\end{abstract}


\titlepgskip=-15pt

\maketitle
\section{Introduction}
\label{sec:intro}
{\let\thefootnote\relax\footnote{* This research work was conducted while S.S. was affiliated with Kyushu University; S.S. is currently with Fujitsu Limited, Japan.}}
Fault-tolerant quantum computing (FTQC) with superconducting qubits is expected to require millions of physical qubits to achieve practical utility \cite{mohseni2024build,croot2025enabling}. Current cryogenic platforms typically support around 100 physical qubits (PQs), barely sufficient for single-digit logical qubits (LQs) using surface-code based error correction \cite{2025acharyaQuantumErrorCorrection}. Experimental demonstrations have shown logical error rates can be suppressed with increasing code distance \cite{2025acharyaQuantumErrorCorrection}. For example, the 105-qubit system in \cite{2025acharyaQuantumErrorCorrection} realizes a distance-7 surface code and demonstrates a logical qubit capable of preserving quantum information over extended durations. 
To support practical fault-tolerant algorithms, systems must scale to tens to hundreds of logical qubits.
Even before accounting for routing space, lattice-surgery workspace, and magic-state distillation, this requirement already points toward cryogenic platforms that can support thousands to 10k physical qubits. 
Consequently, achieving 10k-qubit systems has become a near-term milestone for many research efforts \cite{ibmRoadmap,quantumaiRoadmapGoogle,globalFujitsuStarts,quantwareQuantWareAnnounces}. 

At this scale, scalability depends not only on qubit fabrication but also on the cryogenic infrastructure required to remove heat while delivering control and readout signals \cite{mohseni2024build,croot2025enabling}. A system-level understanding of the heat loads that limit scalability is therefore necessary to guide design choices and prioritize engineering efforts.

Prior works have established that passive heat conduction through control and readout lines, attenuator dissipation and limited millikelvin cooling capacities limit the scalability of large-scale quantum systems \cite{krinner2019engineering,joshi2023scaling,min2023qisim,raicu2025cryogenic}. However, they do not consistently include the combined active, passive, and ohmic heat associated with the cryogenic readout-amplifier chain. While this omission is relatively benign in earlier systems, it becomes consequential as modern lower-PHL interconnects and higher-capacity refrigerators shift the scalability bottleneck towards the 4\,K stage. As a result, analyses using previous models can misidentify bottleneck stages and overestimate qubit capacity.

Previous models are also tightly coupled to specific technology assumptions, making it difficult to explore design options with alternative and emerging technologies. It is therefore not clear which design choices and strategies provide the largest system-level scalability improvements. We therefore require a flexible thermal scalability model that includes the major heat loads while enabling fair comparison across both legacy and emerging technologies under a unified framework.

Furthermore, previous models only estimate the PQ yield of the system. However, physical qubit capacity is only an intermediate metric. The more relevant question is how many logical qubits the system can ultimately support. Existing models do not provide a mechanism to relate the PQ yield obtained from thermal scalability analysis to the resulting LQ yield under given physical error rates and target logical error rate requirements.

In this work, we build upon previous scalability models by introducing a system-level heat estimation framework for large-scale cryogenic superconducting transmon systems (\Cref{sec:model}). Our model evaluates passive, active, and ohmic heat loads, including previously overlooked amplifier-associated heat sources, enabling estimation of PQ capacity from a thermal perspective. Our unified modeling framework supports fair cross-technology system-level comparison of the thermal implications of various technology options. Furthermore, this framework also integrates logical-error analysis to estimate the resulting LQ capacity of FTQC systems.

Using our model, we re-evaluate prior scalability studies under modern system assumptions (\Cref{sec:bottlenecks}). We expose how advances in low heat conduction cabling technologies have shifted the heat landscape, giving rise to new bottlenecks. We  demonstrate that amplifier-related heat is the dominant heat source limiting system scalability in modern systems. We further show how omitting these loads leads to misidentification of bottleneck stages and overestimation of system scalability.

We then analyze various solutions for reducing amplifier-induced thermal bottlenecks and their system-level tradeoffs (\Cref{sec:solutions}). We compare wiring materials (copper, Manganin, and Yttrium Barium Copper Oxide (YBCO)) and readout-amplifier technologies (High Electron Mobility Transistor (HEMT) and Superconductor-Insulator-Superconductor (SIS) mixer amplifiers), and evaluate their passive, active, and ohmic heat contributions as well as tradeoffs. Our analysis reveals that reducing active dissipation alone does not necessarily reduce total heat load because passive or ohmic contributions can become dominant. In some cases, the same amplifier can exhibit both the best and worst thermal performance depending on the choice of biasing wire material. From this design-space exploration, we identify amplifier configurations that 
can significantly increase the system PQ capacity.

Finally, we use these insights to evaluate hypothetical future systems motivated by emerging trends, including larger refrigeration platforms and optical approaches (\Cref{sec:pathways}). The purpose of this analysis is to investigate various architectural and technological choices of forward-looking cryogenic single-fridge 10k-qubit systems.

In summary, this paper makes the following contributions:
\begin{itemize}
    \item We build upon previous approaches to develop a flexible, extensible, cross-technology system-level heat estimation model for large-scale cryogenic superconducting qubit systems (\Cref{sec:model}). The model is integrated with a logical error model that enables designers to estimate both PQ and LQ yields for given design configurations and parameters.
    
    \item Using this model, we identify that previously unaccounted readout amplifier-associated heat (active dissipation, bias-line passive conduction and ohmic dissipation) constitutes a dominant thermal bottleneck (\Cref{sec:bottlenecks}) and how omitting these heat loads can lead to incorrect identification of bottleneck stages and overestimation of qubit capacity.
    
    \item We explore mitigation strategies that reduce amplifier-associated heat loads. We quantify tradeoffs across wiring materials and amplifier technologies, and identify configurations that substantially improve system scalability (\Cref{sec:solutions}).
    
    \item Using these insights, we evaluate forward-looking system-level pathways toward single-fridge 10k-qubit cryogenic systems (\Cref{sec:pathways}).
\end{itemize}

The remainder of this paper is organized as follows.
\Cref{sec:background} summarizes cryogenic-system components and the FTQC context relevant to heat modeling.
\Cref{sec:model} defines our modeling methodology and its integration with logical-error analysis.
\Cref{sec:bottlenecks} re-evaluates prior scalability studies and identifies amplifier-associated heat as the critical overlooked bottleneck.
\Cref{sec:solutions} analyzes mitigation strategies and their tradeoffs across wiring and amplifier technology choices.
\Cref{sec:pathways} discusses some forward-looking scaling pathways motivated by emerging trends.
In \Cref{sec:scope}, we discuss the modeling assumptions and scope of this work and finally conclude in \Cref{sec:conclusion}.
\section{Background and Motivation}
\label{sec:background}

\Figure[!ht](topskip=0pt, botskip=0pt, midskip=0pt)[width=0.95\textwidth]
{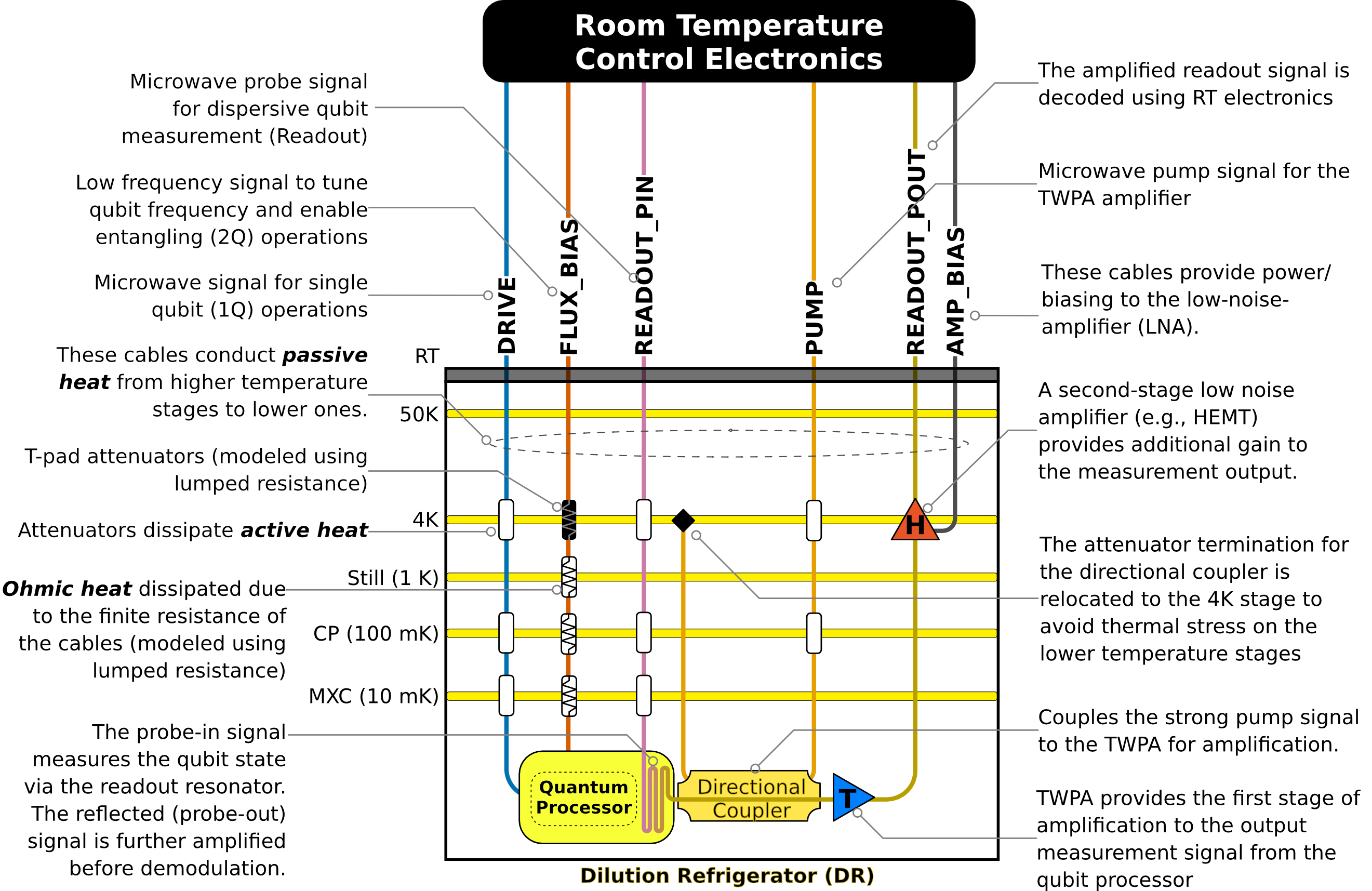}
{\parbox[t]{0.95\linewidth}{Simplified schematic of the cryogenic system in \cite{krinner2019engineering} consisting of the dilution refrigerator that houses the quantum processor and the supporting infrastructure. The major heat loads arise from heat conduction via the metallic cables (passive heat loads), heat dissipation through attenuators and amplifiers (active heat loads), and Joule heating (ohmic heat loads). Components with negligible heat footprint (e.g., circulators, filters, isolators) are not shown for simplicity.}
\label{fig:cryogenic_system}
}

\subsection{Cryogenic System Overview}
\label{sec:background_cryo}

A typical dilution refrigerator (DR) includes a room-temperature stage (RT$\sim$300\,K), followed by intermediate stages such as 50\,K and 4\,K cooled by a pulse-tube cryocooler. Colder stages typically include the Still ($\sim$1\,K), Cold Plate (CP$\sim$100 \unit{\milli\kelvin}), and Mixing Chamber (MXC$\sim$10 \unit{\milli\kelvin}). Each stage provides a finite cooling power. The quantum processor is located at the mixing chamber where temperatures reach a few tens of millikelvin. Such extreme operating temperatures are necessary to suppress thermal noise and minimize operational errors.  In addition to the qubit chip, the DR also houses supporting infrastructure required for operation, including cables, attenuators, amplifiers and filters. A simplified schematic of a representative cryogenic system (fixed-frequency fixed-coupling) is shown in \Cref{fig:cryogenic_system}.

Multiple microwave signal lines connect the qubits to RT control electronics. 
\leg{DRIVE} cables deliver high-frequency control pulses for single-qubit (1Q) operations. 
\leg{FLUX\_BIAS} microwave cables carry low-frequency signals used to tune qubit frequencies. In some architectures like \cite{krinner2019engineering}, the same line is also used to mediate entangling two-qubit (2Q) operations. Modern architectures use tunable couplers, which require additional \leg{COUPLER} cables, to dynamically enable or disable inter-qubit coupling. 
For readout, a probe tone is injected through the \leg{READOUT\_PIN} cable into a resonator coupled to the qubit. 
The reflected signal propagates through an amplification chain that typically includes a quantum-limited amplifier (e.g., Traveling Wave Parametric Amplifier (TWPA)) for first-stage amplification and a subsequent low-noise amplifier (LNA), typically a HEMT, at 4\,K before being routed to RT electronics for demodulation via the \leg{READOUT\_POUT} cable. The TWPA amplifies the weak readout signal by transferring energy from a strong external pump tone delivered through a separate \leg{PUMP} line and coupled via the directional coupler.

The heat loads in the cryogenic system can be categorized into three types. 
\textit{Passive heat load} (PHL) arises from thermal conduction along metallic cables (control, readout, pump, and biasing) that span temperature gradients between adjacent stages. 
\textit{Active heat load} (AHL) originates from power dissipated by components thermally anchored to a stage, such as attenuators and amplifiers. 
\textit{Ohmic heat load} (OHL) arises from resistive (Joule) heating in conductors that carry DC or low-frequency currents, especially when the conductors have significant electrical resistance. 
For a given stage, the \textit{total heat load} (THL) is the sum of all PHL, AHL, and OHL contributions.

To compare thermal stress across stages with different cooling powers, we normalize each heat load contribution by the cooling power of the stage where that heat is absorbed. 
We refer to this quantity as the normalized heat load (NHL) \cite{krinner2019engineering,raicu2025cryogenic}. 
For a heat load contribution $P_i$ absorbed at stage $s$ with cooling power $P_{\mathrm{cool},s}$, the corresponding NHL is
\begin{equation}
    \mathrm{NHL}_{i,s} = \frac{P_i}{P_{\mathrm{cool},s}} .
\end{equation}
The total NHL of stage $s$ is therefore the sum of the normalized contributions from all heat sources at that stage:
\begin{equation}
    \mathrm{NHL}_{\mathrm{THL},s}
    =
    \frac{\mathrm{THL}_s}{P_{\mathrm{cool},s}}
    =
    \sum_i \mathrm{NHL}_{i,s}.
\end{equation}
This convention allows us to compare not only total stage-level thermal stress, but also the relative importance of individual PHL, AHL, and OHL contributors within the same stage \cite{krinner2019engineering,joshi2023scaling,raicu2025cryogenic}.

\subsection{The Need for system-level thermal scalability analysis}
As the number of qubits increases, the number of required interconnects and supporting components also increases. 
Consequently, PHL, AHL, and OHL generally increase with qubit count. 
A temperature stage approaches its cooling limit when its total NHL approaches unity i.e., the THL at that stage nearly exhausts the available cooling power. 
As the system scales, the first stage whose total NHL reaches unity defines the \textbf{thermal scalability bottleneck}. 
At this point, the corresponding stage can no longer absorb additional heat and maintain steady-state operation. 
This bottleneck determines the maximum physical qubit capacity supported by the cryogenic platform.

To scale the system further, the thermal bottlenecks must first be resolved. 
This requires three steps: (i) identifying the bottleneck temperature stage, (ii) determining the dominant heat contributors at that stage, and (iii) evaluating system-wide effects of alternative design choices that reduce the thermal footprint of the system.

These steps require a \textit{system-level} thermal analysis model that decomposes heat contributions by component and by temperature stage. Such a model identifies which stages are thermally constrained and which heat sources ultimately limit system scalability. Once these constraints are known, candidate solutions can be evaluated to reduce the relevant heat loads. However, a solution that resolves a bottleneck locally is not sufficient. A design change that reduces one heat load at one stage can unintentionally increase another heat load elsewhere in the cryogenic system and even degrade the overall thermal scalability of the system. Therefore, the heat model must capture the system-level thermal implications of each evaluated solution, rather than only its local effect. With this capability, designers can evaluate and compare architectural and technological strategies at a system level. This enables informed design decisions that reduce limiting heat loads and improve scalability. 

\subsection{Why Thermal Scalability Must Be Revisited}
\label{subsec:revisit}
\Figure[!ht](topskip=0pt, botskip=0pt, midskip=0pt)[width=\linewidth]
{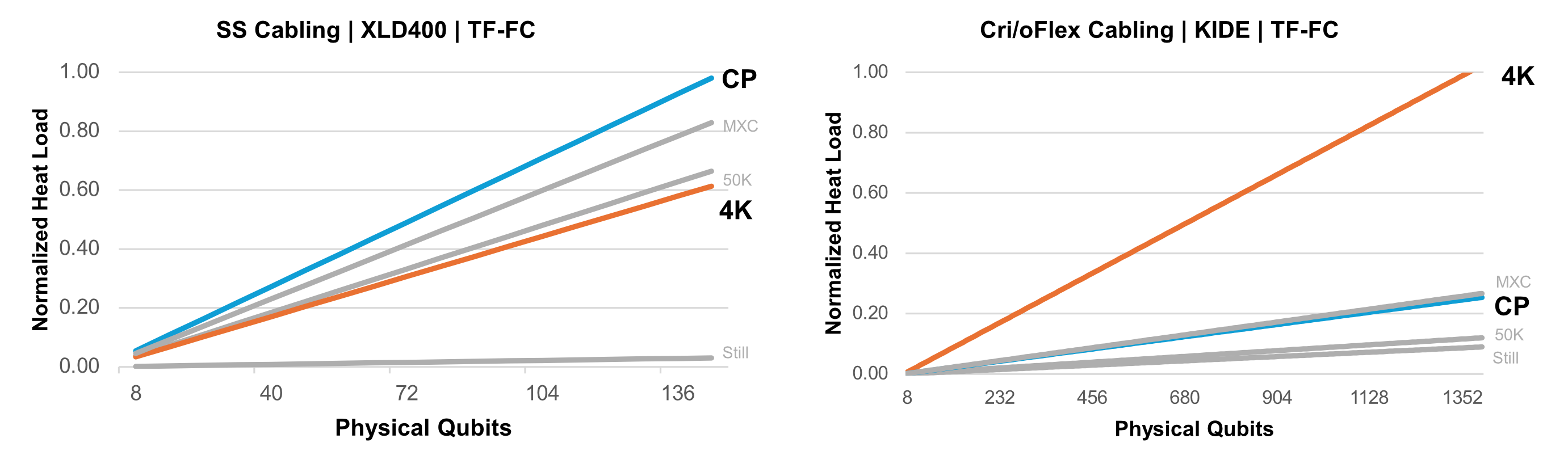}
{Normalized heat loads at different temperature stages under two wiring assumptions for the system in \cite{krinner2019engineering}. 
(left) Reproduced scalability trend for the stainless-steel wiring architecture considered in \cite{krinner2019engineering}. 
In this case, the millikelvin stages dominate because stainless-steel drive and flux-bias cables introduce large passive and ohmic heat loads. 
(right) Hypothetical replacement of stainless-steel cables with Cri/oFlex wiring \cite{delftcircuitsCryogenicCables} and upgrading the fridge to KIDE. 
Higher-capacity refrigeration and lower-PHL cabling increase the qubit count, but more importantly they change the scaling trend and shift the dominant bottleneck toward the 4\,K stage. This shift motivates explicit modeling of readout-amplifier heat loads and amplifier-bias wiring, which become critically important in the current technological context.
\label{fig:krinner-trends}
}

Prior works have established important baselines for evaluating the thermal scalability of superconducting quantum systems. 
A representative example is \cite{krinner2019engineering}, which analyzes a 50-qubit tunable-frequency fixed-coupling (TF-FC) architecture implemented with stainless-steel (SS) coaxial cables in a Bluefors XLD400 dilution fridge.
Their analysis reveals that the CP is the dominant thermal bottleneck primarily due to the large PHL conducted by the SS microwave lines relative to the available cooling power at CP. \Cref{subsec:krinner} discusses this in more detail. \Cref{fig:krinner-trends}(left) shows how the NHL of different temperature stages in the system scales with increasing qubit counts. Heat loads increase at the highest rate in CP (followed by MXC) and therefore CP is the first stage to exhaust its cooling power. In this configuration, the 4\,K heat loads have little effect on the system scalability.

Since the study in \cite{krinner2019engineering}, several components of the cryogenic stack have changed. 
Modern systems increasingly use interconnects with lower PHL, such as cupronickel coaxial cables \cite{raicu2025cryogenic,blueforsHighDensityWiring} and microwave striplines \cite{delftcircuitsCryogenicCables}, instead of SS coaxial cables. Refrigerator platforms are improving, with larger cooling powers being reported (or targeted). This reduces the thermal stress in millikelvin stages and the dominant stress shifts to other temperature stages. \Cref{fig:krinner-trends}(right) shows how the heat stress at 4\,K rises at a much higher rate than the other stages when the system in \cite{krinner2019engineering} is hypothetically rewired using modern Cri/oFlex microstrip cables \cite{delftcircuitsCryogenicCables} and a more powerful modern KIDE fridge is used, with all other assumptions unchanged. This behavior is an example of \textit{bottleneck migration}: improving one part of the cryogenic stack can expose a different stage or component as the new scalability limit.

A critical limitation of prior scalability models \cite{krinner2019engineering,joshi2023scaling,raicu2025cryogenic,min2023qisim} is that they cannot capture this migration because they omit or incompletely account for the heat loads at 4\,K. In particular, two contributions at the 4\,K stage are critical: (i) the active dissipation of cryogenic readout amplifiers (e.g., HEMTs), and (ii) the PHL conducted through amplifier bias lines (typically copper). For the original system in \cite{krinner2019engineering}, these terms remain secondary (and can therefore be omitted without significant implications) because the 4\,K heat load scales weakly with qubit count compared to CP and MXC (\Cref{fig:krinner-trends}(left)). However, this balance changes when better wiring and refrigeration technologies are introduced and amplifier-related heat becomes a first-order contributor.

Additionally, alternatives to traditional readout amplifiers and biasing lines have been recently introduced. For instance, ultra-low-power HEMT (ULP-HEMT) and SIS-Mixer amplifiers have much lower active heat dissipation than conventional HEMTs. Alternative materials to copper such as Manganin can also be used for biasing wires. However, the choice of amplifier and associated bias wiring creates a tradeoff between AHL, PHL and OHL. Thus, thermal analysis of modern systems with currently available amplifier and wiring technologies is necessary to understand their impact on system scalability.

The need for re-analysis is not limited to improvements in readout amplifier and wiring choices. 
The available design space has expanded across multiple dimensions. 
Qubit architectures now include fixed-frequency (FF) and tunable-frequency (TF) qubits, as well as fixed-coupling (FC) and tunable-coupling (TC) schemes. 
Each choice changes the required number of drive, flux-bias, and coupler-bias lines. Optical interconnects further change the tradeoff by reducing PHL while introducing new active heat sources, such as photodetectors at cryogenic stages. 
These changes make it difficult to estimate present-day scalability limits using previously proposed models.

These observations motivate a modern re-evaluation of thermal scalability. 
A useful model must explicitly include PHL, AHL, and OHL from all major control, readout, and supporting infrastructures. 
It must also be flexible enough to evaluate different qubit modalities, wiring technologies, amplifier technologies, and refrigerator cooling profiles under a unified accounting framework. 
Without such a model, the dominant bottleneck may be misidentified, and design effort may be directed toward components that are no longer the true scalability limit.

In this work, we explicitly model amplifier-associated heat loads, including amplifier AHL and amplifier-bias PHL/OHL during thermal scalability analysis. 
We then compare legacy and emerging technology choices (e.g., amplifiers, amplifier biasing, fridges) using a consistent and common framework to determine how bottlenecks migrate as individual components improve. 
This re-analysis is necessary for identifying credible pathways toward 10k-qubit cryogenic systems and for understanding which engineering improvements most directly increase useful fault-tolerant computing capacity.

\subsection{Estimating the logical qubit yield}
The previous subsection defines scalability in terms of the number of physical qubits that can be thermally supported by the cryogenic platform. 
However, the ultimate objective is to maximize the number of logical qubits that a system can support, i.e., the logical qubit yield. A higher PQ capacity does not automatically translate into a higher LQ yield. The conversion from PQ capacity to LQ yield depends on the code distance required to achieve the maximum tolerable logical error rate given the physical error rates of the system. The physical error rates of the system include 1Q gate errors, 2Q gate errors, measurement errors, reset errors, and idling errors. Higher physical error rates require larger code distances, which reduce the number of logical qubits that can be encoded within the same physical qubit budget. \Cref{{subsec:ler}} discusses how we can estimate LQ yield in more detail.

\subsection{Acronyms and Heat-Load Taxonomy}
The cryogenic system is a complex multi-stage thermal platform. 
Each temperature stage absorbs heat from multiple components, and these heat loads arise under different operating conditions. 
To make the modeling framework clear and readable, we first establish the heat load taxonomy used throughout this work. 

Throughout this work, each heat contribution is labeled using a structured tag that identifies (i) the cable, (ii) the physical origin of the heat, and (iii) the operational context. PHLs are written as \leg{CABLE|PASSIVE|IDLE}, which denotes the steady conductive heat associated with a given \leg{CABLE} and is treated as present regardless of workload activity. The label \leg{IDLE} indicates that the associated heat is continuously present during system operation. Active heat loads are written as \leg{CABLE|COMPONENT|OPERATION}, which indicates heat dissipated by a specific \leg{COMPONENT} on that \leg{CABLE} when executing a particular \leg{OPERATION}. The list of all heat loads is shown in \Cref{tab:heat_loads}.

To assist the reader, we have also summarized all major acronyms used throughout this work  in \Cref{tab:acryonms}.

\section{Heat Modeling Methodology}
\label{sec:model}

In \Cref{subsec:revisit}, we outlined the requirements for a model suitable for large-scale cryogenic systems. Specifically, the model must
\begin{itemize}
    \item capture dominant heat sources that emerge at scale, such as amplifier-related contributions,
    \item support fair comparison across legacy and emerging technology options, and 
    \item connect thermal constraints to logical qubit yield under realistic workloads.
\end{itemize}

To address these requirements, we improve upon prior system-level thermal models by incorporating previously neglected heat sources and by standardizing the representation of different system components. Our model captures passive, active, and ohmic heat loads across all temperature stages and supports a consistent comparison across wiring technologies, amplifier architectures, and cryogenic configurations.

We further integrate a workload-aware formulation and a logical error model to enable direct estimation of logical qubit yield from the underlying physical and thermal constraints. As a result, our framework links device-level parameters, system architecture, and cryogenic limitations to estimate the computational capacity of the system.

Note that the framework is a steady-state architectural model, intended to \textit{compare} heat load trends and \textit{identify candidate bottlenecks}, not to predict the exact operating point of a specific installed refrigerator.

\subsection{Dilution Refrigerators}
\begin{table*}[t]
\centering
\caption{Dilution fridge specifications used for scalability analysis in this work. We assume that the listed cooling powers are the available residual cooling powers.}
\label{tab:all_fridges}

\begin{threeparttable}
\begin{tabular}{lcccccc}

\toprule
\textbf{Fridge} 
& {\textbf{50\,K}} 
& {\textbf{4\,K}} 
& {\textbf{2\,K}} 
& {\textbf{Still}} 
& {\textbf{CP}} 
& {\textbf{MXC}}  \\
\midrule

\textbf{XLD400} \cite{krinner2019engineering} \\
\quad{\textit{Cooling Power}} \tnote{$\dagger$} &
\qty{30}{\watt} & 
\qty{1.5}{\watt} & 
N/A & 
\qty{40}{\milli\watt} & 
\qty{200}{\micro\watt} & 
\qty{19}{\micro\watt} 
\\

\quad{\textit{Operating Temp.}} &
\qty{45}{\kelvin} & 
\qty{4.2}{\kelvin} & 
N/A & 
\qty{1.2}{\kelvin} & 
\qty{140}{\milli\kelvin} & 
\qty{20}{\milli\kelvin} 
\\
\quad{\textit{Flange Separation}} & 
\qty{20}{\centi\meter} & 
\qty{29}{\centi\meter} &
N/A & 
\qty{25}{\centi\meter} &
\qty{17}{\centi\meter} &
\qty{14}{\centi\meter}
\\
\midrule

\textbf{XLD1000sl} \cite{raicu2025cryogenic} \\
\quad{\textit{Cooling Power}} & 
\qty{30}{\watt} & 
\qty{0.7}{\watt} & 
N/A & 
\qty{7}{\milli\watt} & 
\qty{1}{\milli\watt} & 
\qty{30}{\micro\watt}
\\

\quad{\textit{Operating Temp.}} &
\qty{40}{\kelvin} & 
\qty{3.5}{\kelvin} & 
N/A & 
\qty{1.4}{\kelvin} & 
\qty{200}{\milli\kelvin} & 
\qty{20}{\milli\kelvin} 
\\
\quad{\textit{Flange Separation}} & 
\qty{30.53}{\centi\meter} &
\qty{31.55}{\centi\meter} &
N/A & 
\qty{27.75}{\centi\meter} &
\qty{19.65}{\centi\meter} &
\qty{19.65}{\centi\meter}
\\ 
\midrule

\textbf{KIDE} \cite{blueforsKIDECryogenic} \\
\quad{\textit{Cooling Power}} \tnote{$\ddagger$} & 
\qty{90}{\watt} & 
\qty{6}{\watt} & 
N/A & 
\qty{90}{\milli\watt} & 
\qty{3}{\milli\watt} & 
\qty{90}{\micro\watt} 
 \\

\quad{\textit{Operating Temp.}} &
\qty{40}{\kelvin} & 
\qty{4.5}{\kelvin} & 
N/A & 
\qty{1.2}{\kelvin} & 
\qty{200}{\milli\kelvin} & 
\qty{20}{\milli\kelvin} 
\\
\quad{\textit{Flange Separation}} & 
\qty{30.53}{\centi\meter} &
\qty{31.55}{\centi\meter} &
N/A & 
\qty{27.75}{\centi\meter} &
\qty{19.65}{\centi\meter} &
\qty{19.65}{\centi\meter}
\\ 
\midrule

\textbf{Colossus} \cite{hollister2022large,hollister2024update}\\
\quad{\textit{Cooling Power}} & 
\qty{9}{\kilo\watt} & 
\qty{200}{\watt} & 
\qty{10}{\watt} & 
\qty{100}{\milli\watt} & 
\qty{3}{\milli\watt} & 
\qty{300}{\micro\watt} 
\\

\quad{\textit{Operating Temp.}} &
\qty{80}{\kelvin} & 
\qty{5}{\kelvin} & 
\qty{2}{\kelvin} & 
\qty{1}{\kelvin} & 
\qty{100}{\milli\kelvin} & 
\qty{20}{\milli\kelvin} 
\\
\quad{\textit{Flange Separation}} & 
\qty{24.64}{\centi\meter} &
\qty{39.88}{\centi\meter} &
\qty{54.61}{\centi\meter} &
\qty{46.23}{\centi\meter} &
\qty{54.61}{\centi\meter} &
\qty{54.61}{\centi\meter}
\\

\textbf{Colossus-CP} \tnote{$\star$}\\
\quad{\textit{Cooling Power}} & 
\qty{9}{\kilo\watt} & 
\qty{200}{\watt} & 
\qty{10}{\watt} & 
\qty{100}{\milli\watt} & 
\qty{6}{\milli\watt} & 
\qty{240}{\micro\watt} 
\\

\quad{\textit{Operating Temp.}} &
\qty{80}{\kelvin} & 
\qty{5}{\kelvin} & 
\qty{2}{\kelvin} & 
\qty{1}{\kelvin} & 
\qty{100}{\milli\kelvin} & 
\qty{20}{\milli\kelvin} 
\\
\quad{\textit{Flange Separation}} & 
\qty{24.64}{\centi\meter} &
\qty{39.88}{\centi\meter} &
\qty{54.61}{\centi\meter} &
\qty{46.23}{\centi\meter} &
\qty{54.61}{\centi\meter} &
\qty{54.61}{\centi\meter}
\\

\bottomrule
\end{tabular}

\begin{tablenotes}
\item[$\dagger$] Measured cooling powers at 50\,K and 4\,K already account for the total PHL due to 24 pairs of AWG35 copper wires and 24 pairs of AWG36 phosphor-bronze wires (less than \qty{50}{\milli\watt}).
\item[$\ddagger$] Estimated values (see text for details).
\item[$\star$] We consider a hypothetical modification in which two dilution units originally serving the MXC stage are reassigned to the CP stage. We refer to this configuration as \textbf{Colossus-CP}.

\end{tablenotes}

\end{threeparttable}
\end{table*}

The choice of fridge determines (i) the available cooling power and (ii) the operating temperatures of various temperature stages as well as (iii) the distance between the temperature flanges. The operating temperatures and the inter-flange separation directly affect the PHL (as well as the electrical resistance) of various wires. 

\Cref{tab:all_fridges} summarizes the operating temperatures and the corresponding residual cooling powers of the different DRs considered in this work. These include legacy systems such as XLD400 \cite{krinner2019engineering,joshi2023scaling}, currently deployed systems such as XLD1000sl \cite{raicu2025cryogenic} and KIDE \cite{blueforsKIDECryogenic}, and emerging large-scale systems such as Colossus \cite{hollister2022large,hollister2024update}.

We assume, under steady-state conditions, the temperature stages maintain the constant operating temperature as long as the total heat load entering a stage remains below the available cooling power i.e., all incoming heat is fully absorbed. We assume this constant temperature condition to simplify modeling because varying the operating temperatures can impact the available cooling power of the stages \cite{manifold2025thermal} as well as the PHL and resistance of the cables.

The cooling powers for XLD400 and XLD1000sl are taken from reported measurements \cite{krinner2019engineering,raicu2025cryogenic}. Bluefors KIDE \cite{blueforsKIDECryogenic} is a state-of-the-art cryogenic fridge designed for modular large-scale quantum computing. It is composed of three dilution units. Since its measured cooling powers have not been publicly reported, we estimate its cooling power to be roughly $3\times$ that of XLD1000sl at CP and MXC, with larger cooling powers at the warmer stages. For more discussion, refer to \Cref{app:bluefors-kide}.

Colossus \cite{hollister2022large,hollister2024update} represents a large-scale approach that integrates ten dilution units within a single system. It is being developed by Fermilab and achieves significantly higher cooling power at 50\,K, 4\,K, and MXC compared to other systems. The values in \Cref{tab:all_fridges} are projected residual cooling powers. Due to its large size, Colossus includes an additional 2\,K stage between the 4\,K and Still. However, its CP cooling power remains comparable to KIDE. This is because Colossus has dedicated dilution units that intercept heat at CP before it reaches MXC, which limits the effective cooling power at CP.

\subsection{Passive Heat Loads}
Passive heat load (also referred to as static heat load) originates from thermal conduction due to the temperature gradient along interconnects that span two adjacent temperature stages. In our model, the PHL attributed to a given stage is defined as the conductive heat flowing \textit{into} that stage from the immediately higher-temperature flange. Contributions from nonadjacent higher-temperature stages and thermal radiation are omitted. Microwave interconnects required for qubit control and readout introduce PHL into all intermediate stages. These include drive lines, flux or coupler bias lines (when applicable), readout feed-in and feed-out lines, and pump lines. In addition, as discussed before, the readout amplifiers located at 4\,K and 50\,K require DC biasing and control wiring which also introduce significant PHL. To our knowledge, this is the first work to explicitly account for the PHL due to readout amplifier biasing wires during scalability analysis. 

\subsubsection{Qubit Control and Readout Lines}

\begin{table*}[t]
\centering
\caption{Passive heat loads and DC resistances for different signal wiring options. PHL values indicate heat flow rate into the temperature stage from the higher temperature flange (assuming the inter-flange distances and operating temperatures corresponding to XLD1000sl).}
\label{tab:signal_cables}

\begin{threeparttable}
\begin{tabular}{llcccccc}
\toprule
\textbf{Cable}
&\textbf{Name}   
& \textbf{50\,K} 
& \textbf{4\,K} 
& \textbf{Still} 
& \textbf{CP}    
& \textbf{MXC} 
& \textbf{Reference} \\
\midrule

UT-085-SS-SS (Drive)
& SS Drive              
& \qty{30.06}{\milli\watt}       
& \qty{822.3}{\micro\watt}        
& \qty{2.52}{\micro\watt}           
& \qty{391.76}{\nano\watt}    
& \qty{13.89}{\nano\watt}      
& \cite{krinner2019engineering} \\

UT-085-SS-SS (Flux)
& SS Flux            
& \qty{37.4}{\milli\watt}       
& \qty{986.76}{\micro\watt}        
& \qty{1.26}{\micro\watt}           
& \qty{293.82}{\nano\watt}    
& \qty{30.99}{\nano\watt}      
& \cite{krinner2019engineering} \\

\quad
\textit{DC Resistance}
&
& N/A
& N/A
& N/A
& \qty{420}{\milli\ohm}
& \qty{150}{\milli\ohm}
& \cite{krinner2019engineering} \\

UT-085-NbTi (coax)
& NbTi (coax)
& --       
& --        
& \qty{630.63}{\nano\watt}        
& \qty{293.82}{\nano\watt} 
& \qty{21.37}{\nano\watt}   
& \cite{krinner2019engineering} \\

SC-086-NbTi (coax) \tnote{$\dagger$}
& SC086-NbTi (coax)
& --       
& --        
& \qty{157.13}{\nano\watt}         
& \qty{22.01}{\nano\watt}     
& \qty{135.24}{\pico\watt}     
& \cite{coax2022} \\

\addlinespace
\midrule

HDW
&              
& \qty{7.57}{\milli\watt}       
& \qty{315.7}{\micro\watt}        
& \qty{2.37}{\micro\watt}           
& \qty{549.1}{\nano\watt}    
& \qty{11.32}{\nano\watt}      
& \cite{raicu2025cryogenic} \\

\quad
\textit{DC Resistance}
&
& \qty{278.16}{\ohm}
& \qty{96.67}{\milli\ohm}
& \qty{84.5}{\milli\ohm}
& \qty{60.27}{\milli\ohm}
& \qty{60.27}{\milli\ohm}
& \cite{raicu2025cryogenic} \\

\addlinespace
\midrule

Cri/oFlex (Ag)
& Ag   
& \qty{2.94}{\milli\watt}       
& \qty{643.84}{\micro\watt}        
& \qty{1.68}{\micro\watt}            
& \qty{495.88}{\nano\watt}      
& \qty{12.01}{\nano\watt}        
& \cite{delftcircuitsCryogenicCables} \\

Cri/oFlex (NbTi)
& NbTi
& --                            
& --                             
& \qty{168.27}{\nano\watt}             
& \qty{49.59}{\nano\watt}       
& \qty{1.2}{\nano\watt}        
& \cite{delftcircuitsCryogenicCables} \\

\addlinespace
\midrule

Optical Fiber \tnote{$\dagger$} 
& Fiber    
& \qty{34.03}{\micro\watt}       
& \qty{1.13}{\micro\watt}     
& \qty{17.98}{\nano\watt}            
& \qty{4.40}{\nano\watt}      
& \qty{21.87}{\pico\watt}     
& \cite{smith1978effect} \\  

\bottomrule
\end{tabular}
\begin{tablenotes}
\item[$\dagger$] Estimated values using the thermal conductivity models in \Cref{app:phl_details}. All other values are derived from reported measurements/data as described in \Cref{app:phl_details}.
\end{tablenotes}

\end{threeparttable}

\end{table*}
\Cref{tab:signal_cables} summarizes the PHLs for the qubit control and readout cables we use in our scalability analysis. Since the PHL depends on the wire length and the temperature gradient across the cable, the values in \Cref{tab:signal_cables} assume the inter-flange separations and operating temperatures corresponding to XLD1000sl (See \Cref{app:phl_details} for more details). 

Stainless-steel (SS) coaxial cables were historically used for both high-frequency drive and low-frequency flux lines \cite{krinner2019engineering}. For readout feed-out lines below 4\,K, NbTi coaxial cables are typically used because they become superconducting, which reduces attenuation and eliminates ohmic dissipation. UT-085 NbTi coaxial cables are much thicker than SC-086-NbTi and hence conduct more passive load.
More recent systems adopt lower-PHL cupronickel-based coaxial cables such as those used in the Bluefors High Density Wiring (HDW) platform \cite{raicu2025cryogenic,blueforsHighDensityWiring}. These cables reduce conductive heat flow while maintaining compatibility with conventional coax-based routing.

Microwave striplines with lower PHL and higher routing density \cite{monarkha2024equivalence,paluch2025thermalization} have also been recently proposed. In these structures, a planar conductor is embedded in a dielectric, which reduces the effective cross-sectional area and therefore the conductive heat flow. Delft Circuits implements this approach in the Cri/oFlex family \cite{delftcircuitsCryogenicCables}. Ag striplines are typically used for drive lines, while NbTi variants are used below 4\,K for flux-bias and readout RF lines (\leg{READOUT\_POUT}) due to their superconducting behavior. In addition to lower PHL, striplines also thermalize faster compared to SS coaxial cables \cite{paluch2025thermalization}. 

We also model systems that use optical fibers as an alternative to metallic coaxial cables. Optical fibers offer very low thermal conductivity and support for multiplexing \cite{lecocq2021control,joshi2023scaling}. As shown in \Cref{tab:signal_cables}, their PHL is orders of magnitude lower than metallic wiring, particularly at higher temperature stages. Additional modeling details of their PHL are provided in \Cref{app:phl_details}.

\subsubsection{Amplifier Biasing Lines}
Readout amplifiers for second-stage amplification require DC biasing lines to supply operating power. These lines introduce significant PHL into the 50\,K and 4\,K stages. In this work, we consider common implementations that use copper wires (e.g., AWG30 or AWG35) \cite{lownoisefactoryLNFNANO9MNoise,krinner2019engineering} as well as alternative materials such as Manganin (Mn) \cite{ismael2021development} and high-temperature superconducting YBCO-on-Kapton cables \cite{solovyov2021ybco}.

\begin{table*}[tbp]
\centering
\caption{Passive heat load and equivalent resistance of bias wires between RT--50\,K and 50\,K--4\,K segments for different materials assuming operating temperatures and inter-flange distances of XLD1000sl.}
\label{tab:biasing_cables}

\begin{threeparttable}

\begin{tabular}{crrrrrr}
\toprule
Cable 
& Name
& \textbf{PHL (RT--50\,K)}        
& \textbf{PHL (50\,K--4\,K)}        
& \textbf{Resistance (RT--50\,K)} 
& \textbf{Resistance (50\,K--4\,K)} 
& \textbf{Ref.}      
\\
\midrule
\textit{\textbf{Copper AWG35}}
& Cu-35   
& \qty{10.02}{\milli\watt}  
& \qty{1.23}{\milli\watt}  
& -   
& -   
& \cite{krinner2019engineering}  \\

\textit{\textbf{Copper AWG30}}
& Cu-30   
& \qty{19.88}{\milli\watt}  
& \qty{6.81}{\milli\watt}  
& \qty{48.37}{\milli\ohm}   
& \qty{2.31}{\milli\ohm}   
& \cite{nistCryogenicMaterial,copperCryogenicProperties}  \\

\textit{\textbf{Manganin AWG30}}
& Mn 
& \qty{693.92}{\micro\watt} 
& \qty{25.47}{\micro\watt} 
& \qty{2.89}{\ohm}         
& \qty{2.77}{\ohm}         
& \cite{twire,xiang2020characterization,lakeshoreCryogenicWire} \\

\textit{\textbf{YBCO-on-Kapton}}
& YBCO     
& \qty{3.97}{\micro\watt}   
& \qty{3.84}{\micro\watt}  
& -                         
& 0 (assumed)                       
& \cite{solovyov2021ybco} \\ 

\bottomrule
\end{tabular}

\begin{tablenotes}
\item YBCO cables are modeled with cross-sectional dimension of \qty{1}{\milli\meter} $\times$ \qty{101}{\micro\meter}.
\end{tablenotes}

\end{threeparttable}
\end{table*}

\Cref{tab:biasing_cables} summarizes the PHL and resistance of representative bias-wire materials used in our analysis. The system in \cite{krinner2019engineering} uses custom AWG35 copper wires. We use this specification only when modeling that particular system, as only the PHL values are reported and the corresponding resistance data is not available. For all other configurations, we use commercially available AWG30 copper wires \cite{lownoisefactoryLNFNANO9MNoise}, since both thermal and electrical properties are well characterized.

A key observation is that a \emph{single} AWG30 copper wire introduces approximately \qty{7}{\milli\watt} of PHL into the 4\,K stage, which is comparable to the active dissipation of a HEMT amplifier (see \Cref{tab:amp_options}). Since each amplifier requires three bias wires, copper biasing alone can account for a large fraction of the per-amplifier heat load.

Manganin reduces PHL by more than an order of magnitude relative to copper, improving scalability at the cost of higher electrical resistance. We assume AWG30 diameter for consistency. Moreover, its high resistance requires a higher RT supply voltage to compensate for the voltage drop due to wire resistance.

YBCO cables are superconducting below 77\,K making them very attractive for biasing 4\,K stage amplifiers. They exhibit the lowest PHL due to the material and geometry. However, only prototypes have been demonstrated to date \cite{solovyov2021ybco}.

These differences in PHL of various cables strongly influence the 4\,K heat budget and are a key factor in determining scalability bottlenecks in later sections.
Details of the PHL estimation methodology of the cables are provided in \Cref{app:phl_details}.

\subsection{Ohmic Heat Loads}
Ohmic heat loads arise from Joule heating in cables that carry DC or low-frequency currents. This includes amplifier-bias lines, flux-bias or coupler-bias lines that remain continuously electrically energized during operation. To our knowledge, prior scalability analyses have not systematically accounted for the ohmic loads in amplifier bias wires. 

These cables span multiple temperature stages and the resistivity of the material can change significantly with temperature. 
To estimate the OHL, we therefore model each cable segment between adjacent stages using an effective lumped resistance $R_J$ whose value is determined using the temperature-dependent resistivity model $\rho(T)$ as follows. (See \Cref{app:ohl_details} for more details.)

The rate of Joule heating $P_J$ due to a wire segment carrying current $I$ between temperature stages with temperature $T_H$ and $T_L$ (assuming a linear temperature profile approximation)is 
\begin{equation}
    P_J = I^2 \frac{L}{A(T_H - T_L)} \int_{T_L}^{T_H} \rho(T)\, dT,
\label{eq:ohmicload}
\end{equation}
where $L$ is the wire length and $A$ is the cross-sectional area. 

The equivalent effective resistance $R_J$ over that temperature span also dissipates ohmic heat at the same rate and is thus given by 
\begin{equation}
R_J = \frac{L}{A (T_H - T_L)} \int_{T_L}^{T_H} \rho(T)\, dT
\label{eq:eqv_resistance}
\end{equation}

\subsubsection{Amplifier Bias Lines}
The lumped resistance values to estimate OHL for amplifier biasing lines are summarized in \Cref{tab:biasing_cables}. Copper exhibits low resistivity and therefore low OHL, but introduces high PHL. In contrast, Manganin significantly reduces PHL but has resistance that is orders of magnitude higher, leading to substantial OHL at typical bias currents. This trade-off between thermal and resistive heat loads is a key factor in determining scalable wiring choices. YBCO bias wiring is assumed to have zero resistance (even with very low cross-sectional area) over the 50\,K--4\,K segment due to superconductivity below 77\,K.

Ohmic heat is generated continuously along the wire. Since bias wires are not perfectly thermalized at intermediate stages, we assume that the generated heat is equally shared between the two temperature stages connected by the segment. For a segment between 50\,K and 4\,K, the heat attributed to 4\,K for a single wire is $0.5 I^2 R_J$. Accounting for both forward and return conductors, the total heat assigned to 4\,K becomes $I^2 R_J$.

\subsubsection{Qubit Control Lines}
At microwave frequencies, conductor loss is governed by the frequency-dependent AC resistance, which increases because current is confined near the conductor surface by the skin effect. However, for the low microwave powers and duty cycles considered in this work, this dissipation is generally small compared with other heat loads. We therefore do not model skin-effect-induced OHL explicitly.

In contrast, flux-bias and coupler-bias signals are DC or low-frequency pulses, for which Joule heating in the cable conductors can be estimated from their DC resistance. This conductor OHL is generally small at the warmer cryogenic stages relative to the available cooling power and dominant loads from attenuators and amplifiers. However, it can consume a non-negligible fraction of the limited cooling power at the millikelvin stages \cite{krinner2019engineering,raicu2025cryogenic}. We therefore model the resistance offered by each cable segment at the millikelvin stages using an effective lumped resistance derived from the reported measurements for stainless-steel and HDW coaxial cables and summarized in \Cref{tab:signal_cables}. Following the thermal-anchoring convention in \cite{raicu2025cryogenic}, we assign the Joule heat generated within each segment to the lower-temperature stage at which the segment is anchored.

\subsection{Active Heat Loads}
Active heat loads typically originate from RF attenuators, readout amplifiers, and, in optical architectures, photodetectors. Other cryogenic components such as filters, TWPAs, circulators, isolators, and directional couplers may dissipate power, but their contributions are small relative to the dominant terms and are not included in our scalability analysis.

\subsubsection{RF Attenuators}
\begin{table}[tbp]
\centering
\caption{Attenuator distribution across temperature stages used in this work, adapted from \cite{krinner2019engineering}. Values denote attenuation placed at each stage.}
\label{tab:attenuator}

\resizebox{\columnwidth}{!}{%
\begin{threeparttable}

\begin{tabular}{lccccl}
\toprule
\textbf{Cable}
& \textbf{4\,K} 
& \textbf{Still} 
& \textbf{CP} 
& \textbf{MXC} 
& \textbf{Comments}     
\\ 
\midrule

\textbf{\leg{DRIVE}}        
& \qty{20}{\decibel}                          
& -               
& \qty{20}{\decibel}              
& \qty{20}{\decibel}               
& For system in \cite{krinner2019engineering}                                        
\\

\textbf{\leg{DRIVE}}        
& \qty{20}{\decibel}                   
& \qty{10}{\decibel}            
& \qty{10}{\decibel}              
& \qty{20}{\decibel}               
&                                
\\

\textbf{\leg{FLUX\_BIAS}}         
& \qty{20}{\decibel} \tnote{$\dagger$}                         
& -                                  
& -                               
& -                                
& For TF systems                      
\\

\textbf{\leg{COUPLER}}      
& \qty{20}{\decibel} \tnote{$\dagger$}                          
& -                                  
& -                               
& -                                
& For TC systems                      
\\

\textbf{\leg{PUMP}}         
& \qty{20}{\decibel}                           
& -                                  
& \qty{10}{\decibel}              
& \qty{20}{\decibel}\tnote{$\ddagger$}               
&                
\\

\textbf{\leg{READOUT\_PIN}} 
& \qty{20}{\decibel}                           
& -                 
& \qty{20}{\decibel}              
& \qty{20}{\decibel}               
& For system in \cite{krinner2019engineering}
\\
\textbf{\leg{READOUT\_PIN}} 
& \qty{20}{\decibel}                           
& \qty{10}{\decibel}                 
& \qty{10}{\decibel}              
& \qty{20}{\decibel}               
& 
\\
\bottomrule
\end{tabular}

\begin{tablenotes}
\footnotesize
\item[$\dagger$] Implemented using T-pad resistor network.
\item[$\ddagger$] The \qty{20}{\decibel} attenuation for the pump line is provided by the directional coupler. The impedance-matching termination for the directional coupler may be relocated to 4\,K stage.
\end{tablenotes}

\end{threeparttable}
}
\end{table}

\begin{table}[tbp]
\centering
\caption{Signal power required at the MXC stage for various operations.}
\label{tab:mxc_power}

\begin{threeparttable}

\begin{tabular}{llllr}
\toprule
\textbf{Operation}
& \textbf{Cable}    
& \textbf{$P_{MXC}$}
& \textbf{Reference}
 \\
\midrule
\textbf{1Q}
& \leg{DRIVE}         
& \qty{-71}{\dBm}
& \cite{krinner2019engineering}  
\\

\textbf{2Q}
& \leg{DRIVE}            
& \qty{-66}{\dBm} \tnote{$\dagger$}
& \cite{krinner2019engineering}
\\

\textbf{Readout}
& \leg{READOUT\_PIN} 
& \qty{-120}{\dBm} 
& \cite{bardin2021microwaves,2025acharyaQuantumErrorCorrection}  
\\ 

\textbf{Readout}
& \leg{PUMP}        
& \qty{-55}{\dBm} 
& \cite{krinner2019engineering} 
\\     
\bottomrule          
\end{tabular}

\begin{tablenotes}
\footnotesize
\item[$\dagger$] Only for systems using CX cross-resonance gates.
\end{tablenotes}

\end{threeparttable}

\end{table}

\Figure[!ht](topskip=0pt, botskip=0pt, midskip=0pt)[width=0.97\columnwidth]
{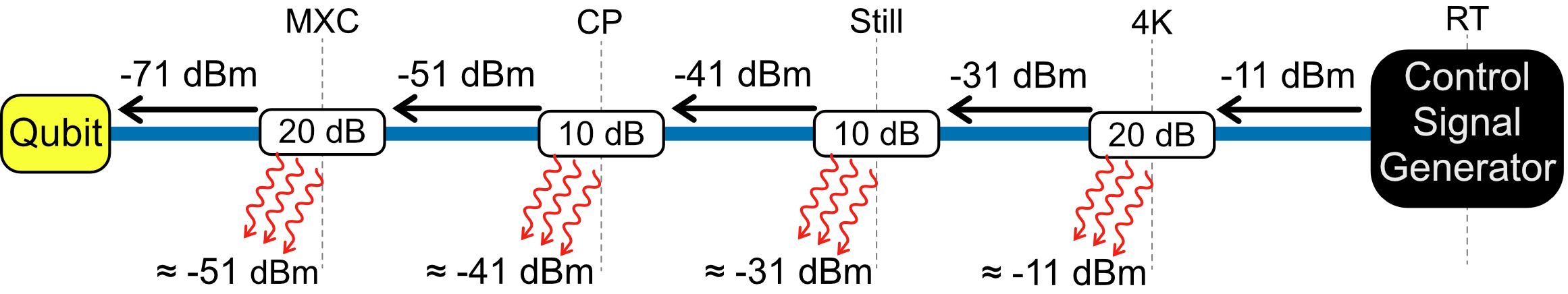}
{Active heat dissipation in attenuators along the \leg{DRIVE} line during a single-qubit (1Q) operation. Each attenuator is assumed to dissipate all incident power as heat (pessimistic estimation). The \leg{DRIVE} line has a total attenuation of \qty{60}{\decibel}, distributed across multiple temperature stages. A 1Q operation requires \qty{-71}{\dBm} of  average power delivered to the qubit at the MXC \cite{krinner2019engineering}. The input power at each attenuator at different temperature stages along the drive line is obtained by propagating this value upstream from the MXC stage. As a result, the control signals generated at RT require an average power of \qty{-11}{\dBm}. 
\label{fig:attenuator_chain}
}

Microwave control and readout lines (\leg{DRIVE}, \leg{PUMP}, \leg{READOUT\_PIN}) include attenuators at multiple temperature stages to thermalize the line and suppress noise from higher-temperature stages. In contrast, \leg{READOUT\_POUT} lines use superconducting cables without attenuation because they carry weak output signals away from the qubit and are isolated by isolators and filters.

Attenuators dissipate 90\%-99\% of the incident signal power as heat depending on the attenuation level. We therefore model the active heat load of an attenuator as equal to its incident power (as a worst-case estimate) and assume that this heat is fully absorbed by the stage where the attenuator is anchored.
Attenuation is distributed across temperature stages to avoid thermally stressing the MXC stage while maintaining adequate noise suppression for qubit operation \cite{krinner2019engineering}. We use the various attenuation schemes from \cite{krinner2019engineering} listed in \Cref{tab:attenuator} for our analysis.
The incident power (or equivalently the dissipated active heat) at each attenuator is determined by starting from the required signal power at MXC ($P_{MXC}$) and propagating it backward through the attenuation chain \cite{krinner2019engineering,raicu2025cryogenic} as shown in \Cref{fig:attenuator_chain}. 

\paragraph*{Signal Power at MXC}
The MXC power levels for various operations used in this work are listed in \Cref{tab:mxc_power}. The values of $P_{MXC}$ are not directly measured in practice. Instead, the input power at room temperature is tuned experimentally, and the effective MXC power is inferred using the known attenuation profile. We use representative values of $P_{MXC}$ from prior work \cite{krinner2019engineering,bardin2021microwaves}. These values are indicative and can be replaced with calibrated measurements for specific systems.

As listed in \Cref{tab:mxc_power}, the readout probe signal at MXC is weak to avoid disturbing the qubit state, whereas the pump signal is strong and used to drive the TWPA. 

The required \leg{DRIVE} signal power at MXC, $P_{DRIVE,MXC}$, depends on the qubit design, coupling strength, and the qubit operation, among many other factors. For 1Q operations, we assume an average power of $P_{DRIVE, MXC}(1Q) = \qty{-71}{\dBm}$ \cite{krinner2019engineering}. For 2Q cross-resonance gates in FF architectures, we use $P_{DRIVE, MXC}(2Q) =\qty{-66}{\dBm}$ because
the flat-topped control pulse remains near peak amplitude for most of its duration \cite{patterson2019calibration,malekakhlagh2020first}.
These values are representative and vary across platforms.

We assume that the TWPA pump signal is injected through a \qty{20}{\decibel} directional coupler located at the MXC. The directional coupler provides the required pump coupling,
while most of the incident pump power exits through the auxiliary port and is absorbed by a \qty{50}{\ohm} termination \cite{krinner2019engineering,raicu2025cryogenic}. This dissipation can impose a nonnegligible active heat load on the MXC.

To evaluate a configuration that transfers this heat load to a stage with greater cooling power, we hypothetically place the termination at the \qty{4}{\kelvin} stage and connect it
to the directional coupler using an NbTi coaxial cable. We denote this additional cable as
\textbf{\leg{DC\_TERMINAL}} and explicitly include its PHL in the heat model.

This configuration assumes that the remote termination remains impedance matched over the pump bandwidth and that sufficient reverse isolation prevents noise from the \qty{4}{\kelvin} termination from reaching the MXC. Refer to \Cref{app:dc_termination} for further discussion.

\subsubsection{Low-Frequency Line Attenuators}

\begin{table}[tbp]
\centering
\caption{DC current values delivered at the output (qubit) for biasing flux and coupler lines.}
\label{tab:DC_currents}

\begin{threeparttable}
\begin{tabular}{@{}lllc@{}}
\toprule
\textbf{System}  
& \textbf{Operation}           
& \textbf{Cable}         
& \textbf{Current (mA)} 
\\ 
\midrule
TF-FC \cite{krinner2019engineering} 
& Flux bias (IDLE)    
& \leg{FLUX\_BIAS}    
& 0.57         
\\
        
& Two-qubit gate (2Q) 
& \leg{FLUX\_BIAS} \tnote{$\dagger$}     
& 0.4          
\\ 
\midrule

TF-TC \cite{raicu2025cryogenic}   
& Flux bias (IDLE)    
& \leg{FLUX\_BIAS}    
& 0.4          
\\

& Two-qubit gate (2Q) 
& \leg{COUPLER} \tnote{$\ddagger$}
& 0.4          
\\ 
\bottomrule
\end{tabular}

\begin{tablenotes}
\footnotesize
\item[$\dagger$] For the TF-FC system in \cite{krinner2019engineering}, 2Q operation involves pulsing the \leg{FLUX\_BIAS} lines to bring the resonant frequencies of the entangling qubits together for a fixed time duration.
\item[{$\ddagger$}] Modern TF-TC systems like \cite{arute2019quantum,2025acharyaQuantumErrorCorrection} use a dedicated \leg{COUPLER} line to activate and deactivate the couplers to enable 2Q operations.
\end{tablenotes}

\end{threeparttable}
\end{table}

Flux and coupler bias lines in TF and TC architectures carry DC currents or low-frequency control signals. These lines typically include a \qty{20}{\decibel} attenuator at the 4\,K stage to suppress thermal noise. The attenuator is impedance-matched to \qty{50}{\ohm} and is commonly implemented as a T-pad network \cite{raicu2025cryogenic}. This is necessary to ensure proper impedance matching with filters and minimize reflections.

For DC or low-frequency bias currents, the same T-pad network behaves as a resistive load that shunts a large fraction of the input current to the ground plane at 4\,K. We model the resulting Joule heating due to the currents flowing through the resistive elements in the attenuator using an effective lumped resistance.
The resistor values reported in \cite{raicu2025cryogenic} for a \qty{20}{\decibel} attenuator correspond to an equivalent lumped resistance of approximately \qty{1.25}{\kilo\ohm}. Refer to \Cref{app:tpad} for more details.

We list the DC currents for flux and coupler biasing used in our modeling in \Cref{tab:DC_currents} based on the reports in \cite{krinner2019engineering,raicu2025cryogenic}. \leg{FLUX\_BIAS} lines carry DC currents to set qubit frequencies and therefore operate continuously. We denote this mode as the \leg{IDLE} operation. In earlier TF-FC architectures such as \cite{krinner2019engineering}, \leg{FLUX\_BIAS} lines were also used to generate pulses for two-qubit (2Q) operations. More recent designs employ tunable couplers, where coupling is controlled using dedicated \leg{COUPLER} lines instead.

\subsubsection{Readout Amplifiers}

\begin{table*}[tbp]
\caption{Readout Amplifiers considered in the scalability analysis.}
\label{tab:amp_options}

\begin{threeparttable}
\centering      
\begin{tabular}{lccccccll}
\toprule
\textbf{Name}
& \textbf{V (Volts)} 
& \textbf{I (mA)} 
& \textbf{AHL (mW)} 
& \textbf{Gain (dB)} 
& \textbf{Noise (K)} 
& \textbf{B/w (GHz)} 
& \textbf{No. of Biasing Lines} \tnote{$\dagger$}                                  
& \textbf{Ref.}                                              
\\ 

\midrule
\textbf{LNF8C}      
& 0.7                
& 15              
& 10.5               
& 42                 
& 1.5                
& 4-8                
& $2\times1 + 1 = 3$                                   
& \cite{lownoisefactoryLNFLNC4_8C}     
\\
\textbf{LNF8G}      
& 0.6                
& 13              
& 7.8               
& 40                 
& 1.5                
& 4-8                
& $2\times1 + 1 = 3$                                   
& \cite{lownoisefactoryLNFLNC4_8GNoise}                     
\\
\textbf{LNF8G (LP)} 
& 0.1                
& 3               
& 0.3               
& 22-24              
& 3                  
& 4-8                
& $2\times1 + 1 = 3$                                   
& \cite{lownoisefactoryLNFLNC4_8GNoise,cha2020300}                     
\\
\textbf{ULP-HEMT}     
& 0.08               
& 2.5             
& 0.2               
& 23.1               
& 2                  
& 4-6                
& $2\times(2\times1 + 1) = 6$ 
& \cite{zeng2023sub}                                        
\\
\textbf{SIS-5w}\tnote{$\ddagger$}       
& -              
& -             
& 0.0081            
& 6-8                
& 11              
& 2-5                
& $2\times1 + 3 = 5$              
& \cite{kojima2023characterization,murayama2024fabrication} 
\\ 
\textbf{SIS-13w}\tnote{$\ddagger$}      
& -              
& -              
& 0.0081            
& 6-8                
& 11              
& 2-5                
& $2\times5 + 3 = 13$    
& \cite{kojima2023characterization,murayama2024fabrication} 
\\ 
\bottomrule
\end{tabular}

\begin{tablenotes}
\footnotesize
\item[$\dagger$] Biasing currents require a pair of wires for incoming and outgoing currents, hence the factor of 2.
\item[$\ddagger$] Refer to \Cref{tab:sis_power} for the power dissipated by the various components of the SIS-Mixer amplifier.
\end{tablenotes}

\end{threeparttable}
\end{table*}

The reflected readout signal from the qubit is very weak, on the order of -120 dBm at the resonator output in the MXC. The first stage amplification by the TWPA (approximately \qty{20}{\decibel} gain) typically has an input saturation power around -100 dBm \cite{macklin2015near} (although there are recent reports that achieve input saturation upto -84 dBm \cite{gaydamachenko2025rf}).  
Further amplification is provided by a low-noise readout amplifier at 4\,K and, in some cases, an additional amplifier at 50\,K. Typically, HEMT amplifiers with 40 dB of gain \cite{lownoisefactoryLNFLNC4_8GNoise} are used to amplify the signal so that it can be processed at room temperature \cite{zurichshhqa}. Thus, in our modeling, we assume that at least 40 dB of additional gain is required after the TWPA. If the amplifier at 4\,K cannot provide this gain, we assume a third amplification stage using an additional HEMT at 50\,K is used to ensure that the signal level meets the requirements for room-temperature I/Q demodulation.  

HEMT amplifiers have large power dissipations that thermally stress the 4\,K stage \cite{raicu2025cryogenic,zeng2023sub}. More recently, ultra-low power HEMT (ULP-HEMT) and SIS-mixer amplifiers with much lower active heat have also been proposed as alternatives to conventional HEMT \cite{zeng2023sub,murayama2024fabrication}. \Cref{tab:amp_options} lists the various amplifiers used in our analysis and their operating parameters.

\paragraph*{HEMT}
The HEMT readout amplifier is assumed to be continuously biased. Its dissipated power is modeled as
\[
P_{HEMT} = V_{ds} \times I_{ds},
\]
where $V_{ds}$ and $I_{ds}$ are the operating voltage and current, respectively. This always-on dissipation is classified as \leg{IDLE} operation. Certain HEMT amplifiers, such as the LNFLNC4\_8C/8G from Low Noise Factory \cite{lownoisefactoryLNFLNC4_8C,lownoisefactoryLNFLNC4_8GNoise}, allow selection among different power modes via the gate-control line $V_{gs}$ \cite{lownoisefactoryLNFLNC4_8GNoise}. These amplifiers require three biasing wires for operation (labelled as \leg{AMP\_BIAS} cables). Two wires are used for power supply (incoming and return current $I_{ds}$ ) and one is used for gate-control. The current flowing through the control line is negligible compared to $I_{ds}$ and is therefore excluded from the active power calculation. \Cref{tab:amp_options} lists the various operating characteristics for the HEMT amplifiers during nominal and low power (LP) operation.

\paragraph*{ULP-HEMT}
Recent work reports a ULP-HEMT amplifier with much lower active power \cite{zeng2023sub}. It has a lower gain and similar noise performance to that of conventional HEMTs (see \Cref{tab:amp_options}). It requires two sets of bias lines (six wires total). Therefore, while AHL decreases, the associated PHL increases due to additional bias wiring. Since the gain is insufficient, an additional LNF8G amplifier at 50\,K for additional amplification is necessary.

\paragraph*{SIS-Mixer}
SIS-mixer amplifiers \cite{kojima2023characterization} provide gain through frequency up-conversion and down-conversion of the input signal. While the frequency conversion process enables amplification, the dominant power dissipation arises from the local oscillator (LO) required to drive the mixing process. SIS amplifiers exhibit higher noise, lower gain, and narrower bandwidth compared to HEMT amplifiers. However, continued advances may allow them to eventually satisfy the requirements of qubit readout. In this work, we treat SIS amplifiers as a representative example of an emerging future readout amplifier technology.

The SIS-mixer considered in this work dissipates approximately \qty{8.1}{\micro\watt} (see \Cref{app:sis_mixer} for details). However, SIS-based operation requires a larger number of bias and control lines compared to HEMT amplifiers. A typical configuration requires five wires (denoted as SIS-5w). To mitigate ohmic heating, especially for lines carrying higher currents, we also consider a configuration where these currents are distributed across additional wires, resulting in a total of 13 wires (SIS-13w). This is equivalent to reducing per-wire resistance by increasing cross-sectional area of the wire.

\subsubsection{Photodetector}
We also model systems that use optical fibers as an alternative to metallic coaxial cables. 
Optical interconnects have been proposed for large-scale superconducting systems due to their significantly lower PHL compared to metallic wiring (see \Cref{tab:signal_cables}) \cite{lecocq2021control,joshi2023scaling}. 
In these architectures, photodetectors (PDs) convert incoming optical signals into microwave signals for driving and measuring qubits. This conversion process dissipates power locally and contributes to the active heat load.

The average dissipated power $P_{PD}$ depends on the required average power at the qubit input $P_{MXC}$, the line impedance $Z$, and the photodetector responsivity $\mathcal{R}$, and is given by \Cref{eq:photodetector} \cite{lecocq2021control}:
\begin{equation}
    P_{PD} = \sqrt{\frac{2P_{MXC}}{Z \mathcal{R}^2}}.
\label{eq:photodetector}
\end{equation}

In optical systems where the PD is installed at the MXC, $P_{PD}$ can become a dominant contributor to the MXC active heat load. Since $P_{PD}$ scales with $\sqrt{P_{MXC}}$, reducing the required qubit input power directly lowers the photodetector dissipation. We therefore also evaluate hypothetical scenarios with reduced $P_{MXC}$ to assess their impact on thermal scalability in \Cref{subsec:optical_kide}.

\subsection{Quantum Workload}
\label{sec:quantum_workload}
The heat loads in a cryogenic system depend, in part, on the quantum workload.
Operation-dependent components, such as attenuators and photodetectors, dissipate heat in proportion to the type and frequency of quantum operations. 
Therefore, their contributions must be modeled explicitly as a function of the workload. To our knowledge, this is the first work that directly maps these operation-level activities to heat dissipation. Prior works approximate such effects using an \emph{activity factor} \cite{krinner2019engineering,raicu2025cryogenic,joshi2023scaling,min2023qisim}, which represents an average duty cycle under an abstract workload rather than explicitly modeling the sequence of operations.

In contrast, workload-independent contributions, such as passive heat load (PHL), are always present and do not depend on the executed operations. Similarly, continuous (\leg{IDLE}) operations, such as flux biasing and amplifier biasing, dissipate AHL and OHL independent of the workload.

\subsection{ESM Workload for FTQC}
\label{sec:ftqc}
\Figure[!ht](topskip=0pt, botskip=0pt, midskip=0pt)[width=0.95\textwidth]
{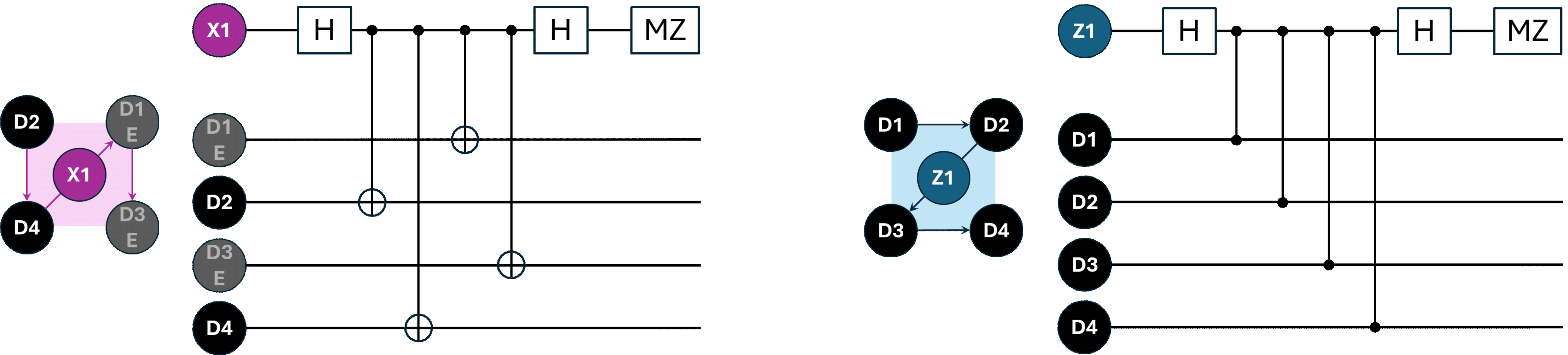}
{\parbox[t]{0.85\linewidth}{ESM operations for an X-ancilla qubit (left) and a Z-ancilla qubit (right). H: Hadamard; MZ: Z-basis measurement}
\label{fig:ESM_basic}}

\Figure[!ht](topskip=0pt, botskip=0pt, midskip=0pt)[width=0.99\textwidth]
{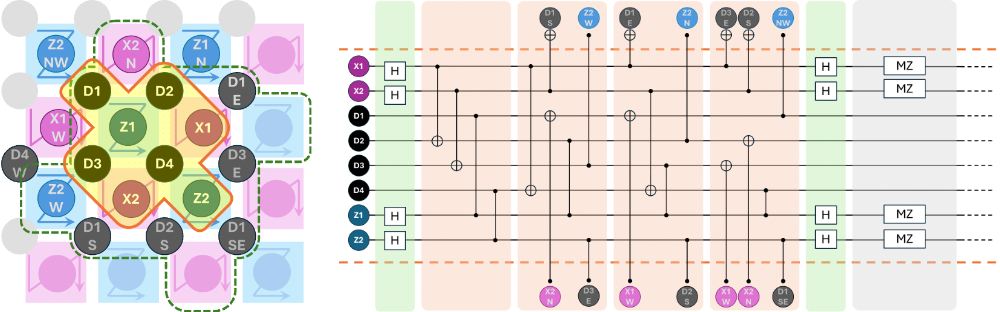}
{\parbox[t]{0.9\linewidth}{A rotated surface-code patch of distance 3 is enclosed in dashed green boundary (left). It consists of a tileable 8-qubit unit cell outlined in orange with yellow fill.
ESM circuit (right) using both CNOT and CZ gates for the 8-qubit unit cell. All gates within the colored shaded regions are performed simultaneously (green: Hadamard gates, orange: entangling CX/CZ gates, gray: measurement operations).}
\label{fig:ESM_original}}

During FTQC, \textit{logical qubits} are encoded using many \textit{physical qubits} and the system employs quantum error correction to suppress errors during computation. 
Among various quantum error correction codes, two-dimensional surface codes \cite{kitaev2003fault, bravyi1998quantum, raussendorf2007fault, fowler2012surface} are widely adopted due to their compatibility with nearest-neighbor connectivity and favorable error thresholds \cite{2025acharyaQuantumErrorCorrection}.

A surface-code logical qubit is implemented as a patch with code distance $d$, where a distance-$d$ patch requires $2d^2 - 1$ physical qubits. Of these, $d^2$ are data qubits and the remainder are ancilla qubits used for syndrome extraction. A distance-3 patch is outlined by a dashed green boundary in \Cref{fig:ESM_original}.

During fault-tolerant operation, the dominant recurring activity is repeated error-syndrome measurement (ESM), also referred to as the \textit{surface-code cycle}. 
This cycle repeatedly triggers control and readout operations across the system and therefore directly determines the dynamic heat loads in the cryogenic infrastructure. \Cref{fig:ESM_basic} illustrates the ESM operations for representative data and ancilla qubits. We will use ESM as the representative FTQC workload throughout this paper for thermal scalability analysis.

We assume a repeating 8-qubit unit-cell layout proposed in \cite{versluis2017scalable} for FTQC ESM, shown in \Cref{fig:ESM_original} (left, outlined in orange with yellow fill). 
Each unit cell consists of four data qubits and four ancilla qubits and serves as the fundamental building block for constructing larger surface-code patches. In tunable coupler architectures, 14 additional couplers are required per unit cell. 
The corresponding ESM circuit for this unit cell is shown in \Cref{fig:ESM_original} (right). 
The circuit is dominated by two-qubit entangling operations, followed by measurement and reset of the ancilla qubits at the end of each cycle.
 
Each ESM cycle consists of a sequence of single-qubit (1Q), two-qubit (2Q), and measurement (MZ) operations. 
Since this sequence repeats identically, the ESM circuit defines a periodic workload. 

Let $L_{ESM}$ denote the total duration of one ESM cycle, which is the sum of the durations of all gate and measurement blocks.
The average active power dissipated by an operation-dependent component over one ESM cycle, $P_{ESM}$, is computed by summing the energy contributions of all operations and normalizing by the cycle duration:
\begin{equation}
P_{ESM} = \frac{1}{L_{ESM}} \sum_{i \in \mathcal{O}} N_i P_i L_i
\end{equation}
where $\mathcal{O} = \{1\text{Q}, 2\text{Q}, \text{MZ}\}$, and
\begin{description}
    \item[$N_i$] number of operations of type $i$ per cycle,
    \item[$P_i$] instantaneous power dissipated during operation $i$,
    \item[$L_i$] duration (latency) of operation $i$.
\end{description}
$P_i$ is estimated using the models described in the previous sections, while $N_i$ and $L_i$ depend on the architecture-specific ESM circuit.

\subsubsection{Architecture-specific ESM circuits}

\Figure[!ht](topskip=0pt, botskip=0pt, midskip=0pt)[width=0.70\textwidth]
{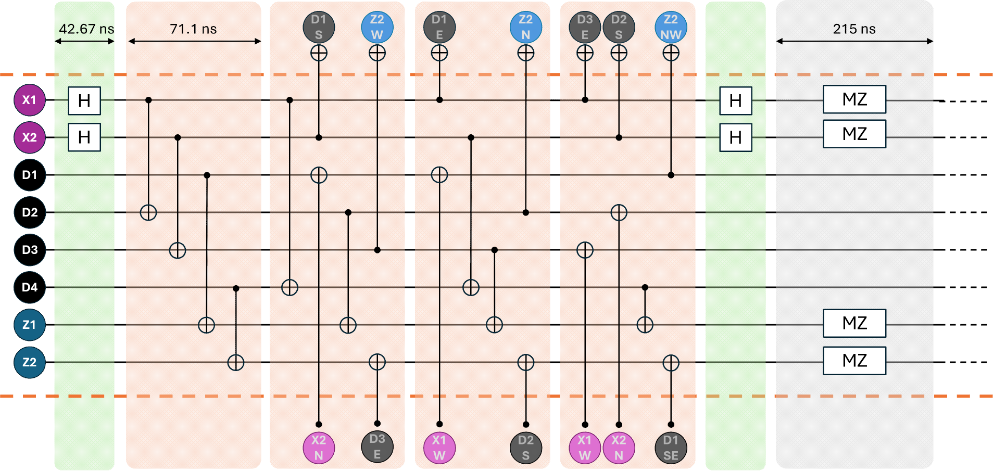}
{\parbox[t]{0.9\linewidth}{Equivalent ESM circuit for a rotated surface code using CX (cross-resonance) entangling gates used in this work to model FF-FC systems.}
\label{fig:ESM_CX}}

\Figure[!ht](topskip=0pt, botskip=0pt, midskip=0pt)[width=0.70\textwidth]
{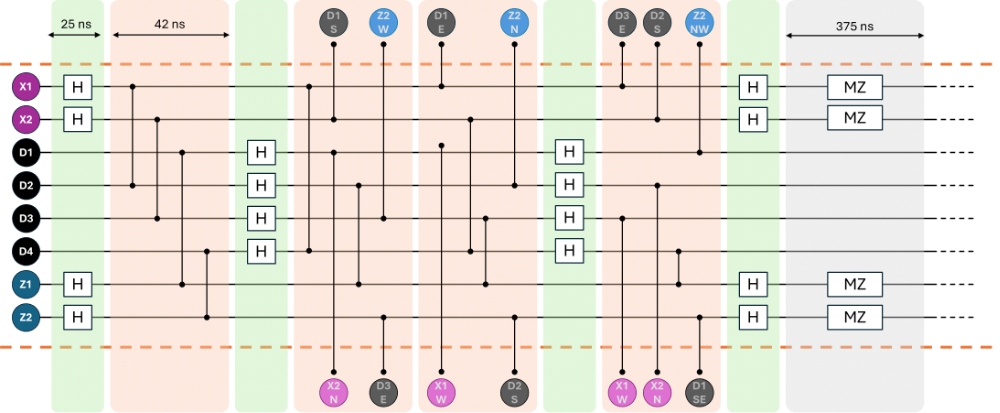}
{\parbox[t]{0.9\linewidth}{Equivalent ESM circuit for a rotated surface code using CZ entangling gates used in this work to model TF systems.}
\label{fig:ESM_CZ}}

The circuit in \Cref{fig:ESM_original} assumes that both CX and CZ gates are available in the system. 
In practice, the available gate set depends on the qubit architecture. 
The representative FF-FC architectures considered in this work implement cross-resonance CX gates\cite{kandala2021demonstration,spring2025fast} for entangling operations, whereas the representative tunable qubit architectures implement CZ gates \cite{arute2019quantum,sung2021realization,2025acharyaQuantumErrorCorrection,krinner2020benchmarking}. 
We therefore consider two functionally equivalent ESM implementations to analyze different qubit architectures: a CX-based circuit (\Cref{fig:ESM_CX}) and a CZ-based circuit (\Cref{fig:ESM_CZ}).

\subsubsection{ESM Operational Parameters}
\begin{table}[tbp]
\caption{Operation parameters for CX- and CZ-based ESM circuits.}
\label{tab:operations_summary}
\centering
\begin{tabular}{lccc}
\toprule
& \textbf{1Q} 
& \textbf{2Q} 
& \textbf{Readout} 
\\
\midrule
\textbf{CX gate set} & & & \\
\quad Duration (ns)     
& 42.67 \cite{underwood2024using}      
& 71.1  \cite{underwood2024using}       
& 215 \cite{spring2025fast} \\
\quad Operations per ESM
& 4           
& 16          
& 4 \\
\quad Multiplexing Ratio         
& 1          
& 1           
& 4 \cite{spring2025fast,krinner2020benchmarking} \\
\midrule
\textbf{CZ gate set} & & &  \\
\quad Duration (ns)     
& 25 \cite{2025acharyaQuantumErrorCorrection}          
& 42 \cite{2025acharyaQuantumErrorCorrection}          
& 375 \cite{2025acharyaQuantumErrorCorrection} \\
\quad Operations per ESM 
& 16          
& 16          
& 4 \\
\quad Multiplexing Ratio         
& 1           
& 1           
& 4 \\
\bottomrule
\end{tabular}
\end{table}

The number of operations per cycle and their latencies depend on the chosen architecture.
For example, in our assumed workloads, the CZ-based circuit requires additional Hadamard gates compared to the CX-based implementation. 
On the other hand, TF-based systems typically support faster gate operations than FF-based systems.

To enable a fair comparison across architectures, we adopt the following assumptions in our modeling. 
First, we assume a fixed ESM cycle duration of $L_{ESM} = \qty{1100}{\nano\second}$ \cite{versluis2017scalable,krinner2020benchmarking,2025acharyaQuantumErrorCorrection} that includes idle/control/reset overhead. This constraint ensures that error correction completes within the coherence time of the qubits. 
Although this bound is motivated by TF systems, we apply it uniformly to all architectures for consistency. 
Second, we assume a four-way multiplexed dispersive readout for ancilla qubits as in \cite{krinner2020benchmarking,spring2025fast} unless stated otherwise.

For CZ-based systems, we use operation latencies from the 105-qubit experimental system reported in \cite{2025acharyaQuantumErrorCorrection}. 
While the actual implementation in \cite{2025acharyaQuantumErrorCorrection} differs from the simplified circuit in \Cref{fig:ESM_CZ}, these values provide a representative baseline.

For CX-based systems, a full ESM implementation has not yet been experimentally demonstrated. 
We therefore construct a hypothetical system using experimentally reported latencies from recent FF-FC platforms \cite{underwood2024using,spring2025fast}. 
Readout latencies of up to \qty{215}{\nano\second} have been reported, and CX gate durations range from \qty{71.1}{\nano\second} to \qty{711}{\nano\second}. 
We use the lower bound of \qty{71.1}{\nano\second} to ensure that the ESM cycle fits within the assumed $L_{ESM}$ constraint.

The resulting operation counts ($N_i$) and latencies ($L_i$) for both CX- and CZ-based systems are summarized in \Cref{tab:operations_summary}. 
These values are indicative and do not correspond to a single validated hardware platform. 
More accurate modeling can be obtained using calibrated measurements from a specific system.

\subsection{Heat Load Calculation}
Our modeling framework is built around the 8-qubit unit cell described in \Cref{sec:ftqc}. We therefore first compute the complete thermal load at each temperature stage by summing all PHL, AHL and OHL associated with cables, attenuators, amplifiers, and other components required to operate the unit cell. 

To evaluate larger systems, we scale this unit-cell heat load linearly. For example, a 10k-qubit system corresponds to 1250 repetitions of the 8-qubit unit cell. All heat contributions are therefore multiplied by 1250 to estimate the thermal profile of the system. This linear scaling assumes uniform tiling of identical unit cells and neglects minor boundary effects arising from the finite geometry of the surface code. 

In practical fault-tolerant computation, not all qubits are continuously engaged in ESM cycles and certain qubit blocks must remain available for lattice surgery, routing, and magic-state distillation. The effective active heat load would be reduced compared to a fully saturated ESM workload. In this work, however, we adopt a conservative assumption: all qubits are continuously engaged in ESM with high activity. This yields an upper bound on steady-state active dissipation and therefore provides a lower bound on the maximum number of qubits that can be supported within the cryogenic heat constraints.

\subsection{Logical Error Model}
\label{subsec:ler}
We integrate a logical error model into the system-level framework to translate physical qubit capacity into logical qubit yield. Given a target LER and a set of physical error probabilities, the logical error model estimates the minimum code distance required to achieve that target. From the code distance, we can compute the number of PQs required per LQ. Combining this with the maximum number of PQs supported by the cryostat yields the total number of LQs that can be realized under the given configuration.

For our evaluations, we use the mean physical error probabilities reported in \cite{2025acharyaQuantumErrorCorrection}, summarized in \Cref{tab:physical_errors} for both CX- and CZ-based systems. These values serve as representative device-level performance parameters and provide a consistent reference point for architectural comparison. In reality, the physical error probabilities vary significantly across systems and architectures.

\begin{table}[h]
\caption{Representative baseline physical qubit error probabilities used in our evaluations. These values are based on \cite{2025acharyaQuantumErrorCorrection} which reports a CZ-based platform. We use its reported mean errors to model both CX- and CZ-based systems as a common assumption.}
\label{tab:physical_errors}
\centering
\begin{tabular}{lc}
\toprule
\textbf{Operation} & \textbf{Error Probability} \\
\midrule
1Q Gate     & $6.2 \times 10^{-4}$ \\
2Q Gate     & $2.8 \times 10^{-3}$ \\
Measurement & $8.0 \times 10^{-3}$ \\
Reset       & $1.5 \times 10^{-3}$ \\
Idle        & $9.0 \times 10^{-3}$ \\
\bottomrule
\end{tabular}
\end{table}

To estimate the LER as a function of code distance, we simulate rotated surface-code stabilizer circuits (ten million shots) using \texttt{stim} \cite{gidney2021stim} and decode syndromes with \texttt{pymatching} \cite{Higgott2025sparseblossom}. For each code distance $d \in \{11, 13, 15, 17, 19, 21,23, 25\}$, we compute the corresponding logical error rate under the assumed physical error rates, establishing a baseline mapping between $d$ and LER.

Using the simulated data points, we perform an exponential fit to interpolate the required code distance for an arbitrary target LER, consistent with the expected exponential suppression of logical errors with increasing $d$. This continuous fit is retained only as a guide to illustrate the error scaling trend. For final physical resource estimates, we do not evaluate the continuous curve directly; instead, we look strictly at discrete, odd code distances. The final required resource count is determined by selecting the smallest odd integer $d$ whose predicted (or simulated) performance successfully achieves the target LER.

By integrating this logical error model with the thermal capacity model, we establish a direct link between cryogenic heat constraints and logical computational capability. This integration allows us to evaluate system configurations not only in terms of PQ capacity, but in terms of the LQ yield achievable at a specified target LER. This process is shown in \Cref{fig:LQ_model}.

In this context, LQ refers specifically to independent rotated memory patches. The calculated LQ yield thus serves as an upper bound for logical memory-patch capacity, omitting the overhead required for routing, lattice surgery, magic-state factories, and spare qubits.

\Figure[!ht](topskip=0pt, botskip=0pt, midskip=0pt)[width=0.97\columnwidth]
{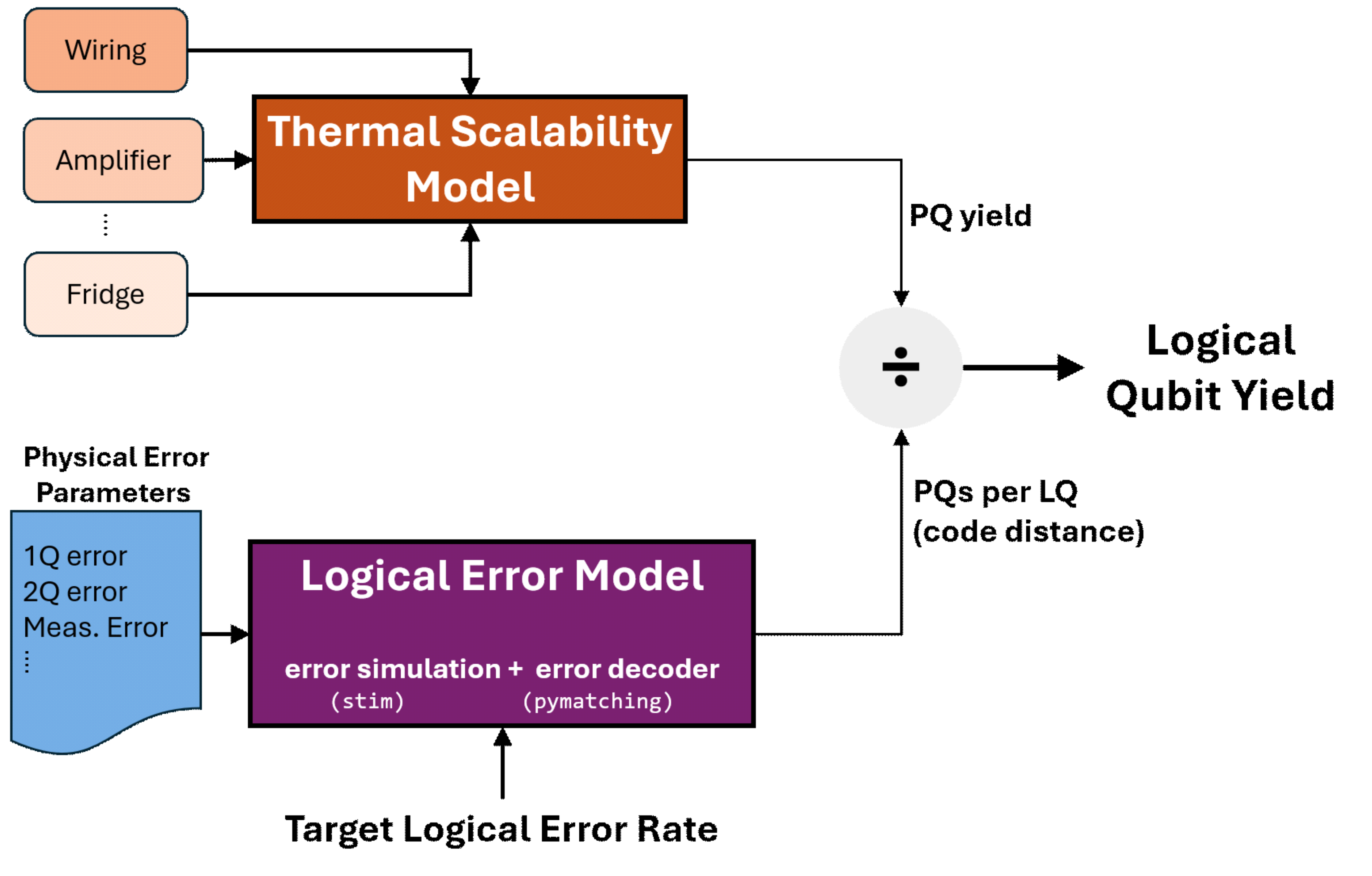}
{Estimation of logical qubit yield. The thermal scalability model gives the PQ yield of the system given the system parameters (e.g., cooling powers, amplifier loads, etc.). The logical error model estimates the code distance required to achieve a target LER given the physical error rates of the system (e.g., 1Q, 2Q, measurement error, etc.). Dividing the PQ yield by the number of PQs required per LQ gives the logical qubit yield of the given system.
\label{fig:LQ_model}
}
\section{Revisiting Previous Scalability Estimates}
\label{sec:bottlenecks}
In this section, we revisit prior scalability analyses under both their original assumptions and updated technological contexts using our modeling methodology. 

\subsection{100-qubit-scale system}
\label{subsec:krinner}

\Figure[!ht](topskip=0pt, botskip=0pt, midskip=0pt)[width=\linewidth]
{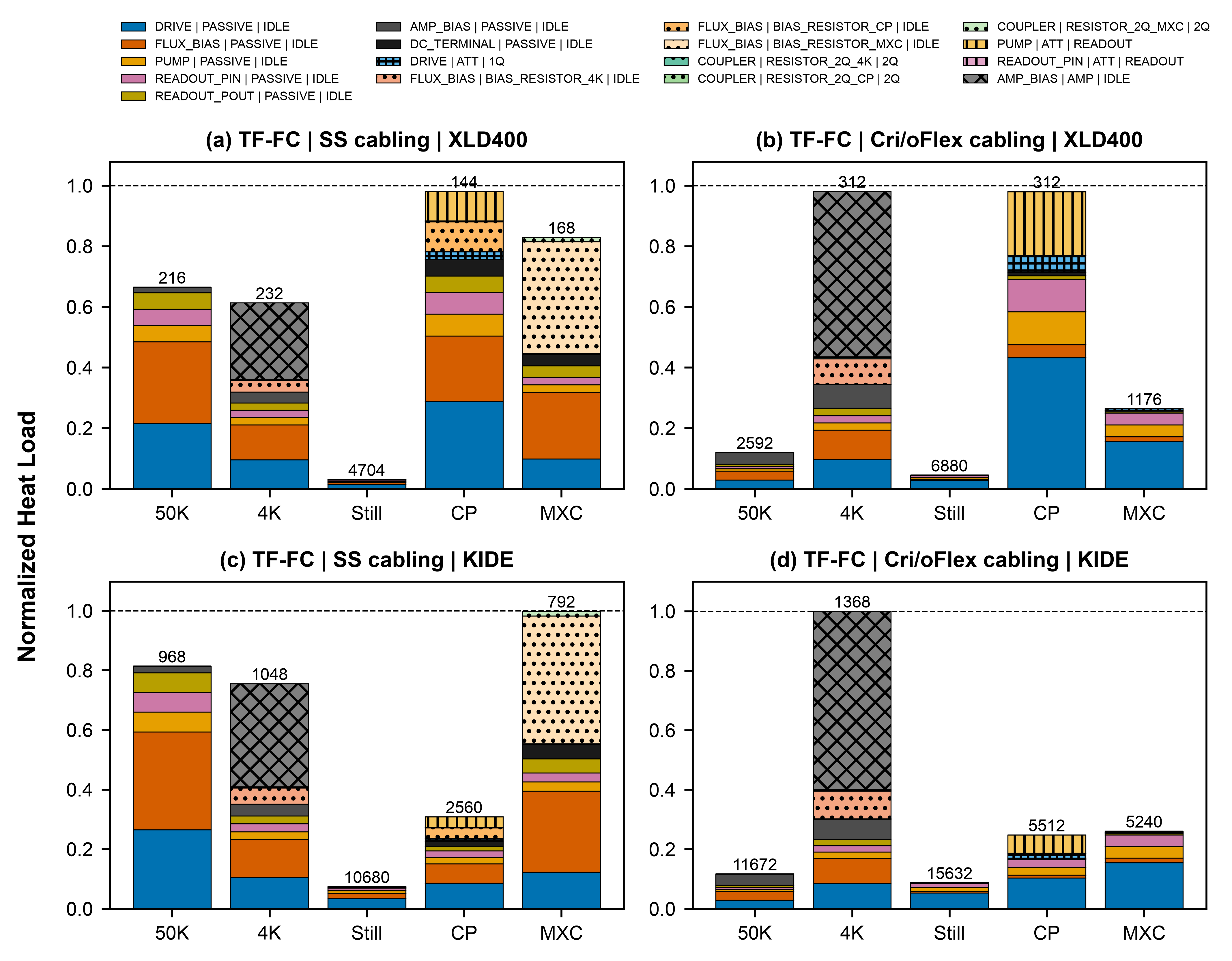}
{Thermal breakdown of the reconstructed TF-FC system across different wiring and refrigerator configurations. Stacked segments represent normalized contributions from individual heat components. Numbers above bars indicate maximum qubit capacity per stage. (a) Original system in \cite{krinner2019engineering}, where CP is the dominant bottleneck. (b) Replacement with microstrip wiring reduces the heat loads at 50\,K and MXC; 4 K becomes co-limiting with CP. (c) Increasing cooling power using KIDE improves capacity but shifts the bottleneck to MXC (due to joule heating of SS \leg{FLUX\_BIAS} cables). (d) Combined improvements expose amplifier heat at 4\,K as the dominant constraint. 
\label{fig:krinner_all}
}

We begin by reconstructing the TF-FC system for the 100-qubit-scale system described in \cite{krinner2019engineering}, which serves as a canonical reference for superconducting qubit cryogenic design. This reconstruction serves two purposes. First, it validates our modeling framework against a well-established real-world system baseline. Second, it provides a controlled setting to expose heat sources that were not explicitly modeled in prior work and to evaluate their impact under modern technological conditions.

\subsubsection{Modeling Assumptions}
The reference system implements a tunable-frequency fixed-coupling (TF-FC) architecture using stainless-steel coaxial cables in an XLD400 dilution refrigerator. Although the heat analysis presented in \cite{krinner2019engineering} assumes 50 qubits, the authors provide design guidelines to extend it to a 100 to 150-qubit scale system.

Each qubit requires a dedicated drive and flux-bias line, while readout is multiplexed across multiple qubits. The original study assumes 6-way multiplexing whereas we assume 4-way multiplexing. 2Q operations are realized through low-frequency flux pulses delivered via \leg{FLUX\_BIAS} lines. Accordingly, we model the workload using CZ-based surface-code cycles (\Cref{fig:ESM_CZ}). The system uses a HEMT amplifier (LNF8C) at the 4\,K stage, along with associated copper-based biasing wiring. Further modeling details are provided in \Cref{app:tf-fc-ss-xld400}.

\subsubsection{Modeling Results}
\Cref{fig:krinner_all}(a) shows the thermal profile of the original system using our modeling methodology. Each stacked bar corresponds to a temperature stage, with the heights representing normalized total heat load. The stacked segments indicate contributions from individual sources. The numbers above each bar denote the maximum number of qubits supportable by that stage under full utilization. The horizontal dashed line denotes the cooling power limit of the fridge for each temperature stage. We will use these conventions throughout the work.

Our results closely reproduce the conclusions in \cite{krinner2019engineering}. The maximum qubit capacity of the system is 144 qubits and the CP is the primary bottleneck, followed by MXC. The dominant heat contributors at the millikelvin stages are the PHL associated with drive and flux-bias lines and the OHL due to resistivity of SS cables of flux-bias lines.

These findings from \cite{krinner2019engineering} motivated two major research directions-- (i) increasing cooling power at millikelvin stages, and (ii) developing wiring technologies with lower thermal conduction and resistance. These directions have led to modern systems with significantly improved refrigeration (e.g., KIDE) and wiring technologies (e.g., HDW, Cri/oFlex).

\subsubsection{Technology Shifts}
\label{subsec:newtech}

We now evaluate how the heat landscapes change under modern technological assumptions, i.e., a higher capacity fridge (KIDE) and newer wiring technologies (Cri/oFlex microstrips).

\paragraph*{Effect of Improved Refrigeration}
Replacing XLD400 with a higher-capacity KIDE refrigerator (\Cref{fig:krinner_all}(c)) (while keeping the SS cabling) increases qubit capacity to 800 qubits ($\approx 5 \times$). However, this improvement is inconsistent with the increase in CP cooling power ($\approx 15\times$). This is because the bottleneck shifts to MXC, where cooling power scales less aggressively ($\approx 5\times$). The dominant contributor at MXC remains the ohmic heat from flux-bias lines. This indicates that SS \leg{FLUX\_BIAS}-induced ohmic heat at MXC remains a critical limitation.

\paragraph*{Effect of Advanced Wiring}
Cri/oFlex wiring uses superconducting NbTi cables below 4\,K for flux biasing which effectively eliminates any joule heating at the millikelvin stages. Furthermore, Cri/oFlex has a much lower PHL at 50\,K than SS which reduces the thermal stress at 50\,K significantly. Thus, the resulting effect of replacing stainless-steel cables with Cri/oFlex microstrip wiring while using the XLD400 fridge (\Cref{fig:krinner_all}(b)) is:
\begin{enumerate}
    \item Qubit capacity doubles and there is an additional bottleneck at 4\,K.
    \item Heat load at MXC and 50\,K is significantly reduced.
\end{enumerate}

\paragraph*{Combined Effect of Improvements in Wiring and Fridge}
When both advanced wiring and improved refrigeration are applied (\Cref{fig:krinner_all}(d)), the system supports $\approx$1300 qubits. Most importantly, the dominant bottleneck has now shifted to  4\,K because the loads at the millikelvin stages are greatly reduced (due to stronger cooling and superconducting cables). The new bottleneck at 4\,K is dominated by the active heat of the readout amplifier. This reveals a critical insight: \textbf{improvements in one part of the system expose previously hidden bottlenecks elsewhere.} This effect would not be visible in the original model because amplifier heat was not explicitly included. This demonstrates that it is necessary to include amplifier-related wiring heat loads during thermal scalability analysis.

This omission was acceptable in the original system because the heat loads at CP and MXC scaled at much higher rates than the heat load at 4\,K. However in modern systems, amplifier heat (and therefore the total heat loads at 4\,K) scales more strongly than the heat loads at CP and MXC (\Cref{fig:krinner-trends}(b)). Without explicitly modeling amplifier heat, one would incorrectly conclude that there are no bottlenecks at 4\,K.

Going forward, even if we were to hypothetically eliminate all 4\,K heat loads in \Cref{fig:krinner-trends}(right), it would expose the bottleneck at CP, limiting scalability to $\approx$5000 qubits which remains significantly below the 10k target. 

Thus to achieve 10k qubit systems, (i) the active heat load of the readout amplifiers must be reduced substantially, as it becomes the primary bottleneck at the 4\,K stage, (ii) the PHL from drive and flux-bias lines must be further suppressed, since it continues to limit scalability even with advanced wiring, and (iii) the cooling capacity of the refrigerator must be increased across temperature stages (CP cooling power should be increased by approximately $2\times$ to avoid becoming the next limiting constraint). 

\subsection{Optical Fiber alternatives}
\label{sec:historical-optical}

As discussed previously, PHL from metallic cables is a major scalability bottleneck. Optical interconnects have been proposed as an alternative for drive and readout feed-in links \cite{lecocq2021control,joshi2023scaling}, offering significantly lower thermal conduction due to their material and geometry. However, optical control requires opto-electrical conversion inside the cryostat using a PD, which introduces a new active heat load at the stage where it is placed.

In \cite{lecocq2021control}, the PD is located at the MXC stage, where its dissipation strongly stresses the limited cooling power, leading to the conclusion that MXC is the bottleneck due to PD heat. To mitigate this, \cite{joshi2023scaling} proposed relocating the PD to the 4\,K stage.

We evaluate both optical configurations using our model: (i) PD at MXC following \cite{lecocq2021control} and (ii) PD at 4\,K following \cite{joshi2023scaling}. Since the two prior works use different cryogenic assumptions and do not fully specify all system parameters, we establish a unified configuration to enable fair comparison while maintaining as much consistency with the original studies as possible.

The work in \cite{lecocq2021control} uses photonic links for both qubit control (XY drive) and readout, whereas \cite{joshi2023scaling} only considers optical qubit control. To provide a common evaluation framework, we assume optical fibers are used for both drive and readout feed-in lines. Since optical links cannot directly support flux or coupler biasing, we assume a FF-FC architecture. All remaining microwave connections use Cri/oFlex cabling which better reflects modern low-PHL wiring practices. The system further employs a modern LNF8G HEMT amplifier with AWG30 copper biasing wires \cite{lownoisefactoryLNFNANO9MNoise}. The fridge is an XLD400 refrigerator consistent with the system in \cite{joshi2023scaling}.

In \cite{joshi2023scaling}, the microwave signals are generated at 4\,K by the PD, which are transmitted to MXC using NbTi coaxial cables. In our analysis, we instead assume modern NbTi microstrip cables. Detailed modeling assumptions are provided in \Cref{app:ff-fc-optical_MXC-xld400,app:ff-fc-optical_4K-xld400}.

\paragraph*{Bottleneck Misidentification}

\Figure[!ht](topskip=0pt, botskip=0pt, midskip=0pt)[width=\linewidth]
{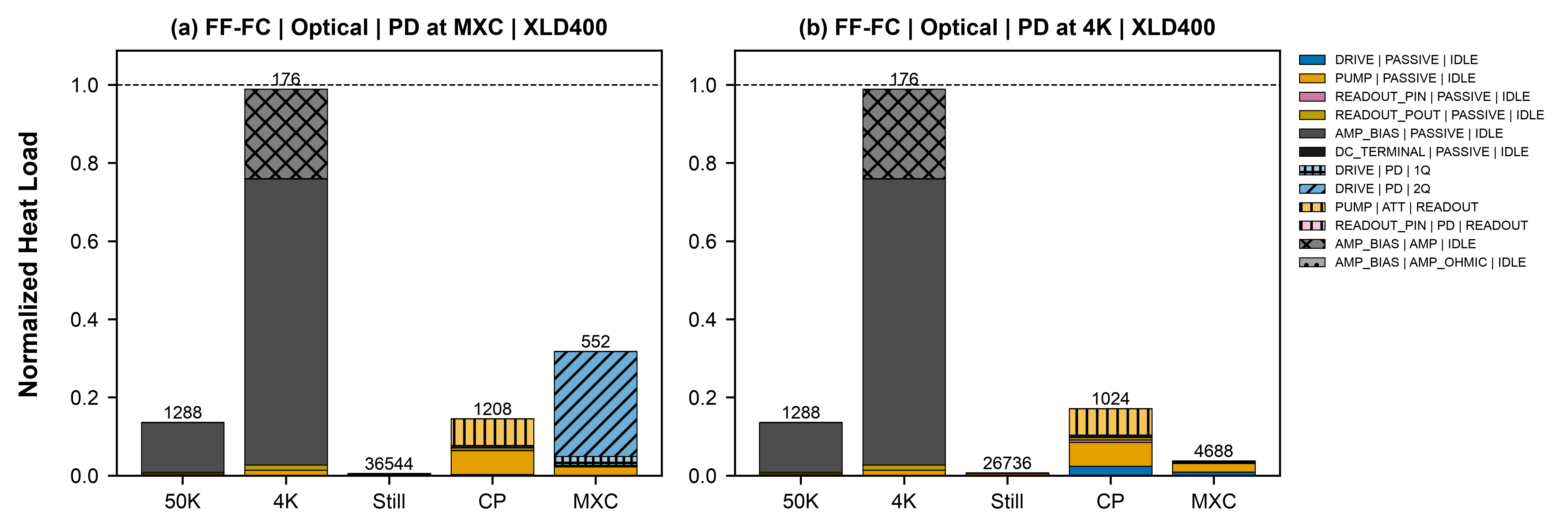}
{Normalized heat load breakdown for an FF-FC system using optical drive and readout feed-in links in an XLD400 refrigerator. (a) PD at MXC \cite{lecocq2021control}. (b) PD relocated to 4\,K \cite{joshi2023scaling}. In both cases, the 4\,K stage becomes the dominant bottleneck once amplifier-related heat is included. The dashed line indicates the cooling-power limit of the refrigerator at every temperature stage.
\label{fig:lecoq_joshi}
}

The heat breakdowns are shown in \Cref{fig:lecoq_joshi}. In both configurations, the system supports only $\approx$170 qubits. This is similar to the qubit capacity of the original SS-based system in \cite{krinner2019engineering}. Importantly, this limit is not caused by PD dissipation. \textbf{Instead, the bottleneck under the modeled configuration is the 4\,K stage due to amplifier-related heat.} More than 90\% of the total 4\,K load originates from the amplifier subsystem, the majority of which originates from passive conduction through amplifier biasing wires.

This behavior is consistent with the observations from the previous section. Once the dominant heat loads at CP and MXC are reduced, the bottleneck shifts to previously secondary heat sources. Here, the newer LNF8G amplifier reduces active dissipation, but the larger AWG30 copper bias wires substantially increase passive heat conduction at 4\,K compared to the system in \cite{krinner2019engineering}.

These results further support our claim that amplifier-related heat must be explicitly modeled during scalability analysis. Previous studies identified PD dissipation at MXC as the primary bottleneck because amplifier-related heat was not included. However, once these loads are modeled, the limiting stage shifts to 4\,K, fundamentally changing the design priorities.

Only after removing the amplifier bottleneck at 4\,K does PD placement become important. In a hypothetical scenario with zero amplifier-related heat, placing the PD at MXC limits scalability to $\approx$ 550 qubits due to PD dissipation at MXC. Relocating the PD to 4\,K removes this limitation but requires additional cabling (we assume NbTi). This shifts the bottleneck to CP and increases the capacity by another $\approx$ 500 qubits.

\paragraph*{Scalability Overestimation}

Prior work \cite{joshi2023scaling} estimated that up to 3k qubits could be supported in an XLD400 system with optical wiring and PD at 4\,K. This estimate is optimistic because it neglects several system-level heat contributions, including passive loads from pump and readout lines, amplifier biasing wires, and attenuator-induced active heat. When all passive, active, and ohmic heat loads are modeled consistently, the achievable capacity is greatly reduced. This highlights that incomplete heat accounting leads to overestimation of scalability and misidentification of bottlenecks.

\paragraph*{Implications}

This case study shows that optical interconnects effectively reduce passive conduction from metallic wiring. However, this reduction is only a minor improvement over the original SS-based cable system from \cite{krinner2019engineering}. This is because once the originally dominant bottleneck heat loads are resolved using optical fibers, the bottleneck migrates to 4\,K (due to amplifier-related heat) which limits further scalability. Moreover, our study also shows that PD heat is not the dominant scalability bottleneck in these systems as previously assumed. 

\subsection{High-Density Wiring}
\label{subsec:hdw}

\Figure[!ht](topskip=0pt, botskip=0pt, midskip=0pt)[width=0.97\columnwidth]
{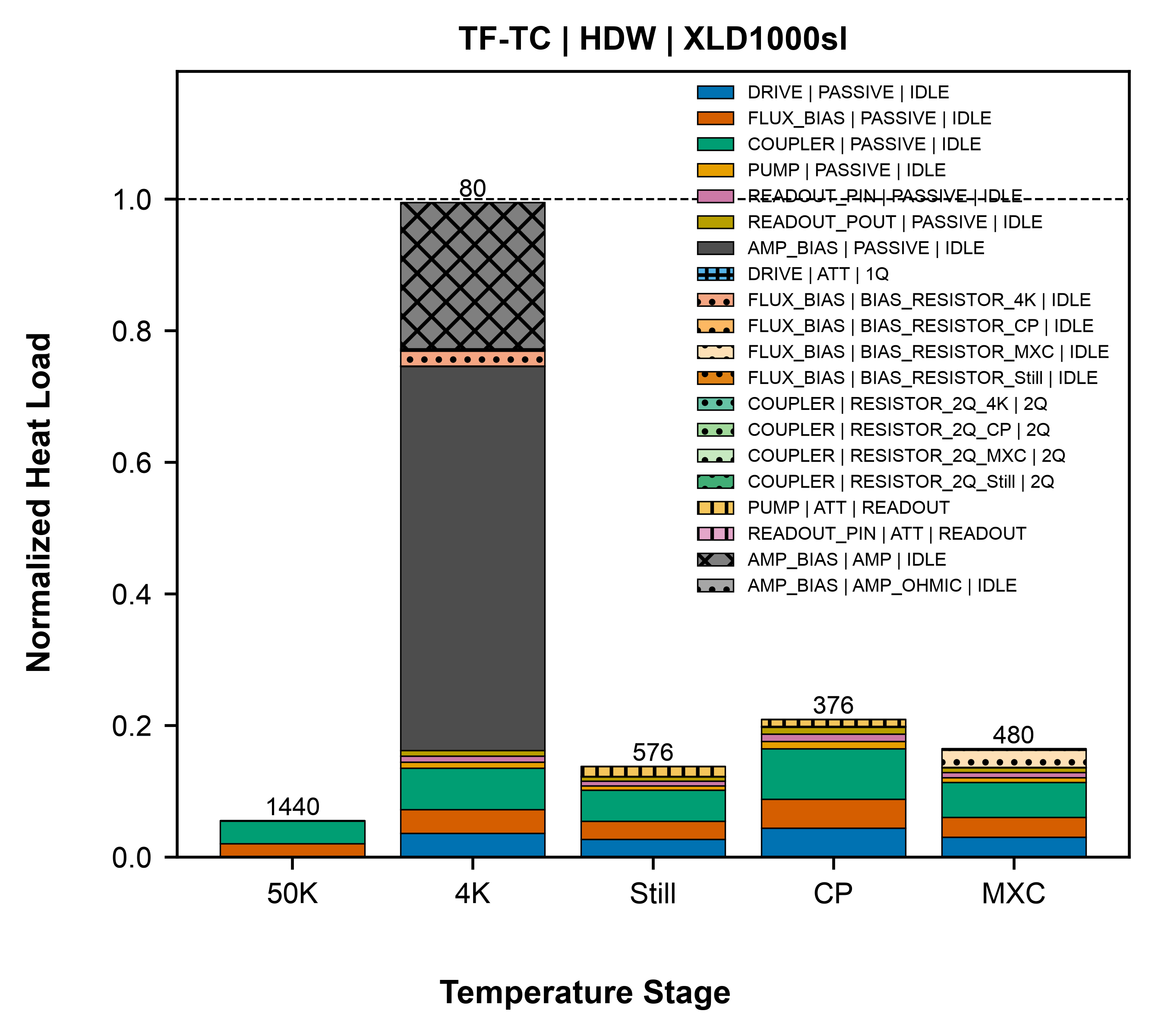}
{Normalized heat load breakdown for a TF-TC architecture using HDW cables in an XLD1000sl refrigerator. Including passive heat conduction from amplifier biasing wires significantly reduces the supported qubit count compared to \cite{raicu2025cryogenic}. The PHL contribution of amplifier biasing wires was omitted in the original study.
\label{fig:raicu}
}

We next revisit a recent scalability study based on Bluefors High-Density Wiring (HDW) \cite{blueforsHighDensityWiring} and the XLD1000sl refrigerator \cite{raicu2025cryogenic}. This work evaluates cupronickel coaxial cables with substantially lower PHL than legacy stainless-steel wiring.

The architecture considered is TF-TC, which requires an additional control line per qubit for tunable coupler biasing in addition to the flux biasing line. The analysis in \cite{raicu2025cryogenic} estimates the system can support $\approx$144 qubits, with the 4\,K stage identified as the primary bottleneck. The study concludes that the dominant contributions at 4\,K are amplifier active heat and the combined PHL of control and readout cables (passive heat conduction through amplifier biasing wires is not included in their analysis).

We model a comparable system using our framework with CZ-based ESM workloads while keeping all other architectural parameters consistent. The key difference is the explicit inclusion of passive heat conduction from amplifier biasing wires between RT and 4\,K (see \Cref{app:tf-tc-hdw-xld1000sl} for more modeling details).

Our analysis (\Cref{fig:raicu}) agrees with \cite{raicu2025cryogenic} that 4\,K is the limiting stage. As observed in the previous analyses, reducing millikelvin-stage PHL shifts the bottleneck to 4\,K. However, our analysis reveals that  the thermal stress at 4\,K is significantly higher than the estimate in \cite{raicu2025cryogenic}. Consequently, the supported qubit count drops to $\approx$80, roughly half of the original estimate. This reduction is primarily due to the inclusion of the PHL of AWG30 copper bias wires. (To isolate this effect, we repeat the analysis excluding bias-line PHL and obtain $\approx$192 qubits, substantially closer to the original estimate of 144 qubits in \cite{raicu2025cryogenic}. This confirms that the discrepancy arises from the previously unmodeled passive conduction of amplifier biasing wires).

Therefore, our analyses demonstrate that amplifier-related heat (AHL and PHL) must be included in scalability analysis. While negligible in earlier systems dominated by stainless-steel wiring \cite{krinner2019engineering}, \textbf{these amplifier heat loads emerge as the dominant source of scalability bottlenecks} in modern systems with better wiring alternatives (like optical fibers, Cri/oFlex) and stronger fridges (KIDE).
\section{Strategies to reduce amplifier heat}
\label{sec:solutions}
In this section, we evaluate several options to reduce amplifier-related heat loads and how these choices affect the PQ and LQ yield. In our analysis, we will assume the operating temperatures and the inter-flange distances (i.e., wire lengths) corresponding to XLD1000sl fridge for consistency. The readout multiplexing factor can greatly influence the amplifier heat loads. For example, increasing the multiplexing from 4$\times$ to 8$\times$ can reduce the number of amplifiers (and the associated heat load by half). However, the multiplexing factor depends on many other factors such as amplifier bandwidth and resonator spacing. Analyzing these effects is beyond the scope of this work. For consistency, we assume 4$\times$ readout multiplexing in all our analyses.

The amplifier-related heat loads depend on the choice of amplifier and the biasing cable. We will analyze the various combinations of amplifiers and biasing cables listed in \Cref{tab:amp_options} and \Cref{tab:biasing_cables} to see their impact on system scalability. The choice of amplifier determines the AHL, the number of biasing wires necessary and whether an additional amplification stage is necessary at 50\,K and the choice of biasing wire determines the PHL and OHL load. 

\paragraph*{Optimal Current Splitting}
While Manganin has a lower PHL than copper, it also offers a higher electrical resistance.
Thus, when choosing between Manganin and copper, it is necessary to determine when Manganin reduces the \textit{total} heat load (passive and ohmic) relative to copper. Ohmic heating grows quadratically with current, so it can be reduced by splitting the current across multiple wire pairs (equivalently, reducing per-wire resistance by increasing cross-sectional area).

Assume the amplifier requires $n = 2m + k$ wires. The bias current $I$ is carried by $m$ supply-return pairs, so each of the $2m$ current-carrying wires carries $I/m$ current. The remaining $k$ wires carry negligible current and contribute negligibly to ohmic heating (e.g., gate-bias wires).

The total heat load into a given stage is
\begin{equation}
\label{eq:total_heat}
    H = A_{amp} + (2m+k)P + \left[0.5\times \left( \frac{I}{m}\right)^2 \times R \right] \times 2m
\end{equation}
where $A_{amp}$ is the amplifier active heat, $P$ is the passive heat per wire into that stage, and $R$ is the (lumped) wire resistance. The factor $0.5$ models equal partition of ohmic dissipation between the two stages connected by the wire segment (i.e., 50\,K and 4\,K stages each absorb half of the ohmic dissipation between them). Increasing $m$ reduces ohmic heat through reduced current per wire, but increases PHL through additional wires.

Differentiating \Cref{eq:total_heat} with respect to $m$ and setting the derivative to zero yields
\begin{equation}
\label{eq:best_split}
    m_{opt} = \sqrt{\frac{I^2R}{2P}}.
\end{equation}
Rounding $m_{opt}$ to the nearest integer that minimizes $H$ gives the practical split factor.

\subsection{Analysis of Different Amplifier--Wire Configurations}
\label{subsec:amp_wire_analysis}

\begin{table*}[t]
\centering
\caption{Heat Load Breakdown by Amplifier and Wiring Material}
\label{tab:heat_breakdown}
\begin{tabular}{llrrr}
\toprule
\textbf{Amplifier} & \textbf{Heat Load} 
& \textbf{Copper (Cu)} 
& \textbf{Manganin (Mn)} 
& \textbf{YBCO} \\
\midrule

\multirow{4}{*}{\textbf{LNF8G}}
& Active & \qty{7.8}{\milli\watt}
      & \qty{7.8}{\milli\watt} 
      & \qty{7.8}{\milli\watt} \\

& Passive & \textbf{\qty{20.27}{\milli\watt}} 
      & \qty{76.1}{\micro\watt} 
      & \qty{11.52}{\micro\watt} \\

& Ohmic & \qty{392.92}{\nano\watt} 
      & \qty{467.83}{\micro\watt} 
      & 0 \\

& \textbf{Total} 
      & \qty{28.07}{\milli\watt} 
      & \qty{8.34}{\milli\watt} 
      & \qty{7.81}{\milli\watt} \\
\midrule

\multirow{4}{*}{\textbf{LNF8G (LP)}}
& Active & \qty{300}{\micro\watt} 
      & \qty{300}{\micro\watt} 
      & \qty{300}{\micro\watt} \\

& Passive & \qty{20.27}{\milli\watt} 
      & \qty{76.1}{\micro\watt} 
      & \qty{11.52}{\micro\watt} \\

& Ohmic & \qty{20.92}{\nano\watt} 
      & \qty{24.91}{\micro\watt} 
      & 0 \\

& \textbf{Total} 
      & \qty{20.57}{\milli\watt} 
      & \qty{401.01}{\micro\watt}
      & \qty{311.52}{\micro\watt} \\
\midrule

\multirow{4}{*}{\textbf{ULP-HEMT}}
& Active & \qty{200}{\micro\watt} 
      & \qty{200}{\micro\watt} 
      & \qty{200}{\micro\watt} \\

& Passive & \textcolor{red}{\qty{40.54}{\milli\watt}} 
      & \qty{152.19}{\micro\watt} 
      & \qty{23.05}{\micro\watt} \\

& Ohmic & \qty{14.53}{\nano\watt} 
      & \qty{17.3}{\micro\watt} 
      & 0 \\

& \textbf{Total} 
      & \textcolor{red}{\qty{40.74}{\milli\watt}} 
      & \textcolor{green}{\qty{369.49}{\micro\watt}} 
      & \qty{223.05}{\micro\watt} \\
\midrule

\multirow{4}{*}{\textbf{SIS-5w}}
& Active & \qty{8.1}{\micro\watt} 
      & \qty{8.1}{\micro\watt} 
      & \qty{8.1}{\micro\watt} \\

& Passive & \qty{33.78}{\milli\watt} 
      & \qty{126.83}{\micro\watt} 
      & \qty{19.21}{\micro\watt} \\

& Ohmic & \qty{1.13}{\micro\watt} 
      & \textcolor{magenta}{\qty{1.34}{\milli\watt}} 
      & 0 \\

& \textbf{Total} 
      & \qty{33.79}{\milli\watt} 
      & \textcolor{magenta}{\qty{1.47}{\milli\watt}} 
      & \textcolor{green}{\qty{27.31}{\micro\watt}} \\
\midrule

\multirow{4}{*}{\textbf{SIS-13w}}
& Active & - 
      & \qty{8.1}{\micro\watt} 
      & - \\

& Passive & - 
      & \textcolor{blue}{\qty{329.75}{\micro\watt}} 
      & - \\

& Ohmic & - 
      & \textcolor{blue}{\qty{268.01}{\micro\watt}} 
      & - \\

& \textbf{Total} 
      & - 
      & \qty{605.85}{\micro\watt} 
      & - \\
\bottomrule
\end{tabular}
\end{table*}

We compare amplifier-wire configurations using the per-amplifier heat breakdown at the 4\,K stage in \Cref{tab:heat_breakdown}. The goal is to understand (i) which heat component dominates (active, passive, or ohmic), (ii) how the dominance changes with amplifier choice and bias-wire material, and (iii) which combination has the lowest heat load.

\subsubsection{PHL Dominates in Copper Wiring}
\textbf{With copper wiring, the total amplifier heat at 4\,K is dominated by its PHL across all amplifier options.} This demonstrates that, with copper bias wiring, reducing AHL of the amplifier alone has limited impact. For amplifiers that require many biasing wires like ULP-HEMT or SIS, the PHL becomes prohibitively large. A clear example is ULP-HEMT. Although its AHL is much lower than that of LNF8G, it requires six biasing wires and therefore doubles the PHL relative to three-wire HEMT configurations. As a result, ULP-HEMT has the largest total amplifier-associated heat when using copper wires (shown in red).

\subsubsection{PHL and OHL Tradeoffs in Manganin Wiring}
Replacing copper with Manganin reduces the PHL but the OHL also increases. For HEMT variants, the AHL is large enough that the increased OHL is not so significant. However, for low-AHL SIS, the increased OHL due to Manganin becomes quite significant. SIS requires high biasing currents for its local oscillator (LO). In the SIS-5w case, all LO current passes through a pair of relatively high resistance AWG30 Manganin wires. The resulting OHL brings the total heat load to the milliwatt range (highlighted in magenta). \textbf{Thus, even though SIS has the lowest AHL, it has a higher total load due to the OHL of its Manganin biasing wires.}

\paragraph*{Splitting Current to Reduce Total Load}
\Figure[!ht](topskip=0pt, botskip=0pt, midskip=0pt)[width=0.95\columnwidth]
{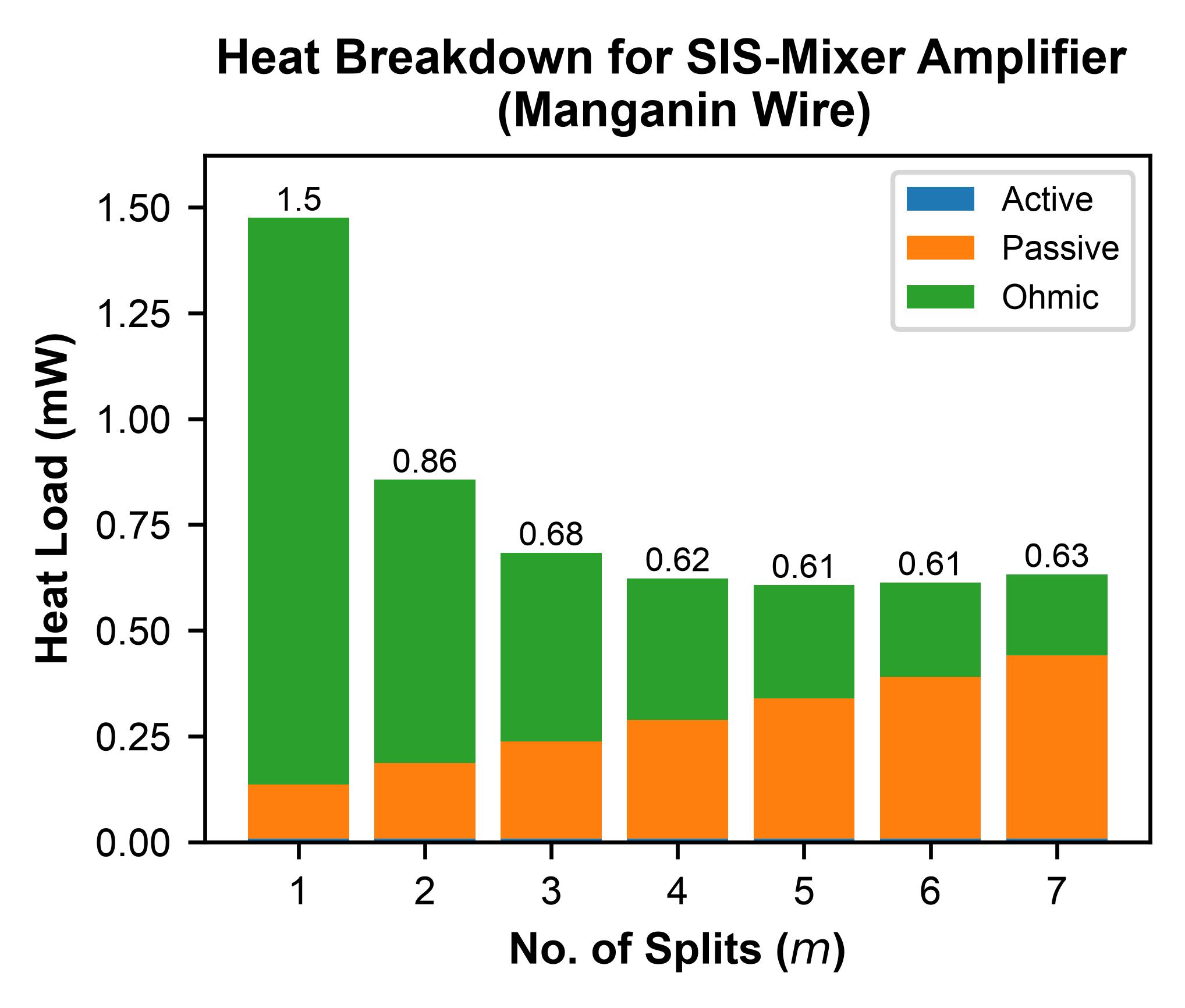}
{Tradeoff between ohmic and passive heat loads when splitting the LO current-carrying wires in SIS amplifier. For $m$ splits, the total number of wires $n = 2m + 3$. As we increase the number of splits, the ohmic load decreases due to less current flowing in the wires while the passive load increases due to larger number of wires. For LO = \qty{22}{\milli\ampere}, we observe that the total heat load is minimized for $m=5 or 6$ splits. AHL is practically negligible compared to the PHL and OHL. We assume the operating temperatures and inter-flange separations corresponding to an XLD1000sl fridge.
\label{fig:sis_split}
}

It is possible to reduce the OHL significantly at the cost of a moderate increase in PHL by splitting the high LO current across additional wires. For an LO current of \qty{22}{\milli\ampere} (see \Cref{tab:sis_power}), \Cref{eq:best_split} gives $m_{opt}=5$. Therefore, the LO current-carrying path should be distributed across $2m=10$ wires (5 supply and 5 return) to minimize the total amplifier-associated heat load. The total wire count then becomes $n=2m+k=13$. This split configuration is denoted as SIS-13w. Compared to the non-split configuration, the PHL increases by approximately $2.5\times$, while the OHL decreases by approximately $5\times$ (blue values in \Cref{tab:heat_breakdown}). Overall, \textbf{the total heat load is reduced to less than half that of the non-split SIS configuration.}

The increase in wire density caused by current-splitting may cause flange crowding. Assuming 4-way readout multiplexing, a 10k-qubit system requires 2500 amplifiers. At 13~AWG30 wires per SIS amplifier, the wiring footprint is comparable to a square with a side of \qty{40}{\centi\meter}. This can be challenging for compact refrigerators such as XLD1000sl and KIDE, especially when combined with the microwave wiring required for drive and readout. In such cases, the number of splits can be decreased at the expense of slightly higher amplifier heat loads. \Cref{fig:sis_split} illustrates how the total heat load as well as the PHL / OHL contributions in SIS-Manganin configuration varies with the number of current splits. As expected, the PHL increases and OHL decreases with the number of splits with the minimum total heat loads at 5-6 splits. However, to minimize crowding, a split of 2-3 also might be acceptable depending upon the available thermal budget.

\subsubsection{Superconducting Biasing Wires}
YBCO wiring has orders of magnitude lower thermal conductivity than copper or Manganin while being superconducting at temperatures below 77K. This makes it an attractive best-case option for biasing lines. Since YBCO bias wiring is not yet a mature, broadly deployed technology, we use it as an optimistic bound to estimate the upper limit of heat scalability benefits from bias-line improvements.

In the YBCO case, the AHL becomes the dominant contributor to the total amplifier-associated heat load because the PHL and OHL are nearly eliminated. For all amplifier variants using YBCO wiring, the total heat is therefore largely determined by the amplifier AHL. This means that further reductions in amplifier operating power directly improve thermal performance. This trend is most clearly visible for the SIS amplifier. Despite requiring more bias wires per amplifier, its extremely low AHL allows it to achieve the lowest total heat load among all configurations.

\subsubsection{Lowest-Heat Amplifier--Wire Configurations}
From \Cref{tab:heat_breakdown}, the SIS $\times$ YBCO combination achieves the lowest total heat load among all configurations. Although this is a hypothetical setup, it represents the best-case scenario because it simultaneously minimizes AHL and PHL while eliminating OHL. At 10k scale, the total amplifier-associated heat consumes only slightly more than 1\% of the available 4\,K cooling power of a KIDE refrigerator.

A more practical solution is the (ULP-HEMT $\times$ Manganin) or (LNF8G(LP) $\times$ Manganin) combination. Reducing the active dissipation of HEMT with low power operation also lowers the current draw, which decreases OHL. Replacing copper with Manganin further reduces the PHL contribution from the bias wires. These configurations consume roughly 20\% of the total 4\,K cooling power of a KIDE refrigerator.

Another important observation is that \textbf{the choice of bias-wire material directly determines the total thermal load.} ULP-HEMT produces the highest heat load when combined with copper, but among the lowest when combined with Manganin. The total heat changes by nearly two orders of magnitude between these two cases. This demonstrates that explicitly modeling the PHL of amplifier bias wires is essential for correctly evaluating thermal scalability.

\subsection{System-Level Impact}

\subsubsection{Physical Qubit Yield}

\begin{table}[t]
\centering
\caption{Physical qubit (PQ) capacity of a KIDE refrigerator under different amplifier--bias wire configurations. The values in green correspond to configurations where the 4\,K stage is no longer the primary bottleneck.}
\label{tab:4k_pq}
\begin{tabular}{lccc}
\toprule
\textbf{Amplifier} 
& \textbf{Copper} 
& \textbf{Manganin} 
& \textbf{YBCO} \\
\midrule
LNF8G       & 776                   & 2152                        & 2264  \\
LNF8G(LP)   & 1024                  & \textcolor{green}{4688}     & \textcolor{green}{4688}  \\
ULP-HEMT    & \textcolor{red}{544}  & \textcolor{green}{4688}     & \textcolor{green}{4688}  \\
SIS-5w      & 656                   & \textcolor{green}{4688}     & \textcolor{green}{4688} \\
SIS-13w     & --                    & \textcolor{green}{4688}     & --   \\
\bottomrule
\end{tabular}
\end{table}

The previous subsection analyzed the amplifier--bias wire combinations in isolation to identify the configurations with the lowest amplifier-associated heat load. However, minimizing the heat load of the amplifier subsystem does not necessarily maximize the thermal scalability of the overall system. Once the amplifier-related heat is sufficiently reduced, the thermal bottleneck may migrate to another temperature stage. Beyond this point, further reductions in amplifier heat provide little or no improvement in PQ capacity. Therefore, it is necessary to evaluate these configurations at the system level to determine how the reduced amplifier heat translates into PQ capacity. We perform this evaluation under the following assumptions.

We assume an FF-FC architecture (and therefore CX-based ESM workloads) because it minimizes the number of control wires per qubit and simplifies analysis. We further assume the system uses Cri/oFlex RF microstrips for control and readout (with 4-way readout multiplexing). Additional modeling details are provided in \Cref{app:ff-fc-delft-hemt_cu-kide}. Since all amplifiers (except LNF8G) provide less than 40\,dB gain, we assume an additional HEMT is installed at 50\,K to achieve the required gain. We assume this HEMT is also LNF8G.

\Cref{tab:4k_pq} reports the PQ capacity for different amplifier--wire configurations in a KIDE refrigerator. The LNF8G $\times$ Copper configuration (\Cref{fig:amp_solutions}(a)) represents current systems and supports only about 700 qubits. As discussed earlier, the primary bottleneck is the 4\,K stage, where the readout amplifier subsystem dominates the thermal budget.

Using lower-AHL amplifiers together with lower-PHL bias wires increases the PQ capacity to more than 4k qubits. Beyond this point, however, further reductions in amplifier-associated heat no longer improve the system PQ yield. For example, the SIS-13w $\times$ Manganin and SIS $\times$ YBCO configurations support the same PQ capacity, even though their per-amplifier heat loads differ by more than $22\times$ (See \Cref{tab:heat_breakdown}).

The heat-load breakdown for the ULP-HEMT $\times$ Manganin configuration in \Cref{fig:amp_solutions}(b) explains this saturation (See \Cref{app:ff-fc-delft-hemt-ulp_mn-kide} for configuration details). Once the amplifier-associated heat is sufficiently reduced, the bottleneck migrates from the 4\,K stage to MXC, limiting the system to 4688 PQs. At this point, after accounting for all other heat loads at 4\,K, approximately 39\% of the available cooling power (about \qty{2.32}{\watt}) remains for the readout amplifier subsystem. With 4-way readout multiplexing, this corresponds to a thermal budget of approximately \qty{2}{\milli\watt} per amplifier module before the MXC reaches its cooling capacity. Consequently, any configuration in \Cref{tab:heat_breakdown} with a total amplifier heat below this threshold no longer limits the system at the 4\,K stage.

After the bottleneck shifts to MXC, the PHL of the microwave cables becomes the primary limitation under this new configuration. The drive cables contribute the largest heat load because each qubit requires a dedicated drive line, while the readout and pump microwave lines contribute approximately one quarter of this load due to 4-way multiplexing. Therefore, once amplifier heat is sufficiently reduced, improving the thermal scalability of the system requires reducing the passive heat load of the microwave wiring.

In systems that require an additional LNF8G stage at 50\,K, the additional thermal stress is not significant. \textbf{This suggests that using lower-gain, lower-power amplifiers at 4\,K and compensating with additional amplification at 50\,K may be a practical strategy.} However, the resulting system-level noise performance must still be evaluated, which is outside the scope of this work.

\Figure[!ht](topskip=0pt, botskip=0pt, midskip=0pt)[width=\linewidth]
{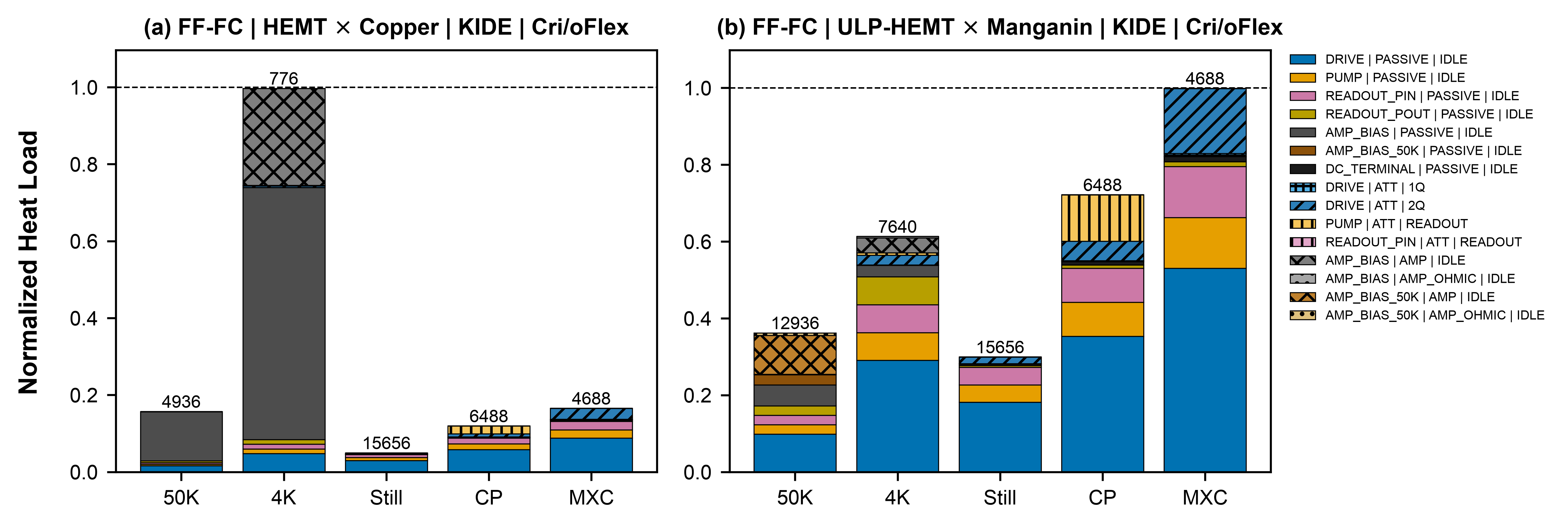}
{Normalized heat load breakdown for FF-FC systems using different amplifier$\times$bias wire configurations. (a) HEMT $\times$ Copper (representing currently available systems) where 4\,K stage remains the primary bottleneck; (b) ULP-HEMT $\times$ Manganin with an additional 50\,K amplification stage. The primary bottleneck shift back to the millikelvin stages (where the major heat contributor is the PHL from the microwave lines). The dashed line indicates the refrigerator cooling-power limit.
\label{fig:amp_solutions}
}

\subsubsection{Logical Qubit Yield}

\Figure[!ht](topskip=0pt, botskip=0pt, midskip=0pt)[width=0.98\columnwidth]
{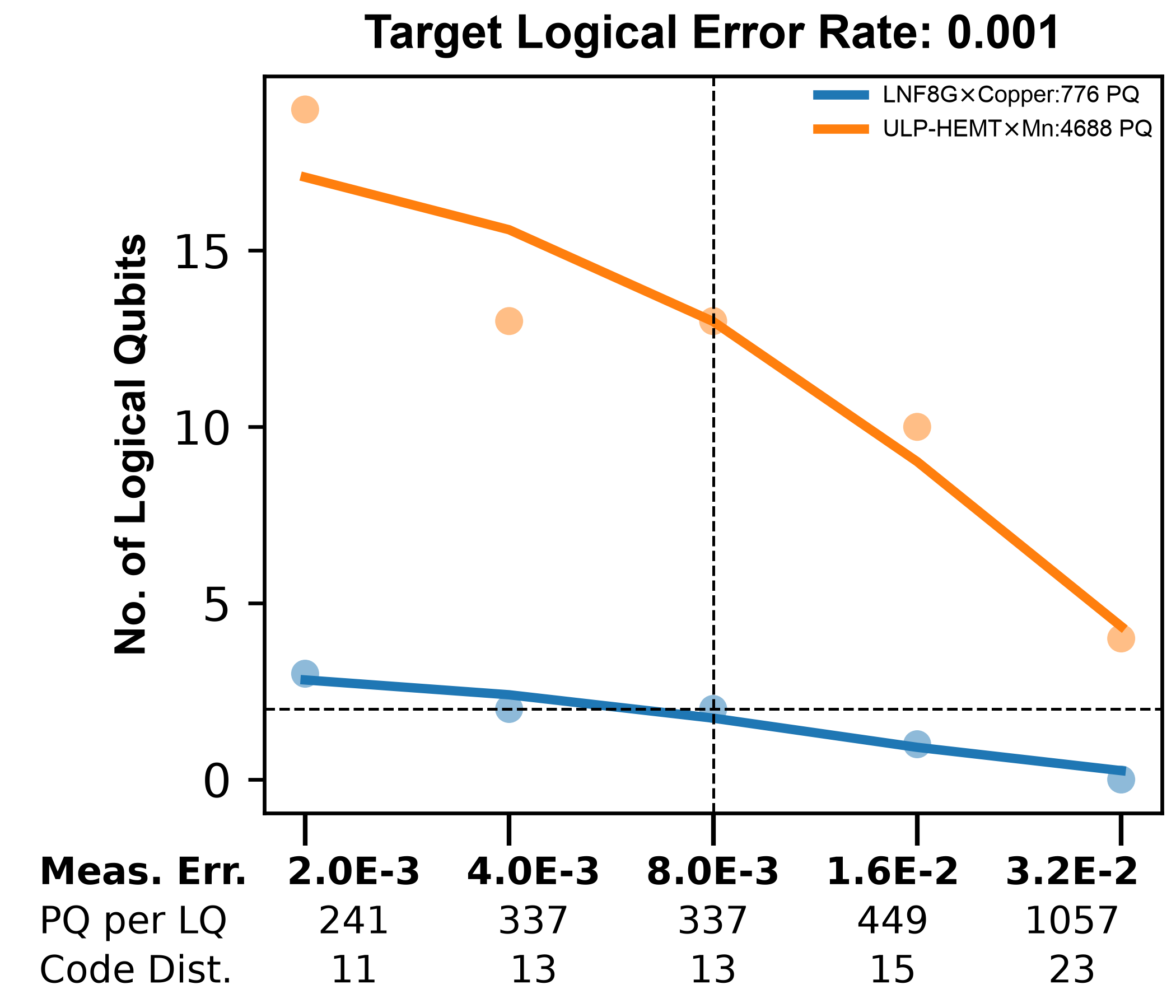}
{LQ yield versus measurement error for different amplifier$\times$bias wire configurations in a KIDE refrigerator. The vertical dashed line indicates the baseline measurement error ($0.008$), while the horizontal dashed line indicates the LQ yield of the baseline LNF8G$\times$Copper system. Circles represent LQ-capacity values derived from the simulated code-distance results. The exponential trend lines are shown as a visual guide.
\label{fig:readout_lq}
}

We also estimate how the amplifier configuration affects the LQ yield. The primary impact of the amplifier is on the measurement error through changes in readout SNR, amplification, demodulation, digitization, and signal processing. Modeling these effects in detail is outside the scope of this work because they depend strongly on implementation-specific parameters.

Instead, we estimate the required code distance for different measurement errors while assuming all other physical error rates (1Q, 2Q, idle, etc.) remain unchanged. We therefore isolate the impact of measurement error on the LQ yield for a given PQ capacity.

We compare two representative configurations:
(i) LNF8G$\times$Copper representing current systems (baseline),
(ii) ULP-HEMT$\times$Manganin representing a practical near-term improvement.

The baseline system uses the physical error rates listed in \Cref{tab:physical_errors}. In \Cref{fig:readout_lq}, the vertical dashed line indicates the baseline measurement error ($0.008$). At this error rate, the required code distance is 13, corresponding to 337 PQs per LQ. Since the baseline system supports about 700 PQs, it can host only 2 LQs (horizontal dashed line). If the ULP-HEMT$\times$Manganin system can achieve the same measurement error as the baseline, the required code distance remains unchanged. Since this system supports roughly 4k PQs, it can host 13 LQs. Note that the exponential trend line is there only to serve as a visual guide.

If the measurement error increases (e.g., due to lower SNR due to additional amplification stages or higher amplifier noise floors) the required code distance also increases and the LQ yield decreases. For example, a $4\times$ increase in measurement error increases the required code distance up to 23, requiring nearly 1k PQs per LQ. In this regime, only 4 LQs are possible with the ULP-HEMT. However, this is still better than the baseline LQ yield. Therefore, even if low-gain amplifier configurations degrade the measurement error, the larger PQ capacity enabled by lower thermal load can still improve the overall LQ yield. This observation is consistent with the qualitative claim in \cite{croot2025enabling}.

Conversely, if the measurement error can be reduced through improved amplifier noise performance, demodulation, or machine-learning-based state discrimination \cite{lienhard2021machine}, the LQ yield increases further. Under such conditions, a single KIDE refrigerator could support up to 19 LQs.
\section{Possible Pathways Toward 10k-Qubit Systems}
\label{sec:pathways}
In this section, we explore several hypothetical and forward-looking system configurations that may enable operation in the 10k-qubit regime. These analyses are intended to identify possible scaling directions and quantify the conditions under which such scaling may become thermally feasible.

\subsubsection{Tradeoffs Between RF Wiring Options}
\label{subsubsec:wire_tradeoffs}

\Figure[!ht](topskip=0pt, botskip=0pt, midskip=0pt)[width=\linewidth]
{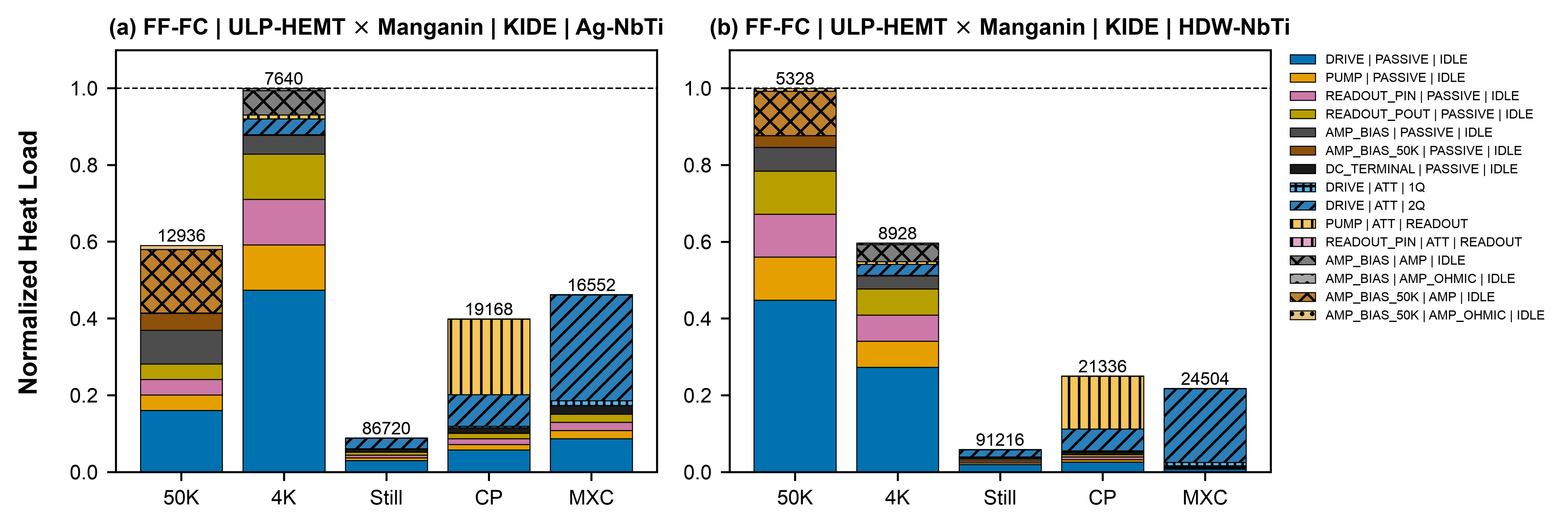}
{Normalized heat load breakdown for FF-FC systems using ULP-HEMT amplifiers with Manganin bias wires and two RF wiring configurations. (a) Cri/oFlex wiring with Ag microstrips above 4\,K and NbTi microstrips below 4\,K. (b) Coaxial wiring with HDW cables above 4\,K and NbTi superconducting coaxial cables below 4\,K. In both configurations, NbTi reduces the thermal load at the millikelvin stages.
\label{fig:wiring_choice}
}

\Figure[!ht](topskip=0pt, botskip=0pt, midskip=0pt)[width=0.97\columnwidth]
{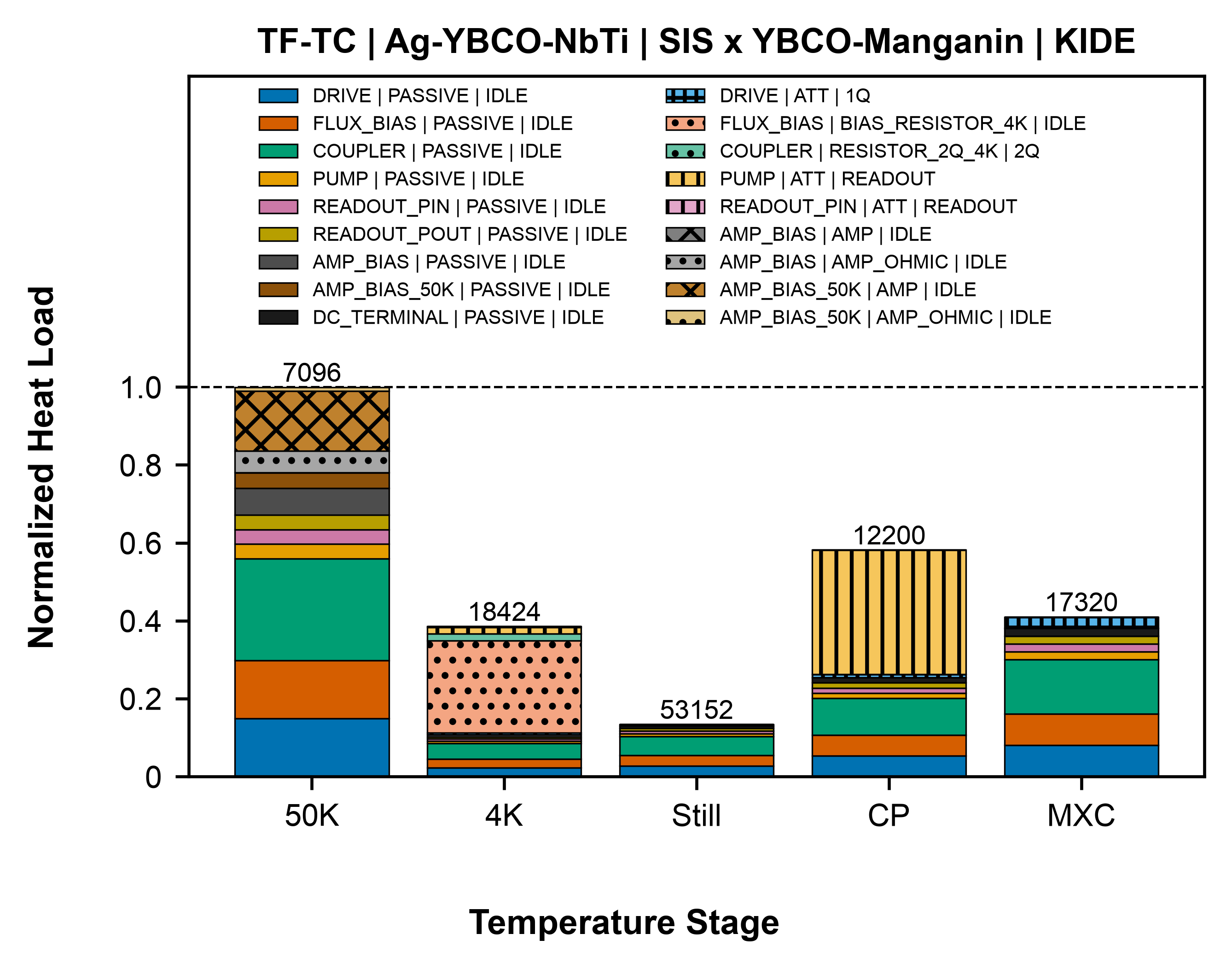}
{Normalized heat load breakdown for a hypothetical KIDE-based system using Ag-YBCO-NbTi microwave striplines, TF-TC qubits, and SIS amplifiers. 
The reduced microwave line PHL shifts the primary bottleneck from 4\,K to 50\,K where we have conservatively estimated the cooling power to be \qty{90}{\watt} (\Cref{app:bluefors-kide}). 
If we resolve the bottleneck at 50\,K (e.g., by increasing the cooling power at 50\,K by a little more than 70\%), the next bottleneck is at CP and the system can support approximately 12k physical qubits. When this happens, pump line attenuation dominates the CP heat load, while continuous flux biasing dominates the remaining heat load at 4\,K.
\label{fig:kide-ybco-delft}
}

In \Cref{fig:amp_solutions}, reducing the amplifier-associated heat causes the PHL of the microwave RF wiring to become the dominant thermal contribution especially at the millikelvin stages. We therefore evaluate three wiring configurations that use low PHL NbTi conductors below 4\,K but differ in the conductors and transmission line structures used at higher temperature stages. We will use ULP-HEMT$\times$Manganin readout amplifier configuration to prevent any 4\,K bottlenecks due to readout amplifiers with an additional LNF8G at 50\,K to achieve the required gain.

The first configuration uses Ag Cri/oFlex microstrips above 4\,K and NbTi microstrips below 4\,K (\Cref{fig:wiring_choice}(a)). The NbTi microstrips have approximately 10\% of the PHL of Ag microstrips. They therefore substantially reduce the heat load at CP and MXC and remove the bottlenecks at these stages. The resulting system supports approximately 7.5k PQs. However, the Ag sections above 4\,K continue to impose a large heat load on the 4\,K stage, which becomes the primary bottleneck. Additional modeling details are provided in \Cref{app:ff-fc-delft_nbti-ulp_mn-kide}.

The second configuration uses HDW coaxial cables above 4\,K and SC-086 NbTi coaxial cables below 4\,K (\Cref{fig:wiring_choice}(b)). Compared with Ag Cri/oFlex microstrips, HDW cables impose a lower PHL on the 4\,K stage (see \Cref{tab:signal_cables}), while the superconducting SC-086 NbTi sections reduce the PHL at the lower-temperature stages, particularly CP and MXC. This configuration therefore alleviates the thermal constraints at both the 4\,K and millikelvin stages. Consequently, the primary bottleneck shifts to the 50\,K stage, limiting the system capacity to approximately 5k PQs.

We conservatively estimate the available cooling power at 50\,K to be \qty{90}{\watt} (\Cref{app:bluefors-kide}). The actual capacity may be higher because KIDE employs functionally separated pulse-tube cryocoolers for operational, payload, and radiation-shield cooling. This separation may increase the available cooling powers. In order for the bottleneck to shift away from 50\,K to 4\,K we would require the available 50\,K cooling power to be increased by $\approx$70\% from \qty{90}{\watt} to \qty{153}{\watt}. If this is achieved, the primary bottleneck shifts to the 4\,K stage, and the estimated system capacity increases to approximately 8k PQs. Additional modeling details are provided in \Cref{app:ff-fc-hdw_nbti-ulp_mn-kide}.

In either of the above cases, the major heat load responsible is the PHL from the control and readout cables and \textit{not} the readout amplifier chain-related heat loads. Using an amplifier configuration with a lower thermal footprint (e.g., SIS \(\times\) YBCO) would only provide limited gains. This indicates that reducing the PHL at 4\,K should be the next focus to increase system scalability.    

The third configuration attempts this by retaining the Cri/oFlex stripline structure but replacing the Ag conductor between 50\,K and 4\,K with a high-temperature superconductor such as YBCO \cite{solovyov2021ybco} due to its superconducting nature and low PHL. Although YBCO was previously considered only for DC amplifier bias wiring, in principle, it can also support microwave transmission \cite{solovyov2021ybco}.

At the time of writing, YBCO microwave interconnects have only been demonstrated as laboratory-scale prototypes. Further advances in materials, fabrication, and packaging are required to produce long, low-loss YBCO striplines with suitable impedance, crosstalk, and signal integrity characteristics.

We model a shielded YBCO stripline with a structure similar to Cri/oFlex, in which the signal conductor is placed between two ground planes. We assume that each YBCO ground plane has a width equal to twice the signal line pitch. Since the YBCO planes dominate thermal conduction, we estimate the PHL of the complete shielded stripline as five times the PHL of a single YBCO bias wire.

YBCO is used only between 50\,K and 4\,K because it remains superconducting below its critical temperature of 77\,K. Ag is retained between RT and 50\,K. Below 4\,K, NbTi is preferred because it has substantially lower PHL than YBCO at temperatures below 4\,K.

We evaluate this configuration using a TF-TC architecture and a CZ-based workload. TF-TC requires the largest number of cables per qubit cell among the architectures considered in this work. It therefore represents a conservative case for evaluating RF interconnect PHL. To isolate the effect of reducing microwave line PHL, we also assume SIS amplifiers with YBCO bias wiring at 4\,K. This minimizes amplifier-associated heat. Additional modeling details are provided in \Cref{app:tf-tc-ybco_delft-sis_ybco-kide}.

The resulting heat-load breakdown is shown in \Cref{fig:kide-ybco-delft}. The substantial reduction in the PHL of the microwave lines alleviates the thermal constraints at both the 50\,K and 4\,K stages. The bottleneck is at 50\,K with a 7.2k PQ capacity despite the additional flux and coupler-bias lines required by the TF-TC architecture. If the 50\,K cooling constraint is removed, as discussed previously, CP becomes the next bottleneck, and the system can support approximately 12k PQs.

Several observations follow from this result:
\begin{itemize}
    \item The primary bottleneck shifts from 4\,K to 50\,K due to the limited cooling power assumed at 50\,K for the KIDE fridge. The next bottleneck is at CP where pump line attenuation is the dominant heat source. This load may be reduced by moving part of the pump attenuation to the Still stage.

    \item Amplifier associated heat at 4\,K becomes nearly negligible because SIS amplifiers have low AHL and YBCO bias wires have low PHL. The remaining cooling margin indicates that other low power amplifier configurations, including ULP-HEMT and LNF8G(LP), may also be feasible.

    \item The dominant heat source at 4\,K is the ohmic dissipation of the T-pad attenuator used for continuous flux biasing of TF qubits. Coupler bias attenuators produce less ohmic heat because they are active only during 2Q operations. If couplers also require static bias, this will increase the heat stress further at 4\,K.

    \item TF-TC architectures incur larger PHL because each unit cell requires additional flux and coupler bias lines. Under otherwise identical assumptions, an FF-FC architecture can support up to 12k PQs when 50\,K is the bottleneck. If this bottleneck is resolved, the qubit count increases to 16k as shown in \Cref{app:ff-fc-ybco_delft-sis_ybco-kide}. This result illustrates the thermal tradeoff between qubit control architectures.
\end{itemize}

The Ag-YBCO-NbTi configuration represents a hypothetical upper bound. Long, low-loss YBCO microwave striplines suitable for large-scale quantum systems have not yet been demonstrated. The result should therefore be interpreted as an exploratory design point that shows how substantial reductions in microwave line PHL can shift the system bottleneck and may enable qubit counts beyond 10k.

Overall, using NbTi below 4\,K reduces the millikelvin heat load in both the microstrip and coaxial cable families. HDW with NbTi coaxial cables provides greater scalability than Ag Cri/oFlex with NbTi microstrips because HDW imposes a lower PHL at 4\,K. Replacing the Ag section of the stripline with YBCO provides a further reduction and enables approximately 12k PQs for TF-TC and 16k PQs for FF-FC systems under optimistic assumptions. However, all hybrid configurations may introduce practical challenges related to thermalization, signal integrity, routing, connector integration, fabrication, and cost.

\subsubsection{KIDE Fridge with Optical Fiber Interconnects}
\label{subsec:optical_kide}

\Figure[!ht](topskip=0pt, botskip=0pt, midskip=0pt)[width=\linewidth]
{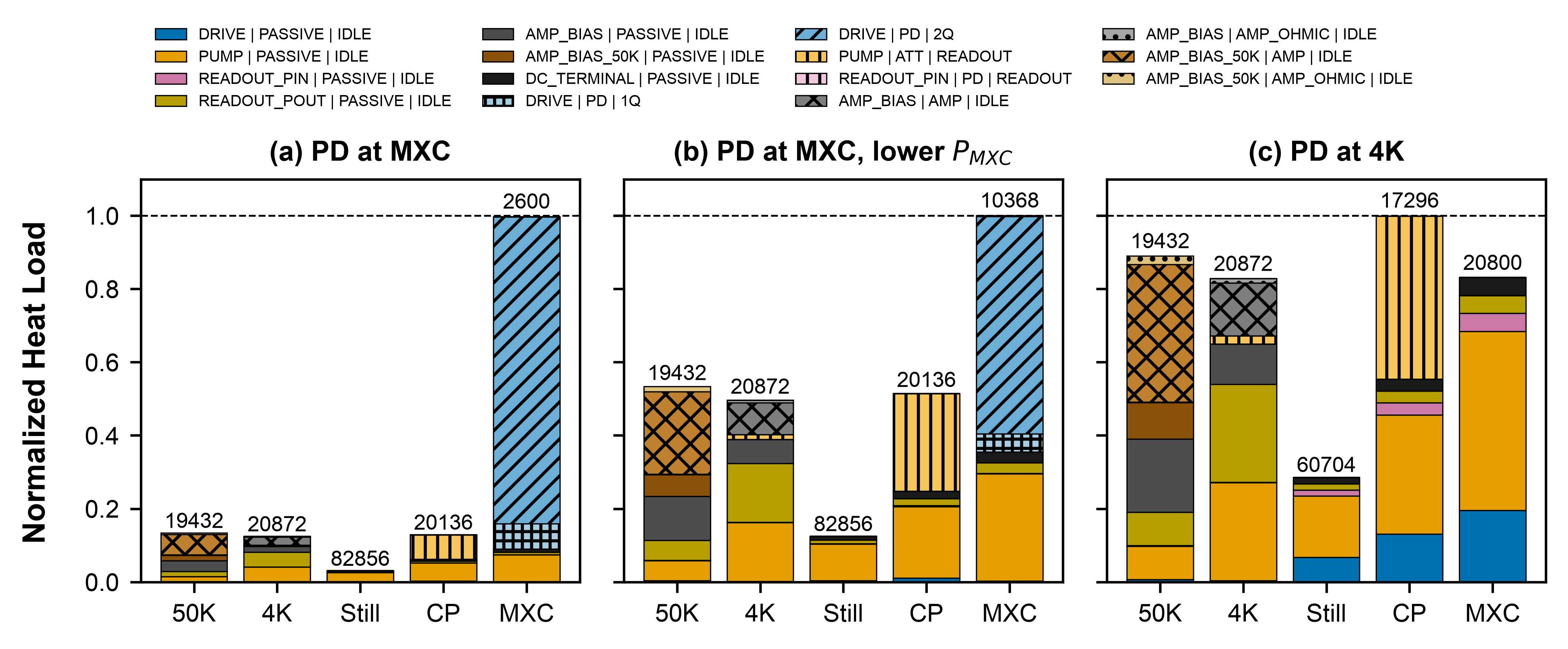}
{Heat-load breakdown for KIDE-based optical-interconnect architectures using ULP-HEMT amplifiers with Manganin bias lines. 
(a) \textbf{Optical-Baseline}: Optical architecture with the PD located at MXC. After amplifier heat is mitigated, the dominant bottleneck shifts to MXC due to large PD dissipation, limiting the system to approximately 2.6k physical qubits. 
(b) \textbf{Optical-LP}: Hypothetical architecture where qubit drive powers are reduced by \qty{15}{\decibel}. Reduced PD dissipation supports slightly more than 10k PQs under optimistic assumptions. 
(c) \textbf{Optical-PD-at-4\,K}: Optical architecture with the PD relocated to 4\,K as proposed in \cite{joshi2023scaling}. The MXC bottleneck is removed and the system exceeds 10k physical qubits despite reintroducing NbTi microwave lines between 4\,K and MXC.
\label{fig:kide-optical}
}

\Figure[!ht](topskip=0pt, botskip=0pt, midskip=0pt)[width=0.98\columnwidth]
{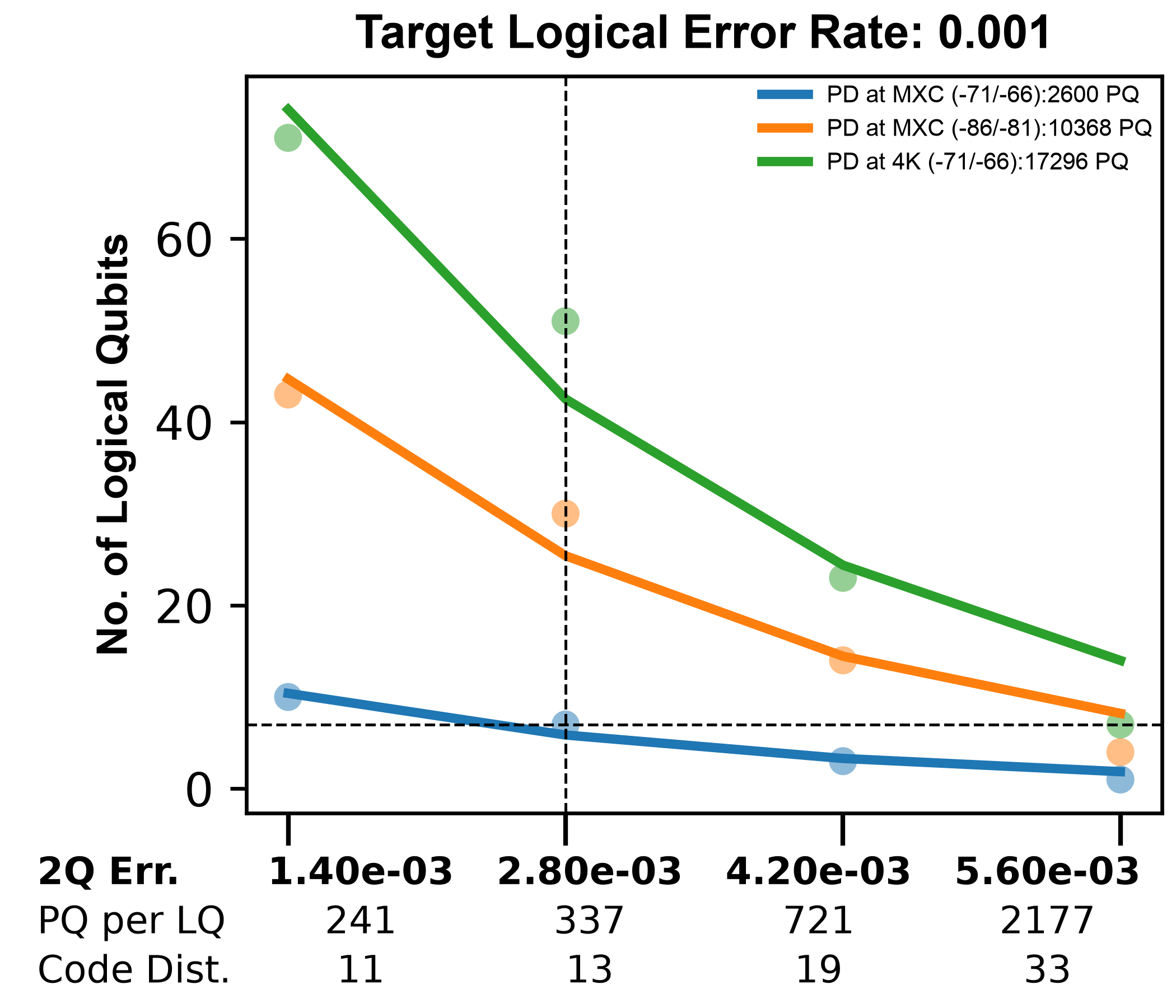}
{LQ yield sensitivity to 2Q error rate for KIDE-based optical architectures. Circles represent simulation data points and solid lines represent exponential trends. Although reduced qubit drive power lowers PD dissipation and increases PQ capacity, the LQ yield remains strongly dependent on 2Q fidelity. Even with 50--100\% higher 2Q error rates, the Optical-LP system maintains LQ yields comparable to Optical-Baseline due to its larger PQ capacity. Results assume all other error rates remain unchanged.
\label{fig:optical-2q-ler}
}

In \Cref{sec:historical-optical}, we discussed optical interconnects as an alternative to metallic microwave lines due to their extremely low PHL. However, our earlier analysis showed that optical fibers alone provide only limited scalability benefits because amplifier-associated heat at 4\,K remained the dominant bottleneck.

Here, we reevaluate optical fiber-based architectures assuming that amplifier heat has been mitigated using ULP-HEMT amplifiers with Manganin bias lines as discussed in \Cref{sec:solutions}. All other assumptions remain unchanged except that the XLD400 fridge is replaced with the higher-cooling-power KIDE fridge (Additional modeling details are provided in \Cref{app:ff-fc-kide-optical-baseline}). 

\paragraph*{Optical Baseline}
\Cref{fig:kide-optical}(a) shows the resulting heat breakdown for the baseline optical configuration where the PD is located at MXC \cite{lecocq2021control}. We refer to this configuration as \textbf{Optical-Baseline}. In this configuration, the dominant bottleneck shifts from 4\,K to MXC due to the large active heat dissipated by the PD as reported in \cite{lecocq2021control,joshi2023scaling}. The majority of this dissipation originates from 2Q operations because they require higher drive power and occur more frequently in the ESM workload. As a result, the system is limited to 2.6k physical qubits. We explore two approaches to mitigate this MXC bottleneck.

\paragraph*{Optical-LP: Reducing Qubit Drive Power}
The first approach is to reduce PD dissipation directly by lowering the qubit drive power $P_{MXC}$. From \Cref{eq:photodetector}, reducing $P_{MXC}$ lowers the optical-to-microwave conversion power and therefore the PD heat dissipation. In \Cref{fig:kide-optical}(b), we evaluate a hypothetical low-power configuration, denoted \textbf{Optical-LP}, where the required 1Q/2Q drive powers are reduced by \qty{15}{\decibel}, from \qty{-71}{\dBm}/\qty{-66}{\dBm} to \qty{-86}{\dBm}/\qty{-81}{\dBm}, respectively (Modeling details are provided in \Cref{app:ff-fc-kide-optical-lp}).

Such reductions in drive power would require substantial device-level improvements. Lower drive powers generally require stronger coupling between the qubit and drive line, which may increase physical errors due to leakage, Purcell decay, and decoherence. Therefore, this scenario should be interpreted as a sensitivity study with an optimistic theoretical limit rather than a validated self-consistent operating regime.

Under these assumptions, the PD heat decreases sufficiently that the system can approach 10k physical qubits. However, PD heat at MXC remains the primary scalability bottleneck. To evaluate the impact on LQ yield, we estimate the LQ yield for different 2Q error rates as shown in \Cref{fig:optical-2q-ler}. We focus primarily on 2Q error rates because preliminary experiments indicate that variations in 1Q error rates produce comparatively small changes in LQ yield, whereas modest increases in 2Q error rates can severely degrade logical performance. This behavior is expected because ESM workloads are dominated by entangling operations.

For the physical error rates in \Cref{tab:physical_errors}, the Optical-Baseline system yields 7 LQs, requiring a code distance of 13 for a target LER of 0.001. Under the same error assumptions, Optical-LP yields 30 LQs due to its substantially larger PQ capacity.

We also observe that even if lowering the drive power increases the 2Q error rate by 50--100\%, the resulting LQ yield remains comparable to Optical-Baseline with the lower error rates. Although higher 2Q error rates require larger code distances, the larger PQ capacity of Optical-LP compensates for the increased overhead.

\paragraph*{PD-at-4\,K}

The second approach is to relocate the PD to 4\,K as proposed in \cite{joshi2023scaling}. We refer to this configuration as \textbf{PD-at-4\,K}. The resulting heat breakdown is shown in \Cref{fig:kide-optical}(c) (Refer to \Cref{app:ff-fc-kide-optical-pd4k} for more modeling details). Relocating the PD removes the dominant heat source from MXC and transfers it to 4\,K, where the substantially larger cooling power can comfortably absorb the dissipation.

This relocation requires NbTi microwave lines between 4\,K and MXC, reintroducing metallic interconnects and their associated PHL. However, because NbTi has relatively low PHL, the additional heat remains manageable. Under these assumptions, the system exceeds 10k physical qubits.

Noise analysis in \cite{joshi2023scaling} suggests that relocating the PD to 4\,K should not significantly degrade system fidelity. Therefore, assuming the physical error rates in \Cref{tab:physical_errors}, the resulting LQ yield is  51 logical qubits, due to its large PQ capacity.

These results indicate that optical interconnects can provide meaningful scalability benefits only if the PD thermal stress at MXC is mitigated without significantly degrading logical fidelity. Otherwise, increases in PQ capacity may not translate into proportional gains in logical scalability.

\subsubsection{Colossus Fridge}
\label{subsec:colossus}

\Figure[!ht](topskip=0pt, botskip=0pt, midskip=0pt)[width=\linewidth]
{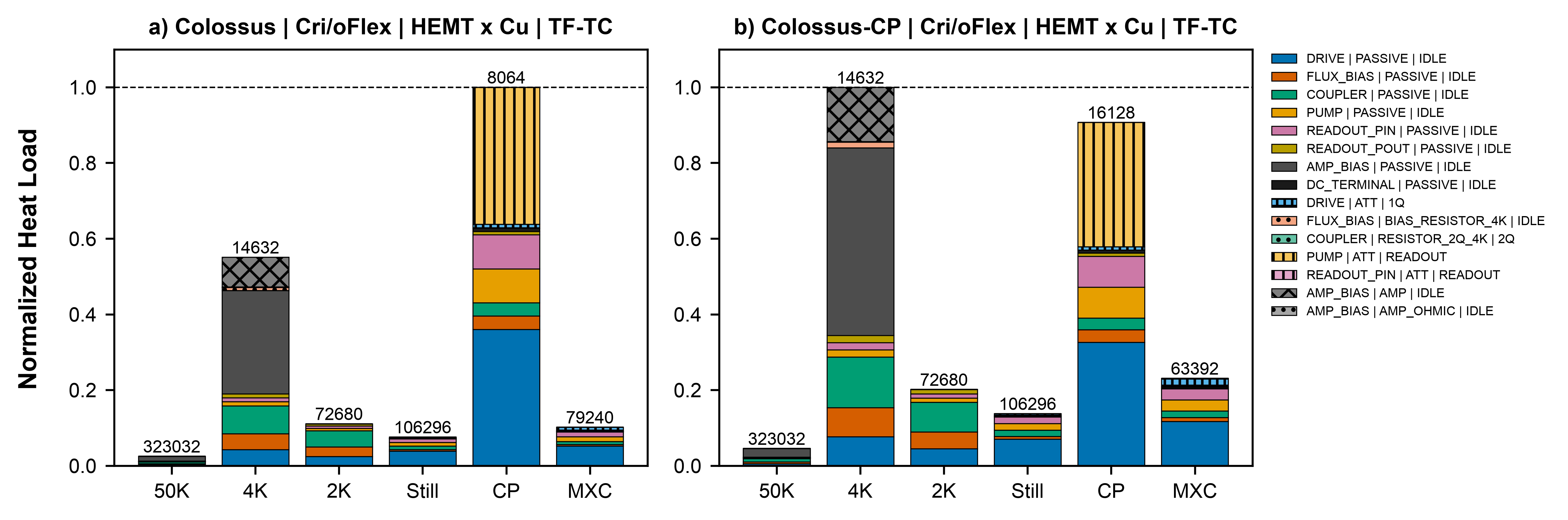}
{Heat-load breakdown for a Colossus-based system under the TF-TC configuration with Cri/oFlex cables. 
(a) Despite large cooling margins at 50\,K and 4\,K, the limited CP cooling capacity constrains scalability to approximately 8k physical qubits, with CP as the dominant bottleneck. 
(b) In the hypothetical Colossus-CP configuration, two dilution units are reassigned from MXC to CP, increasing CP cooling capacity. Under this assumption, the system can exceed 10k physical qubits, with 4\,K and CP operating near their cooling limits. These results assume ideal redistribution of dilution capacity and do not account for potential thermodynamic coupling constraints.
\label{fig:colossus_10k}
}

We next evaluate a Colossus-class cooling profile as a forward-looking sensitivity case. This calculation is not intended to predict the realized performance of the Colossus platform. Instead, it evaluates how the modeled bottleneck changes if a single integrated cryostat provides the Colossus-class cooling capacities. 
Colossus provides substantially higher cooling power at 50\,K and 4\,K compared to KIDE but the cooling power available at CP is limited by the number of installed dilution units (see \Cref{tab:all_fridges}).

We assume a \textbf{Colossus} fridge that uses a TF-TC architecture with Cri/oFlex microwave cables with conventional LNF8G HEMT-based readout chains with copper wires. This configuration represents how currently available systems would immediately scale with a large Colossus-like fridge. See \Cref{app:tf-tc-delft-hemt_cu-colossus} for more modeling details.

\Cref{fig:colossus_10k}(a) shows the thermal analysis for this configuration assuming TF-TC architecture. The amplifier-related heat loads, which were previously dominant bottlenecks in KIDE-based systems, no longer impose critical constraints due to the large cooling powers at 4\,K. Instead, the primary bottleneck shifts to the CP stage where the PHL from microwave lines and the pump line attenuation overwhelms the CP stage and limits the PQ count to around 8k qubits. With FF-FC architecture, the flux and coupler biasing cables are not required and the qubit count increases to around 9k (See \Cref{app:ff-fc-delft-hemt_cu-colossus}). 

One potential mitigation strategy is to redistribute dilution units within the system to increase the cooling power at CP at the expense of MXC. Specifically, we consider a hypothetical modification in which two dilution units originally serving the MXC stage are reassigned to the CP stage. This configuration is referred to as \textbf{Colossus-CP} (see \Cref{tab:all_fridges}). 

The resulting heat breakdown is shown in \Cref{fig:colossus_10k}(b) (See \Cref{app:tf-tc-delft-hemt_cu-colossus_cp} for more modeling details).
With the increased CP cooling capacity, the system can accommodate approximately 15k physical qubits even with the TF-TC architecture. The bottleneck now shifts to 4\,K (with the HEMT \(\times\) Copper module being the dominant heat contributor). If we use an alternative amplifier configuration with a smaller thermal footprint, the system would be able to accommodate an additional 1.5k PQs and the bottleneck would shift back to CP.
Our evaluations show that an FF-FC architecture would be able to support as much as 18k physical qubits where 4\,K and CP are co-limiting stages (See \Cref{app:ff-fc-delft-hemt_cu-colossus_cp}). 

The Colossus-CP case is an idealized redistribution of usable cooling capacity from MXC to CP. The values discussed previously are a conditional thermodynamic upper bound rather than a demonstrated platform capability. We note that reallocating dilution units within a Colossus-class system may not be straightforward. The thermodynamic performance of dilution refrigerators depends on coupled fluid flow, heat exchangers, and stage interactions. Modifying dilution distribution could introduce nontrivial stability and efficiency challenges that are not captured in our steady-state heat model.
Furthermore, the power consumption, physical footprint, and capital cost associated with Colossus-class systems are substantial. Whether operating such a large cryogenic platform for a 10k-qubit system is practical or commercially viable remains an open question. Therefore, this analysis should be interpreted as an exploration of thermodynamic feasibility rather than a recommendation of a deployable architecture.

\section{Model Assumptions and Scope}
\label{sec:scope}

The objective of our system-level heat estimation model is not to predict the exact thermal behavior of a specific experimental installation, but to identify dominant heat sources, compare technology options within a unified framework, and determine scalability trends toward 10k systems. The reported PQ and LQ yields should therefore be interpreted as architectural feasibility indicators under steady-state thermal constraints rather than realizable qubit counts for a specific implementation. Real-world implementations would have lower yields due to necessary engineering margins. We are also not recommending any particular system design, as deployment-level evaluation would require significantly more detailed modeling using measured hardware parameters and operational constraints.

To enable architectural-level comparison, we adopt several simplifying assumptions. Cooling power at each temperature stage is assumed constant, and no temperature rise is modeled. In practice, dilution refrigerators are thermodynamically coupled systems in which excessive loading at one stage can influence neighboring stages through helium-flow dynamics and parasitic heat transfer. Accurate modeling of these effects requires detailed manufacturer data and dynamic thermal characterization, as explored in \cite{manifold2025thermal}, and is beyond the scope of this work.

Minor dissipative elements including circulators, isolators, insertion-loss dissipation in filters and directional couplers, and package-level dissipation are also omitted because their reported heat contributions are typically small relative to amplifiers, attenuators, and bias-line dissipation.

We further assume instantaneous and uniform thermalization at each flange. Transient thermal gradients, localized hotspots, and finite heat diffusion dynamics are not modeled. The workload model is based on repeated ESM cycles and does not include implementation-specific auxiliary operations such as echo-X gates or leakage-removal procedures like in \cite{2025acharyaQuantumErrorCorrection}.

We have included reset latency in the ESM cycle duration following \cite{2023acharyaSuppressingQuantumErrors}, but reset-induced active heat loads are not explicitly modeled. Although several experiments report reset times and fidelities, the corresponding microwave powers or flux-current waveforms are generally unavailable. These quantities depend strongly on the reset protocol and device architecture \cite{jerger2024dispersive,mcewen2021removing,sunada2022fast,reed2010fast,chen2024fast}. This prevents a consistent comparison across the various qubit architectures. We therefore defer reset-power modeling until experimentally calibrated waveform data are available. Consequently, the reported thermal capacities are optimistic with respect to reset-induced dissipation.

The error rates, operation powers, and gate latencies used in both the thermal and logical-error models are inspired by reported experimental systems. However, these parameters can vary significantly across architectures and device implementations. In this work, they are used primarily to provide a consistent configuration for fair cross-technology comparison rather than to represent any specific hardware platform.

For scalability analysis, we assume that the 8-qubit unit cell scales linearly to larger systems. This abstraction ignores geometric crowding, flange area limits, routing congestion, chip-layout constraints, crosstalk, and frequency crowding. The purpose of this assumption is to isolate thermal feasibility. If an architecture violates cooling limits even under idealized linear scaling, it is thermally infeasible irrespective of subsequent geometric or fabrication optimizations.

Finally, this work does not model detailed helium thermodynamics, dynamic inter-stage coupling, mechanical packaging constraints, electromagnetic signal integrity, or qubit variability. These effects are important for full-system realization but are orthogonal to the thermal bottleneck analysis performed in this study.
\section{Conclusion}
\label{sec:conclusion}

Designing cryogenic systems for 10k qubits requires careful management of infrastructure-induced heat loads. Existing thermal scalability models do not fully capture several heat sources that have become critical in present-day systems, leading to bottleneck misidentification and overly optimistic scalability estimates. 

In this work, we built upon previous thermal scalability models by incorporating active, passive, and ohmic heat associated with the cryogenic readout chain that were previously omitted. Our modeling framework also enables fair cross-technology comparison across legacy and emerging designs, and integrates thermal analysis with logical-error modeling to estimate both PQ and LQ yields. Using this framework, we showed that amplifier-associated heat loads are the dominant heat contributors in several evaluated modern-system configurations. In particular, the passive, active and ohmic heat associated with readout amplifiers and biasing cables severely constrain the 4\,K stage.

To mitigate this bottleneck, we explored the tradeoffs between amplifier technologies and biasing-wire materials, and showed that maximizing scalability requires jointly minimizing passive, active, and ohmic heat loads. Our analysis also revealed several non-intuitive results, including cases where reducing amplifier active power alone does not improve overall thermal scalability due to increased passive or ohmic contributions.

Finally, we showed that even with current state-of-the-art refrigerators and wiring technologies, achieving 10k scalability remains difficult. Additional advances in refrigeration, wiring, and qubit operation are still required. Our evaluations suggest that high-temperature superconducting cables such as YBCO and optical interconnect architectures are promising directions for future scaling. We also showed that ultra-large refrigerators such as Colossus fundamentally alter the heat-load landscape and may provide a thermal envelope for approximately 10k physical qubits under the stated assumptions.
\appendices
\clearpage
\section{Modeling Methodology}

\subsection{Estimating the cooling powers of Bluefors KIDE}
\label[appsec]{app:bluefors-kide}
To estimate the cooling capacities of the Bluefors KIDE platform, we model it as three independent XLD1000sl-equivalent dilution-cooling sections. We assume that each section uses the enhanced dual-PT425-RM configuration. Bluefors reports that an XLD1000sl system with this configuration can operate its nominal 4\,K flange at 4.5\,K without degrading the dilution-unit performance, providing approximately 2\,W of cooling power available to external loads \cite{blueforsProductsEnhanced}. We therefore assign the KIDE platform a total available cooling power of 6\,W at the nominal 4\,K stage.

Bluefors states that KIDE incorporates nine pulse-tube cryocoolers that are functionally separated among operational cooling, customer-payload cooling, and radiation-shield cooling \cite{blueforsKIDECryogenic}. However, the publicly available specifications neither identify the cryocooler models nor describe the thermal coupling or functional assignment of the individual units. In our model, we treat six of the nine cryocoolers as three dual-pulse-tube cooling sections, one associated with each dilution unit. We do not credit the remaining three cryocoolers toward the cooling power available to user-installed loads, effectively reserving their capacity for radiation-shield cooling and other internal system loads. This allocation is a modeling assumption rather than a manufacturer-specified configuration.

The system in \cite{raicu2025cryogenic} uses an XLD1000sl refrigerator equipped with two PT420 cryocoolers. The authors report that the 50\,K stage provides 30\,W of cooling power available to experimental loads when operated at 40\,K. For three dilution units, the available cooling power is therefore estimated to be \qty{90}{\watt}.
We therefore assign an available cooling power of \qty{90}{\watt} to the nominal 50\,K stage of KIDE, which we assume also operates at 40\,K. This represents a conservative modeling estimate because it applies the residual cooling power reported for an older XLD1000sl configuration with weaker PT420 cryocoolers and does not credit the three additional KIDE cryocoolers toward the user-accessible cooling budget. Nevertheless, because the detailed KIDE cryocooler configuration and internal thermal loads are not publicly available, this value should be interpreted as an approximate cooling budget.

For the lower-temperature stages, we assume the same operating temperatures as those adopted for the XLD1000sl system in \cite{raicu2025cryogenic}. Because KIDE contains three independent dilution-cooling units, we estimate its CP and MXC cooling powers by multiplying the corresponding XLD1000sl values by three. This scaling is consistent with the KIDE specification of more than 90\,$\mu$W at 20\,mK and more than 3000\,$\mu$W at 100\,mK, divided among its three cooling sections \cite{blueforsKIDECryogenic}. For the Still stage, we instead use the vendor-provided estimate of 90\,mW at 1.2\,K for the complete KIDE system.

\subsection{Passive Heat Load Estimation}
\label[appsec]{app:phl_details}

For a cable of cross-sectional area $A$ and length $L$ connecting two stages at $T_H$ and $T_L$, the conductive heat flow is given by
\Cref{eq:phl}, where $k(T)$ is the temperature-dependent thermal conductivity of the cable material:
\begin{equation}
    P = \frac{A}{L} \int_{T_L}^{T_H} k(T)\, dT.
\label{eq:phl}
\end{equation}
This formulation is applied to every cable segment between adjacent stages (e.g., 50\,K to 4\,K, 4\,K to Still, etc.). The total PHL at a stage is obtained by summing the contributions from all cables entering that stage from the warmer side.

Throughout this work, we use the term \textit{linear thermal conductance}, different from thermal conductivity $k(T)$, to denote the quantity $k(T)A=G_L(T)$. This quantity has the SI units of \unit{\watt\meter\per\kelvin}. It represents the thermal conductivity for a unit cable length of cross-sectional area $A$ and is used to compare the passive heat transfer characteristics of different wiring technologies.

Heat transfer is assumed to be dominated by conduction through the metallic components of the wire. Conduction through insulation and polymer coatings is neglected because their thermal conductivity is orders of magnitude lower than that of the metallic conductor \cite{krinner2019engineering,raicu2025cryogenic}. Heat flow is assumed to be one-dimensional along the wire length, which is the dominant direction of heat transfer. Radial conduction and radiative contributions are ignored \cite{raicu2025cryogenic}.

The PHL values reported in \cite{krinner2019engineering,raicu2025cryogenic,delftcircuits,lecocq2021control} correspond to different cryogenic platforms with varying inter-flange distances and operating temperatures. For a fair comparison, we normalize these values to the XLD1000sl by scaling them according to the corresponding inter-flange distances and operating temperatures. During this scaling, we assume that the thermal conductivity of each cable remains constant between adjacent temperature stages when complete material-property data is unavailable. The resulting PHL values used in this work are summarized in \Cref{tab:signal_cables}.

For the Colossus refrigerator, which includes an additional 2\,K stage, we assume that the average thermal conductivity of cables in the 4\,K--2\,K segment is equal to the average conductivity in 2\,K--Still segment.

\subsubsection{PHL of SS Coaxial Cables}
The PHL values for UT-085-SS-SS and UT-085-NbTi in \Cref{tab:signal_cables} are obtained by linearly scaling the mean \emph{measured} values reported in \cite{krinner2019engineering} to the inter-flange distances and operating temperatures of the XLD1000sl refrigerator. This scaling assumes that the effective thermal conductivity remains approximately constant for the relatively small differences in stage temperatures and cable lengths between the two fridges.

The UT-085-NbTi PHL at the Still stage is the only value that was not directly measured. We therefore use the estimate reported in \cite{krinner2019engineering}. Since the Still-stage contribution is small relative to the other heat loads, variations in this value do not affect the main results or conclusions.

The SS Drive and Flux lines have similar PHL from RT to  Still. Below Still, however, the flux line does not include an attenuator at CP. The absence of this thermal anchor allows more heat to propagate toward MXC, resulting in a higher MXC PHL than that of the drive line.

\subsubsection{SC-086 NbTi PHL Calculation}
We estimate the passive conductive load of SC-086-NbTi coaxial cables using the thermal conductivity values reported in \cite{coax2022} over the temperature range of 5\,K to 1\,K. We use a linear fit over the log-log values to estimate the conductivity below 1\,K as in \cite{krinner2019engineering}.

\Figure[!ht](topskip=0pt, botskip=0pt, midskip=0pt)[width=0.95\columnwidth]
{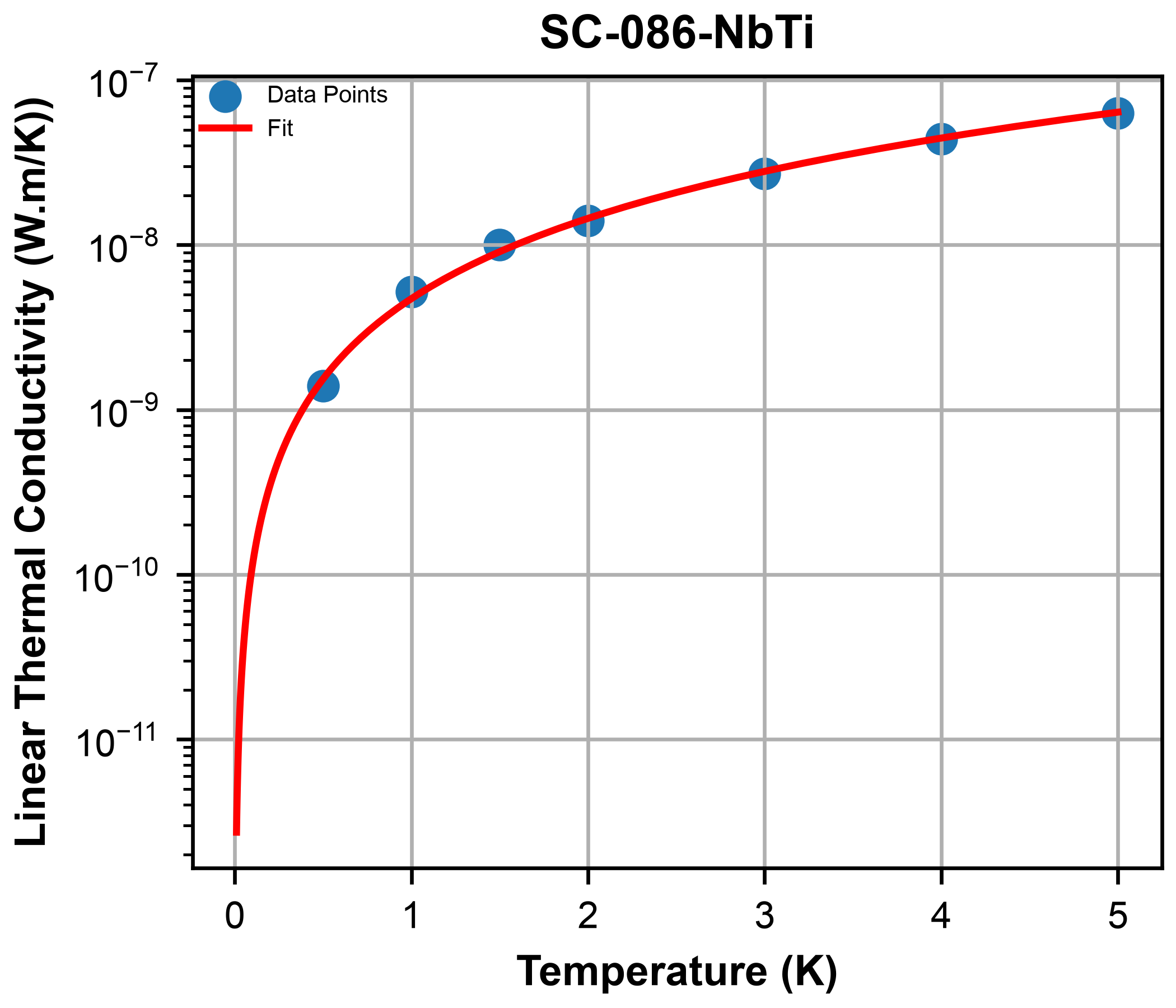}
{Linear thermal conductance of SC-086-NbTi reproduced using data from \cite{coax2022} and a log-log linear fit.
\label{fig:sc_086_nbti_thermal_conductivity}
}

\subsubsection{PHL estimation for Optical Fiber}
\Figure[!ht](topskip=0pt, botskip=0pt, midskip=0pt)[width=0.95\columnwidth]
{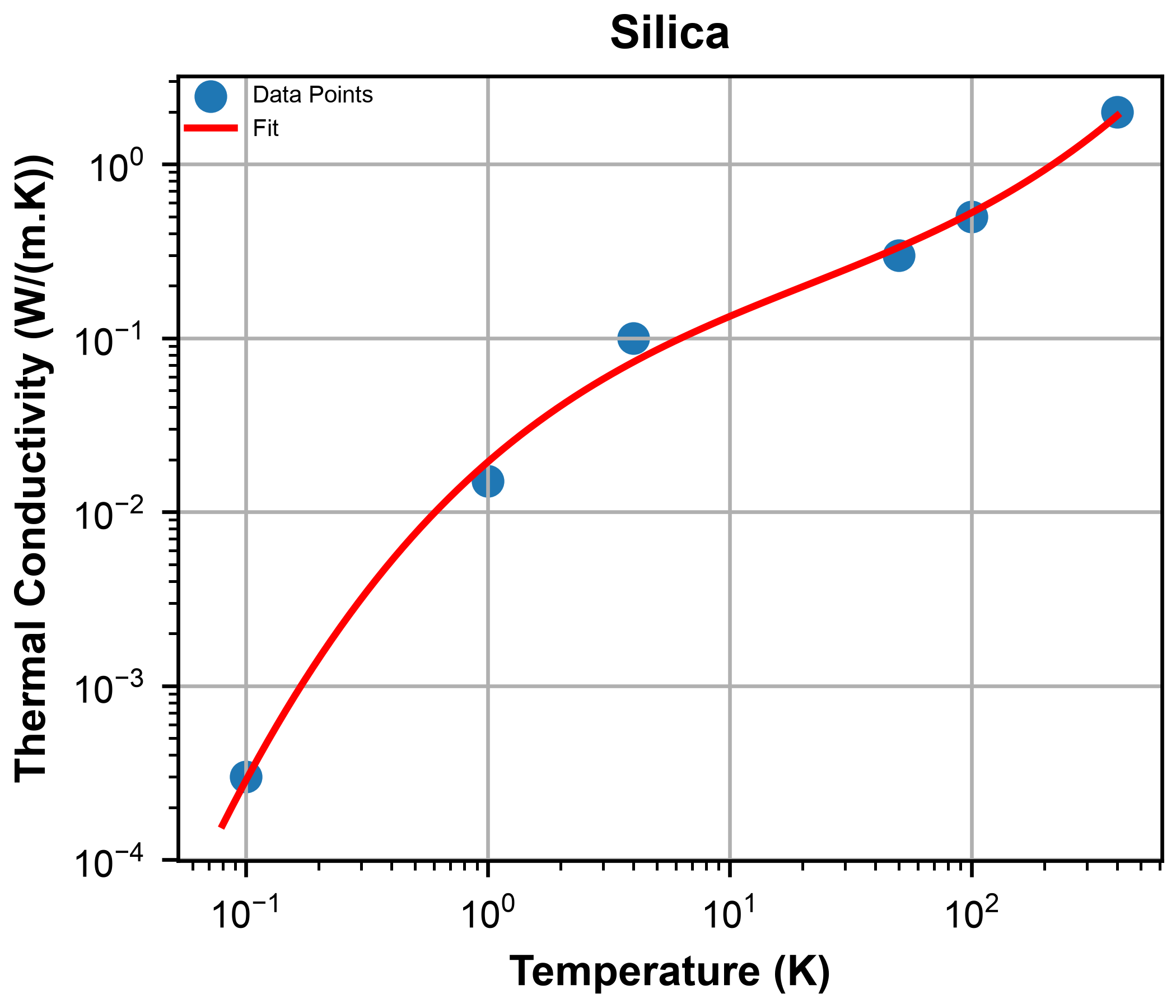}
{Thermal conductivity of silica reproduced using data from \cite{smith1978effect} and a log-log cubic spline fit.
\label{fig:silica_k}
}
We assume the optical fiber consists of a silica core and cladding as in \cite{lecocq2021control}.
To estimate the PHL, we use thermal conductivity data for silica from \cite{smith1978effect,lecocq2021control}, shown in \Cref{fig:silica_k}. A log-log cubic spline fit is applied to the reported data. We assume a fiber diameter of \qty{250}{\micro\meter} \cite{lecocq2021control,joshi2023scaling}, where the core and cladding are both silica, to calculate the PHL.

\subsubsection{PHL for Biasing Wires}
Here we estimate the PHL due to the biasing wires for 4\,K amplifiers. We consider copper, Manganin and YBCO-on-Kapton cables. Since thermal conductivity $k(T)$ is temperature dependent, material-specific models of $k(T)$ are required. In the following, we construct such models for copper, Manganin, and YBCO-on-Kapton and compute their PHL contributions into the 50\,K and 4\,K stages.

\Figure[!ht](topskip=0pt, botskip=0pt, midskip=0pt)[width=0.95\columnwidth]
{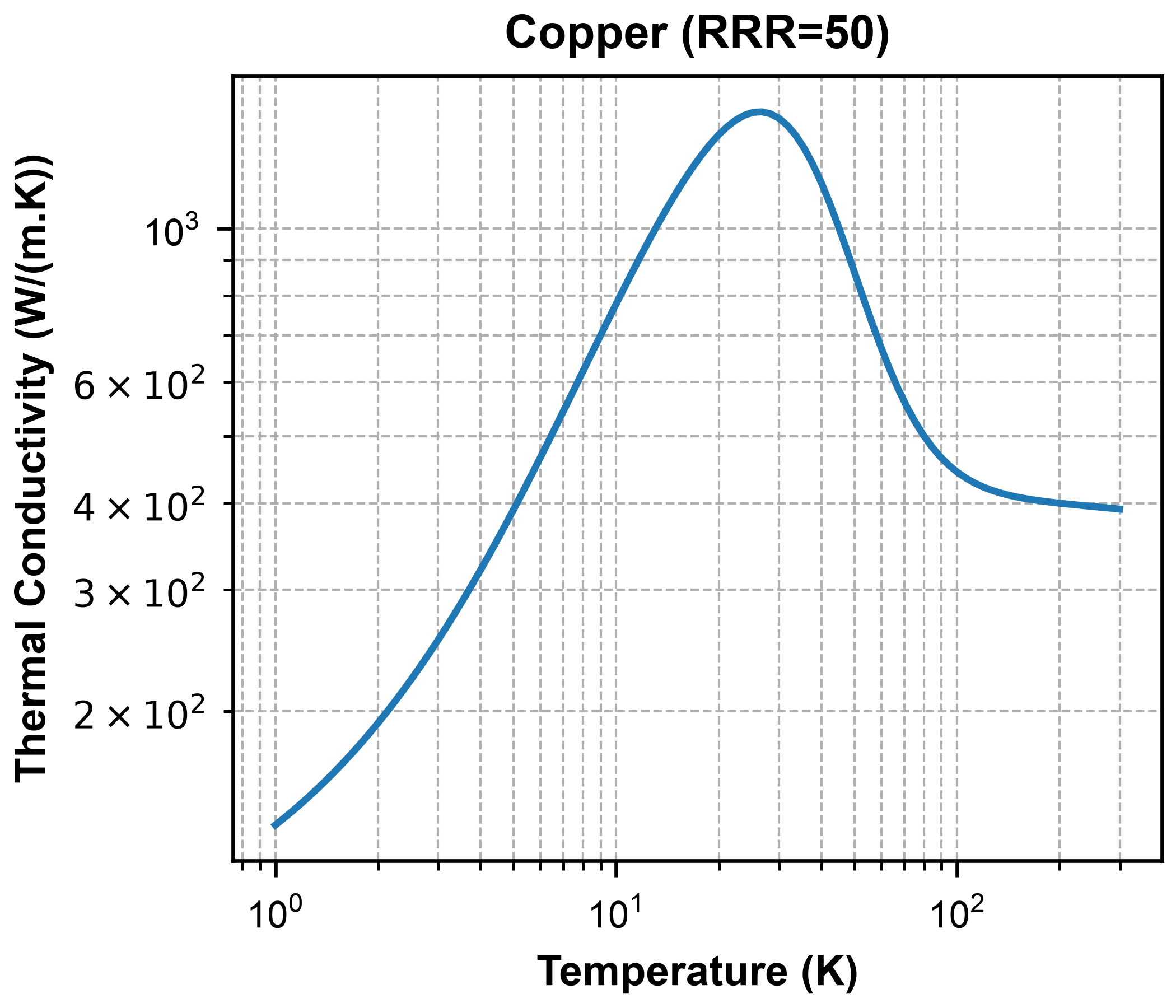}
{Thermal conductivity of copper (RRR=50) using the model provided by NIST \cite{nistCryogenicMaterial}.
\label{fig:cu_k}
}

The heat loads for AWG35 are reported in \cite{krinner2019engineering} for their assumed system. To calculate the PHL of AWG30 copper wires (Cu), we assume a residual-resistance ratio (RRR) of 50, consistent with \cite{green2017connection}. The temperature-dependent thermal conductivity $k_{Cu}$ is obtained from the NIST model \cite{nistCryogenicMaterial}, shown in \Cref{fig:cu_k}. Copper exhibits high thermal conductivity across the full temperature range, with a pronounced increase at \SI{20}{\kelvin}. As a result, copper bias wires conduct significant heat into the 4\,K stage. When multiplied by the large number of bias wires in a 10k-qubit system, this effect becomes critical.

\subsubsection{Manganin Bias Wires}
\Figure[!ht](topskip=0pt, botskip=0pt, midskip=0pt)[width=0.999\columnwidth]
{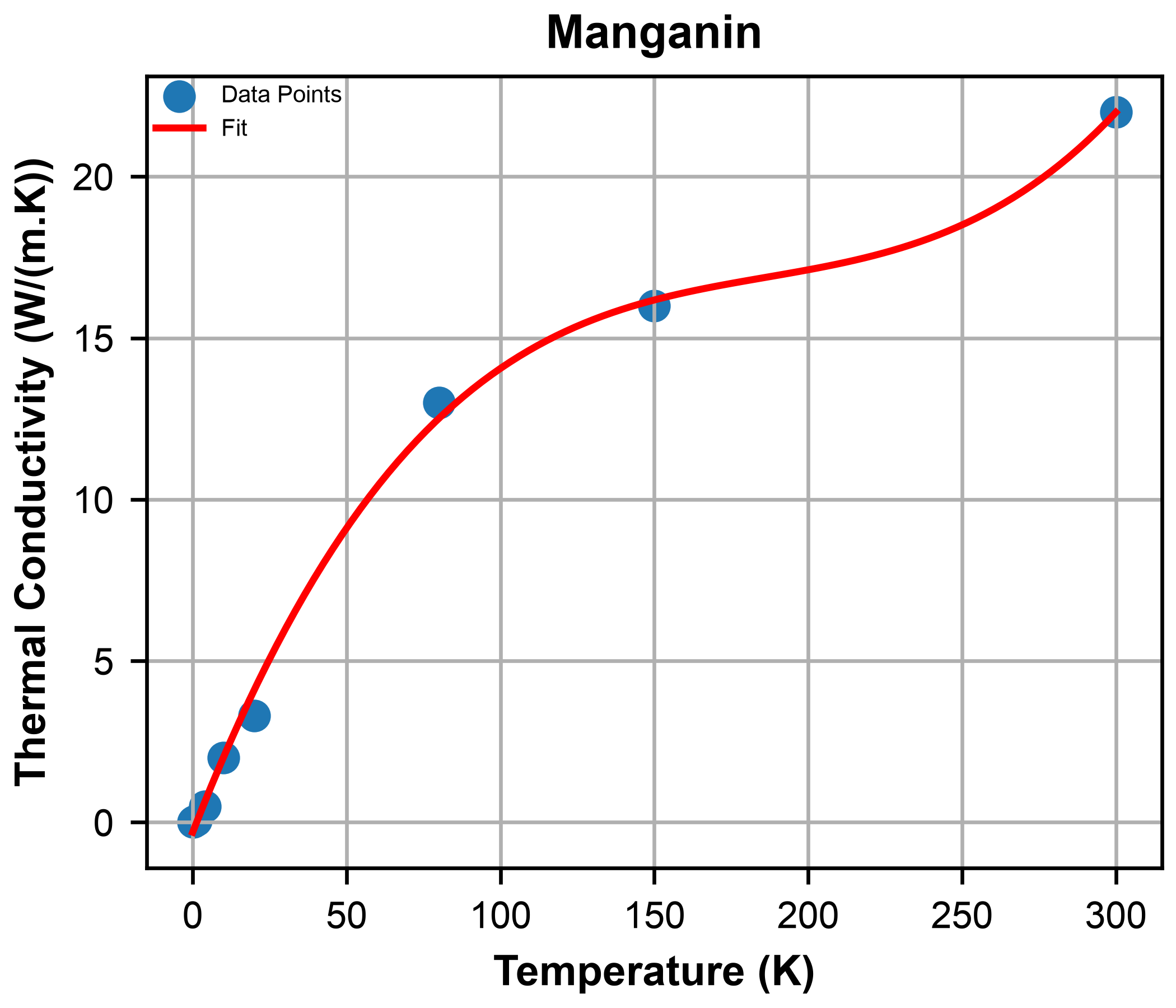}
{Thermal conductivity of Manganin (83\% Copper, 13\% Manganese, 4\% Nickel). Cubic fit using values from \cite{lakeshoreCryogenicWire}. The fit is consistent with trends in \cite{united1966thermal}.
\label{fig:mn_k}
}

Manganin (Mn) is frequently used in cryogenic instrumentation because of its lower thermal conductivity compared to copper \cite{ismael2021development}. We model its thermal conductivity $k_{Mn}$ using data reported in \cite{lakeshoreCryogenicWire} and apply a third-degree polynomial fit, as shown in \Cref{fig:mn_k}. The resulting curve matches the qualitative behavior reported in \cite{united1966thermal}. Across the relevant temperature range, $k_{Mn}$ is more than an order of magnitude lower than $k_{Cu}$. Consequently, Manganin bias wires substantially reduce conductive heat flow into both 50\,K and 4\,K stages relative to copper.

\subsubsection{YBCO-on-Kapton Bias Wires}
High-temperature superconducting YBCO cable interconnects provide an alternative to conventional metallic bias wires. 
We adopt a conservative estimate of the effective thermal conductance based on worst-case values reported in \cite{solovyov2021ybco}. In the most pessimistic case, the effective thermal conduction heat of YBCO-on-Kapton is reported as \qty{12}{\watt\per\meter\squared} for a \qty{1}{\meter} length. We assume the same cable geometry as in \cite{solovyov2021ybco} with width \qty{1}{\milli\meter} and thickness approximately \qty{101}{\micro\meter}. The conductive heat flow (in the most pessimistic case) is estimated using \Cref{eq:phl}. The heat conduction is dominated by the YBCO layer, whose thermal conductivity is orders of magnitude higher than that of the Kapton substrate \cite{solovyov2021ybco}. 

For consistency, all the values in \Cref{tab:biasing_cables} are computed assuming the inter-flange distances of the XLD1000sl refrigerator. 

\subsection{Electrical Resistivity Estimation}
\label[appsec]{app:ohl_details}

To estimate the OHL in wires carrying DC or low-frequency currents, we need to first estimate the effective resistance of these wires. Since the resistivity of these wires can vary significantly over the operating temperature range, we require a temperature-dependent resistivity model to calculate the effective resistance.

\subsubsection{Stainless Steel Coaxial Wires}
For the SS \leg{FLUX\_BIAS} cable, we use the effective lumped resistance values that were experimentally determined 
in \cite{krinner2019engineering}. In their experiments, they report that OHLs from RT to Still are negligible but significant on CP and MXC. Hence, we only consider the OHL due to SS \leg{FLUX\_BIAS} cable in CP and MXC and assume the effective resistances at higher temperature stages to be zero.

\subsubsection{Cupronickel Coaxial Wires}
For the cupronickel HDW cables, we use the temperature-dependent resistivity model reported in \cite{raicu2025cryogenic} to calculate equivalent lumped resistance values for temperatures above $\approx$ 4\,K. For temperatures at or below $\approx$ 4\,K, we assume the resistivity of the cables to be constant at \qty{9.928}{\nano\ohm-\meter} following the convention in \cite{raicu2025cryogenic}. Only the inner-conductor contribution of the coaxial cable is included when estimating DC ohmic loads because the outer conductor has a much lower resistance.

\subsubsection{Copper Resistivity Model}
\Figure[!ht](topskip=0pt, botskip=0pt, midskip=0pt)[width=0.95\columnwidth]
{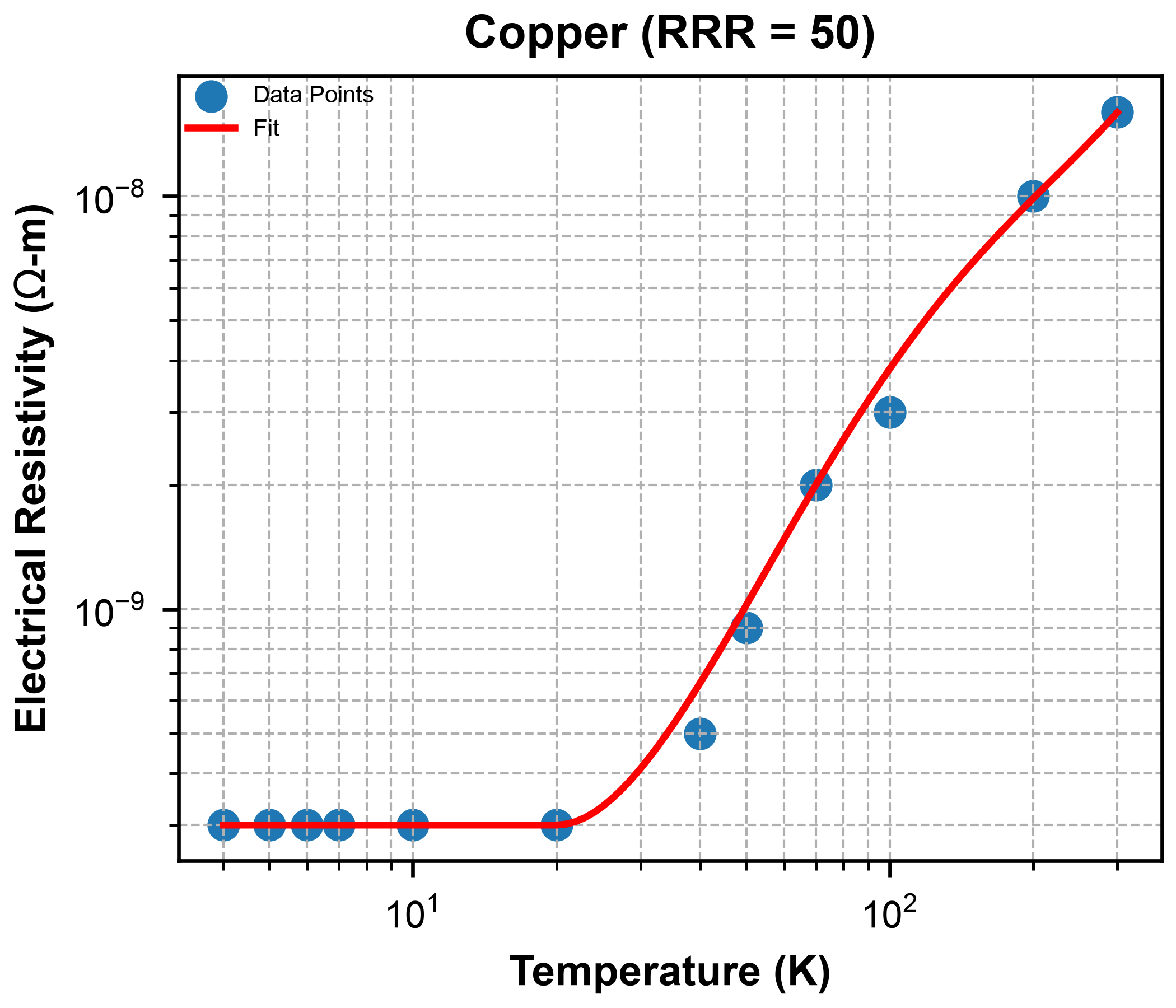}
{Modeled electrical resistivity of copper (RRR=50) based on data extracted from \cite{copperCryogenicProperties}. A piecewise approximation is used assuming constant resistivity below 20\,K.
\label{fig:cu_rho_v2}
}

We estimate the temperature-dependent resistivity of copper with RRR$=50$ using data extracted from \cite{copperCryogenicProperties}. Because the original data are not tabulated in a directly usable form, we construct an approximate model. The resistivity is treated as constant below \SI{20}{\kelvin}, and a fourth-degree polynomial fit in log-log space is applied above \SI{20}{\kelvin} over the range relevant to bias lines between 50\,K and 4\,K. The resulting model is shown in \Cref{fig:cu_rho_v2}. 

\subsubsection{Manganin Resistivity Model}
\Figure[!ht](topskip=0pt, botskip=0pt, midskip=0pt)[width=0.999\columnwidth]
{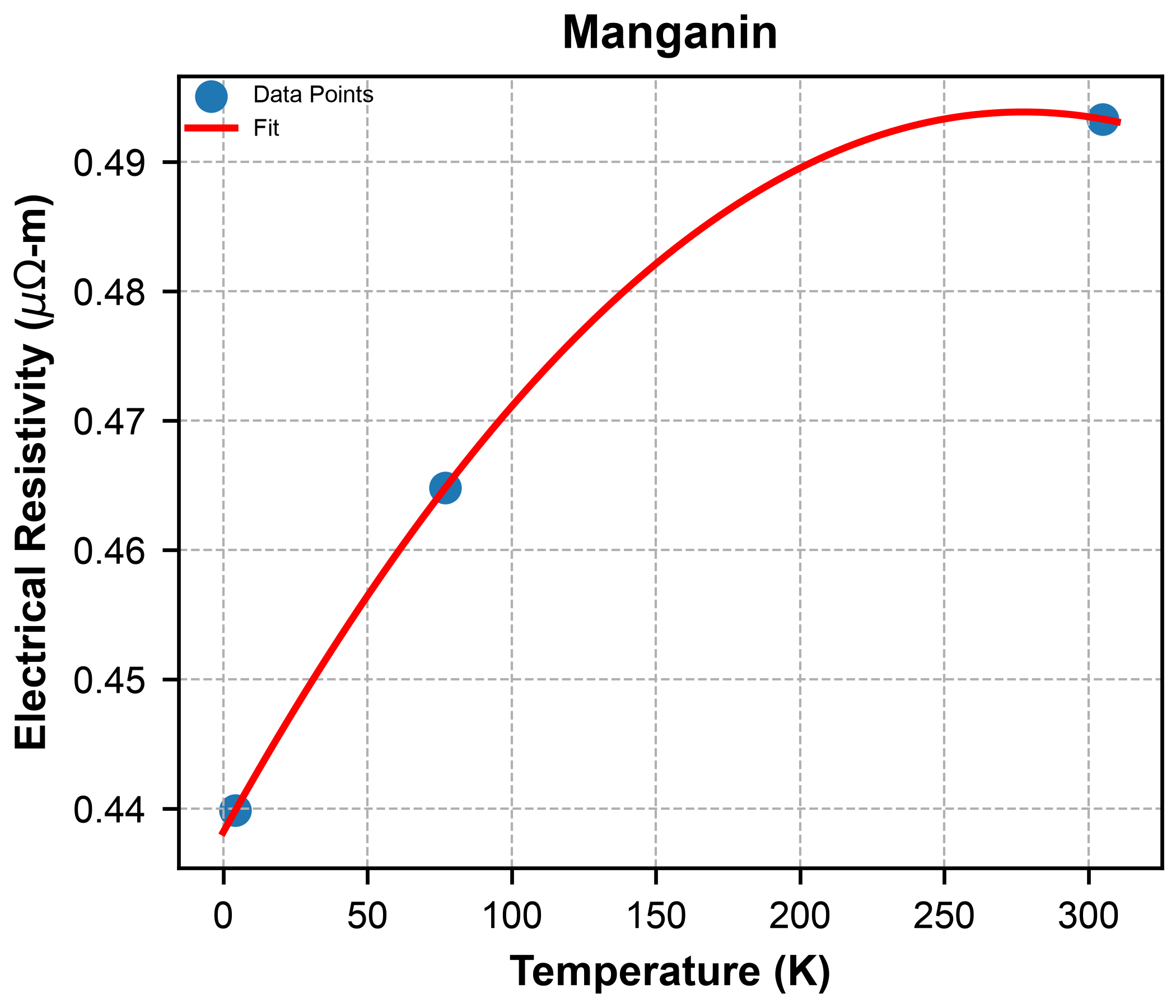}
{Electrical resistivity of Manganin as a function of temperature. Markers denote measured values from \cite{lakeshoreCryogenicWire}; the solid curve is the quadratic fit used in this work.
\label{fig:mn_rho}
}

Manganin is characterized by weak temperature dependence of resistivity. We model $\rho(T)$ using a quadratic fit based on values reported in \cite{lakeshoreCryogenicWire}. The fitted curve is shown in \Cref{fig:mn_rho}. The modeled trend is consistent with reports in \cite{xiang2020characterization,twire}.

\subsection{Relocation of the Directional-Coupler Termination}
\label[appsec]{app:dc_termination}

The pump tone required for TWPA operation is injected into the readout-signal path through a directional coupler. In our model, the pump power incident on the directional coupler is
\(P_{\mathrm{pump,in}}=-35\,\mathrm{dBm}\). For a coupling factor of \(C=20\,\mathrm{dB}\), the power delivered to the coupled port and, subsequently, to the TWPA is $P_{\mathrm{TWPA}}=-55\,\mathrm{dBm}$ consistent with the pump power assumed in \Cref{tab:mxc_power}.

Neglecting coupler insertion loss and impedance mismatch, approximately \(1-10^{-C/10}=0.99\) of the incident pump power exits the through port and is dissipated in its matched \(50\,\Omega\)
termination. The dissipated power is therefore

\begin{equation}
    P_{\mathrm{term}}
    =
    P_{\mathrm{pump,in}}
    \left(1-10^{-C/10}\right)
    \approx
    0.313\,\mu\mathrm{W}.
\end{equation}

For a matched \(50\,\Omega\) load, this power corresponds to an RMS current of

\begin{equation}
    I_{\mathrm{rms}}
    =
    \sqrt{\frac{P_{\mathrm{term}}}{50\,\Omega}}
    \approx
    79\,\mu\mathrm{A}.
\end{equation}

This dissipation constitutes an active heat load at the stage to which the termination is thermally anchored. For an XLD1000sl refrigerator with \(30\,\mu\mathrm{W}\) of nominal MXC cooling power, a single termination dissipating \(0.313\,\mu\mathrm{W}\). 
This termination load alone would be approximately $2500 \times \qty{0.313}{\micro\watt}  = \qty{0.78}{\milli\watt}$ for 10k qubits at four-way multiplexing. This is far above KIDE's assumed \qty{90}{\micro\watt} MXC budget.

The thermal load at the MXC may be reduced by relocating the termination to a warmer stage away from the MXC \cite{raicu2025cryogenic,gaydamachenko2025rf}. In this work, we evaluate a configuration in which the termination is thermally anchored to the 4\,K stage and connected via a superconducting cable \cite{ranadive2022kerr} to the TWPA at the MXC. Note that this configuration is treated here as a system-level design assumption rather than as a demonstrated configuration.

A commercial cryogenic directional coupler such as that reported in \cite{quantummicrowave} is supplied with an external \(50\,\Omega\) termination, making a remotely terminated configuration physically conceivable. The through port of the directional coupler is connected to the remote
termination using a NbTi superconducting transmission line. This additional line is denoted as \textbf{\leg{DC\_TERMINAL}}, and its passive heat load is explicitly included in the model. The line may be implemented using a coaxial or microstrip, provided that it maintains an approximately \(50\,\Omega\) characteristic impedance over the relevant pump-frequency range.

Relocating the termination introduces a potential thermal-noise path from the 4\,K stage toward the MXC. The directional coupler considered in \cite{quantummicrowave} has a nominal coupling factor of
\(20\,\mathrm{dB}\) and a minimum directivity of \(15\,\mathrm{dB}\). For the assumed port orientation, the isolation between the termination port and the qubit-facing coupled port is therefore approximately $35\,\mathrm{dB}$.

The mean thermal-photon occupation at temperature \(T\) and frequency \(f\) is given by the Bose-Einstein distribution

\begin{equation}
    \bar{n}_{\mathrm{th}}(T,f)
    =
    \frac{1}
    {\exp\left(hf/k_{\mathrm{B}}T\right)-1},
\label{eq:bose-einstein}
\end{equation}

At \(T=4.5\,\mathrm{K}\) (KIDE) and \(f=12\,\mathrm{GHz}\) \cite{gaydamachenko2025rf}, the thermal occupation at 4\,K stage is \(\bar{n}_{\mathrm{th}}\approx 7\). With \(35\,\mathrm{dB}\) of isolation, the photon leakage to the MXC is

\begin{equation}
    \bar{n}_{\mathrm{leak}}
    =
    \bar{n}_{\mathrm{th}} 10^{-35/10}
    \approx
    2.3\times10^{-3}.
\end{equation}

This value is below one photon on average but higher than the lower bound requirements estimated in \cite{krinner2019engineering}. It should therefore be interpreted as a first-order estimate rather than as evidence that the relocated termination is intrinsically compatible with qubit operation. Infrared filtering near the base-temperature stage is also required to suppress broadband high-frequency radiation propagating along the termination line. 

\subsection{DC Dissipation in a T-Pad Attenuator}
\label[appsec]{app:tpad}

\Figure[!ht](topskip=0pt, botskip=0pt, midskip=0pt)
[width=0.7\columnwidth]
{figures/T-pad.png}
{Circuit diagram of the symmetric T-pad attenuator used on low-frequency flux- and coupler-bias lines. Each series resistor has resistance $R_1$, and the shunt resistor connected to ground has resistance $R_2$.
\label{fig:tpad}
}

A symmetric T-pad attenuator consists of two identical series resistors, each with resistance $R_1$, and a shunt resistor $R_2$ connected to ground, as shown in \Cref{fig:tpad}. For an attenuator designed to provide an insertion loss of $A~\unit{\decibel}$ between equal source and load reference impedances $Z_0$, the corresponding linear voltage attenuation factor is $ k = 10^{A/20}$.

The resistor values required to obtain the specified attenuation while maintaining an input and output characteristic impedance of $Z_0$ under matched conditions are

\begin{align}
    R_1
    &=
    Z_0
    \left(
        \frac{k-1}{k+1}
    \right),
    \\
    R_2
    &=
    Z_0
    \left(
        \frac{2k}{k^2-1}
    \right)
\end{align}

We model the flux- and coupler-bias lines as using a \qty{20}{\decibel} T-pad designed for a reference impedance of $Z_0=\qty{50}{\ohm}$. The corresponding attenuation factor is $k=10$, giving $R_1 \approx \qty{40.91}{\ohm}$ and $R_2 \approx \qty{10.10}{\ohm}$.

The resistor values above describe the attenuator under matched AC conditions. In the steady-state DC limit, we approximate the second series resistor $R_1$ to be connected to ground through a short circuit because the bias line terminates in a superconducting on-chip bias structure (whose resistance and reactance is assumed negligible relative to $R_1$ and $R_2$).

Let $I_{in}$ denote the current entering the T-pad and $I_{out}$ the current delivered through the output branch. Applying current division at the central node gives

\begin{equation}
\label{eq:current_ratio}
    \frac{I_{out}}{I_{in}}
    =
    \frac{R_2}{R_1+R_2}
    =
    \frac{2k}{k^2+1}
\end{equation}

For $k=10$,

\begin{equation}
    \frac{I_{out}}{I_{in}}
    =
    \frac{20}{101}
    \approx
    0.198
\end{equation}

Thus, approximately 19.8\% of the current entering the attenuator is
delivered to the output load, whereas approximately 80.2\% is shunted
to ground through $R_2$. 

The total Joule power dissipated by the three resistors is

\begin{equation}
\begin{split}
    P_{\mathrm{Tpad}}
    ={}&
    I_{in}^2 R_1
    +
    \left(I_{in}-I_{out}\right)^2 R_2
    +
    I_{out}^2 R_1
\end{split}
\label{eq:tpad_power}
\end{equation}

Substituting the current relation from \Cref{eq:current_ratio} gives

\begin{equation}
\begin{split}
    P_{\mathrm{Tpad}}
    = I_{out}^2
    \Bigg[
        &
        \left(
            \frac{R_1+R_2}{R_2}
        \right)^2 R_1
        +
        \left(
            \frac{R_1}{R_2}
        \right)^2 R_2
        +
        R_1
    \Bigg]
\end{split}
\end{equation}

We define an effective lumped resistance $R_{\mathrm{eff}}$ such that

\begin{equation}
    P_{\mathrm{Tpad}}
    =
    I_{\mathrm{out}}^2 R_{\mathrm{eff}}.
\end{equation}

For a symmetric T-pad, this effective resistance can also be written directly in terms of $Z_0$ and $k$ as

\begin{equation}
\label{eq:tpad_reff}
    R_{\mathrm{eff}}
    =
    Z_0
    \frac{k^4-1}{4k^2}.
\end{equation}

For $Z_0=\qty{50}{\ohm}$ and $k=10$,

\begin{equation}
    R_{\mathrm{eff}}
    =
    \qty{1249.875}{\ohm}
    \approx
    \qty{1.25}{\kilo\ohm}.
\end{equation}

For a required output bias current of $I_{\mathrm{out}}=\qty{0.4}{\milli\ampere}$, as listed in \Cref{tab:DC_currents}, the corresponding input current is approximately \qty{2.02}{\milli\ampere}, and the total power dissipated by the T-pad is \qty{0.20}{\milli\watt}. For continuously applied flux bias, this value represents a steady heat load at the 4\,K stage. 

\subsection{SIS-Mixer Amplifier}
\label[appsec]{app:sis_mixer}
\begin{table}[t]
\centering
\caption{Power consumption of SIS-Mixer amplifier components. Total power is dominated by the local oscillator.}
\label{tab:sis_power}
\begin{tabular}{lcccl}
\toprule
\textbf{Component} 
& {\textbf{Voltage (\unit{\milli\volt})}} 
& {\textbf{Current (\unit{\milli\ampere})}} 
& {\textbf{Power (\unit{\micro\watt})}} 
& \textbf{Ref.} \\
\midrule
SIS-up   
& 9.3   
& 0.025    
& 0.23  
& \cite{kojima2023characterization} \\
SIS-down 
& 10.5  
& 0.11   
& 1.16  
& \cite{kojima2023characterization} \\
LO       
& 0.305  
& 22.0  
& 6.71  
& \cite{murayama2024fabrication} \\
\midrule
\textbf{Total} 
& {} 
& {} 
& \textbf{8.10} 
& {} \\
\bottomrule
\end{tabular}
\end{table}

SIS-Mixer amplifiers \cite{kojima2023characterization} use two cascaded SIS frequency converters and a local oscillator (LO). The SIS-up stage upconverts the signal to about \qty{87.5}{\giga\hertz}, and the SIS-down stage downconverts it back to the original band. This frequency conversion provides gain. Both SIS converters and the LO are continuously biased during operation. The active power of each component is calculated as the product of its operating voltage and current, as summarized in \Cref{tab:sis_power}. Most of the dissipation is from the local oscillator (LO) needed for frequency conversion. Although the total active power of the SIS-Mixer amplifier is significantly lower than that of a HEMT, it requires a larger number of bias lines which increase PHL. In addition, the LO lines carry relatively high currents (approximately \qty{22}{\milli\ampere}), which can introduce substantial ohmic heating in the associated bias wiring.

\begin{table}[tbp]
\centering
\caption{Acronyms used in this work.}
\label{tab:acryonms}
\begin{tabular}{ll}
\hline
\textbf{Acronym} & \textbf{Definition} \\
\hline
1Q & Single-Qubit \\
2Q & Two-Qubit \\
Ag & Silver \\
AHL & Active Heat Load \\
AWG & American Wire Gauge \\
CP & Cold Plate \\
Cu & Copper \\
DR & Dilution Refrigerator \\
ESM & Error-Syndrome Measurement \\
FC & Fixed-Coupling \\
FF & Fixed-Frequency \\
FTQC & Fault-Tolerant Quantum Computing \\
HDW & High-Density Wiring \\
HEMT & High Electron Mobility Transistor \\
IR & Infrared \\
LNA & Low-Noise Amplifier \\
LO & Local Oscillator \\
LQ & Logical Qubit \\
MXC & Mixing Chamber \\
NbTi & Niobium-Titanium \\
NHL & Normalized Heat Load \\
OHL & Ohmic Heat Load \\
PD & Photodetector \\
PHL & Passive Heat Load \\
PQ & Physical Qubit \\
RRR & Residual-Resistance Ratio \\
RT & Room Temperature \\
SIS & Superconductor-Insulator-Superconductor \\
SNR & Signal-to-Noise Ratio \\
SS & Stainless Steel \\
TC & Tunable-Coupling \\
TF & Tunable-Frequency \\
THL & Total Heat Load \\
TWPA & Traveling Wave Parametric Amplifier \\
ULP-HEMT & Ultra-Low-Power High Electron Mobility Transistor \\
YBCO & Yttrium Barium Copper Oxide \\
\hline
\end{tabular}
\end{table}

\begin{table*}[pt]
\centering
\caption{Taxonomy of cryogenic heat loads modeled in this work.}
\label{tab:heat_loads}

\resizebox{\textwidth}{!}{%
\begin{threeparttable}
\begin{tabular}{@{} >{\raggedright\arraybackslash}p{3.25cm} c l >{\raggedright\arraybackslash}p{10cm} @{}}
\toprule
\textbf{Cable} / Heat Source & 
  \textbf{Operation} & 
  \textbf{Label} & 
  \textbf{Description} \\ \midrule

\textbf{DRIVE} & & & \\
\quad Thermal Conduction & - & DRIVE | PASSIVE | IDLE & Microwave signals for qubit control \\
\addlinespace
\quad Attenuator Dissipation & 1Q & DRIVE | ATT | 1Q & \\
                             & 2Q & DRIVE | ATT | 2Q & FF-FC systems only \\
\addlinespace
\quad Photodetector & 1Q & DRIVE | PD | 1Q & Optical Systems only \\
                     & 2Q & DRIVE | PD | 2Q & \\ 
\midrule

\textbf{FLUX\_BIAS} & & & \\
\quad Thermal Conduction & - & FLUX\_BIAS | PASSIVE | IDLE & Tune qubit frequency (TF systems only) \\
\addlinespace
\quad Attenuator Dissipation & - & FLUX\_BIAS | BIAS\_RESISTOR\_4K | IDLE & 20 dB attenuation at 4\,K (modeled using lumped resistance) \tnote{$\star$} \\
\addlinespace
\quad Joule Heating & - & FLUX\_BIAS | BIAS\_RESISTOR\_Still | IDLE & Cable resistance to flux biasing currents (modeled using lumped resistances) \\
                    & - & FLUX\_BIAS | BIAS\_RESISTOR\_CP | IDLE & \\
                    & - & FLUX\_BIAS | BIAS\_RESISTOR\_MXC | IDLE & \\ 
\midrule

\textbf{COUPLER} & & & \\
\quad Thermal Conduction & - & COUPLER | PASSIVE | IDLE & Activate coupling between qubits (TC systems only) \tnote{$\dagger$} \\ \\
\addlinespace
\quad Attenuator Dissipation & 2Q & COUPLER | RESISTOR\_2Q\_4K | 2Q & 20 dB attenuation at 4\,K (modeled using lumped resistance) \tnote{$\star$} \\
\addlinespace
\quad Joule Heating & 2Q & COUPLER | RESISTOR\_2Q\_Still | 2Q & Cable resistance to coupler biasing currents (modeled using lumped resistances) \\
                    & 2Q & COUPLER | RESISTOR\_2Q\_CP | 2Q & \\
                    & 2Q & COUPLER | RESISTOR\_2Q\_MXC | 2Q & \\ 
\midrule

\textbf{PUMP} & & & \\
\quad Thermal Conduction & - & PUMP | PASSIVE | IDLE & Pump tone for TWPA amplification \\
\addlinespace
\quad Attenuator Dissipation & Readout & PUMP | ATT | READOUT & \\ 
\midrule

\textbf{READOUT\_PIN} & & & \\
\quad Thermal Conduction & - & READOUT\_PIN | PASSIVE | IDLE & Measurement Probe-In signal \\
\addlinespace
\quad Attenuator Dissipation & Readout & READOUT\_PIN | ATT | READOUT & \\ 
\midrule

\textbf{READOUT\_POUT} & & & \\
\quad Thermal Conduction & - & READOUT\_POUT | PASSIVE | IDLE & Measurement output (reflected) signal \\ 
\midrule

\textbf{DC\_TERMINAL} & & & \\
\quad Thermal Conduction & - & DC\_TERMINAL | PASSIVE | IDLE & Cable connecting impedance termination at 4\,K to directional coupler port \\ 
\midrule

\textbf{AMP\_BIAS} & & & \\
\quad Thermal Conduction & - & AMP\_BIAS | PASSIVE | IDLE & Biasing wires for readout amplifiers at 4\,K \\
\addlinespace
\quad Readout Amplifier at 4\,K & - & AMP\_BIAS | AMP | IDLE & Active dissipation from amplifiers at 4\,K \\
\addlinespace
\quad Joule Heating & - & AMP\_BIAS | AMP\_OHMIC | IDLE & OHL (Biasing wires for amplifiers at 4\,K) \\ 
\midrule

\textbf{AMP\_BIAS\_50K} & & & \\
\quad Thermal Conduction & - & AMP\_BIAS\_50K | PASSIVE | IDLE & Biasing wires for readout amplifiers at 50\,K \\
\addlinespace
\quad Readout Amplifier at 50K & - & AMP\_BIAS\_50K | AMP | IDLE & Active dissipation from amplifiers at 50\,K \\
\addlinespace
\quad Joule Heating & - & AMP\_BIAS\_50K | AMP\_OHMIC | IDLE & OHL (Biasing wires for amplifiers at 50\,K) \\ 
\bottomrule
\end{tabular}

\begin{tablenotes}
    \item[$\star$] The Joule heating in the wires can be ignored because it is negligible compared to the heat dissipated by the attenuator and the cooling power of the 4\,K stage.
    \item[$\dagger$] In \cite{krinner2019engineering}, 2Q operations are realized by applying low-frequency flux pulses through the \leg{FLUX\_BIAS} lines. As a result, the same physical line is responsible for both static flux biasing (\leg{IDLE}) and dynamic 2Q gates (\leg{2Q}). The ohmic heat load associated with the currents used during 2Q operations is attributed to a separate \leg{COUPLER}-like abstraction, even though no distinct physical line exists in that system. This conceptual separation applies only to the OHL associated with 2Q operations in \cite{krinner2019engineering}. No additional passive heat load (PHL) is introduced. This approach ensures consistency when comparing architectures that use dedicated \leg{COUPLER} lines with those that reuse \leg{FLUX\_BIAS} lines for 2Q control.
\end{tablenotes}

\end{threeparttable}
}
\end{table*}

\clearpage
\onecolumn
\section{Modeling Parameters and Results}
The complete configuration files and notebooks used to generate each figure are available at \url{https://github.com/shaswot/SCQ_heatmodel}.

\subsection{TF-FC system | SS wires | XLD400 fridge | HEMT \(\times\) Copper}
\label[appsec]{app:tf-fc-ss-xld400}
In \Cref{subsec:krinner}, we evaluate the thermal scalability of the system proposed in \cite{krinner2019engineering}. The original study analyzed the heat breakdown for a 50-qubit system which consumed about one-third of the maximum cooling capacity (due to space constraints of the fridge). The following evaluation scales this system to consume all available cooling power and thereby increase the qubit count to about 150 qubits. 

The system is based on a TF-FC architecture using SS coaxial cables in an XLD400 dilution refrigerator. We adopt the attenuator configuration specific to \cite{krinner2019engineering} from \Cref{tab:attenuator}. The impedance termination for the directional coupler is relocated to 4\,K using the \textbf{\leg{DC\_TERMINAL}} NbTi coaxial cable. We assume a $4 \times$ simultaneous readout (original study assumed $6 \times$). This assumption results in exactly two readout chains for one 8-qubit cell, avoiding fractional allocations and simplifying analysis. Each readout chain requires one LNF8C HEMT with three biasing wires.

Since this is a TF system, we use the CZ-based ESM circuit as the representative workload. 2Q operations are realized by applying low-frequency flux pulses through the \leg{FLUX\_BIAS} lines. As a result, the same physical line is responsible for both static flux biasing (\leg{IDLE}) and dynamic 2Q gates (\leg{2Q}). The currents required for flux biasing and 2Q operations are listed in \Cref{tab:DC_currents}.

The ohmic heat load associated with the currents used during 2Q operations is attributed to a separate \leg{COUPLER}-like abstraction, even though no distinct physical line exists in that system. This conceptual separation applies only to the OHL associated with 2Q operations. No additional PHL is introduced. This approach ensures consistency when comparing architectures that use dedicated \leg{COUPLER} lines with those that reuse \leg{FLUX\_BIAS} lines for 2Q control.

\begin{table}[!ht]
\small
\centering
\caption{System configuration for a unit cell of 8 qubits for the system parameters based on \cite{krinner2019engineering} discussed in \Cref{subsec:krinner}.}

\begin{threeparttable}
\resizebox{\textwidth}{!}{%
\begin{tabular}{lccllllll}
\toprule
\textbf{Cable}
 & \textbf{Operations} 
 & \textbf{Count} 
 & \textbf{50\,K} 
 & \textbf{4\,K} 
 & \textbf{Still} 
 & \textbf{CP} 
 & \textbf{MXC} 
 \\
\midrule
\textbf{DRIVE} 
& 1Q
& 8 
& SS Drive  
& SS Drive (20 dB) 
& SS Drive 
& SS Drive (20 dB) 
& SS Drive (20 dB) 
\\
\textbf{FLUX\_BIAS} 
& Idle, 2Q 
& 8 
& SS Flux  
& SS Flux (20 dB $\approx \qty{1.25}{\kilo\ohm}$) 
& SS Flux  
& SS Flux (\qty{0.42}{\ohm})
& SS Flux (\qty{0.15}{\ohm})
\\
\textbf{PUMP} 
& Readout 
& 2 
& SS Drive  
& SS Drive (20 dB) 
& SS Drive  
& SS Drive (10 dB) 
& SS Drive (20 dB)
\\
\textbf{READOUT\_PIN} 
& Readout 
& 2 
& SS Drive  
& SS Drive (20 dB) 
& SS Drive
& SS Drive (20 dB) 
& SS Drive (20 dB) 
\\
\textbf{READOUT\_POUT} 
& Readout 
& 2 
& SS Drive  
& SS Drive  
& NbTi (coax) 
& NbTi (coax) 
& NbTi (coax) 
\\
\textbf{DC\_TERMINAL} 
& Idle 
& 2 
& -  
& -  
& NbTi (coax) 
& NbTi (coax) 
& NbTi (coax) 
\\

\textbf{AMP\_BIAS} 
& Idle  
& 6
& Cu-35
& Cu-35  
\\

\quad \textit{Amplifier} 
& Idle
& 2
& -
& LNF8C
\\
\addlinespace
\midrule
\textbf{Fridge:} XLD400 \\
\quad \textit{Cooling Power} &
& &
\qty{30}{\watt} & 
\qty{1.5}{\watt} & 
\qty{40}{\milli\watt} & 
\qty{200}{\micro\watt} & 
\qty{19}{\micro\watt} &    
\\

\bottomrule
\end{tabular}
}

\end{threeparttable}
\end{table}

\begin{table}[]
\centering
\caption{Operating Parameters.}
\begin{tabular}{@{}lcccr@{}}
\toprule
\textbf{Cable/Operation} & \textbf{Ops. per ESM cycle} & \textbf{$P_{MXC}$} & \textbf{Current (mA)} & \textbf{Duration (ns)} \\ 
\midrule
\textbf{DRIVE} \\
\quad 1Q                        & 16  & -71 dBm  & -     & 25 \\
\textbf{FLUX\_BIAS}              &     & -        & 0.57   & -   \\
\quad 2Q                        & 16  & -        & 0.4   & 42  \\
\textbf{READOUT\_PIN} \\
\quad Readout                   & 4   & -120 dBm & -     & 375   \\
\textbf{PUMP} \\
\quad Readout                   & 4   & -55 dBm  & -     & 375   \\ 
\bottomrule
\end{tabular}
\end{table}

\Figure[!b](topskip=0pt, botskip=0pt, midskip=0pt)[width=0.99\textwidth]
{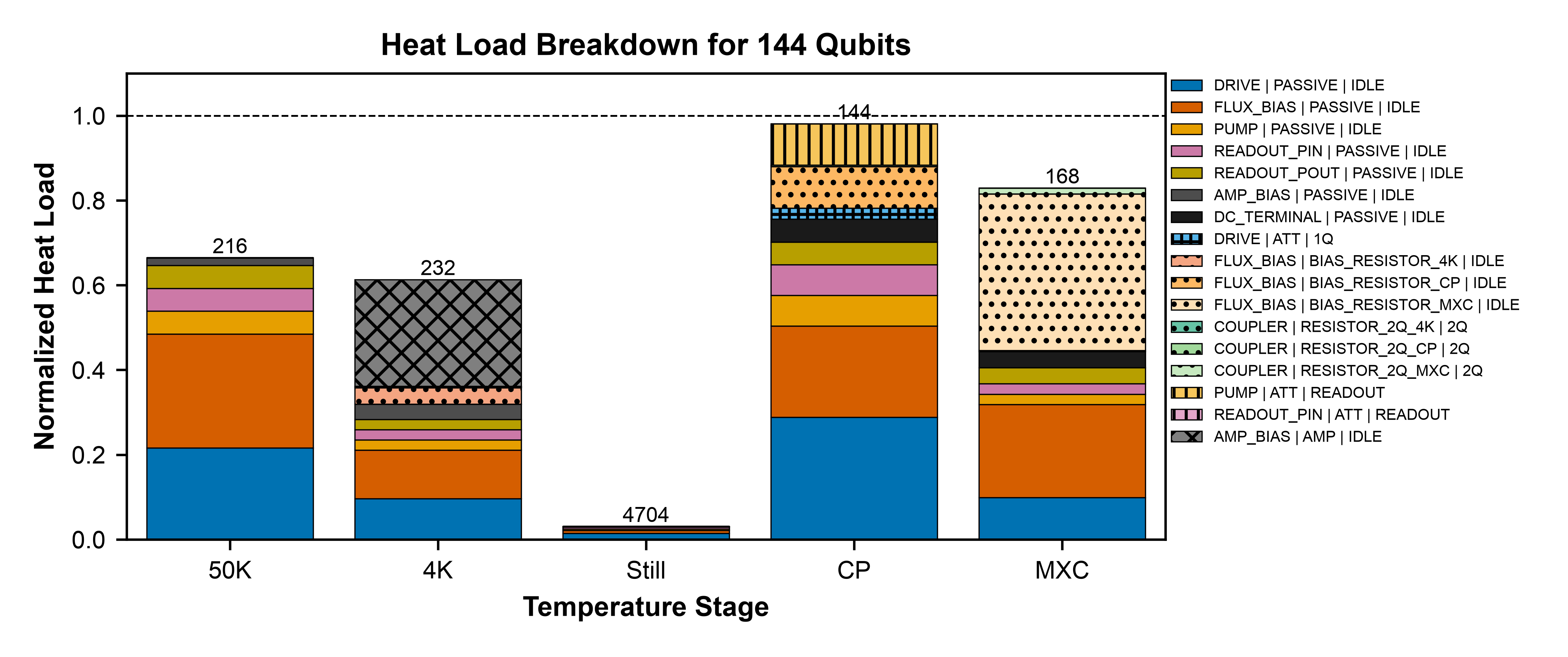}
{\parbox[t]{0.9\linewidth}{Normalized Heat Load Breakdown}
\label{fig:tf-fc-ss-xld400}}
\clearpage

\subsection{TF-FC system | Cri/oFlex wires | XLD400 fridge | HEMT \(\times\) Copper}
\label[appsec]{app:tf-fc-delft-xld400}

In \Cref{subsec:krinner}, we evaluate a system identical to the one in \Cref{app:tf-fc-ss-xld400} but using Delft Cri/oFlex microstrip cables which have lower PHL than SS. The modeling parameters are reported below. 

\begin{table*}[!ht]
\centering
\caption{System Configuration for a unit cell of 8 qubits for the system parameters based on \cite{krinner2019engineering} using Cri/oFlex cabling instead of SS cabling discussed in \Cref{subsec:krinner}.}

\begin{threeparttable}
    
\begin{tabular}{lccllllll}
\toprule
\textbf{Cable}
 & \textbf{Operations} 
 & \textbf{Count} 
 & \textbf{50\,K} 
 & \textbf{4\,K} 
 & \textbf{Still} 
 & \textbf{CP} 
 & \textbf{MXC} 
 \\
\midrule
\textbf{DRIVE} 
& 1Q
& 8 
& Ag  
& Ag (20 dB) 
& Ag 
& Ag (20 dB) 
& Ag (20 dB) 
\\
\textbf{FLUX\_BIAS} 
& Idle, 2Q 
& 8 
& Ag  
& Ag (20 dB $\approx \qty{1.25}{\kilo\ohm}$) 
& NbTi  
& NbTi (\qty{0.42}{\ohm})
& NbTi (\qty{0.15}{\ohm})
\\
\textbf{PUMP} 
& Readout 
& 2 
& Ag  
& Ag (20 dB) 
& Ag  
& Ag (10 dB) 
& Ag (20 dB) 
\\
\textbf{READOUT\_PIN} 
& Readout 
& 2 
& Ag  
& Ag (20 dB) 
& Ag
& Ag (20 dB) 
& Ag (20 dB) 
\\
\textbf{READOUT\_POUT} 
& Readout 
& 2 
& Ag  
& Ag  
& NbTi  
& NbTi  
& NbTi  
\\
\textbf{DC\_TERMINAL} 
& Idle 
& 2 
& -  
& -  
& NbTi  
& NbTi  
& NbTi  
\\

\textbf{AMP\_BIAS} 
& Idle  
& 6
& Cu-35
& Cu-35  
\\

\quad \textit{Amplifier} 
& Idle
& 2
& -
& LNF8C
\\

\addlinespace
\midrule
\textbf{Fridge:} XLD400 \\
\quad \textit{Cooling Power} &
& &
\qty{30}{\watt} & 
\qty{1.5}{\watt} & 
\qty{40}{\milli\watt} & 
\qty{200}{\micro\watt} & 
\qty{19}{\micro\watt} & 
\\
\bottomrule
\end{tabular}


\end{threeparttable}
\end{table*}

\begin{table}[]
\centering
\caption{Operating Parameters.}
\begin{tabular}{@{}lcccr@{}}
\toprule
\textbf{Cable/Operation} & \textbf{Ops. per ESM cycle} & \textbf{$P_{MXC}$} & \textbf{Current (mA)} & \textbf{Duration (ns)} \\ 
\midrule
\textbf{DRIVE} \\
\quad 1Q                        & 16  & -71 dBm  & -     & 25 \\
\textbf{FLUX\_BIAS}              &     & -        & 0.57   & -   \\
\quad 2Q                        & 16  & -        & 0.4   & 42  \\
\textbf{READOUT\_PIN} \\
\quad Readout                   & 4   & -120 dBm & -     & 375   \\
\textbf{PUMP} \\
\quad Readout                   & 4   & -55 dBm  & -     & 375   \\ 
\bottomrule
\end{tabular}
\end{table}

\Figure[!b](topskip=0pt, botskip=0pt, midskip=0pt)[width=0.99\textwidth]
{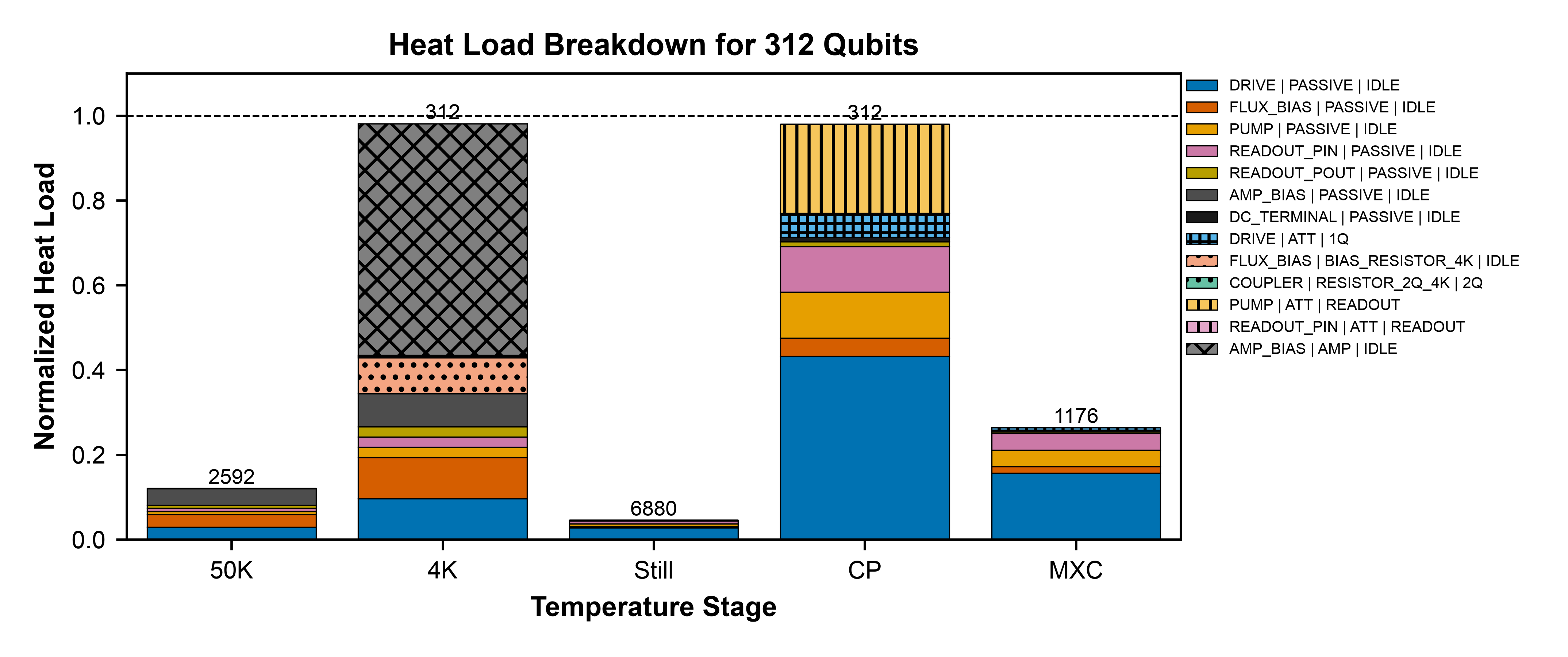}
{\parbox[t]{0.9\linewidth}{Normalized Heat Load Breakdown}
\label{fig:tf-fc-delft-xld400}}
\clearpage

\subsection{TF-FC system | SS wires | KIDE fridge | HEMT \(\times\) Copper}
\label[appsec]{app:tf-fc-ss-kide}

In \Cref{subsec:krinner}, we evaluate a system identical to the one in \Cref{app:tf-fc-ss-xld400} but assume a KIDE fridge with much larger cooling power compared to the original XLD400. The modeling parameters are reported below.

\begin{table*}[!ht]
\centering
\caption{System Configuration for a unit cell of 8 qubits for the system parameters based on \cite{krinner2019engineering} using KIDE fridge discussed in \Cref{subsec:krinner}.}

\begin{threeparttable}
    
\begin{tabular}{lccllllll}
\toprule
\textbf{Cable}
 & \textbf{Operations} 
 & \textbf{Count} 
 & \textbf{50\,K} 
 & \textbf{4\,K} 
 & \textbf{Still} 
 & \textbf{CP} 
 & \textbf{MXC} 
 \\
\midrule
\textbf{DRIVE} 
& 1Q
& 8 
& SS Drive  
& SS Drive (20 dB) 
& SS Drive 
& SS Drive (20 dB) 
& SS Drive (20 dB) 
\\
\textbf{FLUX\_BIAS} 
& Idle, 2Q 
& 8 
& SS Flux  
& SS Flux (20 dB $\approx \qty{1.25}{\kilo\ohm}$) 
& SS Flux  
& SS Flux (\qty{0.42}{\ohm})
& SS Flux (\qty{0.15}{\ohm})
\\
\textbf{PUMP} 
& Readout 
& 2 
& SS Drive  
& SS Drive (20 dB) 
& SS Drive  
& SS Drive (10 dB) 
& SS Drive (20 dB)
\\
\textbf{READOUT\_PIN} 
& Readout 
& 2 
& SS Drive  
& SS Drive (20 dB) 
& SS Drive
& SS Drive (20 dB) 
& SS Drive (20 dB) 
\\
\textbf{READOUT\_POUT} 
& Readout 
& 2 
& SS Drive  
& SS Drive  
& NbTi (coax) 
& NbTi (coax) 
& NbTi (coax) 
\\
\textbf{AMP\_BIAS} 
& Idle  
& 6
& Cu-35
& Cu-35  
\\

\quad \textit{Amplifier} 
& Idle
& 2
& -
& LNF8C
\\

\textbf{DC\_TERMINAL} 
& Idle  
& 2
& -
& -
& NbTi (coax)
& NbTi (coax)
& NbTi (coax)
\\

\addlinespace
\midrule
\textbf{Fridge:} KIDE \\
\quad \textit{Cooling Power} &
& &
\qty{90}{\watt} & 
\qty{6}{\watt} & 
\qty{90}{\milli\watt} & 
\qty{3}{\milli\watt} & 
\qty{90}{\micro\watt} 
\\
\bottomrule
\end{tabular}


\end{threeparttable}
\end{table*}

\begin{table}[]
\centering
\caption{Operating Parameters.}
\begin{tabular}{@{}lcccr@{}}
\toprule
\textbf{Cable/Operation} & \textbf{Ops. per ESM cycle} & \textbf{$P_{MXC}$} & \textbf{Current (mA)} & \textbf{Duration (ns)} \\ 
\midrule
\textbf{DRIVE} \\
\quad 1Q                        & 16  & -71 dBm  & -     & 25 \\
\textbf{FLUX\_BIAS}              &     & -        & 0.57   & -   \\
\quad 2Q                        & 16  & -        & 0.4   & 42  \\
\textbf{READOUT\_PIN} \\
\quad Readout                   & 4   & -120 dBm & -     & 375   \\
\textbf{PUMP} \\
\quad Readout                   & 4   & -55 dBm  & -     & 375   \\ 
\bottomrule
\end{tabular}
\end{table}

\Figure[!b](topskip=0pt, botskip=0pt, midskip=0pt)[width=0.99\textwidth]
{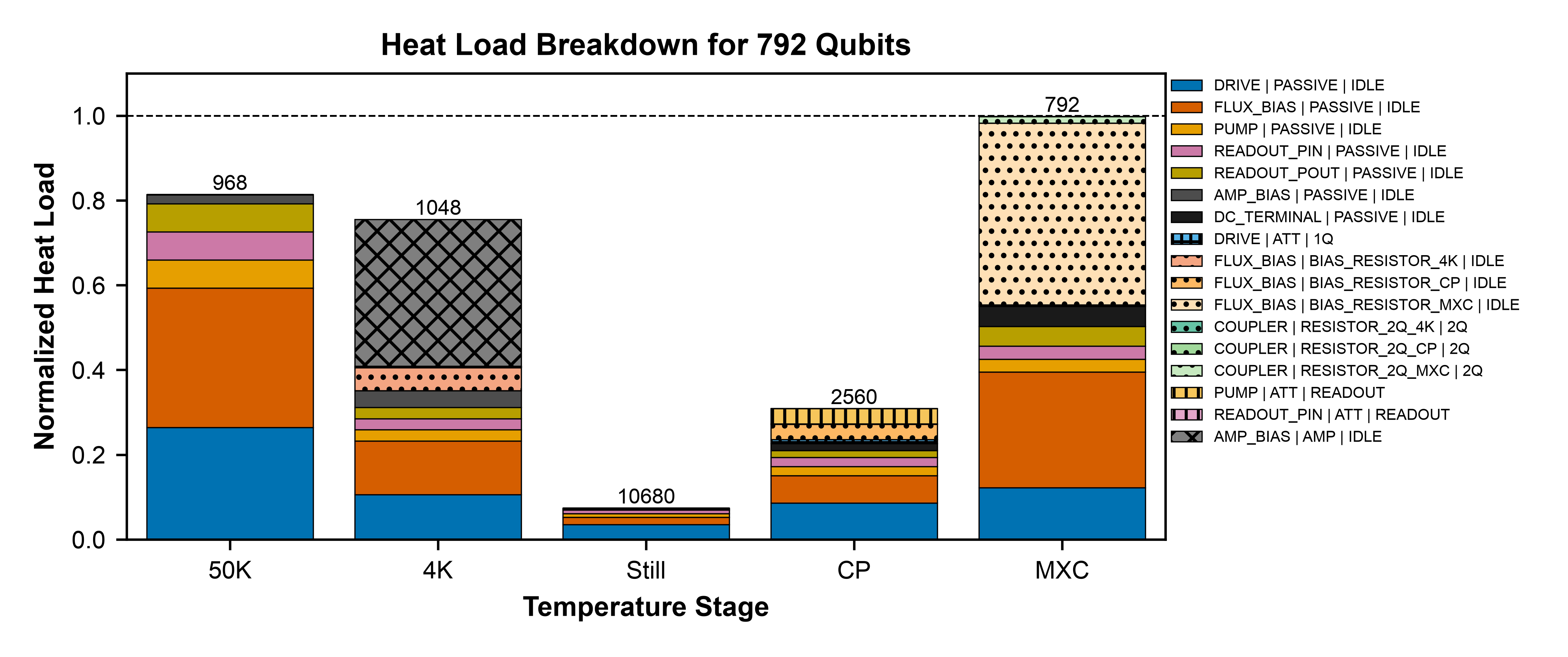}
{\parbox[t]{0.9\linewidth}{Normalized Heat Load Breakdown}
\label{fig:tf-fc-ss-kide}}
\clearpage

\subsection{TF-FC system | Cri/oFlex wires | KIDE fridge | HEMT \(\times\) Copper}
\label[appsec]{app:tf-fc-delft-kide}

In \Cref{subsec:krinner}, we evaluate the combined effects of using lower PHL Delft Cri/oFlex microstrip cabling and a stronger KIDE fridge in the original system in \cite{krinner2019engineering} (\Cref{app:tf-fc-ss-xld400}). The modeling parameters are reported below.

\begin{table*}[!ht]
\centering
\caption{System Configuration for a unit cell of 8 qubits for the system parameters based on \cite{krinner2019engineering} using Cri/oFlex wiring and KIDE Fridge as discussed in \Cref{subsec:krinner}.}

\begin{threeparttable}
    
\begin{tabular}{lccllllll}
\toprule
\textbf{Cable}
 & \textbf{Operations} 
 & \textbf{Count} 
 & \textbf{50\,K} 
 & \textbf{4\,K} 
 & \textbf{Still} 
 & \textbf{CP} 
 & \textbf{MXC} 
 \\
\midrule
\textbf{DRIVE} 
& 1Q
& 8 
& Ag  
& Ag (20 dB) 
& Ag 
& Ag (20 dB) 
& Ag (20 dB) 
\\
\textbf{FLUX\_BIAS} 
& Idle, 2Q 
& 8 
& Ag  
& Ag (20 dB $\approx \qty{1.25}{\kilo\ohm}$) 
& NbTi  
& NbTi (\qty{0.42}{\ohm})
& NbTi (\qty{0.15}{\ohm})
\\
\textbf{PUMP} 
& Readout 
& 2 
& Ag  
& Ag (20 dB) 
& Ag  
& Ag (10 dB) 
& Ag (20 dB)
\\
\textbf{READOUT\_PIN} 
& Readout 
& 2 
& Ag  
& Ag (20 dB) 
& Ag
& Ag (20 dB) 
& Ag (20 dB) 
\\
\textbf{READOUT\_POUT} 
& Readout 
& 2 
& Ag  
& Ag  
& NbTi  
& NbTi  
& NbTi  
\\
\textbf{DC\_TERMINAL} 
& Idle 
& 2 
& -  
& -  
& NbTi  
& NbTi  
& NbTi  
\\

\textbf{AMP\_BIAS} 
& Idle  
& 6
& Cu-35
& Cu-35  
\\

\quad \textit{Amplifier} 
& Idle
& 2
& -
& LNF8C
\\

\addlinespace
\midrule
\textbf{Fridge:} KIDE \\
\quad \textit{Cooling Power} &
& &
\qty{90}{\watt} & 
\qty{6}{\watt} & 
\qty{90}{\milli\watt} & 
\qty{3}{\milli\watt} & 
\qty{90}{\micro\watt} 
\\
\bottomrule
\end{tabular}


\end{threeparttable}
\end{table*}

\begin{table}[]
\centering
\caption{Operating Parameters.}
\begin{tabular}{@{}lcccr@{}}
\toprule
\textbf{Cable/Operation} & \textbf{Ops. per ESM cycle} & \textbf{$P_{MXC}$} & \textbf{Current (mA)} & \textbf{Duration (ns)} \\ 
\midrule
\textbf{DRIVE} \\
\quad 1Q                        & 16  & -71 dBm  & -     & 25 \\
\textbf{FLUX\_BIAS}              &     & -        & 0.57   & -   \\
\quad 2Q                        & 16  & -        & 0.4   & 42  \\
\textbf{READOUT\_PIN} \\
\quad Readout                   & 4   & -120 dBm & -     & 375   \\
\textbf{PUMP} \\
\quad Readout                   & 4   & -55 dBm  & -     & 375   \\ 
\bottomrule
\end{tabular}
\end{table}

\Figure[!b](topskip=0pt, botskip=0pt, midskip=0pt)[width=0.99\textwidth]
{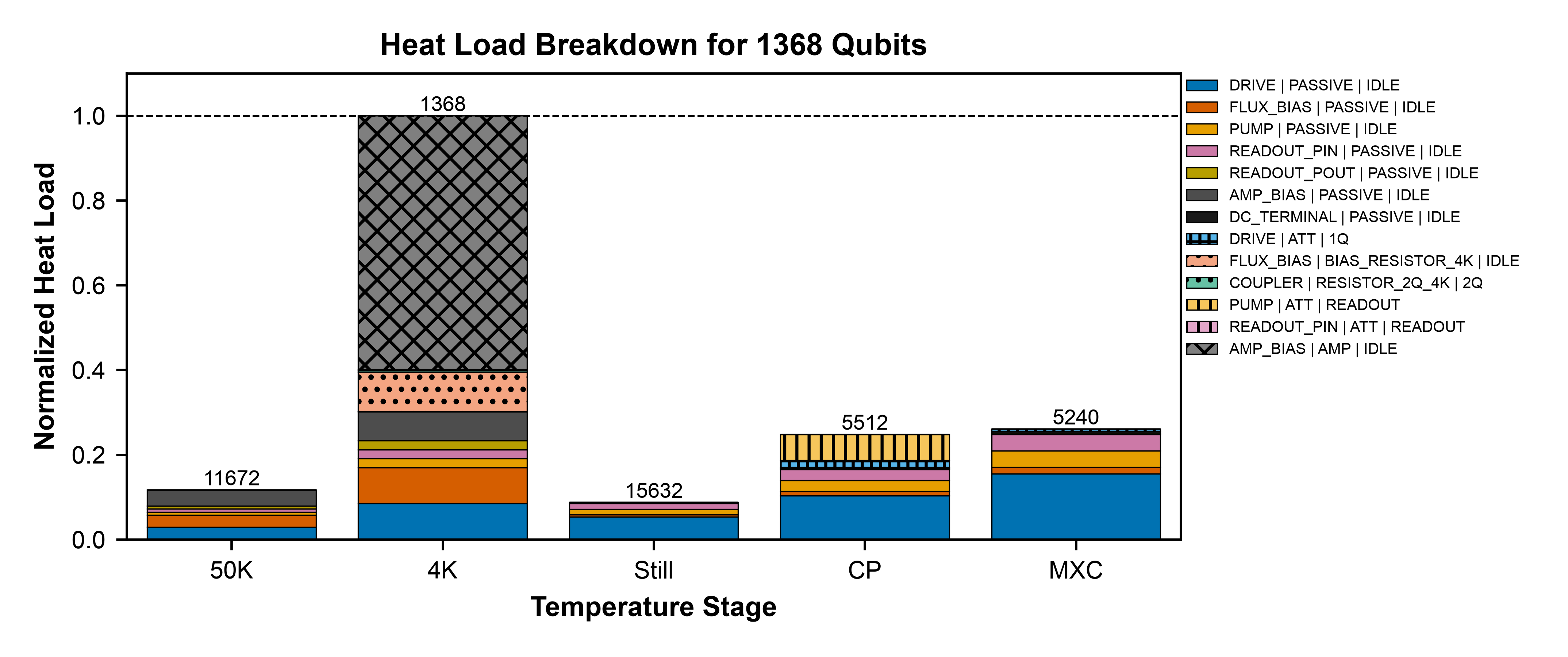}
{\parbox[t]{0.9\linewidth}{Normalized Heat Load Breakdown}
\label{fig:tf-fc-delft-kide}}
\clearpage


\subsection{FF-FC system | Cri/oFlex and Optical Fibers with PD at MXC | XLD400 fridge | HEMT \(\times\) Copper}
\label[appsec]{app:ff-fc-optical_MXC-xld400}

This section provides more details on the thermal scalability of the optical systems discussed in \Cref{sec:historical-optical} based on the original study in \cite{lecocq2021control}. Optical fibers are used for both drive and readout feed-in lines. A PD located at MXC converts the optical signals to microwave control tones. For consistency, we assume a XLD400 fridge and Cri/oFlex wiring for other RF lines. Readout multiplexing is assumed to be 4$\times$.

To model the power dissipation of the PD, we assume an optimistic responsivity $\mathcal{R} = \qty{1}{\ampere\per\watt}$ \cite{joshi2023scaling} and a high control-line impedance $Z = \qty{10}{\kilo\ohm}$. Under these assumptions, the PD dissipates on average approximately \qty{0.141}{\micro\watt} when generating a 1Q control tone with average microwave power $P_{\mu} = -70$\,dBm.

\begin{table*}[!ht]
\centering
\caption{System Configuration for a unit cell of 8 qubits for the system parameters based on \cite{lecocq2021control} as discussed in \Cref{sec:historical-optical}.}

\begin{threeparttable}
    
\begin{tabular}{lccllllll}
\toprule
\textbf{Cable}
 & \textbf{Operations} 
 & \textbf{Count} 
 & \textbf{50\,K} 
 & \textbf{4\,K} 
 & \textbf{Still} 
 & \textbf{CP} 
 & \textbf{MXC} 
 \\
\midrule
\textbf{DRIVE} 
& 1Q, 2Q
& 8 
& Fiber  
& Fiber 
& Fiber 
& Fiber 
& Fiber (PD)
\\

\textbf{PUMP} 
& Readout 
& 2 
& Ag  
& Ag (20 dB) 
& Ag  
& Ag (10 dB) 
& Ag (20 dB)
\\
\textbf{READOUT\_PIN} 
& Readout 
& 2 
& Fiber  
& Fiber 
& Fiber 
& Fiber 
& Fiber (PD)
\\
\textbf{READOUT\_POUT} 
& Readout 
& 2 
& Ag  
& Ag  
& NbTi  
& NbTi  
& NbTi  
\\
\textbf{DC\_TERMINAL} 
& Idle 
& 2 
& -  
& -  
& NbTi  
& NbTi  
& NbTi  
\\

\textbf{AMP\_BIAS} 
& Idle  
& 6
& Cu-30
& Cu-30  
\\

\quad \textit{Amplifier} 
& Idle
& 2
& -
& LNF8G
\\

\addlinespace
\midrule
\textbf{Fridge:} XLD400 \\
\quad \textit{Cooling Power} &
& &
\qty{30}{\watt} & 
\qty{1.5}{\watt} & 
\qty{40}{\milli\watt} & 
\qty{200}{\micro\watt} & 
\qty{19}{\micro\watt} & 
\\
\bottomrule
\end{tabular}

\end{threeparttable}
\end{table*}

\begin{table}[]
\centering
\caption{Operating Parameters.}
\begin{tabular}{@{}lccr@{}}
\toprule
\textbf{Cable/Operation} & \textbf{Ops. per ESM cycle} & \textbf{$P_{MXC}$} & \textbf{Duration (ns)} \\ \midrule
\textbf{DRIVE}        &    &          &       \\
\quad 1Q                    & 4  & -71 dBm  & 42.67 \\
\quad 2Q                    & 16 & -66 dBm  & 71.1  \\
\textbf{READOUT\_PIN} &    &          &       \\
\quad Readout               & 4  & -120 dBm & 215   \\
\textbf{PUMP}         &    &          &       \\
\quad Readout               & 4  & -55 dBm  & 215   \\ \bottomrule
\end{tabular}
\end{table}

\Figure[!b](topskip=0pt, botskip=0pt, midskip=0pt)[width=0.99\textwidth]
{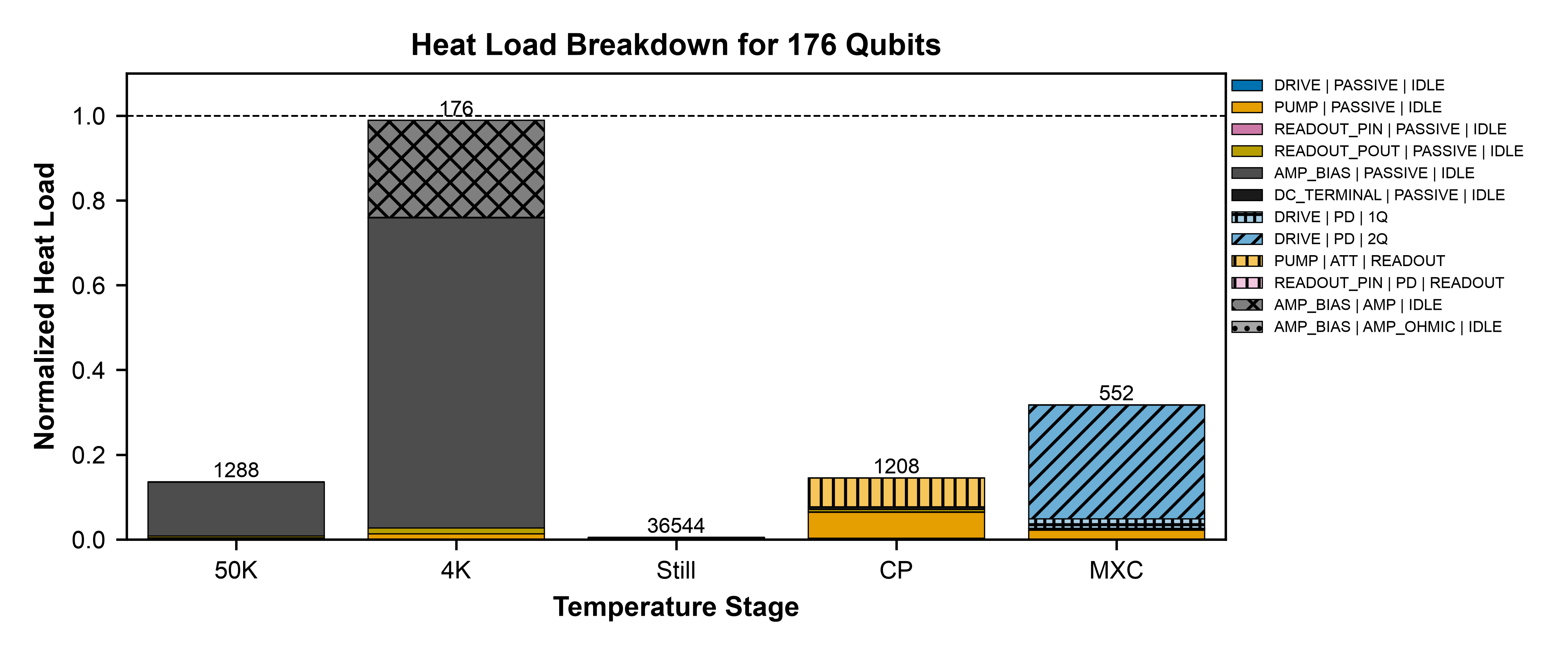}
{\parbox[t]{0.9\linewidth}{Normalized Heat Load Breakdown}
\label{fig:ff-fc-optical_MXC-xld400}}
\clearpage

\subsection{FF-FC system | Cri/oFlex and Optical Fibers with PD at 4\,K | XLD400 fridge | HEMT \(\times\) Copper}
\label[appsec]{app:ff-fc-optical_4K-xld400}

This system is similar to the system in \Cref{app:ff-fc-optical_MXC-xld400} except that the PD is relocated to the 4\,K stage to relieve thermal stress on the MXC as originally proposed in \cite{joshi2023scaling}. In our modeling, we use modern Cri/oFlex microstrip cables for RF signals instead of the coaxial cables assumed in the original system in \cite{joshi2023scaling}. We also include the heat sources from other system components (e.g., amplifiers, pump line and related support lines) that were not included in the original study.

\begin{table*}[!ht]
\centering
\caption{System Configuration for a unit cell of 8 qubits for the system based on \cite{joshi2023scaling} as discussed in \Cref{sec:historical-optical}.}

\begin{threeparttable}
    
\begin{tabular}{lccllllll}
\toprule
\textbf{Cable}
 & \textbf{Operations} 
 & \textbf{Count} 
 & \textbf{50\,K} 
 & \textbf{4\,K} 
 & \textbf{Still} 
 & \textbf{CP} 
 & \textbf{MXC} 
 \\
\midrule
\textbf{DRIVE} 
& 1Q, 2Q
& 8 
& Fiber  
& Fiber (PD)
& NbTi 
& NbTi 
& NbTi 
\\

\textbf{PUMP} 
& Readout 
& 2 
& Ag  
& Ag (20 dB) 
& Ag  
& Ag (10 dB) 
& Ag (20 dB)
\\
\textbf{READOUT\_PIN} 
& Readout 
& 2 
& Fiber  
& Fiber (PD)
& NbTi 
& NbTi 
& NbTi 
\\
\textbf{READOUT\_POUT} 
& Readout 
& 2 
& Ag  
& Ag  
& NbTi  
& NbTi  
& NbTi  
\\
\textbf{DC\_TERMINAL} 
& Idle 
& 2 
& -  
& -  
& NbTi  
& NbTi  
& NbTi  
\\

\textbf{AMP\_BIAS} 
& Idle  
& 6
& Cu-30
& Cu-30  
\\

\quad \textit{Amplifier} 
& Idle
& 2
& -
& LNF8G
\\

\addlinespace
\midrule
\textbf{Fridge:} XLD400 \\
\quad \textit{Cooling Power} &
& &
\qty{30}{\watt} & 
\qty{1.5}{\watt} & 
\qty{40}{\milli\watt} & 
\qty{200}{\micro\watt} & 
\qty{19}{\micro\watt} & 
\\
\bottomrule
\end{tabular}

\end{threeparttable}
\end{table*}

\begin{table}[]
\centering
\caption{Operating Parameters.}
\begin{tabular}{@{}lccr@{}}
\toprule
\textbf{Cable/Operation} & \textbf{Ops. per ESM cycle} & \textbf{$P_{MXC}$} & \textbf{Duration (ns)} \\ \midrule
\textbf{DRIVE}        &    &          &       \\
\quad 1Q                    & 4  & -71 dBm  & 42.67 \\
\quad 2Q                    & 16 & -66 dBm  & 71.1  \\
\textbf{READOUT\_PIN} &    &          &       \\
\quad Readout               & 4  & -120 dBm & 215   \\
\textbf{PUMP}         &    &          &       \\
\quad Readout               & 4  & -55 dBm  & 215   \\ \bottomrule
\end{tabular}
\end{table}

\Figure[!b](topskip=0pt, botskip=0pt, midskip=0pt)[width=0.99\textwidth]
{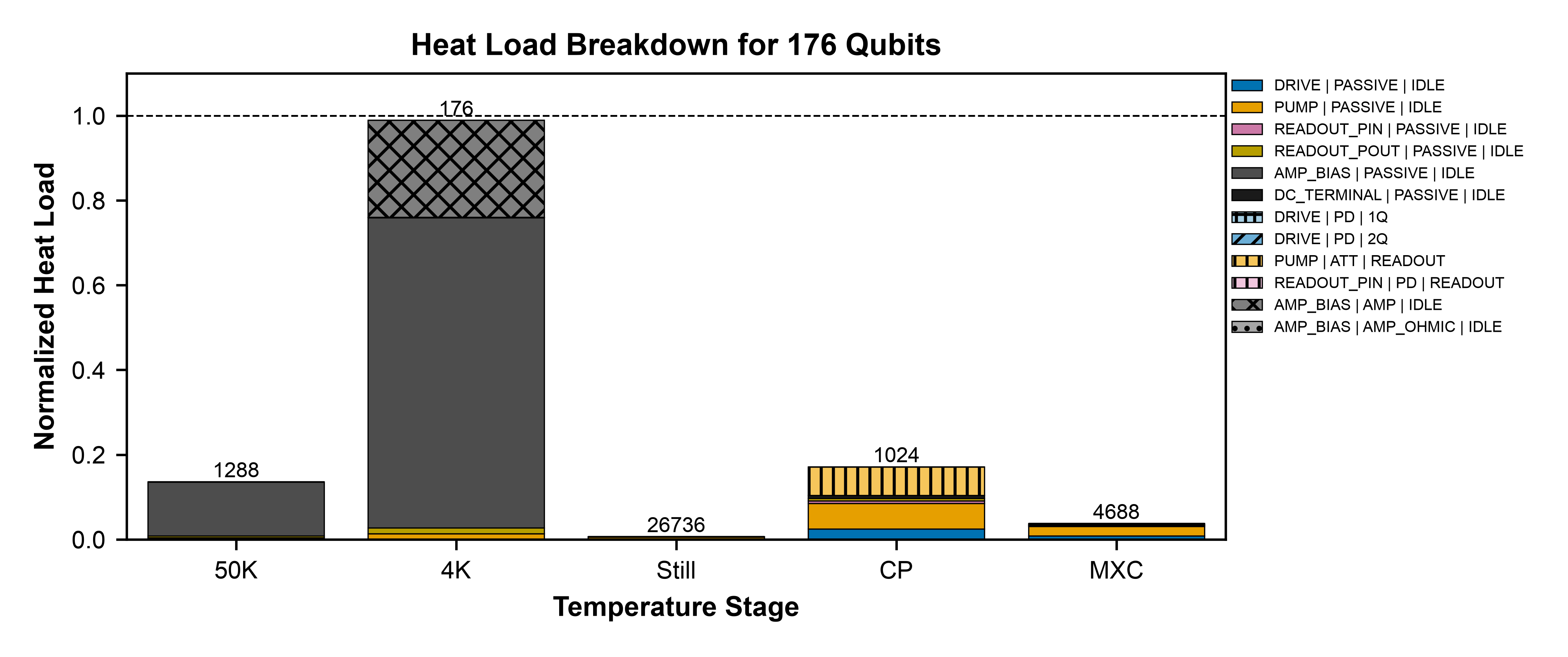}
{\parbox[t]{0.9\linewidth}{Normalized Heat Load Breakdown}
\label{fig:ff-fc-optical_4K-xld400}}
\clearpage


\subsection{TF-TC system | HDW Wiring | XLD1000sl fridge | HEMT \(\times\) Copper}
\label[appsec]{app:tf-tc-hdw-xld1000sl}

This section details the modeling parameters to reproduce the scalability study for a TF-TC system from \cite{raicu2025cryogenic} discussed in \Cref{subsec:hdw}. In this system, the directional coupler port is terminated at the MXC as in the original study and therefore NbTi \leg{DC\_TERMINAL} cables are not necessary. 

\begin{table*}[!ht]
\centering
\caption{System Configuration for a unit cell of 8 qubits for the system parameters based on \cite{raicu2025cryogenic} as discussed in \Cref{subsec:hdw}.}

\begin{threeparttable}
    
\begin{tabular}{lccllllll}
\toprule
\textbf{Cable}
 & \textbf{Operations} 
 & \textbf{Count} 
 & \textbf{50\,K} 
 & \textbf{4\,K} 
 & \textbf{Still} 
 & \textbf{CP} 
 & \textbf{MXC} 
 \\
\midrule
\textbf{DRIVE} 
& 1Q
& 8 
& HDW 
& HDW (20 dB) 
& HDW (10 dB) 
& HDW (10 dB) 
& HDW (20 dB) 
\\
\textbf{FLUX\_BIAS} 
& Idle
& 8 
& HDW 
& HDW (20 dB)
& HDW
& HDW
& HDW
\\
\textbf{COUPLER} 
& 2Q 
& 14 
& HDW 
& HDW (20 dB)
& HDW
& HDW
& HDW
\\

\textbf{PUMP} 
& Readout 
& 2 
& HDW  
& HDW (20 dB) 
& HDW  
& HDW (10 dB) 
& HDW (20 dB)
\\
\textbf{READOUT\_PIN} 
& Readout 
& 2 
& HDW 
& HDW (20 dB) 
& HDW (10 dB) 
& HDW (10 dB) 
& HDW (20 dB) 
\\
\textbf{READOUT\_POUT} 
& Readout 
& 2 
& HDW 
& HDW  
& HDW  
& HDW  
& HDW  
\\

\textbf{AMP\_BIAS} 
& Idle  
& 6
& Cu-30
& Cu-30  
\\

\quad \textit{Amplifier} 
& Idle
& 2
& -
& LNF8G
\\

\addlinespace
\midrule
\textbf{Fridge:} XLD1000sl \\
\quad \textit{Cooling Power} &
& &
\qty{30}{\watt} & 
\qty{0.7}{\watt} & 
\qty{7}{\milli\watt} & 
\qty{1}{\milli\watt} & 
\qty{30}{\micro\watt} & 
\\
\bottomrule
\end{tabular}

\end{threeparttable}
\end{table*}

\begin{table}[]
\centering
\caption{Operating Parameters.}
\begin{tabular}{@{}lcccr@{}}
\toprule
\textbf{Cable/Operation} & \textbf{Ops. per ESM cycle} & \textbf{$P_{MXC}$} & \textbf{Current (mA)} & \textbf{Duration (ns)} \\ 
\midrule
\textbf{DRIVE} \\
\quad 1Q                        & 16  & -71 dBm  & -     & 25 \\
\textbf{FLUX\_BIAS}              &     & -        & 0.4   & -   \\
\textbf{COUPLER} \\
\quad 2Q                        & 16  & -        & 0.4   & 42  \\
\textbf{READOUT\_PIN} \\
\quad Readout                   & 4   & -120 dBm & -     & 375   \\
\textbf{PUMP} \\
\quad Readout                   & 4   & -55 dBm  & -     & 375   \\ 
\bottomrule
\end{tabular}
\end{table}

\Figure[!b](topskip=0pt, botskip=0pt, midskip=0pt)[width=0.99\textwidth]
{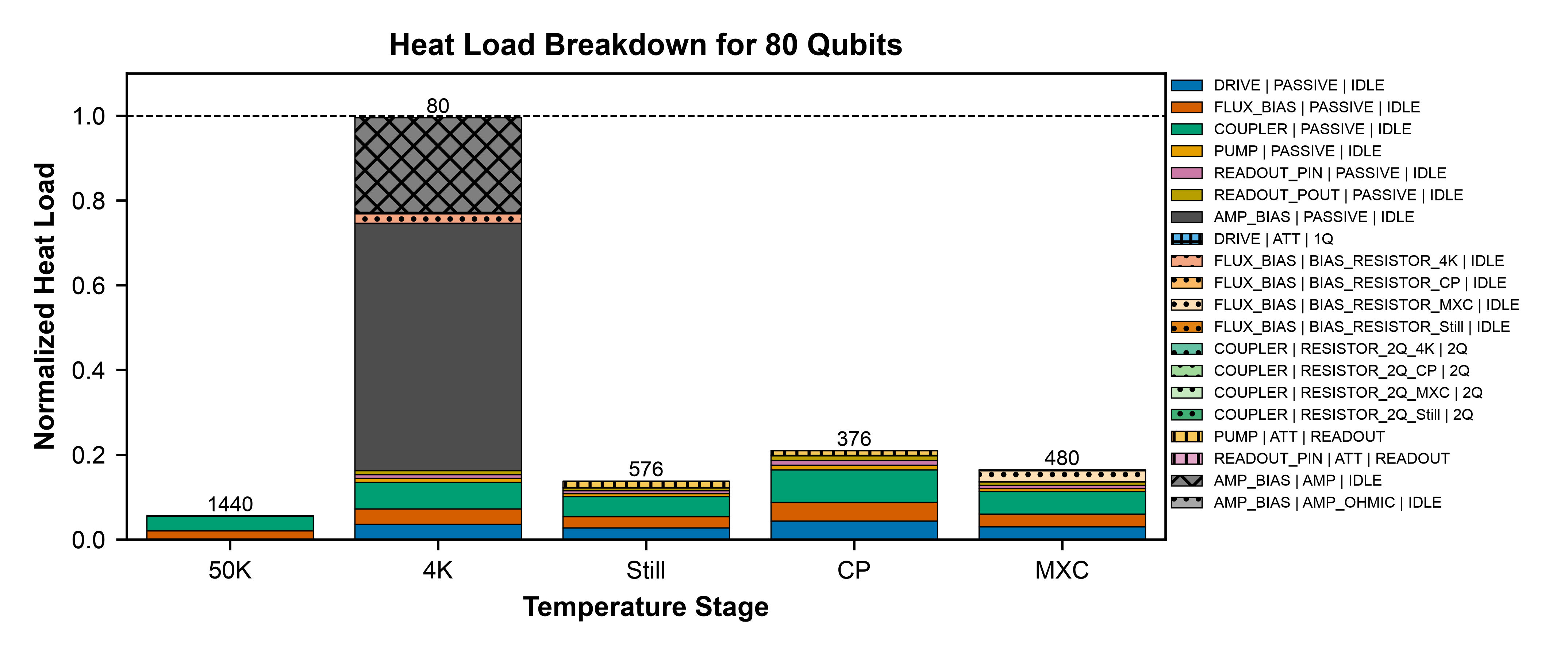}
{\parbox[t]{0.9\linewidth}{Normalized Heat Load Breakdown}
\label{fig:tf-tc-hdw-xld1000sl}}
\clearpage

\subsection{FF-FC system | Cri/oFlex Wiring | KIDE fridge | HEMT \(\times\) Copper}
\label[appsec]{app:ff-fc-delft-hemt_cu-kide}

In \Cref{subsec:amp_wire_analysis}, we evaluate the thermal effects of various combinations of readout amplifiers in \Cref{tab:amp_options} (HEMT, ULP-HEMT, SIS) and biasing wiring materials in \Cref{tab:biasing_cables} (copper, Manganin, YBCO). For a fair and systematic comparison, we assume an FF-FC architecture with Cri/oFlex wiring housed in a KIDE fridge and vary only the amplifier and biasing cables. The directional coupler impedance termination is relocated to the 4\,K stage. The following configuration uses the LNF8G amplifier in which AWG30 copper wires are used. This represents a typical contemporary system:

\begin{table*}[!ht]
\centering
\caption{System Configuration for a unit cell of 8 qubits for the design space exploration conducted in \Cref{subsec:amp_wire_analysis}.}

\begin{threeparttable}
    
\begin{tabular}{lccllllll}
\toprule
\textbf{Cable}
 & \textbf{Operations} 
 & \textbf{Count} 
 & \textbf{50\,K} 
 & \textbf{4\,K} 
 & \textbf{Still} 
 & \textbf{CP} 
 & \textbf{MXC} 
 \\
\midrule
\textbf{DRIVE} 
& 1Q, 2Q
& 8 
& Ag  
& Ag (20 dB)
& Ag (10 dB)
& Ag (10 dB)
& Ag (20 dB)
\\

\textbf{PUMP} 
& Readout 
& 2 
& Ag  
& Ag (20 dB) 
& Ag  
& Ag (10 dB) 
& Ag (20 dB)
\\
\textbf{READOUT\_PIN} 
& Readout 
& 2 
& Ag  
& Ag (20 dB)
& Ag (10 dB)
& Ag (10 dB)
& Ag (20 dB)
\\
\textbf{READOUT\_POUT} 
& Readout 
& 2 
& Ag  
& Ag  
& NbTi  
& NbTi  
& NbTi  
\\
\textbf{DC\_TERMINAL} 
& Idle 
& 2 
& -  
& -  
& NbTi  
& NbTi  
& NbTi  
\\

\textbf{AMP\_BIAS} 
& Idle  
& 6
& Cu-30
& Cu-30  
\\

\quad \textit{Amplifier} 
& Idle
& 2
& -
& LNF8G
\\

\addlinespace
\midrule
\textbf{Fridge:} KIDE \\
\quad \textit{Cooling Power} &
& &
\qty{90}{\watt} & 
\qty{6}{\watt} & 
\qty{90}{\milli\watt} & 
\qty{3}{\milli\watt} & 
\qty{90}{\micro\watt} 
\\
\bottomrule
\end{tabular}

\end{threeparttable}
\end{table*}

\begin{table}[]
\centering
\caption{Operating Parameters.}
\begin{tabular}{@{}lccr@{}}
\toprule
\textbf{Cable/Operation} & \textbf{Ops. per ESM cycle} & \textbf{$P_{MXC}$} & \textbf{Duration (ns)} \\ \midrule
\textbf{DRIVE}        &    &          &       \\
\quad 1Q                    & 4  & -71 dBm  & 42.67 \\
\quad 2Q                    & 16 & -66 dBm  & 71.1  \\
\textbf{READOUT\_PIN} &    &          &       \\
\quad Readout               & 4  & -120 dBm & 215   \\
\textbf{PUMP}         &    &          &       \\
\quad Readout               & 4  & -55 dBm  & 215   \\ \bottomrule
\end{tabular}
\end{table}

\Figure[!b](topskip=0pt, botskip=0pt, midskip=0pt)[width=0.99\textwidth]
{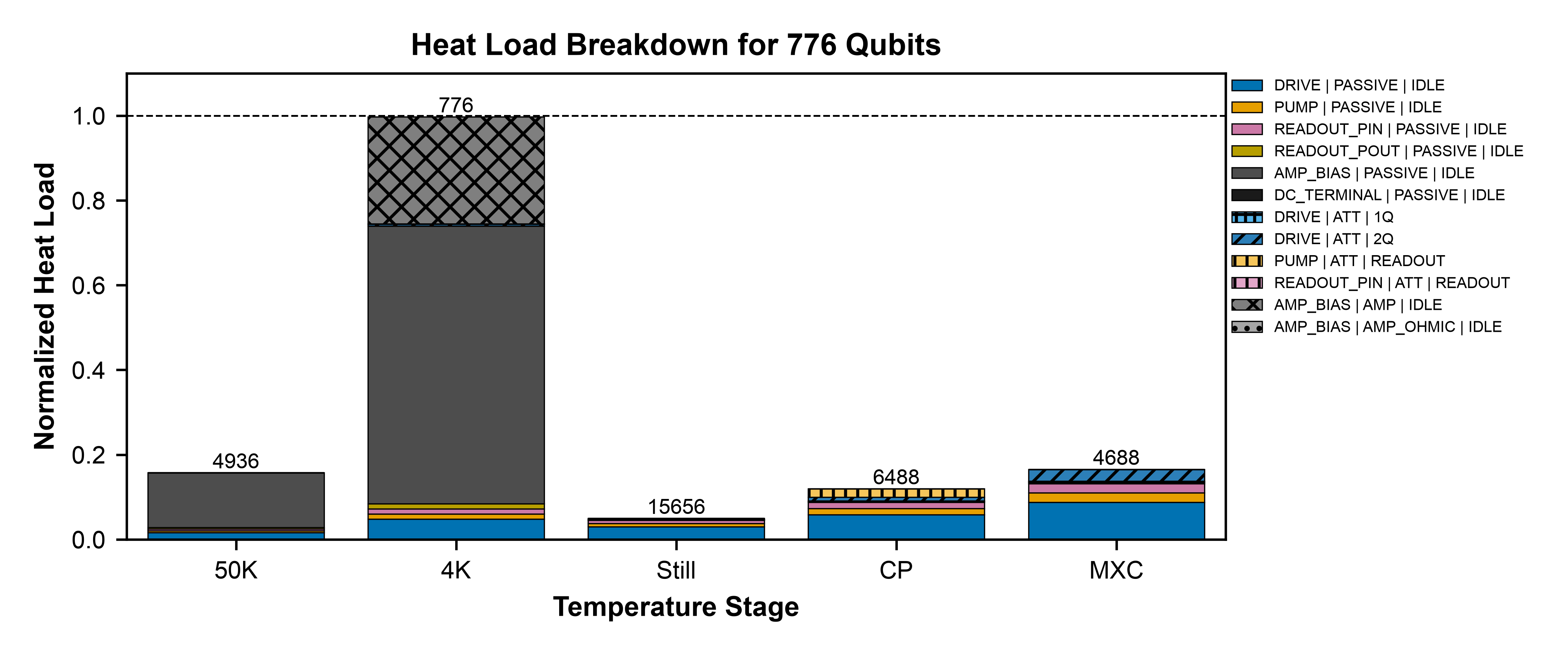}
{\parbox[t]{0.9\linewidth}{Normalized Heat Load Breakdown for the LNF8G-AWG 30 Copper wiring configuration}
\label{fig:ff-fc-delft-hemt_cu-kide}}
\clearpage

\subsection{FF-FC system | Cri/oFlex Wiring | KIDE fridge | ULP-HEMT \(\times\) Manganin}
\label[appsec]{app:ff-fc-delft-hemt-ulp_mn-kide}

The following configuration assumes an FF-FC architecture with Cri/oFlex wiring housed in a KIDE fridge, similar to the previous system. The only difference is that a ULP-HEMT amplifier with AWG30 Manganin wires is being used. Since ULP-HEMT has a limited gain of approximately 23 dB, an additional amplifier stage at 50\,K is necessary. We assume this is a LNF8G HEMT amplifier, also with Manganin biasing wires (designated as \leg{AMP\_BIAS (50\,K)}). This configuration represents a near-term option to remove the amplifier bottleneck at 4\,K.

\begin{table*}[!ht]
\centering
\caption{System Configuration for a unit cell of 8 qubits for the design space exploration conducted in \Cref{subsec:amp_wire_analysis}.}

\begin{threeparttable}
    
\begin{tabular}{lccllllll}
\toprule
\textbf{Cable}
 & \textbf{Operations} 
 & \textbf{Count} 
 & \textbf{50\,K} 
 & \textbf{4\,K} 
 & \textbf{Still} 
 & \textbf{CP} 
 & \textbf{MXC} 
 \\
\midrule
\textbf{DRIVE} 
& 1Q, 2Q
& 8 
& Ag  
& Ag (20 dB)
& Ag (10 dB)
& Ag (10 dB)
& Ag (20 dB)
\\

\textbf{PUMP} 
& Readout 
& 2 
& Ag  
& Ag (20 dB) 
& Ag  
& Ag (10 dB) 
& Ag (20 dB)
\\
\textbf{READOUT\_PIN} 
& Readout 
& 2 
& Ag  
& Ag (20 dB)
& Ag (10 dB)
& Ag (10 dB)
& Ag (20 dB)
\\
\textbf{READOUT\_POUT} 
& Readout 
& 2 
& Ag  
& Ag  
& NbTi  
& NbTi  
& NbTi  
\\
\textbf{DC\_TERMINAL} 
& Idle 
& 2 
& -  
& -  
& NbTi  
& NbTi  
& NbTi  
\\

\textbf{AMP\_BIAS} 
& Idle  
& 12
& Manganin (Mn)
& Manganin (Mn)
\\
\quad \textit{Amplifier} 
& Idle
& 2
& -
& ULP-HEMT
\\

\textbf{AMP\_BIAS (50\,K)} 
& Idle  
& 6
& Manganin (Mn)
& -
\\
\quad \textit{Amplifier} 
& Idle
& 2
& LNF8G
& -
\\

\addlinespace
\midrule
\textbf{Fridge:} KIDE \\
\quad \textit{Cooling Power} &
& &
\qty{90}{\watt} & 
\qty{6}{\watt} & 
\qty{90}{\milli\watt} & 
\qty{3}{\milli\watt} & 
\qty{90}{\micro\watt} 
\\
\bottomrule
\end{tabular}

\end{threeparttable}
\end{table*}

\begin{table}[]
\centering
\caption{Operating Parameters.}
\begin{tabular}{@{}lccr@{}}
\toprule
\textbf{Cable/Operation} & \textbf{Ops. per ESM cycle} & \textbf{$P_{MXC}$} & \textbf{Duration (ns)} \\ \midrule
\textbf{DRIVE}        &    &          &       \\
\quad 1Q                    & 4  & -71 dBm  & 42.67 \\
\quad 2Q                    & 16 & -66 dBm  & 71.1  \\
\textbf{READOUT\_PIN} &    &          &       \\
\quad Readout               & 4  & -120 dBm & 215   \\
\textbf{PUMP}         &    &          &       \\
\quad Readout               & 4  & -55 dBm  & 215   \\ \bottomrule
\end{tabular}
\end{table}

\Figure[!b](topskip=0pt, botskip=0pt, midskip=0pt)[width=0.99\textwidth]
{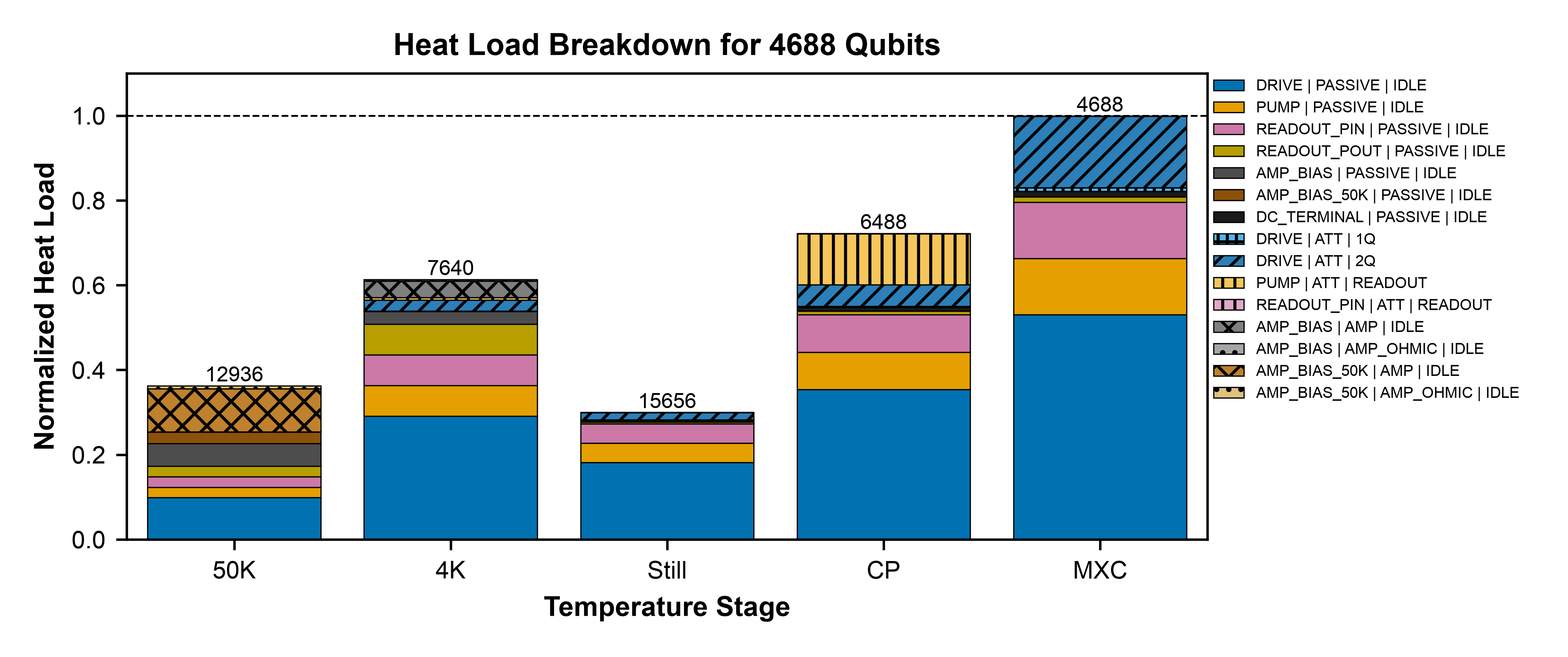}
{\parbox[t]{0.9\linewidth}{Normalized Heat Breakdown for the ULP-HEMT $\times$ Manganin wiring configuration}
\label{fig:ff-fc-delft-hemt-ulp_mn-kide}}
\clearpage

\subsection{FF-FC system | Ag--NbTi Wiring | KIDE fridge | ULP-HEMT \(\times\) Manganin}
\label[appsec]{app:ff-fc-delft_nbti-ulp_mn-kide}

Here, we evaluate a wiring configuration using Cri/oFlex microstrip cables where NbTi is used for \textit{all} cables below the 4\,K stage as discussed in \Cref{subsubsec:wire_tradeoffs}. The readout amplifier at 4\,K is a ULP-HEMT with Manganin biasing wires. This represents a near-term immediate solution to overcome the large heat load at 4\,K due to readout amplifier and its bias lines. Since ULP-HEMT has limited gain, an additional HEMT (also with Manganin biasing wires) is installed at 50\,K.

\begin{table*}[!ht]
\centering
\caption{System Configuration for a unit cell of 8 qubits mentioned in \Cref{subsubsec:wire_tradeoffs}.}
\begin{threeparttable}
\begin{tabular}{lccllllll}
\toprule
\textbf{Cable}
 & \textbf{Operations} 
 & \textbf{Count} 
 & \textbf{50\,K} 
 & \textbf{4\,K} 
 & \textbf{Still} 
 & \textbf{CP} 
 & \textbf{MXC} 
 \\
\midrule
\textbf{DRIVE} 
& 1Q, 2Q
& 8 
& Ag 
& Ag (20 dB) 
& NbTi (10 dB) 
& NbTi (10 dB) 
& NbTi (20 dB) 
\\

\textbf{PUMP} 
& Readout 
& 2 
& Ag  
& Ag (20 dB) 
& NbTi  
& NbTi (10 dB) 
& NbTi (20 dB)
\\
\textbf{READOUT\_PIN} 
& Readout 
& 2 
& Ag 
& Ag (20 dB) 
& NbTi (10 dB) 
& NbTi (10 dB) 
& NbTi (20 dB) 
\\
\textbf{READOUT\_POUT} 
& Readout 
& 2 
& Ag 
& Ag  
& NbTi  
& NbTi  
& NbTi  
\\
\textbf{DC\_TERMINAL} 
& Idle 
& 2 
& -  
& -  
& NbTi  
& NbTi  
& NbTi   
\\

\textbf{AMP\_BIAS} 
& Idle  
& 12
& Manganin (Mn)
& Manganin (Mn)
\\
\quad \textit{Amplifier} 
& Idle
& 2
& -
& ULP-HEMT
\\

\textbf{AMP\_BIAS (50\,K)} 
& Idle  
& 6
& Manganin (Mn)
& -
\\
\quad \textit{Amplifier} 
& Idle
& 2
& LNF8G
& -
\\

\addlinespace
\midrule
\textbf{Fridge:} KIDE \\
\quad \textit{Cooling Power} &
& &
\qty{90}{\watt} & 
\qty{6}{\watt} & 
\qty{90}{\milli\watt} & 
\qty{3}{\milli\watt} & 
\qty{90}{\micro\watt} 
\\
\bottomrule
\end{tabular}
\end{threeparttable}
\end{table*}

\begin{table}[]
\centering
\caption{Operating Parameters.}
\begin{tabular}{@{}lccr@{}}
\toprule
\textbf{Cable/Operation} & \textbf{Ops. per ESM cycle} & \textbf{$P_{MXC}$} & \textbf{Duration (ns)} \\ \midrule
\textbf{DRIVE}        &    &          &       \\
\quad 1Q                    & 4  & -71 dBm  & 42.67 \\
\quad 2Q                    & 16 & -66 dBm  & 71.1  \\
\textbf{READOUT\_PIN} &    &          &       \\
\quad Readout               & 4  & -120 dBm & 215   \\
\textbf{PUMP}         &    &          &       \\
\quad Readout               & 4  & -55 dBm  & 215   \\ \bottomrule
\end{tabular}
\end{table}

\Figure[!b](topskip=0pt, botskip=0pt, midskip=0pt)[width=0.99\textwidth]
{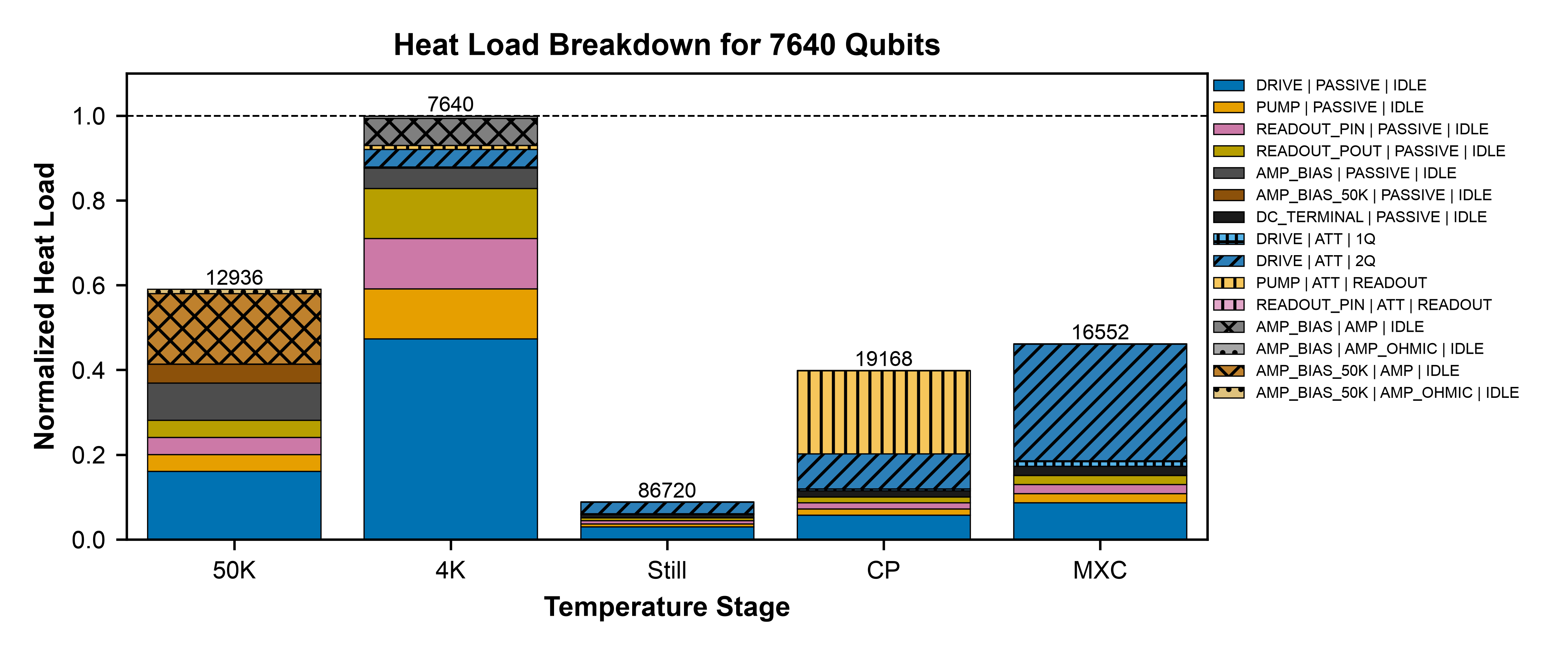}
{\parbox[t]{0.9\linewidth}{Normalized Heat Breakdown for the Ag--NbTi wiring configuration}
\label{fig:ff-fc-delft_nbti-ulp_mn-kide}}
\clearpage

\subsection{FF-FC system | HDW--NbTi Wiring | KIDE fridge | ULP-HEMT \(\times\) Manganin}
\label[appsec]{app:ff-fc-hdw_nbti-ulp_mn-kide}
This configuration is identical to \Cref{app:ff-fc-delft_nbti-ulp_mn-kide} except that we use HDW coaxial cables instead of Cri/oFlex microstrips. We also use SC-086 NbTi coaxial cables to replace the HDW for \textit{all} lines at temperature stages below 4\,K. This has the effect of reducing the thermal stress on the bottleneck CP stage.

\begin{table}[!ht]
\centering
\caption{System Configuration for a unit cell of 8 qubits mentioned in \Cref{subsubsec:wire_tradeoffs}.}
\begin{threeparttable}
\resizebox{\columnwidth}{!}{%
\begin{tabular}{lcclllll}
\toprule
\textbf{Cable}
 & \textbf{Operations} 
 & \textbf{Count} 
 & \textbf{50\,K} 
 & \textbf{4\,K} 
 & \textbf{Still} 
 & \textbf{CP} 
 & \textbf{MXC} 
 \\
\midrule
\textbf{DRIVE} 
& 1Q, 2Q
& 8 
& HDW 
& HDW (20 dB) 
& SC086-NbTi (coax) (10 dB) 
& SC086-NbTi (coax) (10 dB) 
& SC086-NbTi (coax) (20 dB) 
\\

\textbf{PUMP} 
& Readout 
& 2 
& HDW  
& HDW (20 dB) 
& SC086-NbTi (coax)  
& SC086-NbTi (coax) (10 dB) 
& SC086-NbTi (coax) (20 dB)
\\
\textbf{READOUT\_PIN} 
& Readout 
& 2 
& HDW 
& HDW (20 dB) 
& SC086-NbTi (coax) (10 dB) 
& SC086-NbTi (coax) (10 dB) 
& SC086-NbTi (coax) (20 dB) 
\\
\textbf{READOUT\_POUT} 
& Readout 
& 2 
& HDW 
& HDW  
& SC086-NbTi (coax)
& SC086-NbTi (coax) 
& SC086-NbTi (coax)
\\
\textbf{DC\_TERMINAL} 
& Idle  
& 2
& -
& -
& SC086-NbTi (coax)
& SC086-NbTi (coax)
& SC086-NbTi (coax)
\\
\textbf{AMP\_BIAS} 
& Idle  
& 12
& Manganin (Mn)
& Manganin (Mn)
\\
\quad \textit{Amplifier} 
& Idle
& 2
& -
& ULP-HEMT
\\

\textbf{AMP\_BIAS (50\,K)} 
& Idle  
& 6
& Manganin (Mn)
& -
\\
\quad \textit{Amplifier} 
& Idle
& 2
& LNF8G
& -
\\

\addlinespace
\midrule
\textbf{Fridge:} KIDE \\
\quad \textit{Cooling Power} &
& &
\qty{90}{\watt} & 
\qty{6}{\watt} & 
\qty{90}{\milli\watt} & 
\qty{3}{\milli\watt} & 
\qty{90}{\micro\watt} 
\\
\bottomrule
\end{tabular}
}
\end{threeparttable}
\end{table}

\begin{table}[]
\centering
\caption{Operating Parameters.}
\begin{tabular}{@{}lccr@{}}
\toprule
\textbf{Cable/Operation} & \textbf{Ops. per ESM cycle} & \textbf{$P_{MXC}$} & \textbf{Duration (ns)} \\ \midrule
\textbf{DRIVE}        &    &          &       \\
\quad 1Q                    & 4  & -71 dBm  & 42.67 \\
\quad 2Q                    & 16 & -66 dBm  & 71.1  \\
\textbf{READOUT\_PIN} &    &          &       \\
\quad Readout               & 4  & -120 dBm & 215   \\
\textbf{PUMP}         &    &          &       \\
\quad Readout               & 4  & -55 dBm  & 215   \\ \bottomrule
\end{tabular}
\end{table}

\Figure[!b](topskip=0pt, botskip=0pt, midskip=0pt)[width=0.99\textwidth]
{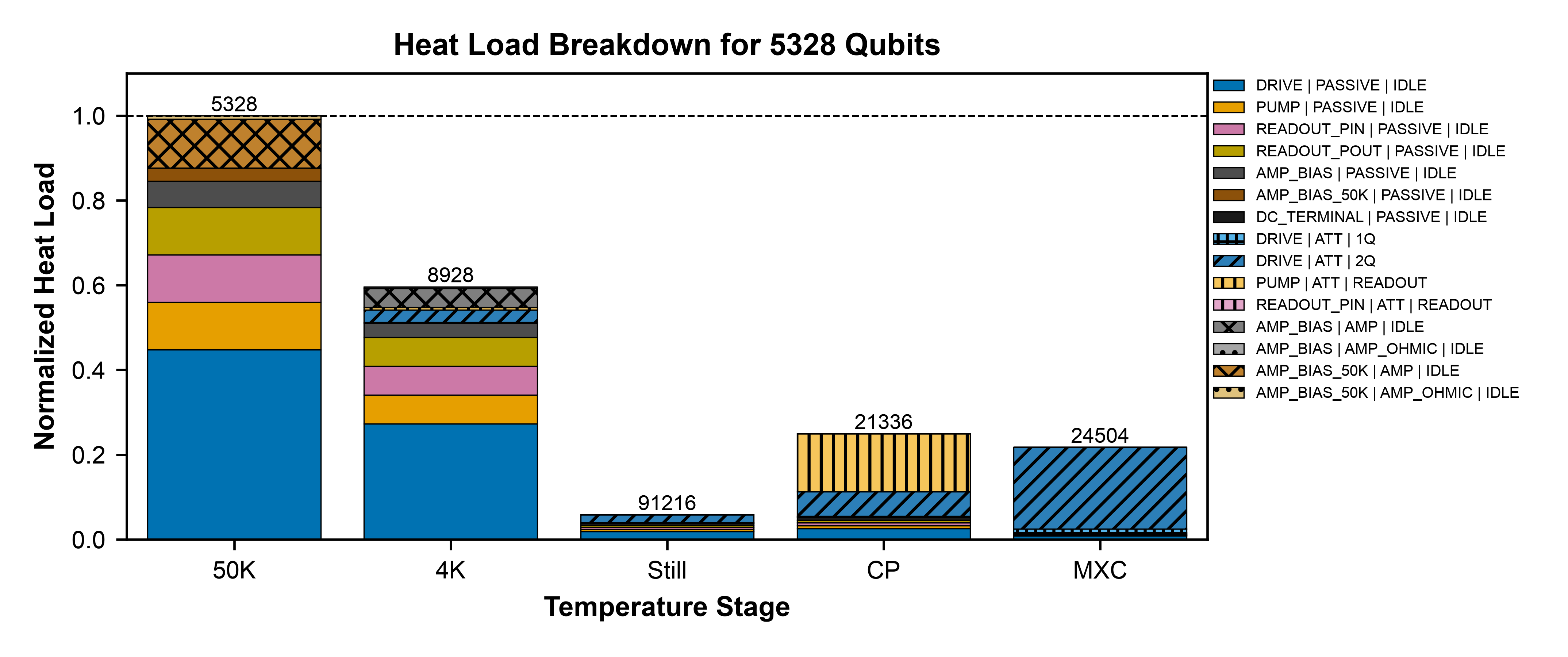}
{\parbox[t]{0.9\linewidth}{Normalized Heat Breakdown for the HDW--NbTi cable configuration}
\label{fig:ff-fc-hdw_nbti-ulp_mn-kide}}
\clearpage

\subsection{TF-TC system | Ag-YBCO-NbTi Wiring | KIDE fridge | SIS \(\times\) YBCO}
\label[appsec]{app:tf-tc-ybco_delft-sis_ybco-kide}

This is a hypothetical system that uses hybrid Ag-YBCO-NbTi microstrip RF cables and SIS-Mixer amplifiers as discussed in \Cref{subsubsec:wire_tradeoffs} to evaluate an exploratory upper-bound scenario where the PHL is minimal. This cabling scheme takes advantage of the low PHL and superconducting nature of YBCO at temperatures between 50\,K and 4\,K. As a result, it has a much lower PHL and negligible ohmic loss. Unlike the previous systems, we assume SIS amplifiers with YBCO biasing wires at 4\,K (to minimize amplifier-related heat) with an additional HEMT at 50\,K biased with Manganin wires. We assume a TF-TC architecture (CZ-based ESM workload) as this requires the most wires per qubit.

\begin{table*}[!ht]
\centering
\caption{Configuration of an 8-qubit unit cell for the system discussed in \Cref{subsubsec:wire_tradeoffs} (TF-TC).}

\begin{threeparttable}
    
\begin{tabular}{lcclllll}
\toprule
\textbf{Cable}
 & \textbf{Operations} 
 & \textbf{Count} 
 & \textbf{50\,K} 
 & \textbf{4\,K} 
 & \textbf{Still} 
 & \textbf{CP} 
 & \textbf{MXC} 
 \\
\midrule
\textbf{DRIVE} 
& 1Q
& 8 
& Ag
& YBCO-Flex (20 dB)
& NbTi (10 dB)
& NbTi (10 dB) 
& NbTi (20 dB) 
\\
\textbf{FLUX\_BIAS} 
& Idle
& 8 
& Ag
& YBCO-Flex (20 dB)
& NbTi 
& NbTi
& NbTi 
\\
\textbf{COUPLER} 
& 2Q 
& 14 
& Ag 
& YBCO-Flex (20 dB)
& NbTi
& NbTi 
& NbTi
\\
\textbf{PUMP} 
& Readout 
& 2 
& Ag  
& YBCO-Flex  (20 dB)
& NbTi
& NbTi (10 dB) 
& NbTi (20 dB)
\\
\textbf{READOUT\_PIN} 
& Readout 
& 2 
& Ag
& YBCO-Flex (20 dB)
& NbTi (10 dB)
& NbTi (10 dB) 
& NbTi (20 dB) 
\\
\textbf{READOUT\_POUT} 
& Readout 
& 2 
& Ag
& YBCO-Flex 
& NbTi 
& NbTi 
& NbTi 
\\

\textbf{AMP\_BIAS} 
& Idle  
& 10
& Manganin (Mn)
& YBCO
\\
\quad \textit{Amplifier} 
& Idle
& 2
& -
& SIS
\\

\textbf{AMP\_BIAS (50\,K)} 
& Idle  
& 6
& Manganin (Mn)
& -
\\
\quad \textit{Amplifier} 
& Idle
& 2
& LNF8G
& -
\\

\addlinespace
\midrule
\textbf{Fridge:} KIDE \\
\quad \textit{Cooling Power} &
& &
\qty{90}{\watt} & 
\qty{6}{\watt} & 
\qty{90}{\milli\watt} & 
\qty{3}{\milli\watt} & 
\qty{90}{\micro\watt} 
\\
\bottomrule
\end{tabular}

\end{threeparttable}
\end{table*}

\begin{table}[]
\centering
\caption{Operating Parameters.}
\begin{tabular}{@{}lcccr@{}}
\toprule
\textbf{Cable/Operation} & \textbf{Ops. per ESM cycle} & \textbf{$P_{MXC}$} & \textbf{Current (mA)} & \textbf{Duration (ns)} \\ 
\midrule
\textbf{DRIVE} \\
\quad 1Q                        & 16  & -71 dBm  & -     & 25 \\
\textbf{FLUX\_BIAS}              &     & -        & 0.4   & -   \\
\textbf{COUPLER} \\
\quad 2Q                        & 16  & -        & 0.4   & 42  \\
\textbf{READOUT\_PIN} \\
\quad Readout                   & 4   & -120 dBm & -     & 375   \\
\textbf{PUMP} \\
\quad Readout                   & 4   & -55 dBm  & -     & 375   \\ 
\bottomrule
\end{tabular}
\end{table}

\Figure[!b](topskip=0pt, botskip=0pt, midskip=0pt)[width=0.99\textwidth]
{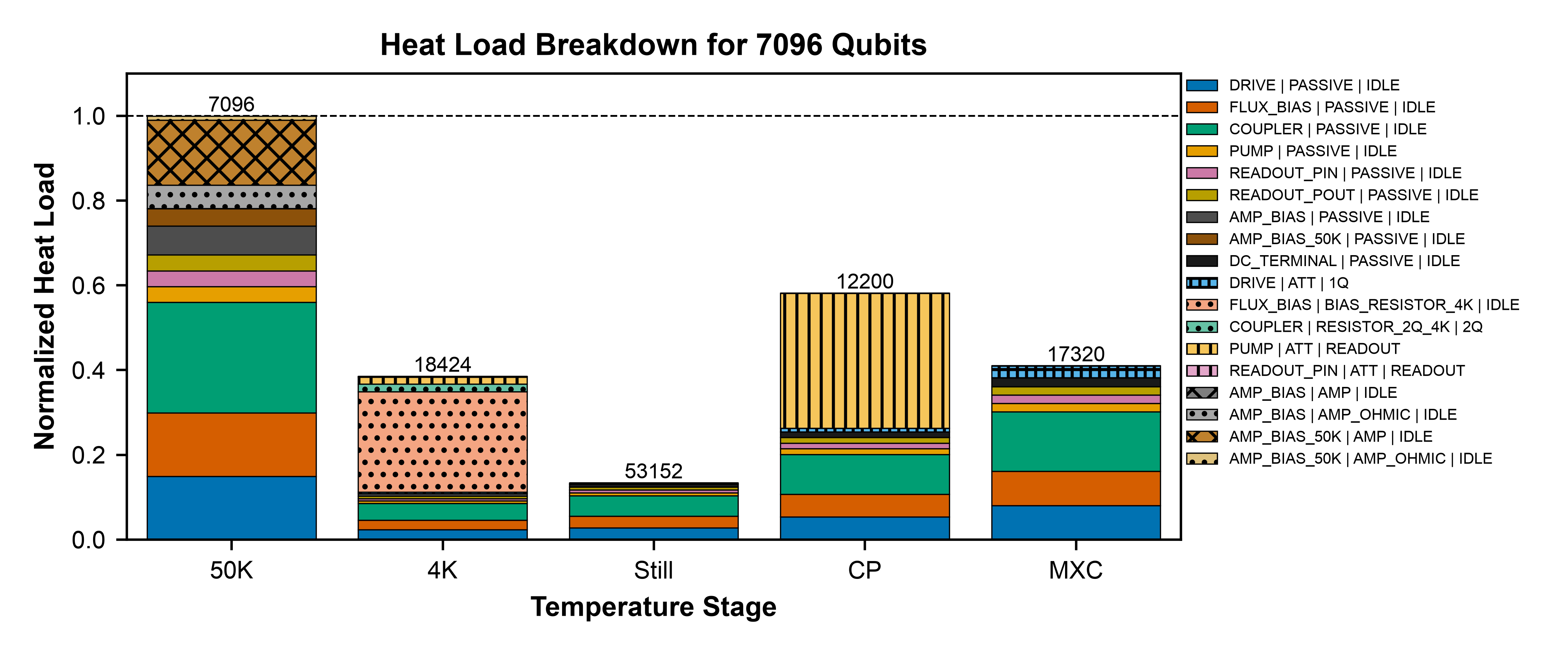}
{\parbox[t]{0.9\linewidth}{Normalized Heat Breakdown assuming a TF-TC system with a hypothetical Ag-YBCO-NbTi microwave wiring and SIS amplifier with YBCO biasing cables.}
\label{fig:tf-tc-ybco_delft-sis_ybco-kide}}
\clearpage

\subsection{FF-FC system | Ag-YBCO-NbTi Wiring | KIDE fridge | SIS \(\times\) YBCO}
\label[appsec]{app:ff-fc-ybco_delft-sis_ybco-kide}
This is a similar system to \Cref{app:tf-tc-ybco_delft-sis_ybco-kide} but for FF-FC qubit architecture (ESM workload with CX gates).

\begin{table*}[!ht]
\centering
\caption{Configuration of an 8-qubit unit cell for the system discussed in \Cref{subsubsec:wire_tradeoffs} (FF-FC).}

\begin{threeparttable}
    
\begin{tabular}{lcclllll}
\toprule
\textbf{Cable}
 & \textbf{Operations} 
 & \textbf{Count} 
 & \textbf{50\,K} 
 & \textbf{4\,K} 
 & \textbf{Still} 
 & \textbf{CP} 
 & \textbf{MXC} 
 \\
\midrule
\textbf{DRIVE} 
& 1Q, 2Q
& 8 
& Ag
& YBCO-Flex (20 dB)
& NbTi (10 dB)
& NbTi (10 dB) 
& NbTi (20 dB) 
\\
\textbf{PUMP} 
& Readout 
& 2 
& Ag  
& YBCO-Flex  (20 dB)
& NbTi
& NbTi (10 dB) 
& NbTi (20 dB)
\\
\textbf{READOUT\_PIN} 
& Readout 
& 2 
& Ag
& YBCO-Flex (20 dB)
& NbTi (10 dB)
& NbTi (10 dB) 
& NbTi (20 dB) 
\\
\textbf{READOUT\_POUT} 
& Readout 
& 2 
& Ag
& YBCO-Flex 
& NbTi 
& NbTi 
& NbTi 
\\
\textbf{AMP\_BIAS} 
& Idle  
& 10
& Manganin (Mn)
& YBCO
\\
\quad \textit{Amplifier} 
& Idle
& 2
& -
& SIS
\\

\textbf{AMP\_BIAS (50\,K)} 
& Idle  
& 6
& Manganin (Mn)
& -
\\
\quad \textit{Amplifier} 
& Idle
& 2
& LNF8G
& -
\\

\addlinespace
\midrule
\textbf{Fridge:} KIDE \\
\quad \textit{Cooling Power} &
& &
\qty{90}{\watt} & 
\qty{6}{\watt} & 
\qty{90}{\milli\watt} & 
\qty{3}{\milli\watt} & 
\qty{90}{\micro\watt} 
\\
\bottomrule
\end{tabular}

\end{threeparttable}
\end{table*}

\begin{table}[]
\centering
\caption{Operating Parameters.}
\begin{tabular}{@{}lcccr@{}}
\toprule
\textbf{Cable/Operation} & \textbf{Ops. per ESM cycle} & \textbf{$P_{MXC}$} & \textbf{Duration (ns)} \\ 
\midrule
\textbf{DRIVE} \\
\quad 1Q                        & 4  & -71 dBm       & 42.67 \\
\quad 2Q                        & 16  & -66 dBm     & 71.1  \\
\textbf{READOUT\_PIN} \\
\quad Readout                   & 4   & -120 dBm      & 215   \\
\textbf{PUMP} \\
\quad Readout                   & 4   & -55 dBm     & 215   \\ 
\bottomrule
\end{tabular}
\end{table}

\Figure[!b](topskip=0pt, botskip=0pt, midskip=0pt)[width=0.99\textwidth]
{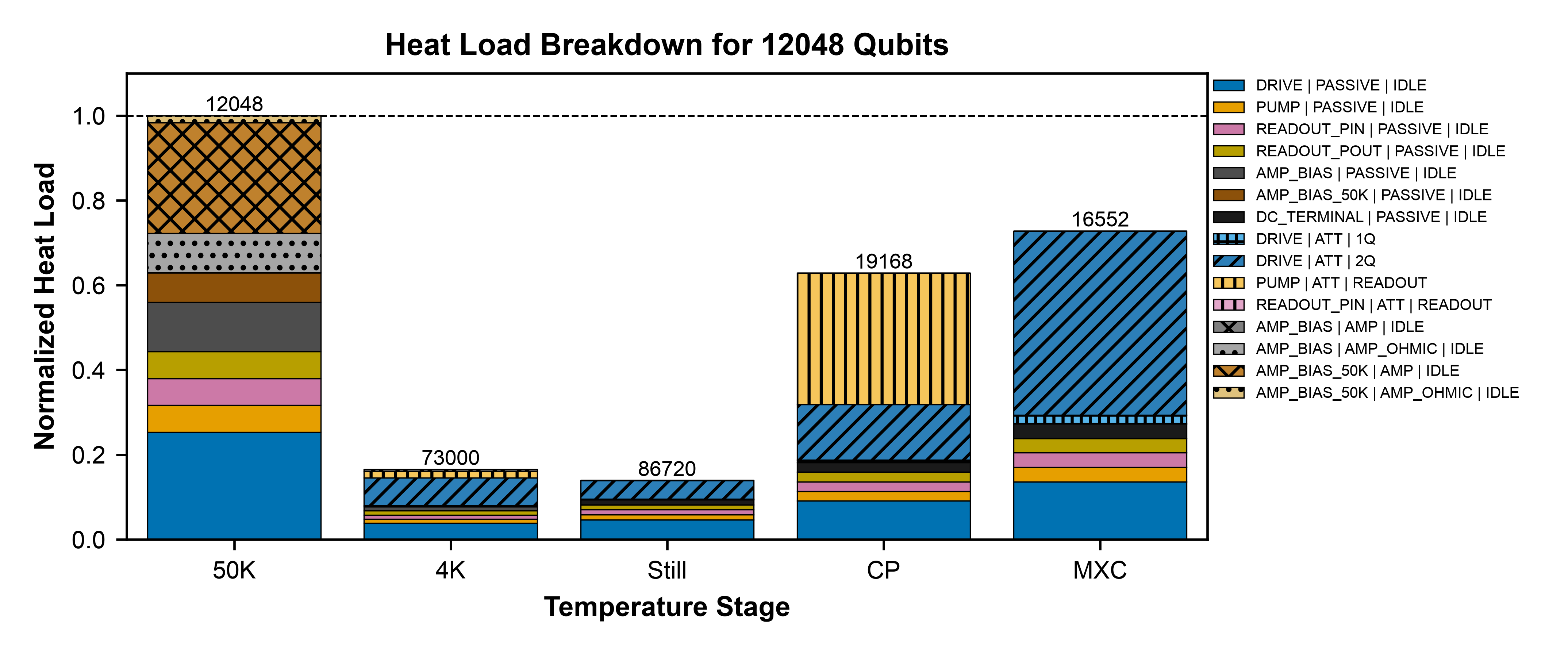}
{\parbox[t]{0.9\linewidth}{Normalized Heat Breakdown assuming an FF-FC system with a hypothetical Ag-YBCO-NbTi microwave wiring and SIS amplifier with YBCO biasing cables.}
\label{fig:ff-fc-ybco_delft-sis_ybco-kide}}
\clearpage

\subsection{FF-FC system | Optical-Baseline | KIDE fridge | ULP-HEMT \(\times\) Manganin}
\label[appsec]{app:ff-fc-kide-optical-baseline}

In \Cref{subsec:optical_kide}, we reevaluate the optical-link-based system proposed in \cite{lecocq2021control} but with a larger KIDE fridge and readout amplifier configuration with a lower thermal footprint (ULP-HEMT $\times$ Manganin). An additional HEMT amplifier is installed at 50\,K. \leg{DRIVE} and \leg{READOUT\_PIN} cables are replaced with optical fibers and the PD is installed at the MXC with the same modeling assumptions as in \Cref{app:ff-fc-optical_MXC-xld400}.

\begin{table*}[!ht]
\centering
\caption{System Configuration for a unit cell of 8 qubits for the system mentioned in \Cref{subsec:optical_kide} where the PD is installed at MXC.}

\begin{threeparttable}
    
\begin{tabular}{lccllllll}
\toprule
\textbf{Cable}
 & \textbf{Operations} 
 & \textbf{Count} 
 & \textbf{50\,K} 
 & \textbf{4\,K} 
 & \textbf{Still} 
 & \textbf{CP} 
 & \textbf{MXC} 
 \\
\midrule
\textbf{DRIVE} 
& 1Q, 2Q
& 8 
& Fiber  
& Fiber 
& Fiber 
& Fiber 
& Fiber (PD)
\\

\textbf{PUMP} 
& Readout 
& 2 
& Ag  
& Ag (20 dB) 
& Ag  
& Ag (10 dB) 
& Ag (20 dB)
\\
\textbf{READOUT\_PIN} 
& Readout 
& 2 
& Fiber  
& Fiber 
& Fiber 
& Fiber 
& Fiber (PD)
\\
\textbf{READOUT\_POUT} 
& Readout 
& 2 
& Ag  
& Ag  
& NbTi  
& NbTi  
& NbTi  
\\
\textbf{DC\_TERMINAL} 
& Idle 
& 2 
& -  
& -  
& NbTi  
& NbTi  
& NbTi  
\\
\textbf{AMP\_BIAS} 
& Idle  
& 12
& Manganin (Mn)
& Manganin (Mn)
\\
\quad \textit{Amplifier} 
& Idle
& 2
& -
& ULP-HEMT
\\

\textbf{AMP\_BIAS (50\,K)} 
& Idle  
& 6
& Manganin (Mn)
& -
\\
\quad \textit{Amplifier} 
& Idle
& 2
& LNF8G
& -
\\

\addlinespace
\midrule
\textbf{Fridge:} KIDE \\
\quad \textit{Cooling Power} &
& &
\qty{90}{\watt} & 
\qty{6}{\watt} & 
\qty{90}{\milli\watt} & 
\qty{3}{\milli\watt} & 
\qty{90}{\micro\watt} 
\\
\bottomrule
\end{tabular}

\end{threeparttable}
\end{table*}

\begin{table}[]
\centering
\caption{Operating Parameters.}
\begin{tabular}{@{}lccr@{}}
\toprule
\textbf{Cable/Operation} & \textbf{Ops. per ESM cycle} & \textbf{$P_{MXC}$} & \textbf{Duration (ns)} \\ \midrule
\textbf{DRIVE}        &    &          &       \\
\quad 1Q                    & 4  & -71 dBm  & 42.67 \\
\quad 2Q                    & 16 & -66 dBm  & 71.1  \\
\textbf{READOUT\_PIN} &    &          &       \\
\quad Readout               & 4  & -120 dBm & 215   \\
\textbf{PUMP}         &    &          &       \\
\quad Readout               & 4  & -55 dBm  & 215   \\ \bottomrule
\end{tabular}
\end{table}

\Figure[!b](topskip=0pt, botskip=0pt, midskip=0pt)[width=0.99\textwidth]
{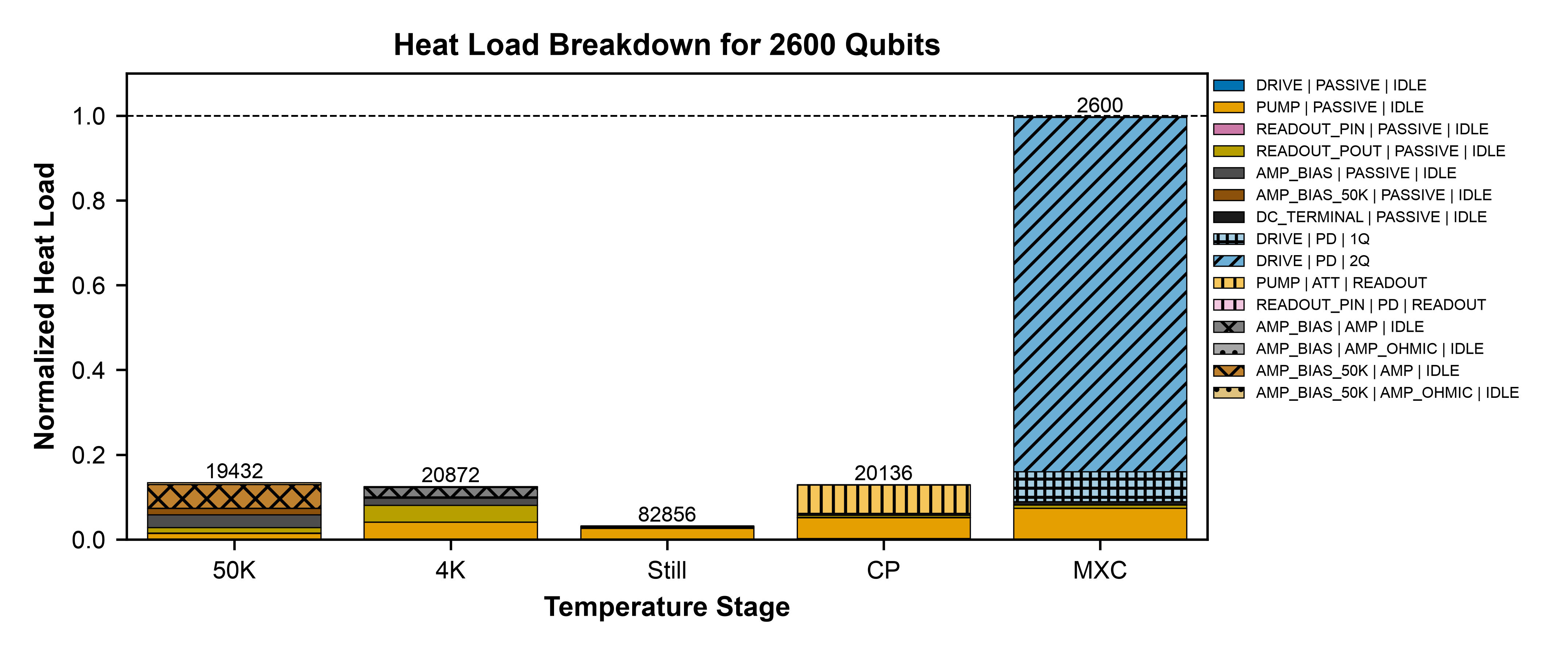}
{\parbox[t]{0.9\linewidth}{Normalized Heat Load Breakdown}
\label{fig:ff-fc-kide-optical-baseline}}
\clearpage

\subsection{FF-FC system | Optical-LP | KIDE fridge | ULP-HEMT \(\times\) Manganin}
\label[appsec]{app:ff-fc-kide-optical-lp}
This system is identical to \Cref{app:ff-fc-kide-optical-baseline} except that the qubit drive powers for 1Q and 2Q operations are reduced by \qty{15}{\decibel}. This reduces the thermal stress at MXC due to PD dissipation enabling the system to accommodate 10k qubits.

\begin{table*}[!ht]
\centering
\caption{System Configuration for a unit cell of 8 qubits for the system mentioned in \Cref{subsec:optical_kide} where the PD is installed at MXC and the qubit drive powers are reduced by \qty{15}{\decibel}.}

\begin{threeparttable}
    
\begin{tabular}{lccllllll}
\toprule
\textbf{Cable}
 & \textbf{Operations} 
 & \textbf{Count} 
 & \textbf{50\,K} 
 & \textbf{4\,K} 
 & \textbf{Still} 
 & \textbf{CP} 
 & \textbf{MXC} 
 \\
\midrule
\textbf{DRIVE} 
& 1Q, 2Q
& 8 
& Fiber  
& Fiber 
& Fiber 
& Fiber 
& Fiber (PD)
\\

\textbf{PUMP} 
& Readout 
& 2 
& Ag  
& Ag (20 dB) 
& Ag  
& Ag (10 dB) 
& Ag (20 dB)
\\
\textbf{READOUT\_PIN} 
& Readout 
& 2 
& Fiber  
& Fiber 
& Fiber 
& Fiber 
& Fiber (PD)
\\
\textbf{READOUT\_POUT} 
& Readout 
& 2 
& Ag  
& Ag  
& NbTi  
& NbTi  
& NbTi  
\\
\textbf{DC\_TERMINAL} 
& Idle 
& 2 
& -  
& -  
& NbTi  
& NbTi  
& NbTi  
\\
\textbf{AMP\_BIAS} 
& Idle  
& 12
& Manganin (Mn)
& Manganin (Mn)
\\
\quad \textit{Amplifier} 
& Idle
& 2
& -
& ULP-HEMT
\\

\textbf{AMP\_BIAS (50\,K)} 
& Idle  
& 6
& Manganin (Mn)
& -
\\
\quad \textit{Amplifier} 
& Idle
& 2
& LNF8G
& -
\\

\addlinespace
\midrule
\textbf{Fridge:} KIDE \\
\quad \textit{Cooling Power} &
& &
\qty{90}{\watt} & 
\qty{6}{\watt} & 
\qty{90}{\milli\watt} & 
\qty{3}{\milli\watt} & 
\qty{90}{\micro\watt} 
\\

\bottomrule
\end{tabular}

\end{threeparttable}
\end{table*}

\begin{table}[]
\centering
\caption{Operating Parameters.}
\begin{tabular}{@{}lccr@{}}
\toprule
\textbf{Cable/Operation} & \textbf{Ops. per ESM cycle} & \textbf{$P_{MXC}$} & \textbf{Duration (ns)} \\ \midrule
\textbf{DRIVE}        &    &          &       \\
\quad 1Q                    & 4  & -86 dBm  & 42.67 \\
\quad 2Q                    & 16 & -81 dBm  & 71.1  \\
\textbf{READOUT\_PIN} &    &          &       \\
\quad Readout               & 4  & -120 dBm & 215   \\
\textbf{PUMP}         &    &          &       \\
\quad Readout               & 4  & -55 dBm  & 215   \\ \bottomrule
\end{tabular}
\end{table}

\Figure[!b](topskip=0pt, botskip=0pt, midskip=0pt)[width=0.99\textwidth]
{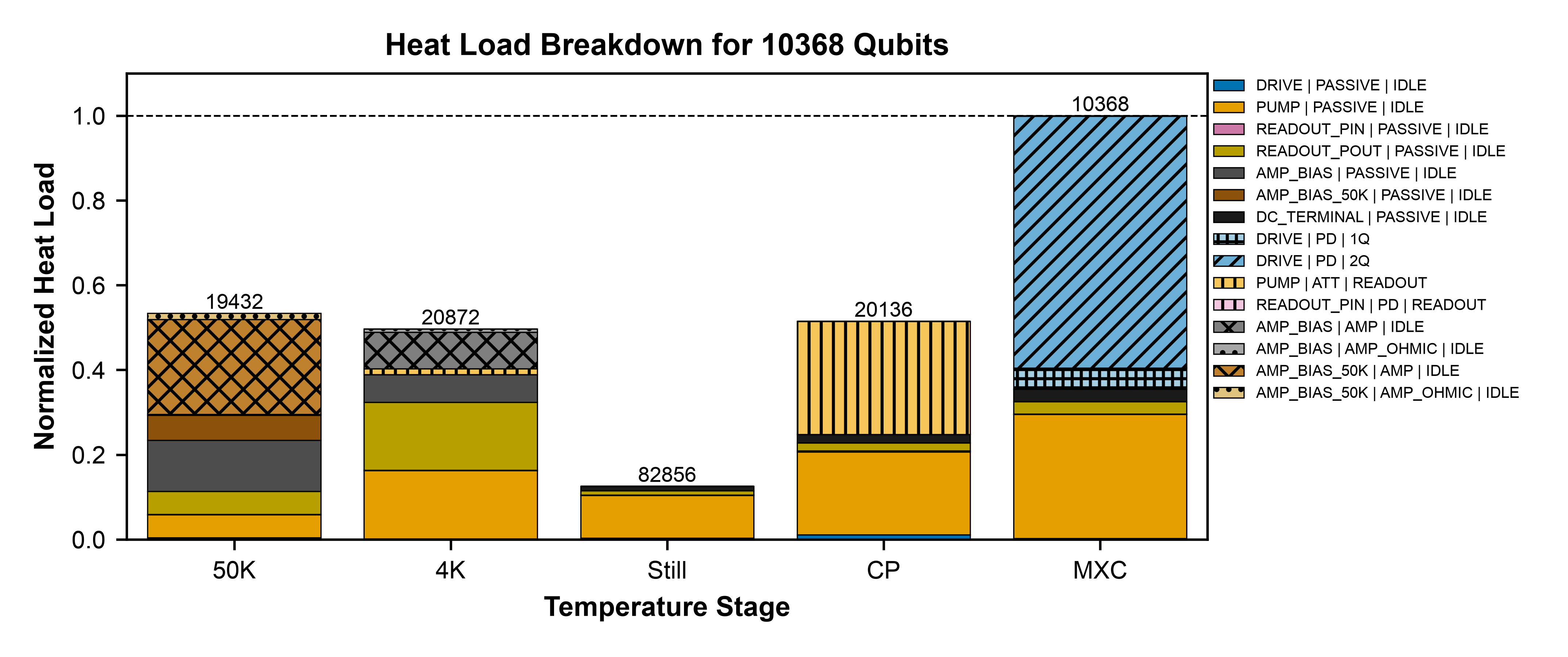}
{\parbox[t]{0.9\linewidth}{Normalized Heat Load Breakdown}
\label{fig:ff-fc-kide-optical-lp}}
\clearpage

\subsection{FF-FC system | Optical-PD at 4\,K | KIDE fridge | ULP-HEMT \(\times\) Manganin}
\label[appsec]{app:ff-fc-kide-optical-pd4k}

This system, discussed in \Cref{subsec:optical_kide}, is similar to the system in \Cref{app:ff-fc-kide-optical-baseline} which uses optical links and a large KIDE fridge but has the PD relocated to 4\,K as in \Cref{app:ff-fc-optical_4K-xld400}.

\begin{table*}[!ht]
\centering
\caption{System Configuration for a unit cell of 8 qubits for the system parameters based on \cite{joshi2023scaling} where the PD is relocated to 4\,K and a higher-capacity KIDE fridge is assumed as mentioned in \Cref{subsec:optical_kide}.}

\begin{threeparttable}
    
\begin{tabular}{lccllllll}
\toprule
\textbf{Cable}
 & \textbf{Operations} 
 & \textbf{Count} 
 & \textbf{50\,K} 
 & \textbf{4\,K} 
 & \textbf{Still} 
 & \textbf{CP} 
 & \textbf{MXC} 
 \\
\midrule
\textbf{DRIVE} 
& 1Q,2Q
& 8 
& Fiber  
& Fiber (PD) 
& NbTi 
& NbTi 
& NbTi 
\\

\textbf{PUMP} 
& Readout 
& 2 
& Ag  
& Ag (20 dB) 
& Ag  
& Ag (10 dB) 
& Ag (20 dB)
\\
\textbf{READOUT\_PIN} 
& Readout 
& 2 
& Fiber  
& Fiber (PD)
& NbTi 
& NbTi 
& NbTi 
\\
\textbf{READOUT\_POUT} 
& Readout 
& 2 
& Ag  
& Ag  
& NbTi  
& NbTi  
& NbTi  
\\
\textbf{DC\_TERMINAL} 
& Idle 
& 2 
& -  
& -  
& NbTi  
& NbTi  
& NbTi  
\\

\textbf{AMP\_BIAS} 
& Idle  
& 12
& Manganin (Mn)
& Manganin (Mn)
\\
\quad \textit{Amplifier} 
& Idle
& 2
& -
& ULP-HEMT
\\

\textbf{AMP\_BIAS (50\,K)} 
& Idle  
& 6
& Manganin (Mn)
& -
\\
\quad \textit{Amplifier} 
& Idle
& 2
& LNF8G
& -
\\

\addlinespace
\midrule
\textbf{Fridge:} KIDE \\
\quad \textit{Cooling Power} &
& &
\qty{90}{\watt} & 
\qty{6}{\watt} & 
\qty{90}{\milli\watt} & 
\qty{3}{\milli\watt} & 
\qty{90}{\micro\watt} 
\\
\bottomrule
\end{tabular}

\end{threeparttable}
\end{table*}

\begin{table}[]
\centering
\caption{Operating Parameters.}
\begin{tabular}{@{}lccr@{}}
\toprule
\textbf{Cable/Operation} & \textbf{Ops. per ESM cycle} & \textbf{$P_{MXC}$} & \textbf{Duration (ns)} \\ \midrule
\textbf{DRIVE}        &    &          &       \\
\quad 1Q                    & 4  & -71 dBm  & 42.67 \\
\quad 2Q                    & 16 & -66 dBm  & 71.1  \\
\textbf{READOUT\_PIN} &    &          &       \\
\quad Readout               & 4  & -120 dBm & 215   \\
\textbf{PUMP}         &    &          &       \\
\quad Readout               & 4  & -55 dBm  & 215   \\ \bottomrule
\end{tabular}
\end{table}

\Figure[!b](topskip=0pt, botskip=0pt, midskip=0pt)[width=0.99\textwidth]
{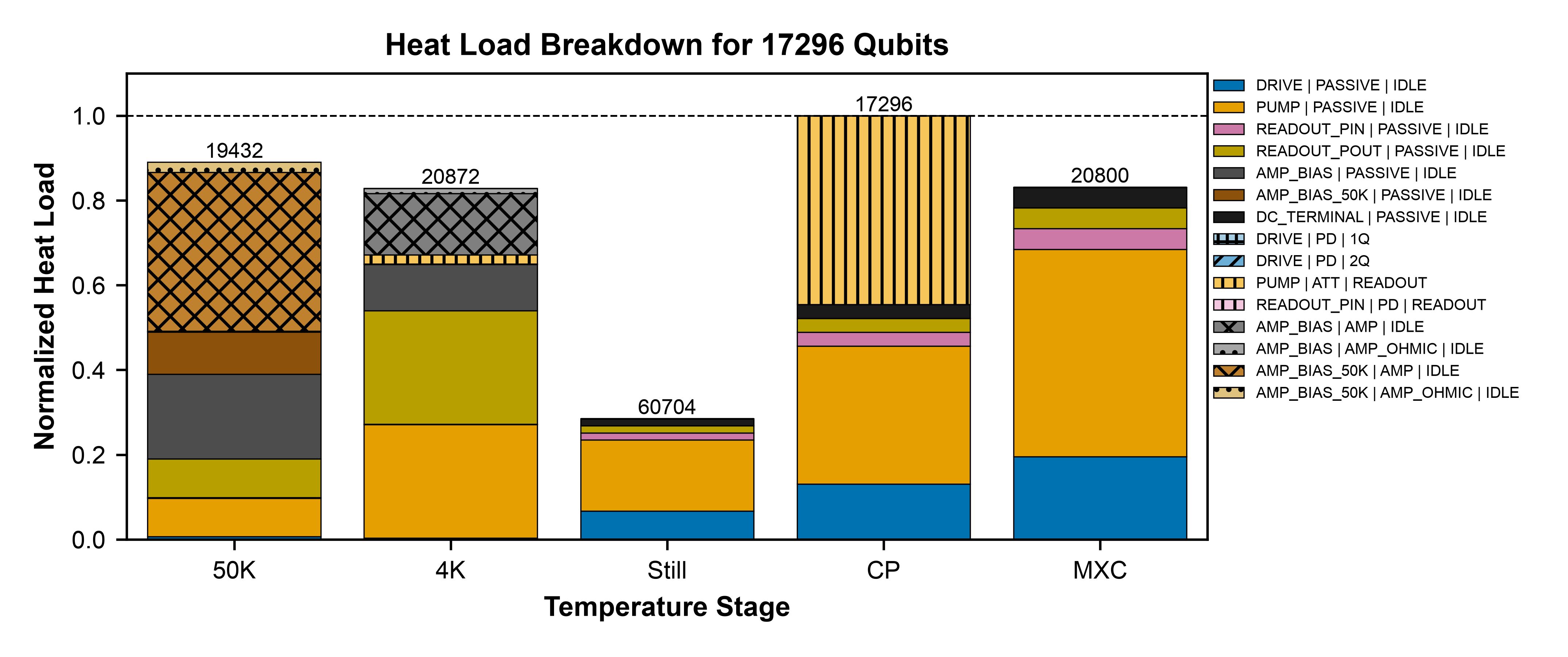}
{\parbox[t]{0.9\linewidth}{Normalized Heat Load Breakdown}
\label{fig:ff-fc-kide-optical-pd4k}}
\clearpage

\subsection{TF-TC system | Cri/oFlex Wiring | Colossus fridge | HEMT \(\times\) Copper}
\label[appsec]{app:tf-tc-delft-hemt_cu-colossus}
Here, we model the thermal performance of a TF-TC system using Cri/oFlex wiring in a much larger Colossus fridge as  mentioned in \Cref{subsec:colossus}. Due to its extremely large cooling power at 4\,K, using conventional LNF8G HEMT-based readout chains with copper wires does not create a thermal bottleneck as it would with other smaller fridges. The fridge also contains an additional 2\,K temperature stage between 4\,K and Still. 

\begin{table*}[!ht]
\centering
\caption{System Configuration for a unit cell of 8 qubits as discussed in \Cref{subsec:colossus}.}

\begin{threeparttable}
    
\begin{tabular}{lcclllllll}
\toprule
\textbf{Cable}
 & \textbf{Operations} 
 & \textbf{Count} 
 & \textbf{50\,K} 
 & \textbf{4\,K} 
 & \textbf{2\,K} 
 & \textbf{Still} 
 & \textbf{CP} 
 & \textbf{MXC} 
 \\
\midrule
\textbf{DRIVE} 
& 1Q
& 8 
& Ag  
& Ag (20 dB)
& Ag
& Ag (10 dB)
& Ag (10 dB)
& Ag (20 dB)
\\
\textbf{FLUX\_BIAS} 
& Idle
& 8 
& Ag
& Ag (20 dB)
& NbTi
& NbTi 
& NbTi
& NbTi 
\\
\textbf{COUPLER} 
& 2Q 
& 14 
& Ag
& Ag (20 dB)
& NbTi
& NbTi
& NbTi 
& NbTi
\\
\textbf{PUMP} 
& Readout 
& 2 
& Ag  
& Ag (20 dB) 
& Ag
& Ag  
& Ag (10 dB) 
& Ag (20 dB)
\\
\textbf{READOUT\_PIN} 
& Readout 
& 2 
& Ag  
& Ag (20 dB)
& Ag
& Ag (10 dB)
& Ag (10 dB)
& Ag (20 dB)
\\
\textbf{READOUT\_POUT} 
& Readout 
& 2 
& Ag  
& Ag  
& NbTi
& NbTi  
& NbTi  
& NbTi  
\\
\textbf{DC\_TERMINAL} 
& Idle 
& 2 
& -  
& -  
& NbTi 
& NbTi  
& NbTi  
& NbTi  
\\

\textbf{AMP\_BIAS} 
& Idle  
& 6
& Cu-30
& Cu-30  
\\

\quad \textit{Amplifier} 
& Idle
& 2
& -
& LNF8G
\\

\addlinespace
\textbf{Fridge:} Colossus \\
\quad \textit{Cooling Power} &
& &
\qty{9}{\kilo\watt} & 
\qty{200}{\watt} & 
\qty{10}{\watt} & 
\qty{100}{\milli\watt} & 
\qty{3}{\milli\watt} & 
\qty{300}{\micro\watt}
\\
\bottomrule
\end{tabular}

\end{threeparttable}
\end{table*}

\begin{table}[]
\centering
\caption{Operating Parameters.}
\begin{tabular}{@{}lcccr@{}}
\toprule
\textbf{Cable/Operation} & \textbf{Ops. per ESM cycle} & \textbf{$P_{MXC}$} & \textbf{Current (mA)} & \textbf{Duration (ns)} \\ 
\midrule
\textbf{DRIVE} \\
\quad 1Q                        & 16  & -71 dBm  & -     & 25 \\
\textbf{FLUX\_BIAS}              &     & -        & 0.4   & -   \\
\textbf{COUPLER} \\
\quad 2Q                        & 16  & -        & 0.4   & 42  \\
\textbf{READOUT\_PIN} \\
\quad Readout                   & 4   & -120 dBm & -     & 375   \\
\textbf{PUMP} \\
\quad Readout                   & 4   & -55 dBm  & -     & 375   \\ 
\bottomrule
\end{tabular}
\end{table}

\Figure[!b](topskip=0pt, botskip=0pt, midskip=0pt)[width=0.99\textwidth]
{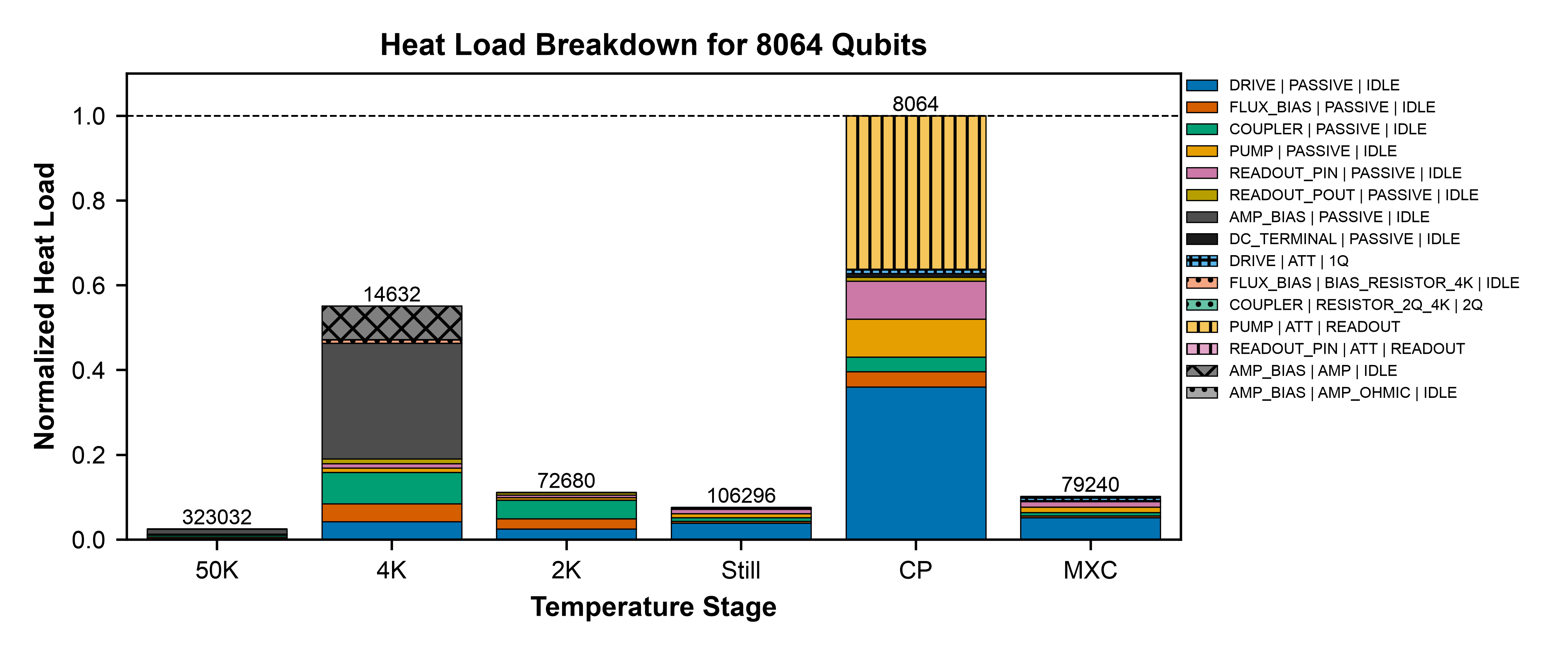}
{\parbox[t]{0.9\linewidth}{Normalized Heat Load Breakdown}
\label{fig:tf-tc-delft-hemt_cu-colossus}}
\clearpage

\subsection{TF-TC system | Cri/oFlex Wiring | Colossus-CP fridge | HEMT \(\times\) Copper}
\label[appsec]{app:tf-tc-delft-hemt_cu-colossus_cp}

This system configuration assumes a hypothetical modification to the original Colossus fridge in which two dilution units originally serving the MXC stage are reassigned to the CP stage (see \Cref{tab:all_fridges}) designated as Colossus-CP. This modification enables the fridge to accommodate more than 10k qubits as discussed in \Cref{subsec:colossus}. 

\begin{table*}[!ht]
\centering
\caption{System Configuration for a unit cell of 8 qubits as discussed in \Cref{subsec:colossus}.}

\begin{threeparttable}
    
\begin{tabular}{lcclllllll}
\toprule
\textbf{Cable}
 & \textbf{Operations} 
 & \textbf{Count} 
 & \textbf{50\,K} 
 & \textbf{4\,K} 
 & \textbf{2\,K} 
 & \textbf{Still} 
 & \textbf{CP} 
 & \textbf{MXC} 
 \\
\midrule
\textbf{DRIVE} 
& 1Q
& 8 
& Ag  
& Ag (20 dB)
& Ag
& Ag (10 dB)
& Ag (10 dB)
& Ag (20 dB)
\\
\textbf{FLUX\_BIAS} 
& Idle
& 8 
& Ag
& Ag (20 dB)
& NbTi
& NbTi 
& NbTi
& NbTi 
\\
\textbf{COUPLER} 
& 2Q 
& 14 
& Ag
& Ag (20 dB)
& NbTi
& NbTi
& NbTi 
& NbTi
\\
\textbf{PUMP} 
& Readout 
& 2 
& Ag  
& Ag (20 dB) 
& Ag
& Ag  
& Ag (10 dB) 
& Ag (20 dB)
\\
\textbf{READOUT\_PIN} 
& Readout 
& 2 
& Ag  
& Ag (20 dB)
& Ag
& Ag (10 dB)
& Ag (10 dB)
& Ag (20 dB)
\\
\textbf{READOUT\_POUT} 
& Readout 
& 2 
& Ag  
& Ag  
& NbTi
& NbTi  
& NbTi  
& NbTi  
\\
\textbf{DC\_TERMINAL} 
& Idle 
& 2 
& -  
& -  
& NbTi 
& NbTi  
& NbTi  
& NbTi  
\\

\textbf{AMP\_BIAS} 
& Idle  
& 6
& Cu-30
& Cu-30  
\\

\quad \textit{Amplifier} 
& Idle
& 2
& -
& LNF8G
\\

\addlinespace
\textbf{Fridge:} Colossus-CP \\
\quad \textit{Cooling Power} &
& &
\qty{9}{\kilo\watt} & 
\qty{200}{\watt} & 
\qty{10}{\watt} & 
\qty{100}{\milli\watt} & 
\qty{6}{\milli\watt} & 
\qty{240}{\micro\watt}
\\
\bottomrule
\end{tabular}

\end{threeparttable}
\end{table*}

\begin{table}[]
\centering
\caption{Operating Parameters.}
\begin{tabular}{@{}lcccr@{}}
\toprule
\textbf{Cable/Operation} & \textbf{Ops. per ESM cycle} & \textbf{$P_{MXC}$} & \textbf{Current (mA)} & \textbf{Duration (ns)} \\ 
\midrule
\textbf{DRIVE} \\
\quad 1Q                        & 16  & -71 dBm  & -     & 25 \\
\textbf{FLUX\_BIAS}              &     & -        & 0.4   & -   \\
\textbf{COUPLER} \\
\quad 2Q                        & 16  & -        & 0.4   & 42  \\
\textbf{READOUT\_PIN} \\
\quad Readout                   & 4   & -120 dBm & -     & 375   \\
\textbf{PUMP} \\
\quad Readout                   & 4   & -55 dBm  & -     & 375   \\ 
\bottomrule
\end{tabular}
\end{table}

\Figure[!b](topskip=0pt, botskip=0pt, midskip=0pt)[width=0.99\textwidth]
{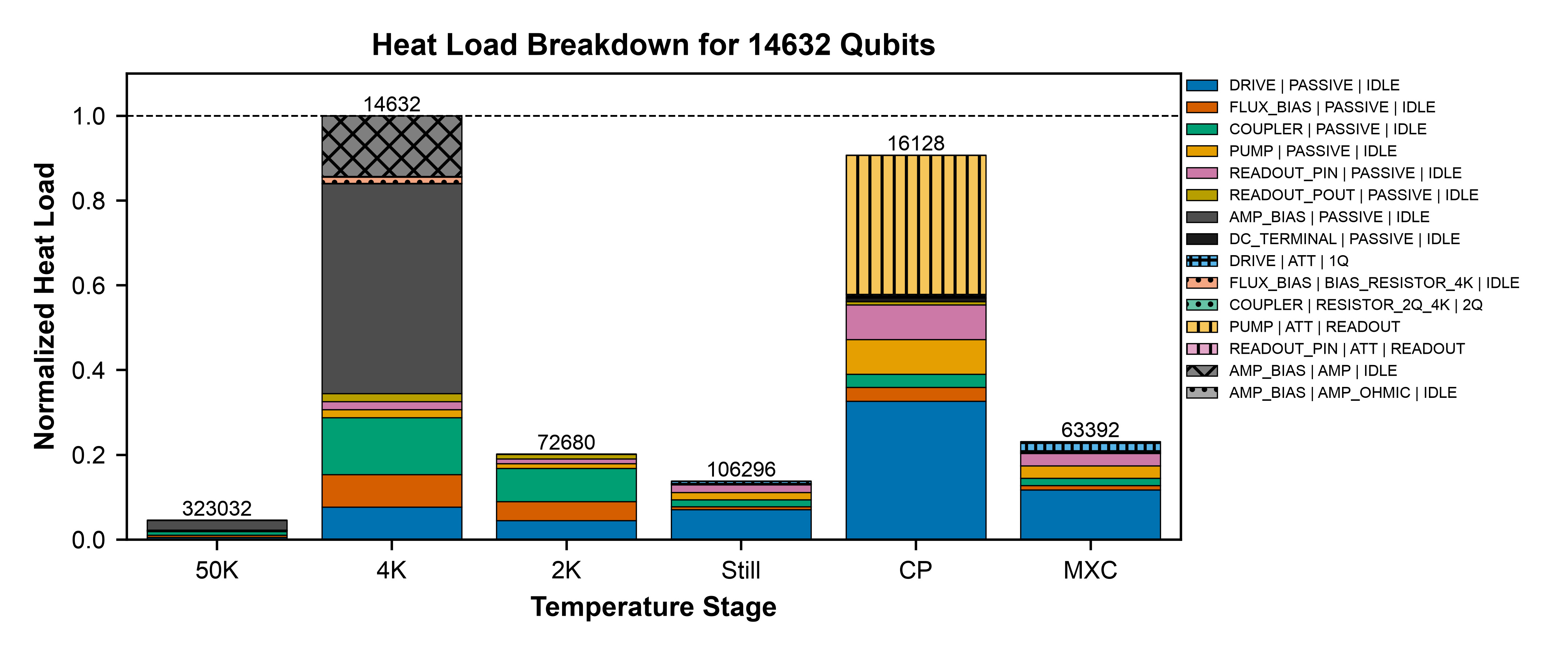}
{\parbox[t]{0.9\linewidth}{Normalized Heat Load Breakdown}
\label{fig:tf-tc-delft-hemt_cu-colossus_cp}}
\clearpage

\subsection{FF-FC system | Cri/oFlex Wiring | Colossus fridge | HEMT \(\times\) Copper}
\label[appsec]{app:ff-fc-delft-hemt_cu-colossus}

This system is similar to \Cref{app:tf-tc-delft-hemt_cu-colossus} that uses the Colossus fridge except that we use the FF-FC architecture and its corresponding workload.

\begin{table*}[!ht]
\centering
\caption{System Configuration for a unit cell of 8 qubits as discussed in \Cref{subsec:colossus}.}

\begin{threeparttable}
    
\begin{tabular}{lcclllllll}
\toprule
\textbf{Cable}
 & \textbf{Operations} 
 & \textbf{Count} 
 & \textbf{50\,K} 
 & \textbf{4\,K} 
 & \textbf{2\,K} 
 & \textbf{Still} 
 & \textbf{CP} 
 & \textbf{MXC} 
 \\
\midrule
\textbf{DRIVE} 
& 1Q, 2Q
& 8 
& Ag  
& Ag (20 dB)
& Ag
& Ag (10 dB)
& Ag (10 dB)
& Ag (20 dB)
\\

\textbf{PUMP} 
& Readout 
& 2 
& Ag  
& Ag (20 dB) 
& Ag
& Ag  
& Ag (10 dB) 
& Ag (20 dB)
\\
\textbf{READOUT\_PIN} 
& Readout 
& 2 
& Ag  
& Ag (20 dB)
& Ag
& Ag (10 dB)
& Ag (10 dB)
& Ag (20 dB)
\\
\textbf{READOUT\_POUT} 
& Readout 
& 2 
& Ag  
& Ag  
& NbTi
& NbTi  
& NbTi  
& NbTi  
\\
\textbf{DC\_TERMINAL} 
& Idle 
& 2 
& -  
& -  
& NbTi 
& NbTi  
& NbTi  
& NbTi  
\\

\textbf{AMP\_BIAS} 
& Idle  
& 6
& Cu-30
& Cu-30  
\\

\quad \textit{Amplifier} 
& Idle
& 2
& -
& LNF8G
\\

\addlinespace
\textbf{Fridge:} Colossus \\
\quad \textit{Cooling Power} &
& &
\qty{9}{\kilo\watt} & 
\qty{200}{\watt} & 
\qty{10}{\watt} & 
\qty{100}{\milli\watt} & 
\qty{3}{\milli\watt} & 
\qty{300}{\micro\watt}
\\
\bottomrule
\end{tabular}

\end{threeparttable}
\end{table*}

\begin{table}[]
\centering
\caption{Operating Parameters.}
\begin{tabular}{@{}lccr@{}}
\toprule
\textbf{Cable/Operation} & \textbf{Ops. per ESM cycle} & \textbf{$P_{MXC}$} & \textbf{Duration (ns)} \\ \midrule
\textbf{DRIVE}        &    &          &       \\
\quad 1Q                    & 4  & -71 dBm  & 42.67 \\
\quad 2Q                    & 16 & -66 dBm  & 71.1  \\
\textbf{READOUT\_PIN} &    &          &       \\
\quad Readout               & 4  & -120 dBm & 215   \\
\textbf{PUMP}         &    &          &       \\
\quad Readout               & 4  & -55 dBm  & 215   \\ \bottomrule
\end{tabular}
\end{table}

\Figure[!b](topskip=0pt, botskip=0pt, midskip=0pt)[width=0.99\textwidth]
{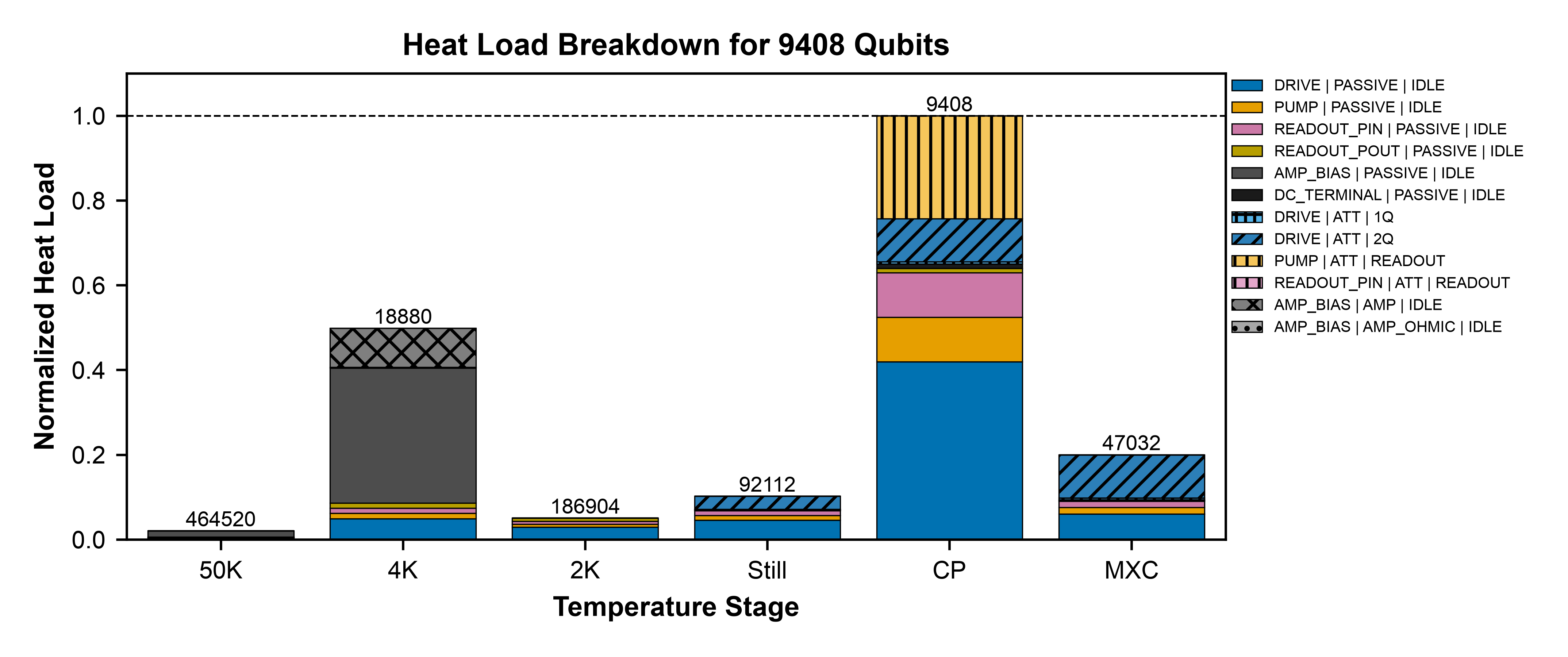}
{\parbox[t]{0.9\linewidth}{Normalized Heat Load Breakdown}
\label{fig:ff-fc-delft-hemt_cu-colossus}}
\clearpage

\subsection{FF-FC system | Cri/oFlex Wiring | Colossus-CP fridge | HEMT \(\times\) Copper}
\label[appsec]{app:ff-fc-delft-hemt_cu-colossus_cp}
This system is similar to \Cref{app:ff-fc-delft-hemt_cu-colossus} except that we assume a hypothetical modification to the original Colossus fridge in which two dilution units originally serving the MXC stage are reassigned to the CP stage (see \Cref{tab:all_fridges}) designated as Colossus-CP.  

\begin{table*}[!ht]
\centering
\caption{System Configuration for a unit cell of 8 qubits as discussed in \Cref{subsec:colossus}.}

\begin{threeparttable}
    
\begin{tabular}{lcclllllll}
\toprule
\textbf{Cable}
 & \textbf{Operations} 
 & \textbf{Count} 
 & \textbf{50\,K} 
 & \textbf{4\,K} 
 & \textbf{2\,K} 
 & \textbf{Still} 
 & \textbf{CP} 
 & \textbf{MXC} 
 \\
\midrule
\textbf{DRIVE} 
& 1Q, 2Q
& 8 
& Ag  
& Ag (20 dB)
& Ag
& Ag (10 dB)
& Ag (10 dB)
& Ag (20 dB)
\\

\textbf{PUMP} 
& Readout 
& 2 
& Ag  
& Ag (20 dB) 
& Ag
& Ag  
& Ag (10 dB) 
& Ag (20 dB)
\\
\textbf{READOUT\_PIN} 
& Readout 
& 2 
& Ag  
& Ag (20 dB)
& Ag
& Ag (10 dB)
& Ag (10 dB)
& Ag (20 dB)
\\
\textbf{READOUT\_POUT} 
& Readout 
& 2 
& Ag  
& Ag  
& NbTi
& NbTi  
& NbTi  
& NbTi  
\\
\textbf{DC\_TERMINAL} 
& Idle 
& 2 
& -  
& -  
& NbTi 
& NbTi  
& NbTi  
& NbTi  
\\

\textbf{AMP\_BIAS} 
& Idle  
& 6
& Cu-30
& Cu-30  
\\

\quad \textit{Amplifier} 
& Idle
& 2
& -
& LNF8G
\\

\addlinespace
\textbf{Fridge:} Colossus-CP \\
\quad \textit{Cooling Power} &
& &
\qty{9}{\kilo\watt} & 
\qty{200}{\watt} & 
\qty{10}{\watt} & 
\qty{100}{\milli\watt} & 
\qty{6}{\milli\watt} & 
\qty{240}{\micro\watt}
\\
\bottomrule
\end{tabular}

\end{threeparttable}
\end{table*}

\begin{table}[]
\centering
\caption{Operating Parameters.}
\begin{tabular}{@{}lccr@{}}
\toprule
\textbf{Cable/Operation} & \textbf{Ops. per ESM cycle} & \textbf{$P_{MXC}$} & \textbf{Duration (ns)} \\ \midrule
\textbf{DRIVE}        &    &          &       \\
\quad 1Q                    & 4  & -71 dBm  & 42.67 \\
\quad 2Q                    & 16 & -66 dBm  & 71.1  \\
\textbf{READOUT\_PIN} &    &          &       \\
\quad Readout               & 4  & -120 dBm & 215   \\
\textbf{PUMP}         &    &          &       \\
\quad Readout               & 4  & -55 dBm  & 215   \\ \bottomrule
\end{tabular}
\end{table}

\Figure[!b](topskip=0pt, botskip=0pt, midskip=0pt)[width=0.99\textwidth]
{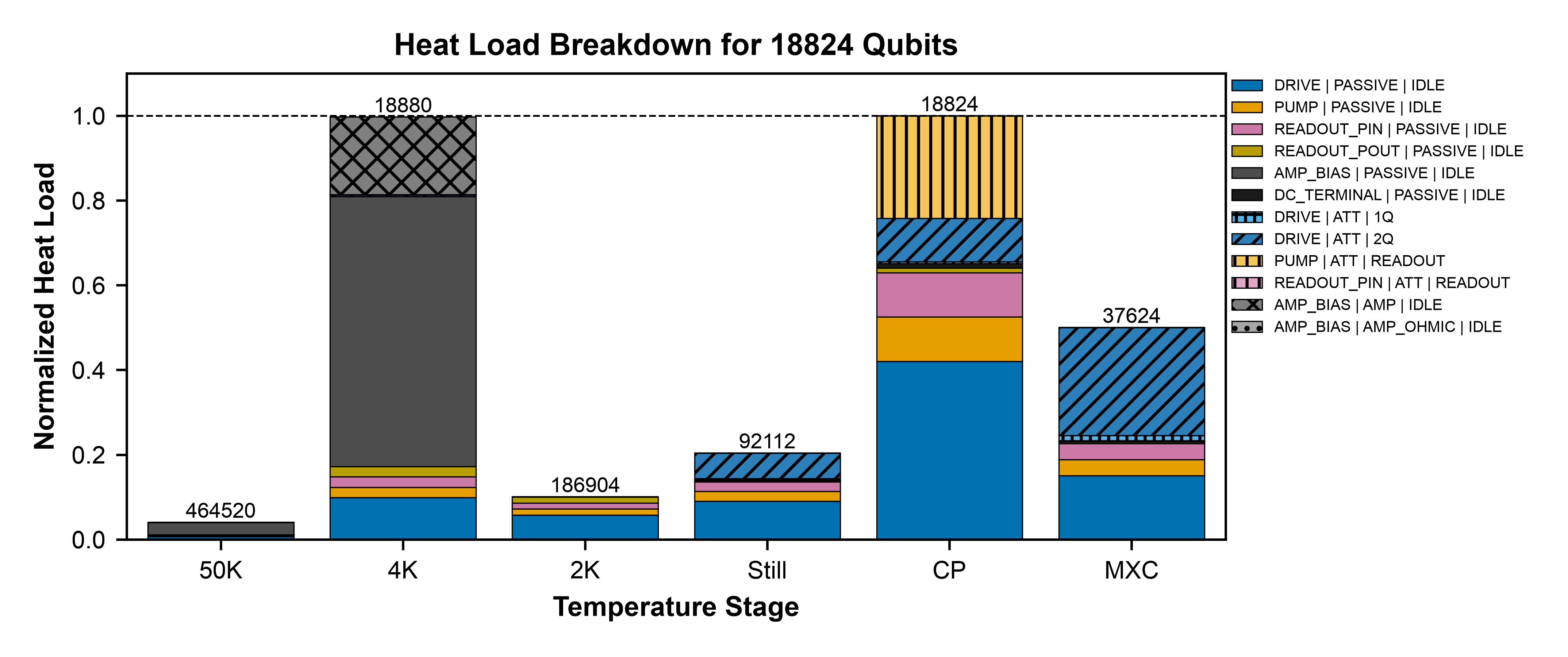}
{\parbox[t]{0.9\linewidth}{Normalized Heat Load Breakdown}
\label{fig:ff-fc-delft-hemt_cu-colossus_cp}}
\clearpage

\twocolumn
\section*{Acknowledgment}
We sincerely thank Yutaka Tabuchi and Shuhei Tamate from RIKEN; Hidehisa Shiomi and Kazuhisa Ogawa from Osaka University; Masamichi Saitoh from ULVAC Cryogenics Inc. and Satoru Masubuchi from Bluefors for their valuable help and discussions.
We acknowledge the use of AI systems for editing to improve readability.
\bibliographystyle{unsrt}
\bibliography{refs}

@article{manifold2025thermal,
  title={Thermal Capacity Mapping of Cryogenic Platforms for Quantum Computers},
  author={Manifold, Scott A and Long, George B and Burnett, Jonathan J},
  journal={arXiv preprint arXiv:2503.10775},
  year={2025}
}

@misc{blueforsKIDECryogenic,
	author = {BlueFors},
	title = {{K}{I}{D}{E} {C}ryogenic {P}latform --- bluefors.com},
	howpublished = {\url{https://bluefors.com/products/kide-cryogenic-platform/}},
	year = {2025},
	note = {[Accessed 07-11-2025]},
}

@inproceedings{hollister2024update,
  title={An update on the Colossus mK platform at Fermilab},
  author={Hollister, Matthew I and Dhuley, Ram C and James, Christopher and Tatkowski, Grzegorz L},
  booktitle={IOP Conference Series: Materials Science and Engineering},
  volume={1302},
  number={1},
  pages={012030},
  year={2024},
  organization={IOP Publishing}
}

@inproceedings{hollister2022large,
  title={A large millikelvin platform at Fermilab for quantum computing applications},
  author={Hollister, Matthew I and Dhuley, Ram C and Tatkowski, Grzegorz L},
  booktitle={IOP Conference Series: Materials Science and Engineering},
  volume={1241},
  number={1},
  pages={012045},
  year={2022},
  organization={IOP Publishing}
}

@misc{globalFujitsuStarts,
	author = {Fujitsu},
	title = {Fujitsu starts official development of plus-10,000 qubit superconducting quantum computer targeting completion in 2030},
	howpublished = {\url{https://global.fujitsu/en-global/pr/news/2025/08/01-01-en}},
	year = {2025},
	note = {[Accessed 04-11-2025]},
}

@misc{delftcircuits,
	author = {Delft Circuits},
	title = {Delft Circuits Product Roadmap},
	howpublished = {\url{https://delft-circuits.com/wp-content/uploads/2025/09/Roadmap_2025.pdf}},
	year = {2025},
	note = {[Accessed 04-11-2025]},
}

@misc{quantumaiRoadmapGoogle,
	author = {Google},
	title = {{R}oadmap | {G}oogle {Q}uantum {A}{I} --- quantumai.google},
	howpublished = {\url{https://quantumai.google/roadmap}},
	year = {2025},
	note = {[Accessed 04-11-2025]},
}

@misc{ibmRoadmap,
	author = {IBM},
	title = {IBM Technology Roadmap},
	howpublished = {\url{https://www.ibm.com/roadmaps/quantum/2025/}},
	year = {2025},
	note = {[Accessed 04-11-2025]},
}

@article{arute2019quantum,
  title={Quantum supremacy using a programmable superconducting processor},
  author={Arute, Frank and Arya, Kunal and Babbush, Ryan and Bacon, Dave and Bardin, Joseph C and Barends, Rami and Biswas, Rupak and Boixo, Sergio and Brandao, Fernando GSL and Buell, David A and others},
  journal={Nature},
  volume={574},
  number={7779},
  pages={505--510},
  year={2019},
  publisher={Nature Publishing Group UK London}
}

@article{jerger2024dispersive,
  title={Dispersive qubit readout with intrinsic resonator reset},
  author={Jerger, M and Motzoi, F and Gao, Y and Dickel, C and Buchmann, L and Bengtsson, A and Tancredi, G and Warren, CW and Bylander, J and DiVincenzo, D and others},
  journal={arXiv preprint arXiv:2406.04891},
  year={2024}
}

@article{mcewen2021removing,
  title={Removing leakage-induced correlated errors in superconducting quantum error correction},
  author={McEwen, Matt and Kafri, Dvir and Chen, Z and Atalaya, Juan and Satzinger, KJ and Quintana, Chris and Klimov, Paul Victor and Sank, Daniel and Gidney, C and Fowler, AG and others},
  journal={Nature communications},
  volume={12},
  number={1},
  pages={1761},
  year={2021},
  publisher={Nature Publishing Group UK London}
}

@article{sunada2022fast,
  title={Fast readout and reset of a superconducting qubit coupled to a resonator with an intrinsic Purcell filter},
  author={Sunada, Yoshiki and Kono, Shingo and Ilves, Jesper and Tamate, Shuhei and Sugiyama, Takanori and Tabuchi, Yutaka and Nakamura, Yasunobu},
  journal={Physical Review Applied},
  volume={17},
  number={4},
  pages={044016},
  year={2022},
  publisher={APS}
}

@article{reed2010fast,
  title={Fast reset and suppressing spontaneous emission of a superconducting qubit},
  author={Reed, Matthew D and Johnson, Blake R and Houck, Andrew A and DiCarlo, Leonardo and Chow, Jerry M and Schuster, David I and Frunzio, Luigi and Schoelkopf, Robert J},
  journal={Applied Physics Letters},
  volume={96},
  number={20},
  year={2010},
  publisher={AIP Publishing}
}

@article{chen2024fast,
  title={Fast unconditional reset and leakage reduction in fixed-frequency transmon qubits},
  author={Chen, Liangyu and Fors, Simon Pettersson and Yan, Zixian and Ali, Anaida and Abad, Tahereh and Osman, Amr and Moschandreou, Eleftherios and Lienhard, Benjamin and Kosen, Sandoko and Li, Hang-Xi and others},
  journal={arXiv preprint arXiv:2409.16748},
  year={2024}
}

@article{kitaev2003fault,
  title={Fault-tolerant quantum computation by anyons},
  author={Kitaev, A Yu},
  journal={Annals of physics},
  volume={303},
  number={1},
  pages={2--30},
  year={2003},
  publisher={Elsevier}
}

@article{bravyi1998quantum,
  title={Quantum codes on a lattice with boundary},
  author={Bravyi, Sergey B and Kitaev, A Yu},
  journal={arXiv preprint quant-ph/9811052},
  year={1998}
}

@article{fowler2012surface,
  title={Surface codes: Towards practical large-scale quantum computation},
  author={Fowler, Austin G and Mariantoni, Matteo and Martinis, John M and Cleland, Andrew N},
  journal={Physical Review A—Atomic, Molecular, and Optical Physics},
  volume={86},
  number={3},
  pages={032324},
  year={2012},
  publisher={APS}
}

@article{raussendorf2007fault,
  title={Fault-tolerant quantum computation with high threshold in two dimensions},
  author={Raussendorf, Robert and Harrington, Jim},
  journal={Physical review letters},
  volume={98},
  number={19},
  pages={190504},
  year={2007},
  publisher={APS}
}

@article{2023acharyaSuppressingQuantumErrors,
  title = {Suppressing Quantum Errors by Scaling a Surface Code Logical Qubit},
  author = {Acharya, Rajeev and Aleiner, Igor and Allen, Richard and Andersen, Trond I. and Ansmann, Markus and Arute, Frank and Arya, Kunal and Asfaw, Abraham and Atalaya, Juan and Babbush, Ryan and Bacon, Dave and Bardin, Joseph C. and Basso, Joao and Bengtsson, Andreas and Boixo, Sergio and Bortoli, Gina and Bourassa, Alexandre and Bovaird, Jenna and Brill, Leon and Broughton, Michael and Buckley, Bob B. and Buell, David A. and Burger, Tim and Burkett, Brian and Bushnell, Nicholas and Chen, Yu and Chen, Zijun and Chiaro, Ben and Cogan, Josh and Collins, Roberto and Conner, Paul and Courtney, William and Crook, Alexander L. and Curtin, Ben and Debroy, Dripto M. and Del Toro Barba, Alexander and Demura, Sean and Dunsworth, Andrew and Eppens, Daniel and Erickson, Catherine and Faoro, Lara and Farhi, Edward and Fatemi, Reza and Flores Burgos, Leslie and Forati, Ebrahim and Fowler, Austin G. and Foxen, Brooks and Giang, William and Gidney, Craig and Gilboa, Dar and Giustina, Marissa and Grajales Dau, Alejandro and Gross, Jonathan A. and Habegger, Steve and Hamilton, Michael C. and Harrigan, Matthew P. and Harrington, Sean D. and Higgott, Oscar and Hilton, Jeremy and Hoffmann, Markus and Hong, Sabrina and Huang, Trent and Huff, Ashley and Huggins, William J. and Ioffe, Lev B. and Isakov, Sergei V. and Iveland, Justin and Jeffrey, Evan and Jiang, Zhang and Jones, Cody and Juhas, Pavol and Kafri, Dvir and Kechedzhi, Kostyantyn and Kelly, Julian and Khattar, Tanuj and Khezri, Mostafa and Kieferov{\'a}, M{\'a}ria and Kim, Seon and Kitaev, Alexei and Klimov, Paul V. and Klots, Andrey R. and Korotkov, Alexander N. and Kostritsa, Fedor and Kreikebaum, John Mark and Landhuis, David and Laptev, Pavel and Lau, Kim-Ming and Laws, Lily and Lee, Joonho and Lee, Kenny and Lester, Brian J. and Lill, Alexander and Liu, Wayne and Locharla, Aditya and Lucero, Erik and Malone, Fionn D. and Marshall, Jeffrey and Martin, Orion and McClean, Jarrod R. and McCourt, Trevor and McEwen, Matt and Megrant, Anthony and Meurer Costa, Bernardo and Mi, Xiao and Miao, Kevin C. and Mohseni, Masoud and Montazeri, Shirin and Morvan, Alexis and Mount, Emily and Mruczkiewicz, Wojciech and Naaman, Ofer and Neeley, Matthew and Neill, Charles and Nersisyan, Ani and Neven, Hartmut and Newman, Michael and Ng, Jiun How and Nguyen, Anthony and Nguyen, Murray and Niu, Murphy Yuezhen and O'Brien, Thomas E. and Opremcak, Alex and Platt, John and Petukhov, Andre and Potter, Rebecca and Pryadko, Leonid P. and Quintana, Chris and Roushan, Pedram and Rubin, Nicholas C. and Saei, Negar and Sank, Daniel and Sankaragomathi, Kannan and Satzinger, Kevin J. and Schurkus, Henry F. and Schuster, Christopher and Shearn, Michael J. and Shorter, Aaron and Shvarts, Vladimir and Skruzny, Jindra and Smelyanskiy, Vadim and Smith, W. Clarke and Sterling, George and Strain, Doug and Szalay, Marco and Torres, Alfredo and Vidal, Guifre and Villalonga, Benjamin and Vollgraff Heidweiller, Catherine and White, Theodore and Xing, Cheng and Yao, Z. Jamie and Yeh, Ping and Yoo, Juhwan and Young, Grayson and Zalcman, Adam and Zhang, Yaxing and Zhu, Ningfeng and {Google Quantum AI}},
  year = 2023,
  month = feb,
  journal = {Nature},
  volume = {614},
  number = {7949},
  pages = {676--681},
  publisher = {Nature Publishing Group},
  issn = {1476-4687},
  doi = {10.1038/s41586-022-05434-1},
  urldate = {2025-01-20},
}

@article{2025acharyaQuantumErrorCorrection,
  title = {Quantum Error Correction below the Surface Code Threshold},
  author = {Acharya, Rajeev and Abanin, Dmitry A. and {Aghababaie-Beni}, Laleh and Aleiner, Igor and Andersen, Trond I. and Ansmann, Markus and Arute, Frank and Arya, Kunal and Asfaw, Abraham and Astrakhantsev, Nikita and Atalaya, Juan and Babbush, Ryan and Bacon, Dave and Ballard, Brian and Bardin, Joseph C. and Bausch, Johannes and Bengtsson, Andreas and Bilmes, Alexander and Blackwell, Sam and Boixo, Sergio and Bortoli, Gina and Bourassa, Alexandre and Bovaird, Jenna and Brill, Leon and Broughton, Michael and Browne, David A. and Buchea, Brett and Buckley, Bob B. and Buell, David A. and Burger, Tim and Burkett, Brian and Bushnell, Nicholas and Cabrera, Anthony and Campero, Juan and Chang, Hung-Shen and Chen, Yu and Chen, Zijun and Chiaro, Ben and Chik, Desmond and Chou, Charina and Claes, Jahan and Cleland, Agnetta Y. and Cogan, Josh and Collins, Roberto and Conner, Paul and Courtney, William and Crook, Alexander L. and Curtin, Ben and Das, Sayan and Davies, Alex and De Lorenzo, Laura and Debroy, Dripto M. and Demura, Sean and Devoret, Michel and Di Paolo, Agustin and Donohoe, Paul and Drozdov, Ilya and Dunsworth, Andrew and Earle, Clint and Edlich, Thomas and Eickbusch, Alec and Elbag, Aviv Moshe and Elzouka, Mahmoud and Erickson, Catherine and Faoro, Lara and Farhi, Edward and Ferreira, Vinicius S. and Burgos, Leslie Flores and Forati, Ebrahim and Fowler, Austin G. and Foxen, Brooks and Ganjam, Suhas and Garcia, Gonzalo and Gasca, Robert and Genois, {\'E}lie and Giang, William and Gidney, Craig and Gilboa, Dar and Gosula, Raja and Dau, Alejandro Grajales and Graumann, Dietrich and Greene, Alex and Gross, Jonathan A. and Habegger, Steve and Hall, John and Hamilton, Michael C. and Hansen, Monica and Harrigan, Matthew P. and Harrington, Sean D. and Heras, Francisco J. H. and Heslin, Stephen and Heu, Paula and Higgott, Oscar and Hill, Gordon and Hilton, Jeremy and Holland, George and Hong, Sabrina and Huang, Hsin-Yuan and Huff, Ashley and Huggins, William J. and Ioffe, Lev B. and Isakov, Sergei V. and Iveland, Justin and Jeffrey, Evan and Jiang, Zhang and Jones, Cody and Jordan, Stephen and Joshi, Chaitali and Juhas, Pavol and Kafri, Dvir and Kang, Hui and Karamlou, Amir H. and Kechedzhi, Kostyantyn and Kelly, Julian and Khaire, Trupti and Khattar, Tanuj and Khezri, Mostafa and Kim, Seon and Klimov, Paul V. and Klots, Andrey R. and Kobrin, Bryce and Kohli, Pushmeet and Korotkov, Alexander N. and Kostritsa, Fedor and Kothari, Robin and Kozlovskii, Borislav and Kreikebaum, John Mark and Kurilovich, Vladislav D. and Lacroix, Nathan and Landhuis, David and {Lange-Dei}, Tiano and Langley, Brandon W. and Laptev, Pavel and Lau, Kim-Ming and Le Guevel, Lo{\"i}ck and Ledford, Justin and Lee, Joonho and Lee, Kenny and Lensky, Yuri D. and Leon, Shannon and Lester, Brian J. and Li, Wing Yan and Li, Yin and Lill, Alexander T. and Liu, Wayne and Livingston, William P. and Locharla, Aditya and Lucero, Erik and Lundahl, Daniel and Lunt, Aaron and Madhuk, Sid and Malone, Fionn D. and Maloney, Ashley and Mandr{\`a}, Salvatore and Manyika, James and Martin, Leigh S. and Martin, Orion and Martin, Steven and Maxfield, Cameron and McClean, Jarrod R. and McEwen, Matt and Meeks, Seneca and Megrant, Anthony and Mi, Xiao and Miao, Kevin C. and Mieszala, Amanda and Molavi, Reza and Molina, Sebastian and Montazeri, Shirin and Morvan, Alexis and Movassagh, Ramis and Mruczkiewicz, Wojciech and Naaman, Ofer and Neeley, Matthew and Neill, Charles and Nersisyan, Ani and Neven, Hartmut and Newman, Michael and Ng, Jiun How and Nguyen, Anthony and Nguyen, Murray and Ni, Chia-Hung and Niu, Murphy Yuezhen and O'Brien, Thomas E. and Oliver, William D. and Opremcak, Alex and Ottosson, Kristoffer and Petukhov, Andre and Pizzuto, Alex and Platt, John and Potter, Rebecca and Pritchard, Orion and Pryadko, Leonid P. and Quintana, Chris and Ramachandran, Ganesh and Reagor, Matthew J. and Redding, John and Rhodes, David M. and Roberts, Gabrielle and Rosenberg, Eliott and Rosenfeld, Emma and Roushan, Pedram and Rubin, Nicholas C. and Saei, Negar and Sank, Daniel and Sankaragomathi, Kannan and Satzinger, Kevin J. and Schurkus, Henry F. and Schuster, Christopher and Senior, Andrew W. and Shearn, Michael J. and Shorter, Aaron and Shutty, Noah and Shvarts, Vladimir and Singh, Shraddha and Sivak, Volodymyr and Skruzny, Jindra and Small, Spencer and Smelyanskiy, Vadim and Smith, W. Clarke and Somma, Rolando D. and Springer, Sofia and Sterling, George and Strain, Doug and Suchard, Jordan and Szasz, Aaron and Sztein, Alex and Thor, Douglas and Torres, Alfredo and Torunbalci, M. Mert and Vaishnav, Abeer and Vargas, Justin and Vdovichev, Sergey and Vidal, Guifre and Villalonga, Benjamin and Heidweiller, Catherine Vollgraff and Waltman, Steven and Wang, Shannon X. and Ware, Brayden and Weber, Kate and Weidel, Travis and White, Theodore and Wong, Kristi and Woo, Bryan W. K. and Xing, Cheng and Yao, Z. Jamie and Yeh, Ping and Ying, Bicheng and Yoo, Juhwan and Yosri, Noureldin and Young, Grayson and Zalcman, Adam and Zhang, Yaxing and Zhu, Ningfeng and Zobrist, Nicholas and {Google Quantum AI and Collaborators}},
  year = 2025,
  journal = {Nature},
  volume = {638},
  number = {8052},
  pages = {920--926},
  publisher = {Nature Publishing Group}
}

@article{versluis2017scalable,
  title={Scalable quantum circuit and control for a superconducting surface code},
  author={Versluis, Richard and Poletto, Stefano and Khammassi, Nader and Tarasinski, Brian and Haider, Nadia and Michalak, David J and Bruno, Alessandro and Bertels, Koen and DiCarlo, Leonardo},
  journal={Physical Review Applied},
  volume={8},
  number={3},
  pages={034021},
  year={2017},
  publisher={APS}
}

@article{croot2025enabling,
  title={Enabling Technologies for Scalable Superconducting Quantum Computing},
  author={Croot, Xanthe and Nowrouzi, Kasra and Spitzer, Christopher and Almudever, Carmen G and Blais, Alexandre and Carroll, Malcolm and Chow, Jerry and Friedman, Daniel and Tokunari, Masao and Charbon, Edoardo and others},
  journal={arXiv preprint arXiv:2512.15001},
  year={2025}
}

@misc{quantwareQuantWareAnnounces,
	author = {QuantWare},
	title = {{Q}uant{W}are announces scaling breakthrough with {V}{I}{O}-40{K}™, delivering 10,000 qubit {Q}uantum {P}rocessors for the first time | {Q}uantware --- quantware.com},
	howpublished = {\url{https://quantware.com/news/quantware-announces-scaling-breakthrough-with-vio-40k}},
	year = {08-12-2025},
	note = {[Accessed 29-01-2026]},
}

@article{raicu2025cryogenic,
  title={Cryogenic thermal modeling of microwave high density signaling},
  author={Raicu, Naomi and Hogan, Tom and Wu, Xian and Vahidpour, Mehrnoosh and Snow, David and Hollister, Matthew and Field, Mark},
  journal={EPJ Quantum Technology},
  volume={12},
  number={1},
  pages={124},
  year={2025},
  publisher={Springer}
}

@article{mohseni2024build,
  title={How to build a quantum supercomputer: Scaling from hundreds to millions of qubits},
  author={Mohseni, Masoud and Scherer, Artur and Johnson, K Grace and Wertheim, Oded and Otten, Matthew and Aadit, Navid Anjum and Alexeev, Yuri and Bresniker, Kirk M and Camsari, Kerem Y and Chapman, Barbara and others},
  journal={arXiv preprint arXiv:2411.10406},
  year={2024}
}

@inproceedings{min2023qisim,
  title={Qisim: Architecting 10+ k qubit qc interfaces toward quantum supremacy},
  author={Min, Dongmoon and Kim, Junpyo and Choi, Junhyuk and Byun, Ilkwon and Tanaka, Masamitsu and Inoue, Koji and Kim, Jangwoo},
  booktitle={Proceedings of the 50th Annual International Symposium on Computer Architecture},
  pages={1--16},
  year={2023}
}

@article{krinner2019engineering,
  title={Engineering cryogenic setups for 100-qubit scale superconducting circuit systems},
  author={Krinner, Sebastian and Storz, Simon and Kurpiers, Philipp and Magnard, Paul and Heinsoo, Johannes and Keller, Raphael and Luetolf, Janis and Eichler, Christopher and Wallraff, Andreas},
  journal={EPJ Quantum Technology},
  volume={6},
  number={1},
  pages={2},
  year={2019},
  publisher={Springer Berlin Heidelberg}
}

@article{joshi2023scaling,
  title={Scaling up superconducting quantum computers with cryogenic RF-photonics},
  author={Joshi, Sanskriti and Moazeni, Sajjad},
  journal={Journal of Lightwave Technology},
  volume={42},
  number={1},
  pages={166--175},
  year={2023},
  publisher={IEEE}
}

@article{lecocq2021control,
  title={Control and readout of a superconducting qubit using a photonic link},
  author={Lecocq, Florent and Quinlan, Franklyn and Cicak, Katarina and Aumentado, Jose and Diddams, SA and Teufel, JD},
  journal={Nature},
  volume={591},
  number={7851},
  pages={575--579},
  year={2021},
  publisher={Nature Publishing Group UK London}
}

@article{ranadive2022kerr,
  title={Kerr reversal in Josephson meta-material and traveling wave parametric amplification},
  author={Ranadive, Arpit and Esposito, Martina and Planat, Luca and Bonet, Edgar and Naud, C{\'e}cile and Buisson, Olivier and Guichard, Wiebke and Roch, Nicolas},
  journal={Nature communications},
  volume={13},
  number={1},
  pages={1737},
  year={2022},
  publisher={Nature Publishing Group UK London}
}

@misc{quantummicrowave,
	author = {Quantum Microwave},
	title = {QMC-CRYOCOUPLER-20NM},
	howpublished = {\url{https://quantummicrowave.com/wp-content/uploads/2023/01/{Q}{M}{C}-{C}{R}{Y}{O}{C}{O}{U}{P}{L}{E}{R}-20{N}{M}-2.pdf}},
	year = {2026},
	note = {[Accessed 21-07-2026]},
}

@article{gaydamachenko2025rf,
  title={rf-SQUID-based traveling-wave parametric amplifier with input saturation power of- 84 dBm across more than one octave in bandwidth},
  author={Gaydamachenko, Victor and Kissling, Christoph and Gr{\"u}nhaupt, Lukas},
  journal={Physical Review Applied},
  volume={23},
  number={6},
  pages={064053},
  year={2025},
  publisher={APS}
}

@phdthesis{lienhard2021machine,
  title={Machine learning assisted superconducting qubit readout},
  author={Lienhard, Benjamin},
  year={2021},
  school={Massachusetts Institute of Technology}
}

@article{macklin2015near,
  title={A near--quantum-limited Josephson traveling-wave parametric amplifier},
  author={Macklin, Chris and O’brien, K and Hover, D and Schwartz, ME and Bolkhovsky, V and Zhang, X and Oliver, WD and Siddiqi, I},
  journal={Science},
  volume={350},
  number={6258},
  pages={307--310},
  year={2015},
  publisher={American Association for the Advancement of Science}
}

@misc{zurichshhqa,
	author = {Zurich Instruments},
	title = {{P}roduct {L}eaflet: {S}{H}{F}{Q}{A} 8.5 {G}{H}z {Q}uantum {A}nalyzer},
	howpublished = {\url{https://www.zhinst.com/sites/default/files/documents/2023-02/zi_shfqa_leaflet.pdf}},
	year = {2023-02},
	note = {[Accessed 09-02-2026]},
}

@misc{lownoisefactoryLNFLNC4_8GNoise,
	author = {Low Noise Factory},
	title = {{L}{N}{F}-{L}{N}{C}4\_8{G} - {L}ow {N}oise {F}actory --- lownoisefactory.com},
	howpublished = {\url{https://lownoisefactory.com/product/lnf-lnc4_8g/}},
	year = {2025},
	note = {[Accessed 04-12-2025]},
}

@misc{lownoisefactoryLNFLNC4_8C,
  author = {Low Noise Factory},
	title = {{L}{N}{F}-{L}{N}{C}4\_8{G} - {L}ow {N}oise {F}actory --- lownoisefactory.com},
	howpublished = {\url{https://lownoisefactory.com/wp-content/uploads/2022/03/lnf-lnc4\_8c.pdf}},
	year = {2022-03-28},
	note = {[Accessed 23-02-2026]},
}

@inproceedings{cha2020300,
  title={A 300-$\mu$W cryogenic HEMT LNA for quantum computing},
  author={Cha, Eunjung and Wadefalk, Niklas and Moschetti, Giuseppe and Pourkabirian, Arsalan and Stenarson, J{\"o}rgen and Grahn, Jan},
  booktitle={2020 IEEE/MTT-S International Microwave Symposium (IMS)},
  pages={1299--1302},
  year={2020},
  organization={IEEE}
}

@article{zeng2023sub,
  title={Sub-mW cryogenic InP HEMT LNA for qubit readout},
  author={Zeng, Yin and Stenarson, J{\"o}rgen and Sobis, Peter and Wadefalk, Niklas and Grahn, Jan},
  journal={IEEE Transactions on Microwave Theory and Techniques},
  volume={72},
  number={3},
  pages={1606--1617},
  year={2023},
  publisher={IEEE}
}

@article{murayama2024fabrication,
  title={Fabrication and Evaluation of waveguide Josephson Array Oscillators at millimeter-wave lengths for SIS mixer-based amplifiers},
  author={Murayama, Yosuke and Kawakami, Akira and Shan, Wenlei and Ezaki, Shohei and Miyachi, Akihira and Masui, Sho and Kojima, Takafumi and Uzawa, Yoshinori},
  journal={IEEE Transactions on Applied Superconductivity},
  year={2024},
  publisher={IEEE}
}

@article{kojima2023characterization,
  title={Characterization of a low-noise superconductor--insulator--superconductor-based microwave amplifier with local oscillator phase-adjusting architecture},
  author={Kojima, T and Masui, S and Shan, W and Uzawa, Y},
  journal={Applied Physics Letters},
  volume={122},
  number={7},
  year={2023},
  publisher={AIP Publishing}
}

@article{spring2025fast,
  title={Fast Multiplexed Superconducting-Qubit Readout with Intrinsic Purcell Filtering Using a Multiconductor Transmission Line},
  author={Spring, Peter A and Milanovic, Luka and Sunada, Yoshiki and Wang, Shiyu and Van Loo, Arjan F and Tamate, Shuhei and Nakamura, Yasunobu},
  journal={PRX Quantum},
  volume={6},
  number={2},
  pages={020345},
  year={2025},
  publisher={APS}
}

@article{underwood2024using,
  title={Using cryogenic CMOS control electronics to enable a two-qubit cross-resonance gate},
  author={Underwood, Devin and Glick, Joseph A and Inoue, Ken and Frank, David J and Timmerwilke, John and Pritchett, Emily and Chakraborty, Sudipto and Tien, Kevin and Yeck, Mark and Bulzacchelli, John F and others},
  journal={PRX Quantum},
  volume={5},
  number={1},
  pages={010326},
  year={2024},
  publisher={APS}
}

@article{patterson2019calibration,
  title={Calibration of a cross-resonance two-qubit gate between directly coupled transmons},
  author={Patterson, AD and Rahamim, J and Tsunoda, T and Spring, PA and Jebari, S and Ratter, K and Mergenthaler, M and Tancredi, G and Vlastakis, B and Esposito, M and others},
  journal={Physical Review Applied},
  volume={12},
  number={6},
  pages={064013},
  year={2019},
  publisher={APS}
}

@article{malekakhlagh2020first,
  title={First-principles analysis of cross-resonance gate operation},
  author={Malekakhlagh, Moein and Magesan, Easwar and McKay, David C},
  journal={arXiv preprint arXiv:2005.00133},
  year={2020}
}

@article{kandala2021demonstration,
  title={Demonstration of a high-fidelity cnot gate for fixed-frequency transmons with engineered zz suppression},
  author={Kandala, Abhinav and Wei, Ken X and Srinivasan, Srikanth and Magesan, Easwar and Carnevale, S and Keefe, GA and Klaus, D and Dial, O and McKay, DC},
  journal={Physical Review Letters},
  volume={127},
  number={13},
  pages={130501},
  year={2021},
  publisher={APS}
}

@article{sung2021realization,
  title={Realization of high-fidelity CZ and ZZ-free iSWAP gates with a tunable coupler},
  author={Sung, Youngkyu and Ding, Leon and Braum{\"u}ller, Jochen and Veps{\"a}l{\"a}inen, Antti and Kannan, Bharath and Kjaergaard, Morten and Greene, Ami and Samach, Gabriel O and McNally, Chris and Kim, David and others},
  journal={Physical Review X},
  volume={11},
  number={2},
  pages={021058},
  year={2021},
  publisher={APS}
}

@article{krinner2020benchmarking,
  title={Benchmarking coherent errors in controlled-phase gates due to spectator qubits},
  author={Krinner, Sebastian and Lazar, Stefania and Remm, Ants and Andersen, Christian K and Lacroix, Nathan and Norris, Graham J and Hellings, Christoph and Gabureac, Mihai and Eichler, Christopher and Wallraff, Andreas},
  journal={Physical Review Applied},
  volume={14},
  number={2},
  pages={024042},
  year={2020},
  publisher={APS}
}

@article{bardin2021microwaves,
  title={Microwaves in quantum computing},
  author={Bardin, Joseph C and Slichter, Daniel H and Reilly, David J},
  journal={IEEE journal of microwaves},
  volume={1},
  number={1},
  pages={403--427},
  year={2021},
  publisher={IEEE}
}

@misc{coax2022,
	author = {coax.co.jp},
	title = {{C}ryogenic {C}atalogue 2022},
	howpublished = {\url{https://www.coax.co.jp/wcaxp/wp-content/uploads/2022/10/Cryogenic_catalogue_2022.pdf}},
	year = {2022},
	note = {[Accessed 15-07-2026]},
}

@misc{blueforsProductsEnhanced,
	author = {Bluefors},
	title = {{N}ew {P}roducts: {E}nhanced {C}ooling, {H}igh-{D}ensity {F}lex {W}iring, and {L}{D}400sl --- bluefors.com},
	howpublished = {\url{https://bluefors.com/news/new-products-enhanced-cooling-high-density-flex-wiring-and-ld400sl/}},
	year = {17-March-2025},
	note = {[Accessed 15-02-2026]},
}

@article{monarkha2024equivalence,
  title={Equivalence of flexible stripline and coaxial cables for superconducting qubit control and readout pulses},
  author={Monarkha, VY and Simbierowicz, Slawomir and Borrelli, M and van Gulik, R and Drobotun, N and Kuitenbrouwer, D and Bouman, D and Datta, Debom and Eskelinen, Patrik and Mannila, Elsa and others},
  journal={Applied Physics Letters},
  volume={124},
  number={22},
  year={2024},
  publisher={AIP Publishing}
}

@article{smith1978effect,
  title={Effect of neutron irradiation on the density of low-energy excitations in vitreous silica},
  author={Smith, Terry Lee and Anthony, PJ and Anderson, AC},
  journal={Physical Review B},
  volume={17},
  number={12},
  pages={4997},
  year={1978},
  publisher={APS}
}

@article{paluch2025thermalization,
  title={Thermalization of a flexible microwave stripline measured by a superconducting qubit},
  author={Paluch, Patrick and Spiecker, Martin and Gosling, Nicolas and Adam, Viktor and Kammhuber, Jakob and Vermeulen, Kiefer and Bouman, Dani{\"e}l and Wernsdorfer, Wolfgang and Pop, Ioan M},
  journal={Applied Physics Letters},
  volume={126},
  number={3},
  year={2025},
  publisher={AIP Publishing}
}

@article{solovyov2021ybco,
  title={YBCO-on-Kapton: Material for high-density quantum computer interconnects with ultra-low thermal loss},
  author={Solovyov, Vyacheslav and Saira, Olli-Pentti and Mendleson, Zachary and Drozdov, Ilya},
  journal={IEEE Transactions on Applied Superconductivity},
  volume={31},
  number={5},
  pages={1--5},
  year={2021},
  publisher={IEEE}
}

@misc{twire,
	author = {Tokyo Wire Works},
	title = {twire.co.jp},
	howpublished = {\url{https://www.twire.co.jp/english/{A}lloy2-en.html}},
	year = {2026},
	note = {[Accessed 10-02-2026]},
}

@misc{delftcircuitsCryogenicCables,
	author = {Delft Circuits},
	title = {Flexible Cryogenic i/o},
	howpublished = {\url{https://photonteck.com/uploads/20241109/b3531c2c55a53a8e034ba1e998f3f7ee.pdf}},
	year = {22-02-2024},
	note = {[Accessed 19-12-2025]},
}

@misc{blueforsHighDensityWiring,
	author = {BlueFors},
	title = {{H}igh-{D}ensity {W}iring --- bluefors.com},
	howpublished = {\url{https://bluefors.com/products/measurement-infrastructure/high-density-wiring/}},
	year = {2025},
	note = {[Accessed 11-12-2025]},
}

@misc{lownoisefactoryLNFNANO9MNoise,
	author = {Low Noise Factory},
	title = {{L}{N}{F}-{N}{A}{N}{O}9{M} - {L}ow {N}oise {F}actory --- lownoisefactory.com},
	howpublished = {\url{https://lownoisefactory.com/product/lnf-nano9m/}},
	year = {2025},
	note = {[Accessed 18-12-2025]},
}

@misc{nistCryogenicMaterial,
	author = {NIST},
	title = {Cryogenic material properties {O}{F}{H}{C} {C}opper --- trc.nist.gov},
	howpublished = {\url{https://trc.nist.gov/cryogenics/materials/OFHC%20Copper/OFHC_Copper_rev1.htm}},
	year = {2010},
	note = {[Accessed 18-12-2025]},
}

@inproceedings{green2017connection,
  title={The connection of refrigeration to a superconducting magnet with a minimum amount of cryogen},
  author={Green, Michael A and Pan, Heng},
  booktitle={IOP Conference Series: Materials Science and Engineering},
  volume={278},
  number={1},
  pages={012180},
  year={2017},
  organization={IOP Publishing}
}

@book{united1966thermal,
  title={Thermal conductivity of selected materials},
  author={United States. National Bureau of Standards and Powell, RW},
  volume={8},
  year={1966},
  publisher={US Government Printing Office}
}

@misc{lakeshoreCryogenicWire,
	author = {Lakeshore Cryotronics},
	title = {{C}ryogenic wire --- lakeshore.com},
	howpublished = {\url{https://www.lakeshore.com/products/categories/specification/temperature-products/cryogenic-accessories/cryogenic-wire}},
	year = {2025},
	note = {[Accessed 18-12-2025]},
}

@misc{copperCryogenicProperties,
	author = {Copper Development Association Inc.},
	title = {{C}ryogenic {P}roperties of {C}opper --- copper.org},
	howpublished = {\url{https://www.copper.org/resources/properties/cryogenic/}},
	year = {2025},
	note = {[Accessed 08-01-2026]},
}

@inproceedings{ismael2021development,
  title={Development of a Cryogenic System for the Characterization of Advanced CMOS technologies down to 350 mK},
  author={Ismael, Mart{\'\i}nez-R and L{\'o}pez-L, Omar and Ferrusca, Daniel and Vel{\'a}zquez, Miguel and Guti{\'e}rrez-D, EA and Durini, D and others},
  booktitle={2021 IEEE International Instrumentation and Measurement Technology Conference (I2MTC)},
  pages={1--6},
  year={2021},
  organization={IEEE}
}

@article{xiang2020characterization,
  title={Characterization of the pressure coefficient of manganin and temperature evolution of pressure in piston-cylinder cells},
  author={Xiang, Li and Gati, Elena and Bud’ko, Sergey L and Ribeiro, Raquel A and Ata, Arif and Tutsch, Ulrich and Lang, Michael and Canfield, Paul C},
  journal={Review of Scientific Instruments},
  volume={91},
  number={9},
  year={2020},
  publisher={AIP Publishing}
}

@article{gidney2021stim,
  doi = {10.22331/q-2021-07-06-497},
  url = {https://doi.org/10.22331/q-2021-07-06-497},
  title = {Stim: a fast stabilizer circuit simulator},
  author = {Gidney, Craig},
  journal = {{Quantum}},
  issn = {2521-327X},
  publisher = {{Verein zur F{\"{o}}rderung des Open Access Publizierens
                in den Quantenwissenschaften}},
  volume = 5,
  pages = 497,
  month = jul,
  year = 2021
}

@article{Higgott2025sparseblossom,
  doi = {10.22331/q-2025-01-20-1600},
  url = {https://doi.org/10.22331/q-2025-01-20-1600},
  title = {Sparse {B}lossom: correcting a million errors per core second with minimum-weight matching},
  author = {Higgott, Oscar and Gidney, Craig},
  journal = {{Quantum}},
  issn = {2521-327X},
  publisher = {{Verein zur F{\"{o}}rderung des Open Access Publizierens in den Quantenwissenschaften}},
  volume = {9},
  pages = {1600},
  month = jan,
  year = {2025}
}

\begin{IEEEbiography}[{\includegraphics[width=1in,height=1.25in,clip,keepaspectratio]{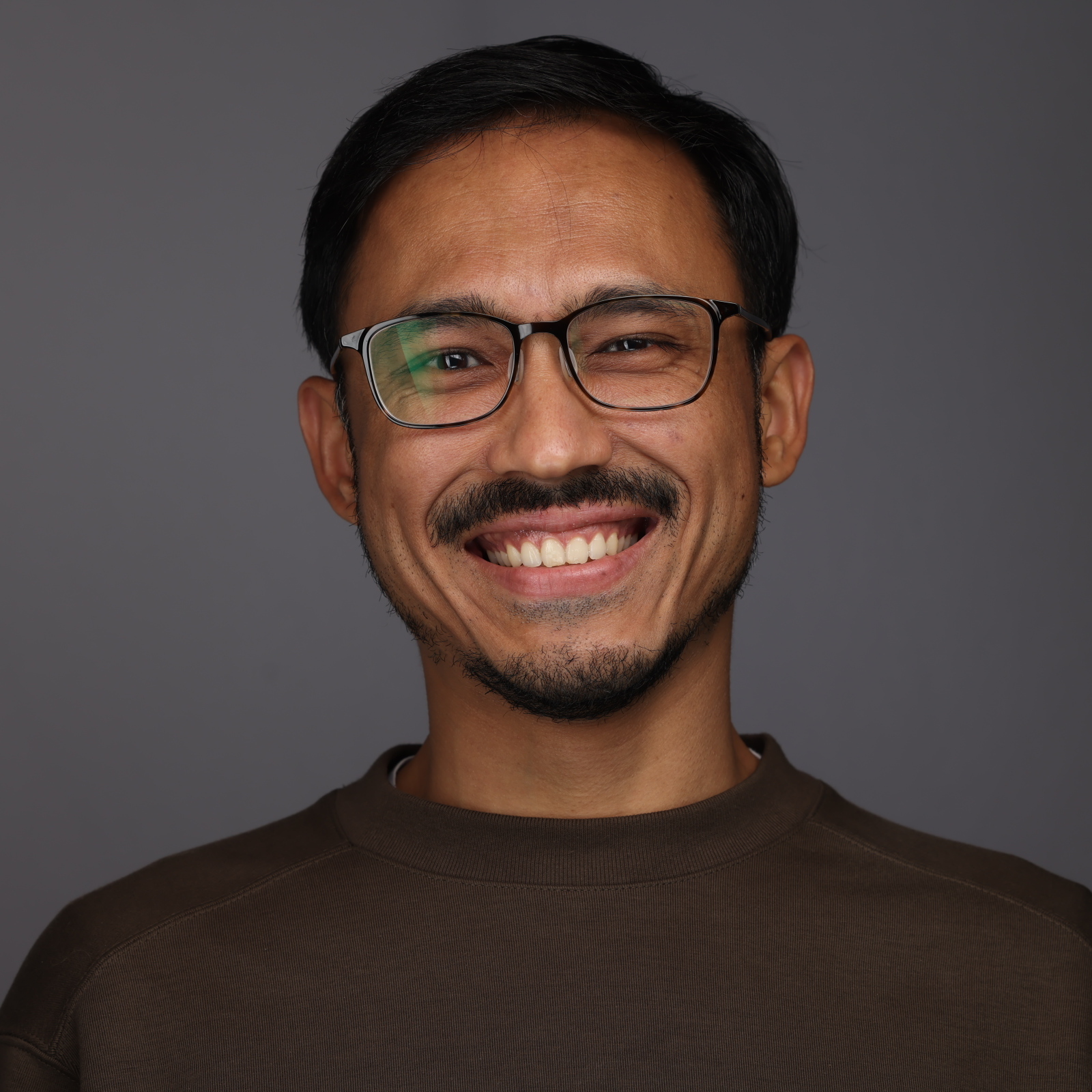}}]{Shaswot Shresthamali} received his B.E. in Electronics and Communication Engineering from Tribhuvan University, Nepal in 2012. He received his M.S. and Ph.D. degrees in Information Science and Technology from The University of Tokyo, Japan in 2018 and 2021, respectively.

From 2021 to 2024, he continued his postdoctoral research at Keio University. From 2024 to 2026, he was a Research Associate Professor with the Department of Advanced Information Technology, Kyushu University. He is currently a researcher at Fujitsu Ltd. His research interests include machine learning, computer architecture, and quantum computing.
\end{IEEEbiography}

\begin{IEEEbiography}[{\includegraphics[width=1in,height=1.25in,clip,keepaspectratio]{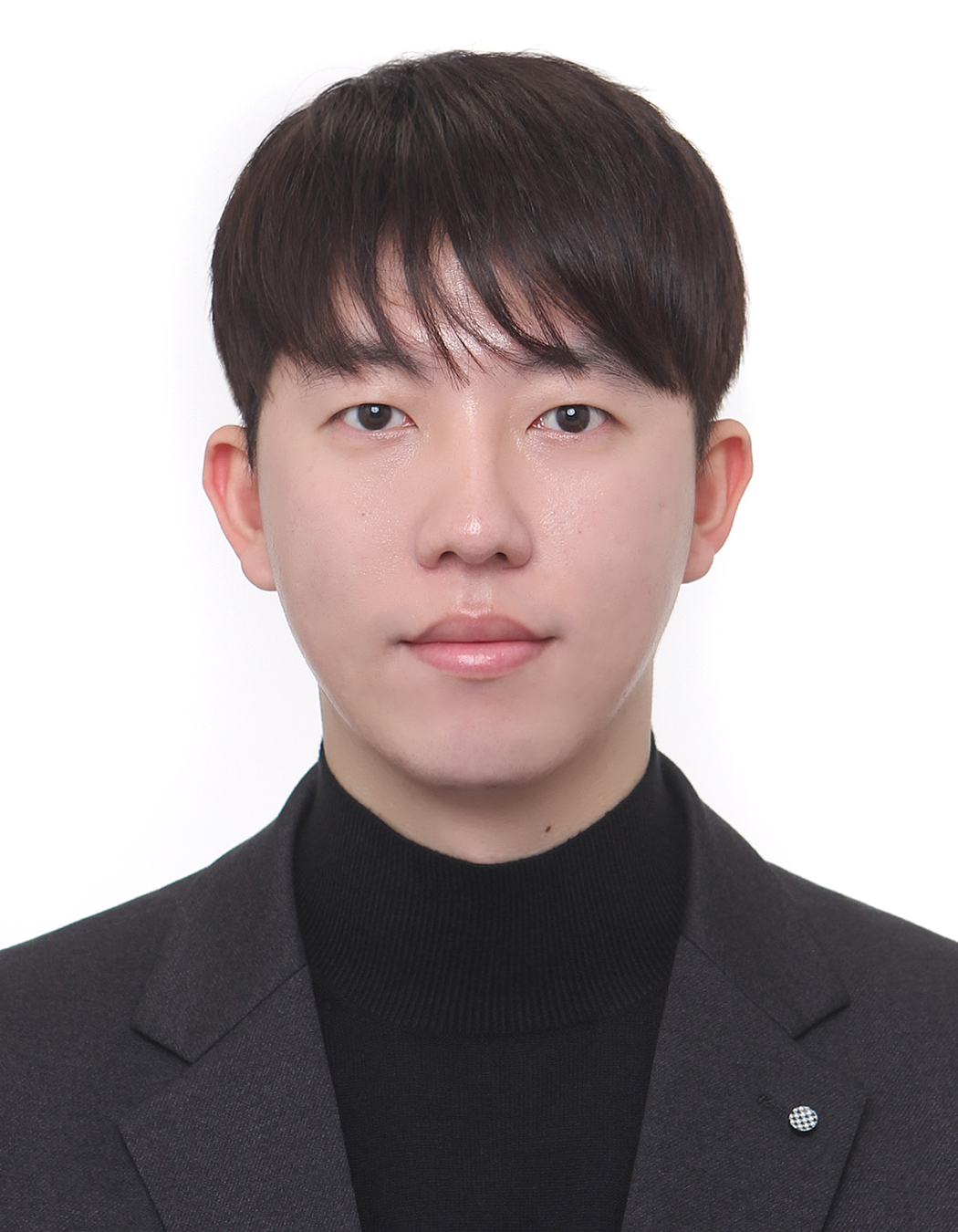}}]{Ilkwon Byun} (Member, IEEE) received his Ph.D. degree in electrical and computer engineering from Seoul National University in 2024. He is currently an Associate Professor with the Department of Advanced Information and Technology, Kyushu University, Fukuoka, Japan.
\end{IEEEbiography}

\begin{IEEEbiography}[{\includegraphics[width=1in,height=1.25in,clip,keepaspectratio]{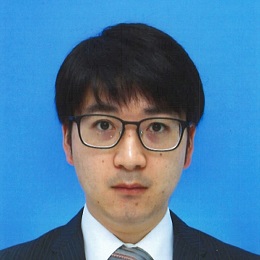}}]{Teruo Tanimoto} (Member, IEEE) received the
Bachelor of engineering and Master of information science and technology degrees from the University of Tokyo, Bunkyo, Japan, in 2010 and 2012, respectively, and the Ph.D. degree in engineering from Kyushu University, Fukuoka, Japan, in 2018.

He is currently an Associate Professor with the Department of Advanced Information and Technology, Faculty of Information Science and
Electrical Engineering, Kyushu University. After working for Fujitsu Laboratories Limited for three years as a Researcher, he joined Kyushu University in 2015. His research interests include quantum computer system architecture, edge computing systems, and secure computer architecture.
\end{IEEEbiography}

\begin{IEEEbiography}[{\includegraphics[width=1in,height=1.25in,clip,keepaspectratio]{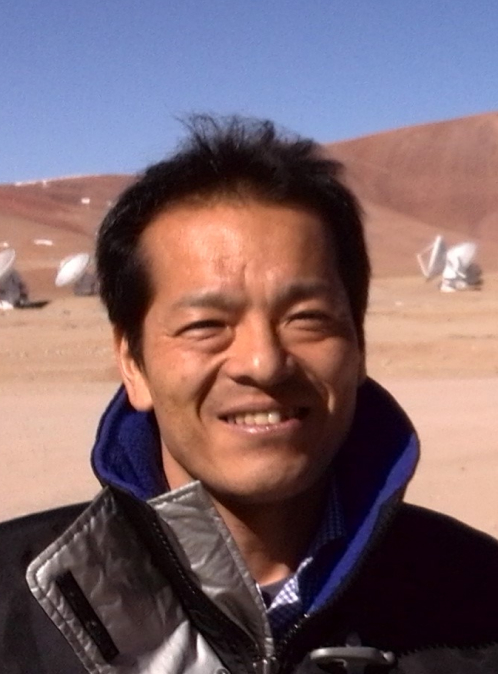}}] {Yoshinori Uzawa} is the Director of Advanced Technology Center of National Astronomical Observatory of Japan (NAOJ) and a professor in The Graduate University for Advanced Studies (SOKENDAI).

He received his M.E. and Ph.D. degrees in applied electronics from the Tokyo Institute of Technology, Tokyo, Japan, in 1991 and 2000, respectively. From 1991 to 2005, he was with the National Institute of Information and Communications Technology (NICT, previously Communications Research Laboratory, Ministry of Posts and Telecommunications), where he worked on the development of quasi-optical submillimeter-wave receivers with NbN SIS junctions. In 2005, he joined the National Astronomical Observatory of Japan (NAOJ) as an Associate Professor, managing the ALMA band 10 receiver development until 2014. After four years with NICT Terahertz Technology Research Center as the Director of Collaborative Research Laboratory of Terahertz Technology, he joined NAOJ as a Professor in 2018 to lead the development of the next generation superconducting receivers. His research interests include the superconducting electronics and terahertz technologies.
\end{IEEEbiography}

\begin{IEEEbiography}[{\includegraphics[width=1in,height=1.25in,clip,keepaspectratio]{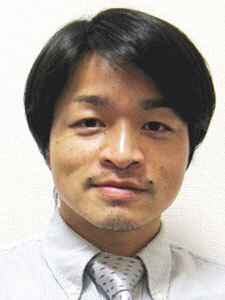}}] {Kunihiro Inomata} received his B.E. and M.E. degrees in electrical engineering and the Ph.D. degree in information sciences from Tohoku University, Japan, in 2001, 2002, and 2005, respectively.
From 2005 to 2016, he was a Research Scientist with RIKEN. Since 2016, he has been with the National Institute of Advanced Industrial Science and Technology (AIST), Japan.
He is currently a Team Leader with the Global Research and Development Center for Business by Quantum-AI Technology (G-QuAT), AIST.
His research interests include low-temperature physics, superconducting quantum devices, and quantum information processing.
\end{IEEEbiography}

\begin{IEEEbiography}[{\includegraphics[width=1in,height=1.25in,clip,keepaspectratio]{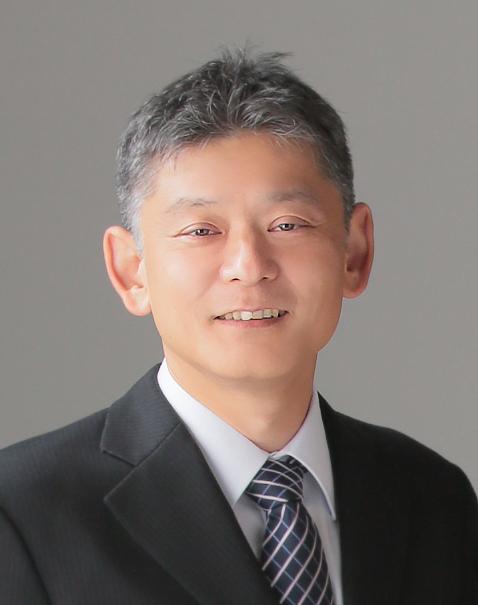}}] {Tsuyoshi Yamamoto} is a Research Fellow in NEC Secure System Platform Research Laboratories and is working on the R\&D of superconducting quantum devices.

He received his Ph.D. degree in applied physics from the University of Tokyo in 2001. In 2001, he joined NEC Corporation, Tsukuba, Japan, where he has been engaged in research on superconducting quantum circuits. From 2009 to 2010, he was a visiting researcher at the University of California, Santa Barbara.
Since 2019 he is a part of the NEC-AIST Quantum Technology Cooperative Research Laboratory, National Institute of Advanced Industrial Science and Technology (AIST), Tsukuba, Japan. 
Since 2026, he has joined Global Research and Development Center for Business by Quantum-AI Technology (G-QuAT), National Institute of Advanced Industrial Science and Technology (AIST).
\end{IEEEbiography}

\begin{IEEEbiography}[{\includegraphics[width=1in,height=1.25in,clip,keepaspectratio]{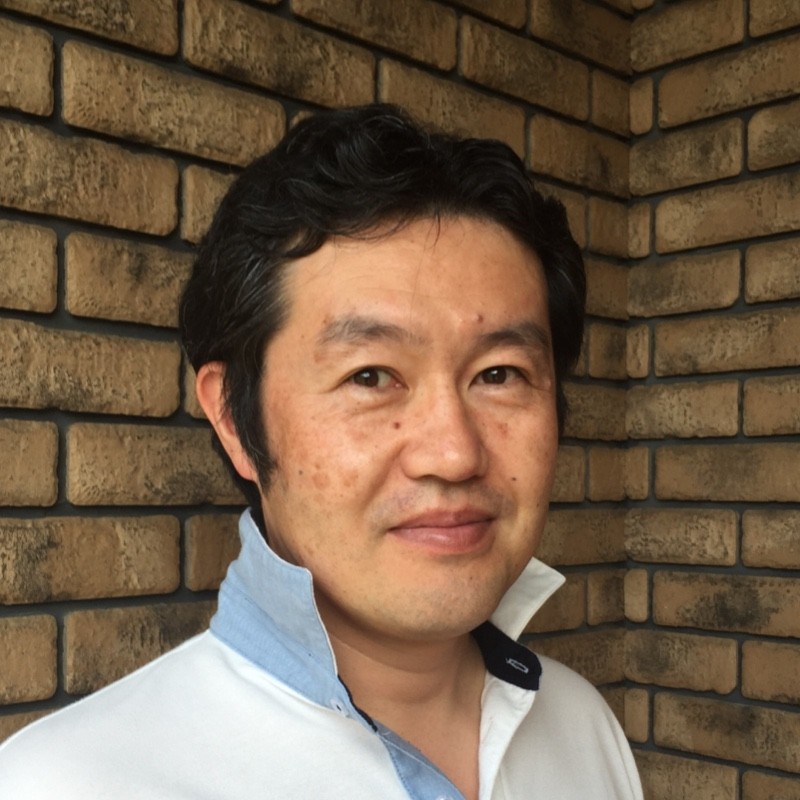}}]{Koji Inoue} (Member, IEEE) received the B.E.
and M.E. degrees in computer science from the Kyushu Institute of Technology, Iizuka, Japan, in 1994 and 1996, respectively, and the Ph.D. degree in computer science from the Department of Computer Science and Communication Engineering, Graduate School of Information Science and Electrical Engineering, Kyushu University, Fukuoka, Japan, in 2001. In 1999, he joined Halo LSI Design and Technology, Inc., New York, NY, USA, as a Circuit Designer. 

He is currently a professor with the Department of Advanced Information Technology, Kyushu University. His research interests include power-aware computing, high-performance computing, secure computer systems, superconductor computing, nanophotonic computing, and quantum computing.
\end{IEEEbiography}

\EOD

\end{document}